Rafael Sfair

# Dinâmica dos anéis de poeira e satélites de Urano e do anel F de Saturno

Tese apresentada à Faculdade de Engenharia do Campus de Guaratinguetá, Universidade Estadual Paulista, como requisito parcial para a obtenção do título de Doutor em Física.

Orientadora: Prof$^a$. Dr$^a$ Silvia M. Giuliatti Winter

Guaratinguetá

2011

**unesp** ⬢

*UNIVERSIDADE ESTADUAL PAULISTA*
*CAMPUS DE GUARATINGUETÁ*

*RAFAEL SFAIR DE OLIVEIRA*

ESTA TESE FOI JULGADA ADEQUADA PARA A OBTENÇÃO DO TÍTULO DE
**"DOUTOR EM FÍSICA"**

PROGRAMA: FÍSICA

APROVADA EM SUA FORMA FINAL PELO PROGRAMA DE PÓS-GRADUAÇÃO

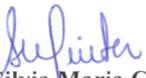

Profª. Drª. Silvia Maria Giuliatti Winter
Coordenadora

*B A N C A   E X A M I N A D O R A:*

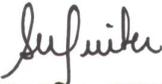

**Profª. Drª. SILVIA MARIA GIULIATTI WINTER**
Orientadora / Unesp-Feg

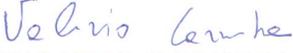

**Prof. Dr. VALERIO CARRUBA**
Unesp-Feg

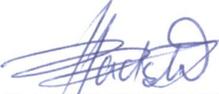

**Prof. Dr. TADASHI YOKOYAMA**
Unesp-Rio Claro

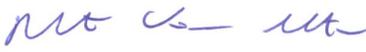

**Prof. Dr. ROBERTO VIEIRA MARTINS**
Observatório Nacional-RJ

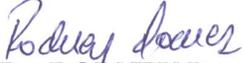

**Prof. Dr. RODNEY DA SILVA GOMES**
Observatório Nacional-RJ

*Fevereiro de 2011*

# Dados Curriculares

Rafael Sfair

NASCIMENTO    30.10.1984 - CURITIBA / PR

FILIAÇÃO    Osni Bento de Oliveira
Ana Rita de Oliveira

FORMAÇÃO

2001 - 2004    Bacharelado e Licenciatura em Física
Departamento de Física
UFPR - Universidade Federal do Paraná

2005 - 2007    Mestrado em Física
Departamento de Física e Química
FEG/UNESP - Universidade Estadual Paulista, Campus de Guaratinguetá

2007 - 2011    Doutorado em Física
Departamento de Física e Química
FEG/UNESP - Universidade Estadual Paulista, Campus de Guaratinguetá

# Resumo

Os anéis $\mu$ and $\nu$ de Urano compõem um segundo sistema de anéis juntamente com os satélites Puck, Mab, Portia e Rosalind. Estes anéis são tênues e compostos por partículas micrométricas, as quais podem ser bastante perturbadas por forças não gravitacionais, como por exemplo a força de radiação solar. Simulações numéricas foram utilizadas para analisar a evolução orbital de um conjunto de partículas deste sistema com raios de 1, 3, 5 e 10 $\mu m$ sob influência da força de radiação e também do achatamento do planeta, combinados com a perturbação gravitacional dos satélites próximos. Como esperado, a componente do arrasto de Poynting-Robertson causa o colapso da órbita, enquando um aumento na excentricidade é verificado devido à componente da radiação solar. A inclusão do achatamento do planeta evita este aumento da excentricidade e confina as partículas na região dos anéis. Encontros com os satélites causam variações no semi-eixo maior das partículas, que podem permanecer na região dos anéis ou colidir com os satélites. Para estas colisões, o resultado mais provável é a deposição na superfície. Como este mecanismo causa a remoção de material do anel, foram investigadas fontes adicionais de partículas. Adotando um valor aproximado para o fluxo de micrometeoritos na órbita de Urano, foi calculada a quantidade de material que pode ser ejetado devido à colisões com projéteis interplanetários. Verificou-se que as ejeções de Mab seriam suficientes para produzir um anel com profundidade óptica comparavável às observações.

Uma análise semelhante dos efeitos da radiação solar foi conduzida para a região de poeira que existe ao redor do anel F de Saturno. O amortecimento causado pelo achatamento do planeta evita as grandes variações da excentricidade, aumentando assim o tempo de vida das partículas de poeira. Além disso, um estudo da fotometria do anel utilizando imagens da sonda Cassini revelou que houve um aumento do brilho do anel nos últimos 25 anos. A forma da curva de fase obtida é semelhante aos dados da Voyager, indicando que, apesar do número de partículas ter aumentado, a distribuição de tamanho dos grãos permanece inalterada.

As regiões de poeira dos anéis de Urano foram observadas no final de 2007 durante o equinócio, quando Sol cruzou o plano dos anéis. Os dados obtidos com o VLT (*Very Large Telescope*) durante quatro noites consecutivas foram tratados e combinados, resultando em imagens com longo tempo de exposição. Para cada imagem, foram extraídos os perfis radiais. Estes perfis serão utilizados para desenvolver um modelo fotométrico.

**Palavras-chave**: Anéis planetários, simulações numéricas, força de radiação solar, análise de imagens.

# Abstract

The $\mu$ and $\nu$ rings of Uranus form a secondary ring-moon system with the satellites Puck, Mab, Portia, and Rosalind. These rings are tenuous and dominated by micrometric particles, which can be strongly disturbed by dissipative forces, such as the solar radiation pressure. We performed a numerical analysis of the orbital evolution of a sample of particles with radii of 1, 3, 5, and 10 $\mu$m under the influence of the solar radiation force and the planetary oblateness, combined with the gravitational interaction with the close satellites. As expected, the Poynting-Robertson component of the solar radiation force causes the collapse of the orbits, while the radiation pressure causes an increase in the eccentricity of the particles. The inclusion of Uranus's oblateness prevents a large variation in the eccentricity, confining the particles in the region of the rings. The encounters with the close satellites produce variations in the semimajor axis of the particles, leading them to move inward and outward within the ring region. These particles can either remain within the region of the rings or collide with a neighbouring satellite, and the most likely result of these collisions is the deposition of particles onto the surface of these satellites. Since this mechanism tends to cause a depletion of material of the rings, we investigate additional sources for these dust particles. Adopting a rough estimative of the flux of interplanetary meteoroids at Uranus' orbit, we calculated the amount of ejecta produced by hypervelocity impacts. We found that the ejections from Mab could generate a ring with optical depth comparable with the observations.

A similar analysis of the effects of the solar radiation force was carried out for the F-ring dust band. The damping due to the Saturn's oblateness prevents the overstated changes of the eccentricity, thus it increases in the lifetime of the particles. Therewithal, a photometric study of the F-ring using Cassini images revealed that substantial secular increase in the brightness of Saturn's F ring has occurred in the last 25 years. The shapes of the phase curves from Cassini and Voyager are similar, suggesting that although the number of dust particles has increased, the overall distribution of sizes is unchanged.

The dust bands that permeate the rings of Uranus were observed late in 2007 during the equinox, when the Sun crossed the ring plane. Images taken with the VLT (*Very Large Telescope*) during four consecutive nights were processed and then combined to result in long-exposure frames. For each frame, the north and south radial profiles were extracted. They will be used to develop a photometric model.

**Keywords**: Planetary rings, numerical simulations, solar radiation pressure, image analysis.

# Lista de Figuras

# Lista de Tabelas

# Sumário

# Capítulo 1

# Introdução

A proposta deste trabalho era analisar a dinâmica dos pequenos satélites da família de Portia e também dos anéis tênues $\mu$ e $\nu$, descobertos em 2005 através de imagens obtidas pelo Telescópio Espacial Hubble (Showalter & Lissauer, 2006). Porém, nos quatro anos de desenvolvimento deste trabalho muitas oportunidades surgiram e o projeto inicial foi adaptado e ampliado, o que está refletido nesta tese. O trabalho se restringia somente a uma análise sobre a dinâmica dos anéis de Urano acabou sendo expandido com trabalhos relacionados à evolução de partículas nos sistemas de Saturno e Plutão, além de estudos relacionados à análise de imagens.

Os trabalhos desenvolvidos foram agrupados de acordo com os respectivos planetas e os capítulos estão organizados de maneira praticamente independente, de forma que o leitor pode acompanhá-los fora de ordem sem grande prejuízo. Cada capítulo possui uma rápida introdução sobre o assunto, a descrição do método utilizado, os resultados obtidos e também a lista com as referências consultadas.

Nos apêndices estão cópias dos artigos publicados que podem servir como uma fonte mais rápida de consulta, uma vez que contém grande parte dos resultados em uma forma condensada.

Para quem deseja seguir linearmente os capítulos, a ordem é a seguinte:

No capítulo 2 é apresentado o estudo da dinâmica dos anéis $\mu$ e $\nu$. Através de simulações numéricas são analisadas não só a interação dos anéis com os satélites próximos como também a influência de forças não gravitacionais. Na última seção é apresentado um estudo sobre um mecanismo de geração de poeira através do impacto de IDPs.

O anel F de Saturno foi objeto dos dois estudos apresentados no capítulo 3. A seção 3.2 contém os resultados referentes a continuação do trabalho iniciado em Sfair (2007) sobre os efeitos da força de radiação solar. Já a análise de imagens do anel F enviadas pela sonda Cassini, um assunto até então inédito no país, é abordada na seção 3.3.

O capítulo 4 descreve as atividades desenvolvidas durante os períodos de estágio realizados no Observatoire de Meudon, sob supervisão do Prof. Bruno Sicardy. Estão descritos os métodos utilizados e os resultados preliminares da análise das imagens dos anéis de Urano obtidas utilizando o VLT (*Very Large Telescope*).

Para os resultados relacionados a Plutão foi reservado o apêndice A, onde pode ser encontrado um estudo sobre as regiões de estabilidade do sistema e também sobre a possibilidade de satélites adicionais.

# Capítulo 2

# Urano

## 2.1 Introdução

Urano é o sétimo planeta em distância ao Sol, cujo semi-eixo maior corresponde a 19.19 unidades astronômicas. Com raio de 25559 km, é o terceiro maior planeta do Sistema Solar e tem densidade de 1.318 g/cm$^3$ (Giorgini et al. 1996). Sua atmosfera é composta primariamente por hidrogênio e hélio, além de uma pequena quantidade de metano, que confere uma coloração azulada ao planeta, e traços de água e amônia (Fegley et al. 1991).

O planeta foi descoberto em 1781 por Willian Herschel enquanto analisava sistematicamente o céu. Urano já havia sido observado em 1690 por John Flamsteed, mas foi erroneamente catalogado como uma estrela, erro cometido também pelo astrônomo frances Pierre Lemonnier em 1750. Mesmo Herschel inicialmente identificou Urano como um cometa e só mais tarde, após a confirmação de outros astrônomos, Herschel reconheceu sua descoberta como sendo um novo planeta (Herschel & Dreyer 1912).

A passagem da sonda Voyager II por Urano em 1986 trouxe inúmeras informações, incluindo a descoberta de novos satélites e anéis (Smith et al. 1986). As imagens obtidas pela sonda em diferentes geometrias e em diferentes comprimentos de onda formam o melhor conjunto de dados disponível sobre o sistema de Urano. Além disso, imagens recentes obtidas principalmente pelo Telescópio Espacial Hubble, pelo VTL (*Very Large Telescope* – ver capítulo 4) e pelo Telescópio Keck complementam as informações sobre os satélites, anéis e sobre o próprio planeta.

Neste capítulo será apresentada uma revisão bibliográfica referente aos satélites e anéis de Urano com maior ênfase aos satélites que formam a chamada família de Portia e aos dois novos anéis recentemente descobertos, $\mu$ e $\nu$. Em seguida será apresentado uma análise sobre as forças perturbativas que atuam neste sistema e por fim um estudo sobre mecanismos de colisão que podem atuar como fonte de partículas para o anel $\mu$.

## 2.2 Satélites

Até o momento Urano possui vinte e três satélites conhecidos (NASA & JPL 2007). Destes, os cinco maiores em ordem crescente de tamanho são: Miranda, Ariel, Umbriel, Oberon e Titânia (tabela 2.1). Estes foram os primeiros satélites a serem descobertos, sendo dois deles (Titânia e Oberon) por Herschel em 1787.

| Satélite | $r$ (km) | $a$ (km) | $e$ ($\times 10^{-3}$) | $i$ (graus) |
|----------|----------|----------|------------------------|-------------|
| Ariel    | 578.9    | 190900   | 1.2                    | 0.041       |
| Umbriel  | 584.7    | 266000   | 3.9                    | 0.128       |
| Titânia  | 788.9    | 436300   | 1.1                    | 0.079       |
| Oberon   | 761.4    | 583500   | 1.4                    | 0.068       |
| Miranda  | 235.8    | 129900   | 1.3                    | 4.338       |

Tabela 2.1: Raio médio ($r$), semi-eixo maior ($a$), excentricidade ($e$) e inclinação ($i$) dos cinco maiores satélites de Urano. Adaptado de Jacobson (1998) e Thomas (1988).

Outros satélites menores foram descobertos através das imagens obtidas pela Voyager II em sua passagem por Urano (Smith et al. 1986). Entre as descobertas estão Cordelia e Ophelia, próximos ao anel $\epsilon$, e um grupo específico de satélites designado família de Portia, formado por Bianca, Cressida, Desdemona, Juliet, Portia, Rosalind, Cupid e Perdita. De fato, Cupid e Perdita foram descobertos posteriormente através de uma reanálise das imagens da Voyager e também com o auxílio do Telescópio Espacial Hubble (HST[1]). A tabela 2.2 resume algumas informações sobre os membros da família de Portia, além de dois outros satélites, Puck e Mab.

A descoberta relativamente recente de novos satélites pode parecer estranha, uma vez que as imagens dos satélites menores de Urano enviadas pela sonda Voyager já tinham sido analisadas. Porém, a maior parte os estudos anteriores se concentrou na detecção de satélites com raio de até 10 km na região próxima ao sistema principal de anéis, enquanto pouco foi feito na região onde os novos satélites foram descobertos. Estudos recentes utilizando dados da Voyager e do HST reduziram a possibilidade de detecção de novos satélites até uma magnitude maior que Cupid, o satélite menos brilhante da família de Portia (Karkoschka 2001c, Showalter & Lissauer 2006).

Pode-se ver que os satélites da família de Portia estão confinados em uma região radial bastante estreita, entre 59166 km e 76417 km, o que corresponde a 2.31-2.99 raios de Urano. Esta proximidade tem implicações na estabilidade do conjunto de satélites, o que será discutido na seção 2.2.6. Além disso, os dois membros externos da família são os menores, enquanto os satélites situados no meio da família são os maiores, configuração que não seria esperada caso a formação da família tenha ocorrido de forma aleatória (Karkoschka 2001a).

---

[1]Do inglês *Hubble Space Telescope*

| Satélite | Raio médio (km) | Semi-eixo maior (km) | Excentricidade ($\times 10^{-3}$) | Inclinação (graus) | Albedo |
|---|---|---|---|---|---|
| Bianca | 27 | 59165.46 | 0.27 | 0.181 | 0.065 |
| Cressida | 41 | 61766.72 | 0.20 | 0.038 | 0.069 |
| Desdemona | 35 | 62658.38 | 0.34 | 0.098 | 0.084 |
| Juliet | 53 | 64358.23 | 0.05 | 0.045 | 0.075 |
| Portia | 70 | 66097.29 | 0.51 | 0.026 | 0.069 |
| Rosalind | 36 | 69926.82 | 0.58 | 0.093 | 0.072 |
| Cupid | 9 | 74392.38 | – | – | 0.067 |
| Belinda | 45 | 75255.61 | 0.28 | 0.028 | 0.070 |
| Perdita | 13 | 76416.73 | 3.29 | 0.068 | 0.070 |
| Puck | 81 | 86004.49 | 0.39 | 0.321 | 0.099 - 0.108 |
| Mab | 12 | 97735.91 | 2.54 | 0.134 | 0.103 |

Tabela 2.2: Raio, elementos orbitais osculadores e albedo geométrico dos membros da família de Portia, Puck e Mab. Os elementos orbitais foram calculados assumindo que cada satélite segue uma órbita kepleriana perturbada pelo achamento do planeta e no caso de Perdita foi incluída uma libração ressonante, como discutido na seção 2.2.3. Adaptado de Showalter & Lissauer (2006).

Os satélites maiores estão distantes do sistema de anéis de Urano e não influenciam a dinâmica dos anéis e da família de Portia. Desta forma será dada mais atenção aos satélites menores, por estarem mais próximos aos anéis, e a interação entre os membros da família de Portia. Nas seções seguintes será feita uma breve descrição dos satélites e de alguns aspectos da dinâmica destes discutidos na literatura.

### 2.2.1 Portia

Portia é o maior satélite entre os membros da família que leva o seu nome. Seu maior tamanho e maior massa, comparados com o outros satélites, faz com que Portia tenha as menores variações nos seus elementos orbitais desde a época da Voyager (Showalter & Lissauer 2006).

Portia está localizado na região interior do anel $\nu$, assim como Puck está próximo à borda do anel $\mu$. Outra semelhança entre os dois sátelites é o indicativo da existência de gelo de água nas suas superfícies, sugerida pelo espectro que mostra absorção na faixa de 2.0 $\mu$m (Dumas et al. 2003).

### 2.2.2 Cupid

Cupid, também designado S/2003 U2 ou Urano XXVII, é o menor satélite da família de Portia. Devido ao seu pequeno tamanho e baixo albedo Cupid não foi detectado nas imagens da Voyager e sua descoberta ocorreu somente em 2003 através da análise de uma sequência de 24 imagens obtidas com o canal de alta resolução da *Advanced Camera for Surveys* do Telescópio Espacial Hubble (Showalter & Lissauer 2003).

A inclinação e excentricidade de Cupid são pequenas, de forma que sua órbita pode ser considerada circular e equatorial em uma primeira aproximação. Entretando deve-se estar atento ao fato de que, devido ao pequeno tamanho do satélite, a quantidade de imagens disponíveis é reduzida e a determinação da posição apresenta uma maior incerteza, o que resulta em uma maior imprecisão na determinação dos elementos orbitais.

A órbita do satélite é bastante próxima à de Belinda, distante apenas 863 km em direção a Urano, e apesar desta proximidade Cupid parece não sofrer nenhuma perturbação e descreve uma órbita bem comportada, ao contrário das órbitas de Mab e Perdita (Showalter & Lissauer 2006).

Entretando, a proximidade entre as órbitas de Cupid e Belinda pode ter implicações na estabilidade destes satélites. O único caso conhecido de um par de satélites com órbitas próximas está em uma ressonância 1 : 1, porém, apesar da ressonância 59 : 58 de Lindblad com Belinda estar 3 km distante da órbita de Cupid, qualquer interação entre os dois satélites parece ser não ressonante.

### 2.2.3 Perdita

Uma busca por satélites interiores à órbita de Puck em 300 imagens da Voyager revelou a existência de duas detecções que não eram um satélite conhecido, um objeto estelar ou algum defeito na imagem (Karkoschka 2001c).

A identificação de um satélite em uma imagem não é um processo deveras complicado, uma vez que efeitos de detecção estelar geram uma mancha (borrão) devido à rotação da sonda, enquanto detecções de satélites geram manchas com diferentes comprimentos e diferentes direções devido ao movimento dos satélites em relação à sonda.

O comprimento e orientação das manchas encontrados nas imagens eram compatíveis com um satélite em órbita equatorial e circular ao redor de Urano. Com a determinação de uma órbita aproximada do satélite, foi possível encontrá-lo em imagens subsequentes, o que permitiu melhorar o cálculo dos elementos orbitais, incluindo uma excentricidade de $0.0012 \pm 0.0005$.

Uma vez confirmada a detecção e determinada a órbita do satélite, este foi designado

S/1986 U10[2] (Karkoschka 1999) e posteriormente recebeu o nome de Perdita. A determinação do tamanho deste satélite está intrinsecamente ligada ao albedo assumido. Considerando o albedo de Perdita similar ao dos demais da família de Portia, o raio do satélite é de 13 km.

Perdita é o membro da família de Portia que possui a maior excentricidade. Isso sugere que o satélite possa estar em uma ressonância de Lindblad, uma vez que um dos efeitos desse tipo de ressonância é o aumento na excentricidade do corpo perturbado (Goldreich & Tremaine 1980). Esta hipótese foi levantada por Karkoschka (2001c) e confirmada por Showalter & Lissauer (2006).

Os dados de Showalter & Lissauer (2006) obtidos com o HST indicavam o valor de $a = 76417.45 \pm 0.03$ km para o semi-eixo maior de Perdita. Quando considerado um modelo que leva em conta também dados da Voyager e uma possível libração devido à ressonância, o melhor ajuste é $a = 76416.731 \pm 0.007$ km, valor que coincide com a ressonância externa $43 : 44$ de Lindblad com Belinda, localizada em 76416.749 $\pm$ 0.015 km.

Para esta ressonância, a amplitude de libração encontrada foi de $7.05°$ com um período de 3.24 anos. Sabe-se que no caso de um satélite com massa $m_{sat}$ e semi-eixo maior $a$, orbitando um planeta de massa $M_p$, a amplitude de libração $L$ é dada por (Murray & Dermott 1999)

$$L = 2.96 \sqrt{\frac{m_{\text{sat}}}{M_p}} a \qquad (2.1)$$

de maneira que o acompanhamento da órbita de Perdita pode oferecer um método alternativo para o cálculo da massa de Belinda e, em última instância, para o albedo deste satélite (Karkoschka 2001c).

A inclusão desta ressonância explica em grande parte o avanço de $57°$ na longitude esperada de Perdita em relação ao esperado pela simples extrapolação da sua posição na época da Voyager. Essa diferença indica que a velocidade média do satélite foi $0.008°$/dia maior do que a velocidade atual, mostrando que Perdita não segue uma órbita kepleriana uniforme.

Uma ressonância $8 : 7$ com Rosalind também está localizada próxima à órbita de Perdita e pode exercer alguma influência na dinâmica do satélite. Desta forma um modelo mais completo para a órbita de Perdita continua em aberto.

## 2.2.4 Mab

A descoberta do satélite S/2003 U1, posteriormente chamado Mab, foi concomitante à descoberta de Cupid (Showalter & Lissauer 2003). Mab, assim como Puck, não faz parte da família de Portia.

---

[2]1986 refere-se ao ano de obtenção das imagens que permitiram a descoberta.

Como o satélite é pequeno e relativamente escuro ele não foi visto nas imagens obtidas pela Voyager II durante sua passagem por Urano. Contudo, a superfície de Perdita é mais escura que a de Mab (c.f. tabela 2.2) e mesmo assim Perdita foi encontrado nas imagens da Voyager. Isso motivou uma reanálise nas imagens da sonda, relevando a presença de Mab em quatro imagens "*backscattered*" (figura 2.1).

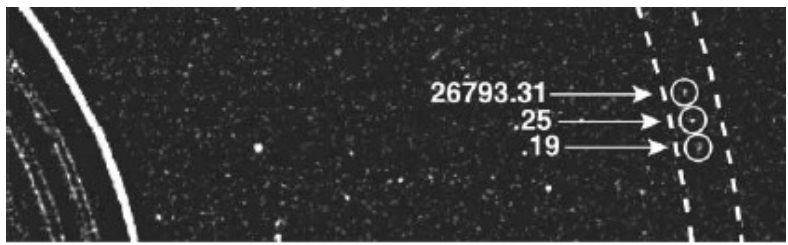

Figura 2.1: Recuperação de Mab em três imagens da Voyager. A figura é formada pela sobreposição das imagens 26793.19, 26793.25 e 26793.31, com 4.8 minutos de diferença entre elas. O satélite está indicado pelos círculos e as linhas tracejadas indicam o raio orbital de Mab ±2000 km. Extraído de Showalter & Lissauer (2006).

Existe uma grande incerteza na determinação do tamanho de Mab, sendo que esta incerteza está relacionada ao albedo assumido para o satélite, cuja órbita está compreendida entre as de Puck e Miranda. Caso Mab possua um albedo similar ao de Puck, seu raio é de aproximadamente 12 km, como apresentado na tabela 2.2. Porém se a superfície de Mab for semelhante a de Miranda, que é muito mais brilhante, o raio de Mab deve ser aproximadamente 6 km, menor que o de Cupid. Desta forma a determinação correta do raio de Mab só será possível através de imagens que permitam comparar semelhanças entre os espectros de Mab e Puck ou Miranda.

Dados obtidos a partir das imagens do HST indicam que a posição do satélite está defasada em 2.2° em relação à posição esperada pela extrapolação da posição de Mab a partir dos dados derivados da Voyager, devido principalmente à incerteza de 0.00034°/dia no movimento médio acumulada durante 18 anos. Com isso a determinação dos elementos orbitais (tabela 2.2) foi feita através do ajuste dos dados do HST juntamente com os dados da Voyager.

Os elementos orbitais apresentam alterações significativas ano a ano (figura 2.2), com diferenças maiores que 1° entre as as posições esperadas em 2004 e 2005, por exemplo. Como nas imagens analisadas o brilho de Mab o destacava e o satélite estava distante de outros, esta diferença na posição não é devido (exclusivamente) a erros no método utilizado, evidenciando a existência de uma perturbação desconhecida atuando sobre Mab. Entretando, nenhuma ressonância de primeira ordem que possa explicar essa perturbação está localizada nas vizinhanças de Mab.

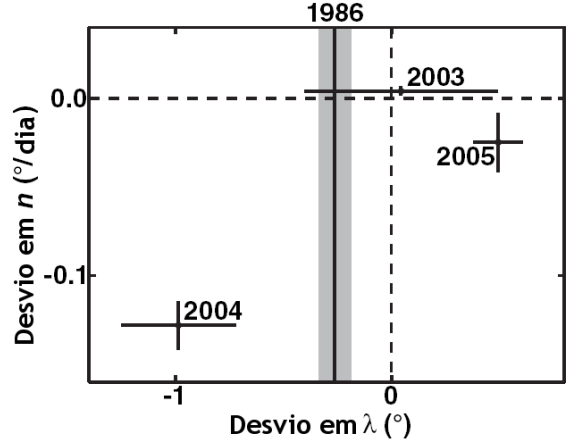

Figura 2.2: Variações ano a ano da órbita de Mab. A origem do gráfico representa o melhor ajuste da longitude $\lambda$ e do movimento médio $n$ para os dados entre 1986 e 2005. As mesmas quantidades baseadas nos ajustes dos dados para cada ano estão indicadas com barras de erro de $\pm 1\sigma$. A incerteza em $n$ para 1986 é grande e somente a variação em $\lambda$ e sua incerteza (região sombreada) está representada. Adaptado de Showalter & Lissauer (2006).

A variação no movimento médio $n$ é de aproximadamente 0.12°/dia, o que corresponde a uma variação de 45 km no semi-eixo maior. Uma distância desta magnitude é muito grande para ser explicada com a existência de um satélite menos massivo coorbital a Mab (Kumar et al. 2010). Já o desvio da longitude média é $\sim 1.5°$, como mostrado na figura 2.2. Uma vez conhecidas as variações do movimento médio e da longitude média, pode-se determinar o período de libração, dado por (Showalter & Lissauer 2006):

$$T \sim 2\frac{\Delta\lambda}{\Delta n}$$ 
(2.2)

Neste caso $T \sim 25$ dias, valor que pode ter sido subestimado devido ao método utilizado no ajuste para a determinação de $\Delta\lambda$ (Showalter & Lissauer 2006).

## 2.2.5 Puck

Puck não faz parte da família de Portia e é externo a todos os membros da família. Ele é o maior dos satélites internos de Urano, com raio de aproximadamente 77 km, facilitando a determinação do seu raio, albedo e elementos orbitais.

Devido a melhor precisão na determinação dos parâmetros de Puck, muitas vezes estes cálculos servem como um teste de consistência para o método utilizado na determinação das características de satélites menores, cuja incerteza é bem maior devido ao reduzido número de imagens ou a resolução obtida (Karkoschka 2001c).

Outra característica interessante de Puck é que seu brilho é maior em um dos lados. Este fato pode estar relacionado à interação entre Puck e as partículas do anel $\mu$, uma vez que o satélite está localizado na borda interna do anel (Showalter & Lissauer 2006). Este fato será discutido na seção 2.3.2.

## 2.2.6 Características gerais e a dinâmica da família de Portia

A análise fotométrica dos membros da família de Portia revela várias similaridades entre os satélites. Karkoschka (2001a) analisou diversas imagens obtidas pela Voyager e pelo HST cobrindo diferentes comprimentos de onda e diferentes ângulos de fase.

Tanto o albedo (tabela 2.2) quando a refletividade mostram claramente que os satélites da família de Portia formam um grupo diferente dos demais objetos próximos (anéis e os satélites regulares). Essa semelhança pode estar eventualmente relacionada à um processo comum na formação e evolução destes satélites (Colwell & Esposito 1990, 1992).

Também verificou-se que, assim como no caso dos anéis principais (seção 2.3.1), o albedo estimado da família de Portia estava subestimado por um fator dois. Esta correção no albedo possibilitou um melhor ajuste nas funções de fase dos satélites menores.

Duncan & Lissauer (1997) realizaram um estudo sobre a estabilidade da família de Portia. A dinâmica desse sistema é complexa, uma vez que todo o conjunto de satélites está confinado em uma região radial bastante estreita. Outra dificuldade está relacionada à determinação das massas dos satélites, pois para alguns deles só é possível obter uma estimativa das massas através de albedos aproximados.

De forma semelhante, não existem muitas informações sobre as órbitas destes satélites. Sabe-se que elas são praticamente coplanares e circulares, como é comum no caso de satélites regulares. Entretanto, várias destas órbitas estão bastante próximas e essa pequena separação levanta a questão de que estas órbitas podem não ser estáveis durante a idade do Sistema Solar (Lissauer 1995).

As simulações numéricas realizadas por Duncan & Lissauer utilizaram o pacote SWIFT (Levison & Duncan 1994), modificado para a inclusão dos coeficientes gravitacionais $J_2$ e $J_4$ que descrevem o achatamento do corpo central. Usando condições iniciais derivadas das imagens da Voyager (Owen & Synnott 1987), foram integrados 400 sistemas baseados na família de Portia, sendo que em algumas simulações também foram incluídos os cinco satélites maiores de Urano. As simulações foram interrompidas quando ocorria cruzamento entre as órbitas de dois satélites quaisquer, ou seja, até a sobreposição radial entre o periapse da órbita de um satélite e o apoapse de outro.

Para cada simulação as massas nominais de todos os satélites foram multiplicadas por um fator $m_f$, em que geralmente $m_f > 1$. Isso permitiu analisar o tempo $t_c$ até o cruzamento de duas órbitas em função do parâmetro $m_f$. Também verificou-se que em um tempo um pouco maior que $t_c$ ocorria a colisão entre os satélites que cruzaram suas órbitas.

Os resultados obtidos mostram que $t_c$ varia com uma potência de $m_f$, e como esperado sistemas com maiores valores de $m_f$ apresentam menores valores de $t_c$. Apesar de não poder prever o comportamento das trajetórias individualmente, poucas simulações divergiram do ajuste com lei de potência.

Verificou-se que sistemas com maiores valores de $m_f$ apresentam um comportamento excitado durante um tempo maior e também que em alguns casos ocorreu um comportamento "acoplado" entre dois satélites, onde a excitação de uma órbita estava relacionada a estabilidade de outra.

A inclusão dos efeitos gravitacionais dos cinco satélites maiores e do momento de quadrupólo de Urano parece não ter afetado as simulações com maiores valores de $m_f$. Nestes casos somente as interações entre Desdemona, Cressida e Juliet parecem ser determinantes.

Para menores valores de $m_f$ a inclusão de $J_2$ e $J_4$ aumentou a estabilidade do sistema, ocasionando um aumento no tempo até o cruzamento. Ao ignorar estes coeficientes, a redução na estabilidade do sistema deve-se a dois efeitos: o momento de quadrupólo altera as taxas de precessão dos satélites, modificando ou até mesmo eliminando ressonâncias seculares; as velocidades iniciais dos satélites utilizadas foram as mesmas nos casos com e sem achatamento, o que gera um erro nos elementos orbitais iniciais (Renner & Sicardy 2006).

Na maior parte dos casos analisados as órbitas dos satélites internos de Urano se cruzaram em um tempo significativamente menor do que a idade do Sistema Solar (entre 4 e 100 milhões de anos). Duncan & Lissauer (1997) argumentam que devido à baixa velocidade relativa, uma possível colisão irá resultar em agregação e as simulações indicam que a colisão mais provável será entre Cressida e Desdemona, o que resultaria em um novo satélite "Cresdemona". O tempo de estabilidade do sistema, formado agora por sete satélites, foi consideravelmente maior e compatível com a idade do Sistema Solar.

Estes resultados indicam que é necessário melhorar a determinação dos parâmetros dos satélites para melhor entender a dinâmica da família de Portia. Além disso, devido à proximidade entre os satélites o sistema é bastante instável, sugerindo que alguns satélites podem ser "novos" em termos geológicos (Colwell & Esposito 1992).

O estudo mais recente acerca da dinâmica dos satélites da família de Portia foi realizado por Showalter & Lissauer (2006). Em comparação com estudos anteriores (Owen & Synnott 1987, Jacobson 1998), a maioria das órbitas sofreu alterações significativas em uma década, principalmente no caso de Belinda (figura 2.3). Entretanto, as variações na energia e no momento angular geralmente estão centradas próximas a zero, consistente com os princípios de conservação. Os elementos orbitais de Cressida e Desdemona sofreram variações em sentidos contrários, como é esperado para órbitas acopladas. Estes dois efeitos já haviam sido sugeridos por Duncan & Lissauer (1997) no estudo envolvendo simulações numéricas citado anteriormente.

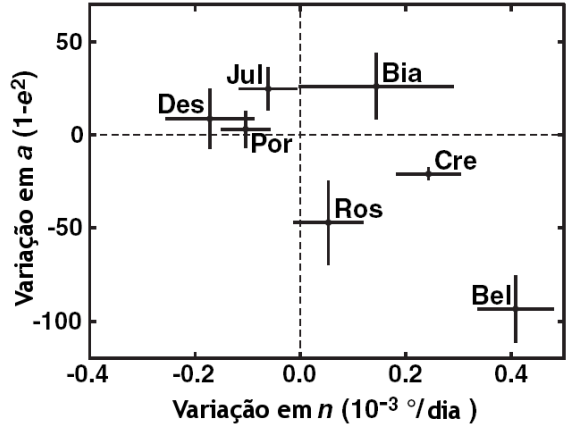

Figura 2.3: Variação dos elementos orbitais dos membros da família de Portia entre 1994 e 2006. A variação em $a$ está relacionada à energia e o produto $a \times (1-e^2)$ ao momento angular. As barras de erro representam a variação de $\pm 1\sigma$. Adaptado de Showalter & Lissauer (2006).

## 2.3 Anéis

Em março de 1977 ocorreu a ocultação da estrela SAO 158687 por Urano. O acompanhamento do brilho desta estrela durante a ocultação revelou a existência de quatro anéis estreitos (anéis $\alpha$, $\beta$, $\gamma$ e $\delta$), com largura de aproximadamente 10 km, praticamente circulares e coplanares, além de um anel mais largo ($\sim 100$ km) e inclinado (anel $\epsilon$) (Elliot et al. 1977). Observações posteriores identificaram mais quatro anéis: $\eta$, 4, 5 e 6. Estes nove anéis, juntamente com os anéis $\zeta$ e $\lambda$ descobertos pela Voyager II (Smith et al. 1986), formam o sistema principal de anéis de Urano.

Urano, seus satélites e anéis foram novamente observados por Showalter & Lissauer (2006) utilizando o Telescópio Espacial Hubble em quatro oportunidades nos anos de 2003, 2004 e 2005. Foram realizadas várias observações com tempo de exposição de 250 segundos utilizando o filtro *clear* e a análise destas imagens revelou a existência de dois novos anéis, provisoriamente chamados R/2003 U1 (ou $\mu$) e R/2003 U2 (R2), depois rebatizados como $\mu$ e $\nu$, respectivamente. A figura 2.4 mostra uma combinação de imagens onde é possível identificar os anéis $\mu$ e $\nu$.

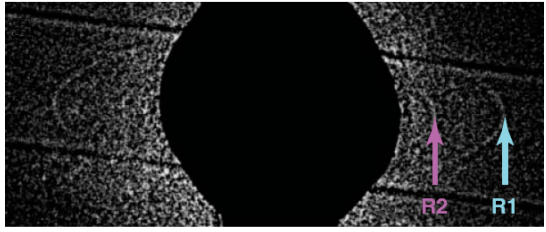

Figura 2.4: Combinação de 24 imagens obtidas com o HST em 2003 onde pode-se ver os anéis $\mu$ e $\nu$. Nestas imagens foram removidos a saturação devido ao planeta e o fundo. O contraste também foi alterado para evidenciar os anéis. Adaptado de Showalter & Lissauer (2006).

Além dos dois anéis, há indícios de outras estruturas similares à anéis tênues em regiões mais próximas ao planeta, mas devido ao ruído na imagem não foi possível obter nenhuma confirmação até o momento. A figura 2.5 mostra um mapa onde é possível ver os novos anéis e alguns satélites da família de Portia, enquanto a tabela 2.3 resume as propriedades dos anéis $\mu$ e $\nu$.

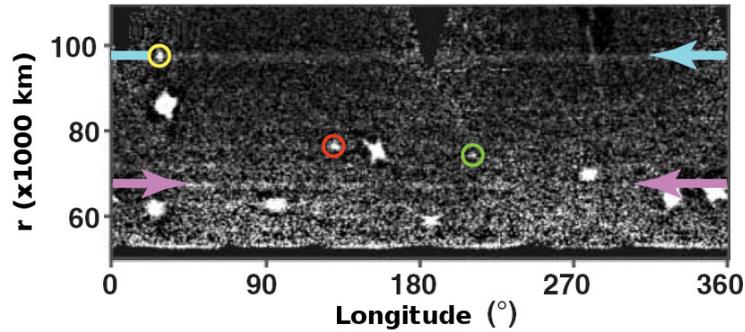

Figura 2.5: Mapa obtido a partir das observações realizadas em 2003. As setas horizontais indicam os anéis $\mu$ (azul) e $\nu$ (violeta). Também são visíveis alguns satélites: Mab (amarelo), Perdita (vermelho) e Cupid (verde). Adaptado (Showalter & Lissauer 2006).

| Anel | Pico radial (km) | Borda interna (km) | Borda externa (km) |
|------|-----------------|--------------------|--------------------|
| $\mu$ ($\mu$) | 97700 | 86000 | 103000 |
| $\nu$ ($\nu$) | 67300 | 66100 | 69900 |

Tabela 2.3: Resumo das informações sobre os anéis $\mu$ e $\nu$. Adaptado de Showalter & Lissauer (2006).

Tanto o anel $\mu$ quanto o $\nu$ estão situados além do anel $\epsilon$, em uma região onde não se esperava encontrar nenhum anel. Este fator, juntamente com a baixa profundidade óptica explica o fato destes anéis não terem sido notados nas imagens da Voyager II. A figura 2.6 mostra de forma esquemática a localização dos anéis e de alguns membros da família de Portia.

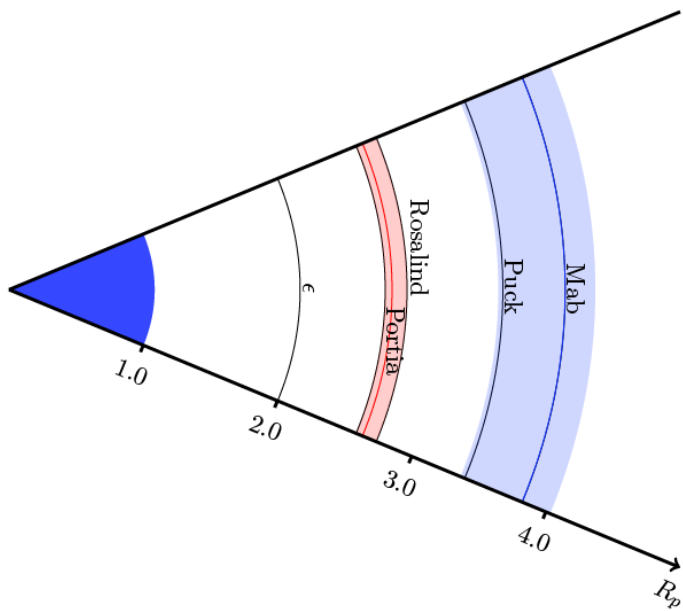

Figura 2.6: Representação esquemática de $\mu$ e $\nu$. Os picos radiais dos anéis estão representados pelas linhas azul ($\mu$) e vermelha ($\nu$) e a extensão radial de cada um pela região sombreada. As posições dos satélites próximos também estão indicadas, assim com o anel $\epsilon$. A escala é dada em unidades de raios de Urano ($R_p$).

Assim como no caso de alguns dos satélites da família de Portia, uma busca nos dados da Voyager permitiu encontrar um conjunto de imagens com grande ângulo de fase $\alpha$, definido como o ângulo formado entre a luz incidente em um objeto e a luz refletida por este. A partir das imagens com $\alpha = 146°$ foi possível fazer uma combinação para então derivar o perfil radial tanto do anel $\mu$ quanto do anel $\nu$. Ambos os perfis são compatíveis com os dados obtidos pelo HST.

Para grandes valores de $\alpha$ os anéis parecem bastante brilhantes, indicando claramente que eles são compostos basicamente por partículas com tamanho da ordem de micrometros. Observações no infravermelho próximo utilizando a óptica adaptativa do telescópio Keck revelaram que o anel $\mu$ é azul, enquanto $\nu$ é extremamente vermelho (de Pater et al. 2006b).

Nas seções seguintes serão discutidos aspectos particulares do sistema principal de anéis e também dos anéis $\mu$ e $\nu$.

### 2.3.1 Sistema principal de anéis

Os anéis internos de Urano apresentam uma série de particularidades. Algumas características destes anéis são apresentadas na tabela 2.4 e a figura 2.7 mostra uma imagem do sistema principal de anéis.

| Anel | Raio orbital médio (km) | Largura (km) | Excentricidade ($\times 10^{-3}$) | Inclinação (graus) |
|------|------------------------|--------------|-----------------------------------|--------------------|
| $\zeta$ | 37850 - 41350 | 3500 | – | – |
| 6 | 41837 | 1.6 - 2.2 | 1.0 | 0.063 |
| 5 | 42234 | 1.9 - 4.9 | 1.9 | 0.052 |
| 4 | 42570 | 2.4 - 4.4 | 1.1 | 0.032 |
| $\alpha$ | 44718 | 4.8 - 10 | 0.8 | 0.014 |
| $\beta$ | 45661 | 6.1 - 11.4 | 0.4 | 0.005 |
| $\eta$ | 47175 | 1.9 - 2.7 | 0? | 0.002 |
| $\gamma$ | 47627 | 3.6 - 4.7 | 0? | 0.011 |
| $\delta$ | 48300 | 4.1 - 6.1 | 0? | 0.004 |
| $\lambda$ | 50023 | 1 - 2 | 0? | 0? |
| $\epsilon$ | 51149 | 19.7 - 96.4 | 7.9 | 0.001 |

Tabela 2.4: Propriedades dos anéis principais de Urano. Adaptado de Karkoschka (2001b), Esposito (2002), Stone & Miner (1986).

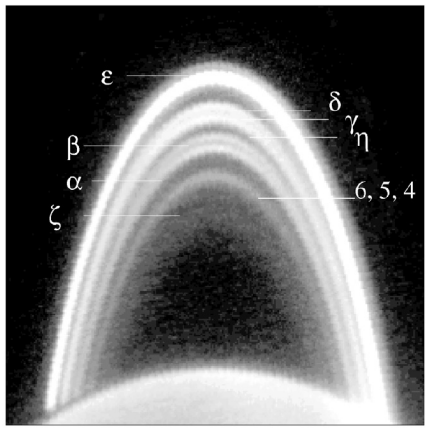

Figura 2.7: Combinação de imagens obtidas com o Telescópio Keck do "*ansae*" norte dos anéis principais de Urano, indicados na figura. Extraído de de Pater et al. (2006a).

O anel $\zeta$ é bastante tênue e está próximo das camadas mais altas da atmosfera de Urano (Smith et al. 1986). Sua existência foi inferida a partir dos dados da Voyager (de fato, o anel é visível em apenas uma imagem) e observações recentes feitas com o telescópio Keck confirmaram a existência do anel e registraram uma mudança significativa na sua localização, mostrando que o anel se estende gradualmente até 32600 km (de Pater et al. 2006a).

Os anéis 6, 5 e 4 são os mais estreitos e menos brilhantes de Urano, enquanto os anéis $\alpha$ e $\beta$ são os mais brilhantes depois do anel $\epsilon$ (Smith et al. 1986). Os anéis $\alpha$ e $\beta$ apresentam variações regulares de brilho e largura, sendo mais brilhantes e largos no apocentro e estreitos e escuros no pericentro (Gibbard et al. 2005). Durante a passagem da Terra pelo plano orbital dos anéis em 2007 estes dois anéis desapareceram, indicando que são verticalmente estreitos e que são compostos por partículas grandes, apesar de ter sido detectado um envelope de poeira envolvendo o anel $\beta$ (de Pater et al. 2007).

O anel $\eta$ é circular, possui uma inclinação muito pequena e pode ser dividido em uma componente opticamente densa e uma região externa com baixa profundidade óptica formada por poeira (de Pater et al. 2007). O próximo anel em distância a Urano é o $\gamma$, também circular e coplanar. Ele apresenta variações significativas de brilho com a longitude (Lane et al. 1986) e especula-se que ao menos uma das bordas do anel é confinada por uma ressonância com Cordélia (Esposito 2002).

Assim como o anel $\eta$, o anel $\delta$ é formado por duas componentes: uma estreita e densa envolvida por outra tênue com baixa profundidade óptica (Karkoschka 1997). Já o anel $\lambda$, descoberto pela Voyager II, é bastante tênue e está localizado entre o anel $\epsilon$ e Cordelia. Observações em diferentes comprimentos de onda e em diferentes geometrias revelaram que o anel é composto por partículas micrométricas e que para algumas geometrias este é o anel mais brilhante de Urano (Smith et al. 1986).

O mais externo anel do sistema principal de Urano, $\epsilon$, é mais largo e brilhante no apocentro de sua órbita e estreito e escuro no pericentro (Karkoschka 2001b). Suas bordas podem ser ajustadas por duas elipses que precessam rigidamente, configuração mantida através da troca de energia e *momentum* entre as partículas do anel e os dois satélites que orbitam próximo ao anel, Cordélia (interno) e Ophelia (externo) (Goldreich & Tremaine 1979).

Acreditava-se que estes anéis principais de Urano estavam entre os objetos mais escuros do Sistema Solar, com albedo de 0.02 (Elliot & Nicholson 1984). Um estudo posterior realizado por Cuzzi (Cuzzi 1985) mostrou que a profundidade óptica dos anéis estava errada por um fator 2 para menos, o que implica em um erro do mesmo fator no brilho dos anéis.

Determinações feitas com imagens da Voyager elevaram o valor do albedo para $0.032 \pm 0.003$ (Ockert et al. 1987). Um estudo mais recente estabeleceu o valor de $0.061 \pm 0.006$ para o albedo do anel $\epsilon$ e valores semelhantes para os outros anéis (Karkoschka 1997). Desta forma, sabe-se agora que os anéis de Urano, que eram parte do grupo dos objetos mais escuros, são mais brilhantes do que a média dos objetos do Sistema Solar.

Mais características destes anéis e observações deles realizadas com o VLT serão apresen-tadas no capítulo 4.

## 2.3.2   Anel $\mu$

O anel $\mu$ apresenta um perfil radial triangular bem definido (Showalter & Lissauer 2006), como pode ser visto na figura 2.8. É interessante notar que o pico do anel coincide com a órbita de Mab, assim como ocorre entre outros anéis e satélites: Pan em Saturno, Galatea em Netuno e os quatro satélites mais internos de Júpiter. Outro fato interessante é a localização de Puck na borda interna do anel, algo semelhante ao que ocorre no anel E de Saturno, cuja borda interna coincide com a órbita de Mimas (figura 2.9) (Hedman et al. 2007).

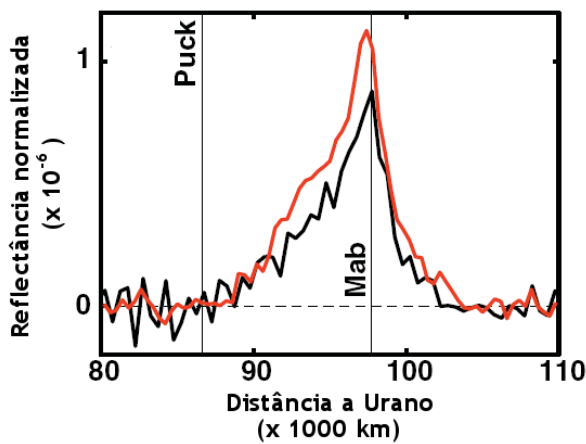

Figura 2.8: Perfis radiais do anel $\mu$ a partir dos dados do HST (preto) e da Voyager (vermelho). Os semi-eixos maiores de Puck e Mab estão indicados com linhas verticais. Adaptado de Showalter & Lissauer (2006).

Um satélite pode limitar a extensão radial de um anel removendo partículas através de colisões. Este mecanismo pode explicar a assimetria do brilho de Puck (apresentada na seção 2.2.5): o material do anel pode colidir preferencialmente em uma determinada região do satélite escurecendo-a, ou pode remover as camadas exteriores de poeira revelando camadas interiores mais brilhantes.

Além do perfil triangular, o anel $\mu$ apresenta várias outras características que o torna similar ao anel E de Saturno. Ambos os anéis estão à distâncias semelhantes dos seus planetas, 3.95 $R_S$ para o caso do anel E e 3.82 $R_U$ para o anel $\mu$, e ambos tem o pico do perfil radial centrado em um satélite, Enceladus e Mab, respectivamente.

Enceladus é tido como a fonte de partículas do anel E. Dados obtidos pela sonda Cassini após um "fly-by" próximo ao satélite revelaram que existem duas contribuições do satélite no processo de criação de partículas (de Pater et al. 2004, Spahn et al. 2006b): a primeira, e mais importante, é a atividade geológica no pólo sul de Enceladus; o segundo fator são as colisões de projéteis interplanetários (micrometeoros) com a superfície do satélite, criando uma nuvem de poeira. Um processo colisional menos eficiente também pode estar ocorrendo em Tethys, o que explicaria um segundo pico menos pronunciado no perfil do anel E.

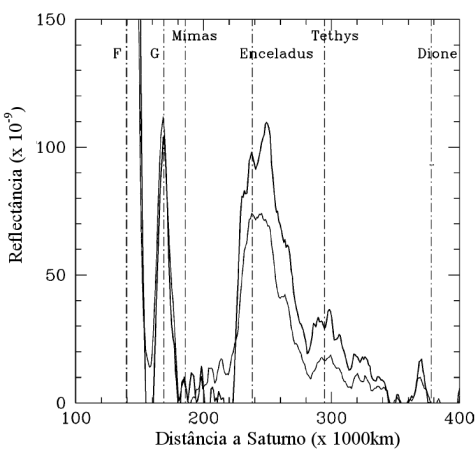

Figura 2.9: Perfis radiais do anel E de Saturno. O perfil superior foi integrado sobre toda a largura do anel, enquanto o inferior apenas sobre o núcleo do anel. As órbitas de satélites próximos estão indicadas por linhas verticais, assim como os anéis F e G. Adaptado de Pater et al. (2004).

A combinação destes dois fatores é responsável pela criação e manutenção do anel E. A contribuição devido às colisões em alta velocidade é menor do que a ejeção de material e gera uma nuvem simétrica de poeira na vizinhança do satélite. Já a atividade geológica do satélite gera mais material e explica a assimetria no perfil radial do anel E (Spahn et al. 2006a).

No caso do anel $\mu$, Mab provavelmente não apresenta nenhuma atividade geológica devido ao seu reduzido tamanho, diferentemente de Enceladus que tem um raio aproximadamente vinte vezes maior. Entretando, um mecanismo semelhante de impactos de micrometeoros, ou mesmo de partículas do próprio anel, pode gerar o material que forma o anel. A taxa de produção de partículas está relacionada à velocidade de impacto, ao fluxo de impactantes e ao tamanho do satélite impactado. Para objetos esféricos a velocidade do material ejetado é comparável com a velocidade de escape e isso faz com que exista um tamanho ótimo para a eficiência na produção de partículas, que corresponde a satélites com raio de $\sim 10$ km (Burns et al. 2001), valor próximo ao diâmetro de Mab. Esta possibilidade será discutida na seção 2.4.

Outra semelhança entre os anéis $\mu$ e E são seus espectros. Imagens do telescópio Keck revelaram que o anel $\mu$ é azul (de Pater et al. 2006b), assim como o anel E (de Pater et al. 2004). Um espectro deste tipo é característico de uma distribuição do tamanho das partículas bastante inclinada e dominada por partículas pequenas, em que o efeito principal é o espalhamento de Rayleigh (Burns et al. 2001).

Para o caso do anel E o espectro azul é explicado através da combinação de forças como o achatamento do planeta, forças eletromagnéticas e a pressão de radiação solar que agem sobre as partículas ejetadas, fazendo com que partículas menores sejam espalhadas tanto verticalmente quanto radialmente, enquanto partículas maiores permanecem na vizinhança de Enceladus e possivelmente recolidem com o satélite após um curto período (de Pater et al. 2006b).

### 2.3.3 Anel $\nu$

O anel $\nu$ está mais próximo à Urano que o anel $\mu$ e também é menos extenso radialmente (tabela 2.3). O anel $\nu$ é acompanhado por dois satélites, Portia próximo à borda interna e Rosalind à externa. Além disso, apesar de não existir nenhum satélite imerso no anel $\nu$ ele apresenta um perfil triangular não simétrico (figura 2.10), similar ao anel $\mu$. Existem ainda evidências de variações de acordo com a longitude: em algumas imagens são vistos aglomerados de partículas (“*clumps*”) em diferentes regiões do anel, fenômeno que pode ser transiente e resultado da colisão entre corpos maiores imersos.

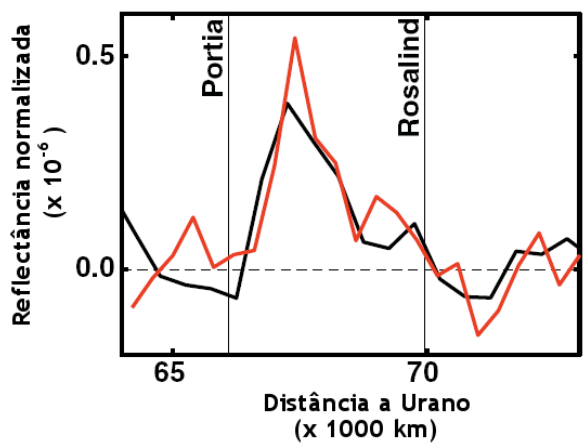

Figura 2.10: Perfis radiais do anel $\nu$ derivado a partir dos dados do HST (preto) e da Voyager (vermelho). As órbitas de Portia e Rosalind estão indicadas por linhas verticais. Adaptado de Showalter & Lissauer (2006).

Assim como o anel $\mu$ apresenta várias similaridades com o anel E de Saturno, o anel $\nu$ tem como seu análogo o anel G de Saturno. O anel G também tem um perfil radial triangular, cujo pico também está deslocado do meio do anel (figura 2.11). A distância dos anéis aos respectivos planetas também é similar: 2.78 raios de Saturno no caso do anel G e 2.64 raios de Urano no caso do anel $\nu$.

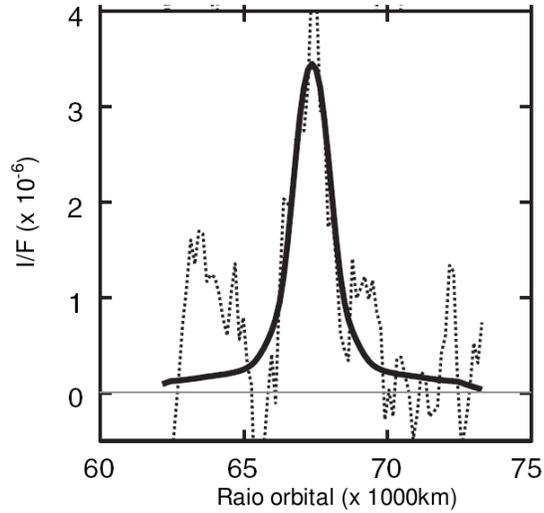

Figura 2.11: Perfil radial da porção sul do anel G de Saturno obtido em agosto de 2005. As medidas de $I/F$ estão indicadas pela linha tracejada e o melhor ajuste dos dados pela linha sólida. Adaptado de de Pater et al. (2006b).

Além da distância e do perfil similares, os dois anéis tem espectros vermelhos (de Pater et al. 2006b), o que está relacionado à partículas de poeira. Partículas pequenas tendem a espalhar a luz incidente cujo comprimento de onda é comparável ao tamanho do grão e esse espalhamento ocorre preferencialmente para grandes ângulos de fase (Burns et al. 2001). Estes dois fatores fazem com que comprimentos de onda maiores, como o vermelho, sejam espalhados com maior eficiência, dominando o espectro.

Novamente as similaridades entre os anéis $\nu$ de Urano e G de Saturno convidam à comparações entre a dinâmica deles. Ao contrário dos anéis E e F, que estão associados à satélites como fonte de material (Enceladus) ou "confinamento" (Prometheus e Pandora), o anel G está 15000 km distante do satélite mais próximo.

A sonda Pionner 11 detectou assinaturas da absorção de prótons com alta energia na região do anel G e sabe-se que grãos de poeira não podem absorver partículas tão energéticas, indicando a presença de uma população de corpos maiores (van Allen et al. 1980). Imagens obtidas pela Cassini em setembro de 2006 revelaram a existência de arcos com 60° de longitude e 250 km de extensão radial, confinados por uma ressonância de corrotação 7 : 6 com Mimas (Porco et al. 2005).

O sistema de medida magnetosférica de baixa energia (LEMMS - *Low Energy Magnetospheric Measurements System*) a bordo da Cassini localizou microassinaturas, quedas localizadas no fluxo de elétrons ao redor do anel G, permitindo estimar a massa dos arcos entre $10^8$ e $10^{10}$ kg (Hedman et al. 2007). Já o pico da reflectância ocorre para ângulos de fase entre 90° e 165°, indicando que a câmera captou a luz espalhada por partículas com tamanhos entre 1 e 10 $\mu$m. A profundidade óptica do anel é estimada em $10^{-5}$, implicando em uma massa entre $10^5$ e $10^6$ para o arco, valor inferior ao determinado pelo LEMMS. Isso mostra que deve existir uma população de partículas maiores, com tamanhos variando entre alguns centímetros a poucos metros. Além disso, recentemente foi descoberto Aegaeon, um satélite de aproximadamente 250 m localizado no arco do anel G (Porco 2009) e Hedman et al. (2010) mostraram que a dinâmica deste satélite é dominada pelas interações com Mimas.

Esta população de corpos maiores, juntamente com Aegaeon, pode ser responsável pela criação de partículas que alimenta todo o anel G. A produção de material seria causada pela colisão entre partículas com estes corpos maiores ou mesmo devido ao impacto de projéteis interplanetários com estes corpos, como proposto por Dikarev et al. (2006) para os anéis de Júpiter. Estes dois fenômenos manteriam a população de partículas do anel G em um regime estacionário (Throop & Esposito 1998, Hedman et al. 2007).

A resolução limite da análise feita por Showalter & Lissauer (2006) é de 5 km, ou seja, provavelmente não há nenhum satélite deste tamanho ou maior na região do anel $\nu$. Porém isso não exclui a possibilidade da existência de inúmeros corpos menores, similar ao que ocorre no anel G. Uma possibilidade para a origem destes corpos menores na região do anel $\nu$ seria a ruptura de um satélite que ali orbitava. Esta região está próxima ao limite de Roche e isso poderia impedir que o satélite se recomponha, ou alternativamente, o processo de aglutinação está em andamento. Assim, o anel seria uma estrutura transiente. Esta teoria de ruptura tem fundamentação na aparente história colisional dos satélites de Urano, pois acredita-se que já existiram outras colisões que causaram fragmentações e estas colisões seriam causadas por impactantes externos (Colwell & Esposito 1990, 1992) ou resultado da própria instabilidade dos satélites internos de Urano (Duncan & Lissauer 1997).

### 2.3.4 Características gerais dos anéis $\mu$ e $\nu$

Os perfis radiais dos anéis $\mu$ e $\nu$ (figuras 2.8 e 2.10) sugerem que ambos possuem uma dinâmica parecida, e sabe-se que partículas circumplanetárias estão sujeitas à várias perturbações, como o achatamento do planeta, forças eletromagnéticas, arrasto atmosférico e força de radiação solar.

Todas as perturbações citadas são resultados de diversas forças de origem diferente e cada uma provoca uma alteração diferente nas órbitas das partículas (Burns et al. 2001). O achatamento do planeta é responsável pela precessão lenta do pericentro da órbita, mantendo o tamanho, forma e inclinação constantes. Já o arrasto de Poynting-Robertson causa uma redução contínua no semi-eixo maior, enquanto partículas elétricamente carregadas sob a ação de forças eletromagnéticas experimentam uma precessão do pericentro das suas órbitas (com intensidade menor do que a causada pelo achatamento). No caso de Urano outra força dissipativa importante é o arrasto atmosférico (Broadfoot et al. 1986): a alta temperatura das camadas mais altas da atmosfera do planeta faz com que esta se expanda a altitudes consideráveis e as colisões das partículas dos anéis com átomos e moléculas da atmosfera causa o colapso das órbitas das partículas.

Porém fica claro que nenhuma dessas forças domina a evolução dos anéis $\mu$ e $\nu$, pois neste caso o perfil radial encontrado não seria triangular. A ação de uma força dissipativa tem uma direção preferencial e este comportamento seria refletido em uma assimetria na distribuição radial do anel. Este assunto será tratado com mais detalhes na próxima seção.

Outros três satélites estão localizados nas bordas dos anéis e provavelmente são responsáveis pela limitação da extensão radial dos anéis $\mu$ e $\nu$. Entretanto, a borda dos anéis não termina abruptamente na órbita dos satélites, indicando a ação de um mecanismo contínuo de remoção de partículas (mas sem a dominação de uma força dissipativa específica, como citado anteriormente). O destino mais provável para o material dos anéis são estes satélites próximos.

Além de Mab, outros satélites tem o tamanho ótimo para a produção de partículas através de impactos. Possíveis partículas ejetadas de Perdita e Cupid seriam removidas rapidamente por Belinda e um anel tênue de poeira gerado por Bianca estaria próximo demais ao planeta para poder ser detectado nas imagens atuais.

## 2.4  Forças perturbativas

Como discutido anteriormente, os anéis $\mu$ and $\nu$ são tênues e apresentam um grande espalhamento de luz para grandes ângulos de fase, indicando que são formados predominantemente por partículas micrométricas (de Pater et al. 2006b). Partículas deste tamanho estão sujeitas à ação de várias forças perturbativas como a maré solar, o achatamento do planeta e a radiação solar. Apesar destas forças serem muito menores que a força gravitacional do planeta, elas alteram a energia orbital da partícula e podem mudar significativamente a evolução ao longo do tempo.

O artigo de Hamilton & Krivov (1996) apresenta um método de comparação da intensidade de cada uma das forças perturbativas mencionadas através de parâmetros adimensionais.

Sendo $n_s$ o movimento médio do planeta ao redor do Sol e $a$ o semi-eixo maior da partícula, o parâmetro $A$ devido à força de maré solar é definido como

$$A \equiv \frac{3}{4} \frac{n_s}{n} \tag{2.3}$$

em que $n$ é o movimento médio da partícula dado por $n = (GM_p/a^3)^{1/2}$ e $GM_p$ é o produto entre a constante universal da gravitação e a massa do planeta.

O parâmetro $C$ relacionado à pressão de radiação pode ser escrito como

$$C \equiv \frac{3}{2} \frac{n}{n_s} \sigma \tag{2.4}$$

onde $\sigma$ é definido pela razão entre a força da pressão de radiação e a força gravitacional do planeta para uma órbita circular. A constante $\sigma$ pode ser escrita como

$$\sigma = \frac{3}{4} Q_{pr} \frac{Fa^2}{GM_p c \rho s} \tag{2.5}$$

Na equação 2.5 $Q_{pr}$ é uma constante relacionada à eficiência da partícula em absorver e espalhar a radiação incidente, $F$ é o fluxo solar que chega ao planeta, $\rho$ é a densidade da partícula cujo raio é $s$ e $c$ é a velocidade da luz.

A intensidade do efeito devido ao achatamento do planeta é representado pelo parâmetro $W$:

$$W \equiv \frac{3}{2} J_2 \left( \frac{R_p}{a} \right)^2 \frac{n}{n_s} \tag{2.6}$$

sendo $J_2$ o primeiro coeficiente da expansão multipolar do campo gravitacional de um planeta de raio $R_p$.

Estes parâmetros foram calculados para uma partícula esférica com densidade de $1 \, \mathrm{g/cm^{-3}}$ e raio de $1 \, \mu\mathrm{m}$ ao redor de Urano, assumindo que a partícula é composta por um material ideal de forma que $Q_{pr} = 1$. O resultado obtido é apresentado na figura 2.12.

É possível notar que a maré solar é relevante apenas para partículas distantes do planeta (acima de 50 raios de Urano), de forma que pode ser tranquilamente ignorada na região dos anéis de poeira analisados. Já os efeitos da radiação solar e do achatamento de Urano são apreciáveis e devem ser levados em consideração. Nas seções seguintes é apresentado o resultado de um estudo numérico sobre a evolução orbital de várias partículas perturbadas por estas duas forças e pela interação gravitacional com os satélites próximos aos anéis.

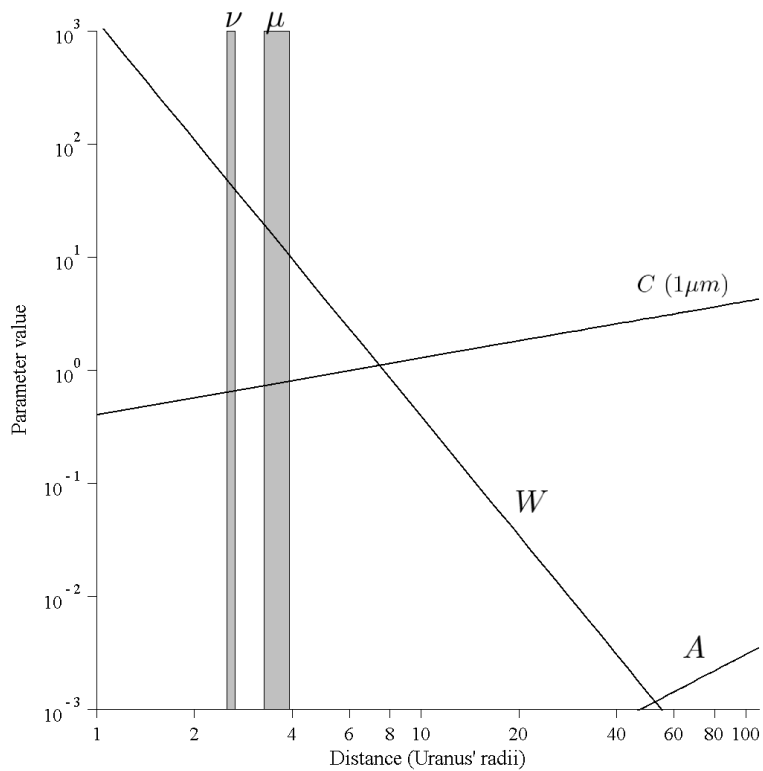

Figura 2.12: Parâmetros adimensionais das forças para uma partícula ao redor de Urano em função da distância ao planeta. As barras verticais indicam a localização e extensão radial dos anéis $\mu$ e $\nu$.

Uma ressalva deve ser feita a respeito da força eletromagnética. Partículas de anéis planetários podem apresentar uma carga elétrica não nula resultante de processos como o *sputtering* causado pelo vento solar ou à fotólise devido à radiação ultravioleta (Burns et al. 2001). Estas partículas carregadas estão sujeitas à influência do campo magnético do planeta, sendo que essa interação muitas vezes é determinante para a evolução orbital das partículas, como ocorre para os anéis de Júpiter (Burns et al. 1985, Horanyi 1996).

O cálculo da força devido às forças eletromagnéticas envolvem os coeficientes $g_{j,k}$ e $h_{j,k}$ do campo magnético do planeta. A determinação destes parâmetros para Urano representa um desafio, considerando as particularidades da orientação do campo magnético do planeta: ele não só está deslocado do centro do planeta como também inclinado em 60° em relação ao eixo de rotação (de Pater & Lissauer 2001).

Assim, o campo magnético não pode ser modelado satisfatoriamente como um dipolo alinhado, sendo necessário incluir outros termos da expansão multipolar. Porém, a única fonte de dados sobre a magnetosfera de Urano são os dados enviados pela sonda Voyager II e o conjunto de informações disponível não permite a determinação precisa dos coeficientes $g_{j,k}$ e $h_{j,k}$ de maior ordem (Holme & Bloxham 1996).

### 2.4.1 Força de radiação solar

Os anéis de Urano estão em órbita equatorial, porém o eixo de rotação do planeta está inclinado 97.86° graus em relação ao plano orbital, fazendo com que Urano seja o planeta com maior obliquidade entre os que possuem um sistema de anéis. Com isso a aproximação de que o Sol está no mesmo plano dos anéis não é válida e o problema não pode ser analisado como sendo bidimensional.

Em um sistema de referência inercial cuja origem coincide com o centro de massa de um planeta (sistema planetocêntrico) a expressão vetorial da força de radiação solar é dada por Mignard (1984)

$$\vec{F} = \beta \left[ \frac{\vec{r}_{sp}}{r_{sp}} \left[ 1 - \frac{\vec{r}_{sp}}{r_{sp}} \left( \frac{\vec{v}_p}{c} + \frac{\vec{v}}{c} \right) \right] - \left( \frac{\vec{v}_p}{c} + \frac{\vec{v}}{c} \right) \right] \tag{2.7}$$

em que $\vec{r}_{sp}$ é o raio vetor que liga o planeta ao Sol e $r_{sp} = |\vec{r}_{sp}|$, $\vec{v}$ é o vetor velocidade da partícula em relação ao planeta, $\vec{v}_p$ o vetor velocidade do planeta ao redor do Sol.

Considerando que o planeta possui uma obliquidade $\gamma$, as componentes da força que atua sobre uma partícula que orbita este planeta são dadas por

$$F_x = \frac{\beta G M_s}{r_{sp}^2} \left[ \cos\left(n_s t\right) - \left( \frac{x_s}{r_{sp}} \right)^2 \left( \frac{v_{xs}}{c} + \frac{v_x}{c} \right) - \left( \frac{v_{xs}}{c} + \frac{v_x}{c} \right) \right] \tag{2.8}$$

$$F_y = \frac{\beta G M_s}{r_{sp}^2} \left[ \cos(\gamma)\sin\left(n_s t\right) - \left( \frac{y_s}{r_{sp}} \right)^2 \left( \frac{v_{ys}}{c} + \frac{v_y}{c} \right) - \left( \frac{v_{ys}}{c} + \frac{v_y}{c} \right) \right] \tag{2.9}$$

$$F_z = \frac{\beta G M_s}{r_{sp}^2} \left[ \sin(\gamma)\sin\left(n_s t\right) - \left( \frac{z_s}{r_{sp}} \right)^2 \left( \frac{v_{zs}}{c} + \frac{v_z}{c} \right) - \left( \frac{v_{zs}}{c} + \frac{v_z}{c} \right) \right] \tag{2.10}$$

onde $(x_s, y_s, z_s)$ são as componentes da posição do Sol e $(v_{xs}, v_{ys}, v_{zs})$ as componentes da sua velocidade. As componentes da velocidade da partícula ao redor do planeta são dadas por $(v_x, v_y, v_z)$. Nestas equações os termos que dependem da velocidade correspondem ao arrasto de Poynting-Robertson (PR) enquanto os demais correspondem à pressão de radiação solar (RP).

Para partículas esféricas que obedecem uma óptica geométrica, o valor de $\beta$ é dado por (Burns et al. 1979)

$$\beta = 5.7 \times 10^{-5} \frac{Q_{\text{pr}}}{\rho s} \tag{2.11}$$

A definição da constante $\beta$ é dada pela razão entre a força devido à radiação e a força gravitacional do Sol. Como ambas as forças apresentam uma dependência com o inverso do quadrado da distância à fonte, a constante é função apenas das características físicas da partícula.

### 2.4.2 Simulações numéricas

As simulações numéricas foram realizadas utilizando o integrador Bulirsch-Stoer disponível no pacote Mercury (Chambers 1999). Este pacote permite a inclusão de uma força definida pelo usuário, tal como a dada pelas equações 2.8-2.10. O Mercury também possibilita a análise para o caso de um corpo central não esférico através da inclusão dos termos da expansão multipolar do campo gravitacional. No presente caso limitou-se ao primeiro termo desta expansão (termo $J_2$) para Urano. Esta é uma aproximação razoável, uma vez que os termos seguintes ($J_4$, $J_6$,...) são pelo menos duas ordens de grandeza menores que $J_2$.

No modelo estudado o planeta está em uma órbita circular ao redor do Sol, de forma que $n_s$ é constante, assim como o fluxo solar. Não foi levada em conta a sombra causada pelo planeta, assim como a reflexão de luz pela atmosfera, uma vez que a contribuição destes dois efeitos é pelo menos duas ordens de grandeza menor do que a perturbação causada pela iluminação direta (Hamilton & Krivov 1996). Pelo mesmo motivo foram ignorados os efeitos devido à rotação das partículas de poeira (efeito Yarkovsky).

Os elementos orbitais e o raio dos satélites analisados estão listados na tabela 2.4.2. Para todos eles assumiu-se a densidade uniforme de 1.3 g/cm$^3$, igual a do satélite Miranda. As informações sobre os anéis $\mu$ e $\nu$, assumidos inicialmente planos e circulares, foram apresentas na tabela 2.3.

|  | Mab | Puck | Rosalind | Portia |
|---|---|---|---|---|
| $a$ (km) | 97735 | 86004 | 69926 | 66097 |
| $e$ ($\times 10^{-3}$) | 2.54 | 0.39 | 0.58 | 0.51 |
| $i$ (°) | 0.134 | 0.321 | 0.093 | 0.18 |
| $\omega$ (°) | 240.30 | 331,88 | 257,28 | 84.41 |
| $\Omega$ (°) | 350.74 | 199.48 | 157.64 | 77.71 |
| $r$ (km) | 12 | 81 | 36 | 70 |

Tabela 2.5: Elementos orbitais e parâmetro físico dos satélites utilizados nas simulações numéricas: semi-eixo maior ($a$), excentricidade ($e$), inclinação ($i$), argumento do pericentro ($\omega$), longitude do nodo ascendente ($\Omega$) e raio ($r$). Todos os elementos orbitais referem-se a época $JED = 2453243.0$. Adaptado de Showalter & Lissauer (2006).

Nas simulações numéricas não foram incluídos outros satélites, uma vez que os satélites internos de Urano são pequenos e suas órbitas estão distantes dos anéis $\mu$ e $\nu$. Desta forma a perturbação gravitacional causada por eles na região dos anéis é pequena e não altera a evolução orbital das partículas.

Não há nenhuma informação detalhada sobre a distribuição dos tamanhos das partículas dos anéis $\mu$ e $\nu$, porém sabe-se que estes anéis são formados predominantemente por partículas micrométricas (de Pater et al. 2006b). Assim, foram escolhidas partículas com tamanhos de 1 $\mu$m, 3 $\mu$m, 5 $\mu$m e 10 $\mu$m com densidade de 1 g/cm$^3$ (gelo puro sólido) como valores representativos da população presente no anel.

Na região de cada um dos anéis foram distribuídas aleatoriamente 10000 partículas de cada um dos tamanhos. A evolução de cada partícula foi acompanhada durante 1000 anos, exceto nos casos onde ocorreu a colisão com algum dos satélites.

### 2.4.3  Resultados

**Efeitos da força de radiação solar**

Se considerarmos que toda a massa do planeta está concentrada em um ponto (caso sem achatamento), a componente da radiação solar causa uma variação de poucos quilômetros no semi-eixo maior das partículas, mas causa grandes variações na excentricidade. A figura 2.13 mostra como varia a excentricidade de partículas de diferentes tamanhos e condições iniciais idênticas devido à esta componente.

Para as maiores partículas o período de oscilação da excentricidade é aproximadamente igual ao período orbital do planeta ($\sim 84$ anos). A excentricidade de partículas de 1 $\mu$m atinge valores maiores que 0.5, que é suficiente para causar a colisão com o anel $\epsilon$. Entretanto, antes de chegar à região do sistema principal de anéis, a partícula cruza a órbita de vários outros satélites da família de Portia e pode eventualmente colidir com algum deles.

As concavidades vistas na figura 2.13 estão relacionadas à grande inclinação do plano equatorial de Urano. A figura 2.14 mostra a influência de $\gamma$ na evolução temporal da excentricidade. Neste caso foram simulados numericamente casos hipotéticos mantendo constantes os parâmetros de uma partícula de 3 $\mu$m e também os parâmetros físicos de Urano, variando apenas o valor da obliquidade.

É possível ver que o aumento da obliquidade causa o aparecimento de um mínimo local próximo a metade do período orbital ($\sim 42$ anos), instante que corresponde a passagem do Sol pelo plano dos anéis.

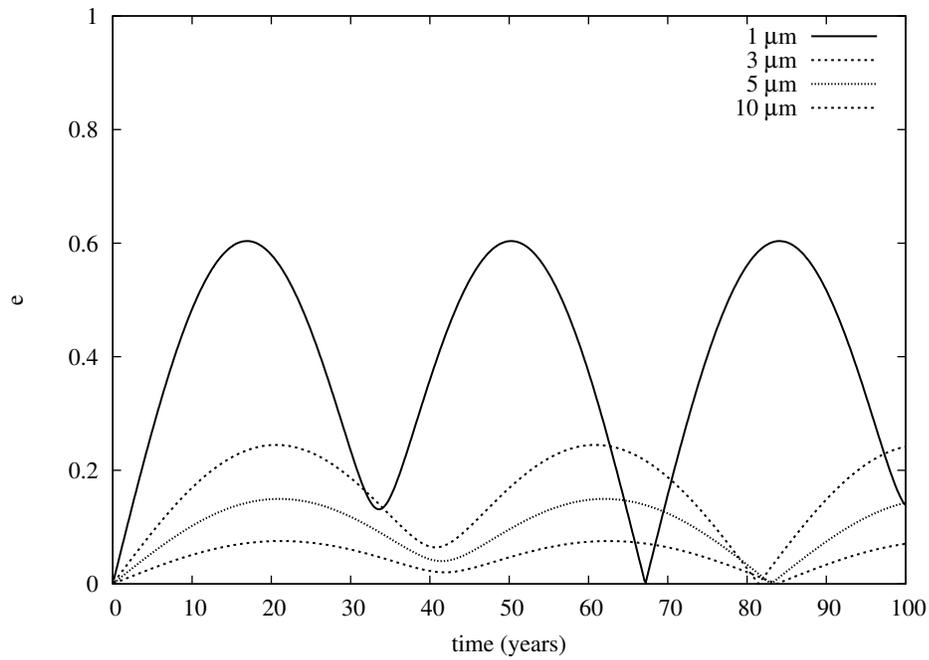

Figura 2.13: Variação da excentricidade devido à pressão de radiação para o caso sem achatamento. As partículas de diferentes tamanhos estão inicialmente localizadas próximo ao pico do perfil radial do anel $\mu$ ($a = 97500$ km).

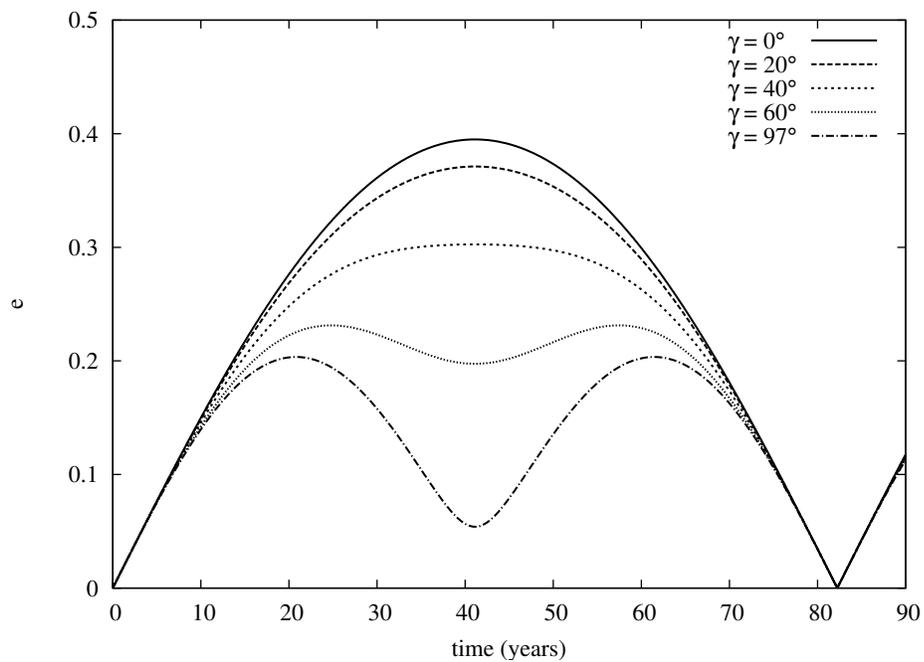

Figura 2.14: Efeito da obliquidade na variação da excentricidade devido à pressão de radiação para o caso sem achatamento. A partícula analisada tem raio de 3 $\mu$m e está inicialmente localizada próximo ao pico do anel $\mu$.

Ao contrário da pressão de radiação, o arrasto de Poynting-Robertson não provoca variações na excentricidade da partícula. O principal efeito desta componente é a redução contínua do semi-eixo maior causada pela perda de energia sofrida pela partícula. No caso de uma partícula de 1 $\mu$m com semi-eixo maior inicial de $a = 97500$ km (próximo ao pico do anel $\mu$), ela decai 23 km em 100 anos, enquanto a variação do semi-eixo maior de uma partícula de 10 $\mu$m é de apenas 2 km no mesmo intervalo de tempo (figura 2.15). Se assumirmos que a taxa de decaimento se mantém constante, o tempo até o colapso da órbita varia de $3.1 \times 10^5$ a $3.6 \times 10^6$ anos, dependendo do tamanho da partícula analisada.

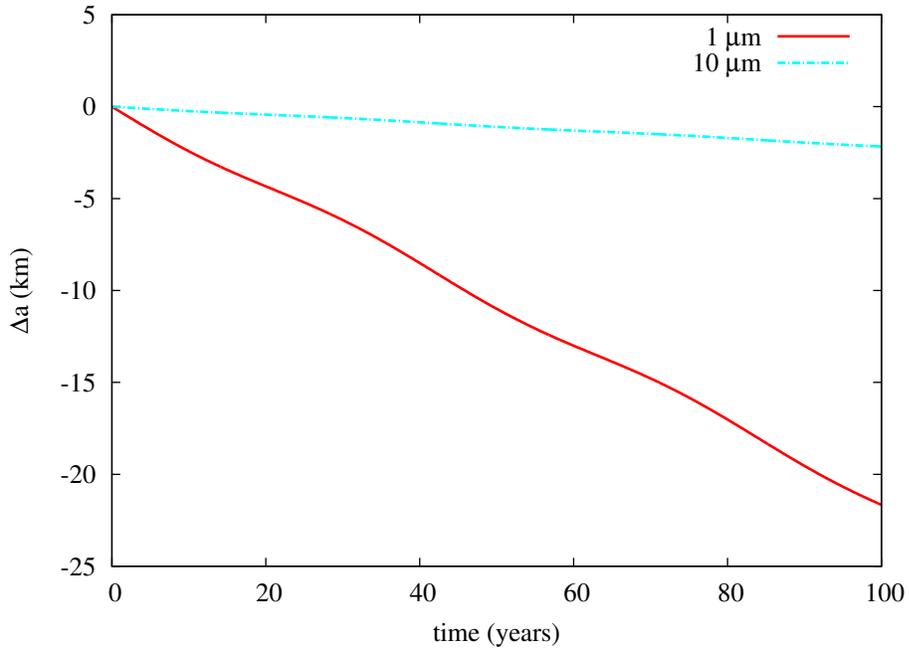

Figura 2.15: Variação do semi-eixo maior de partículas de 1 $\mu$m e 10 $\mu$m devido ao arrasto de Poynting-Robertson para o caso sem achatamento. $\Delta a = 0$ corresponde ao semi-eixo maior inicial da partícula. A região em detalhe é uma ampliação onde pode-se ver as variações de curto período.

O tempo de decaimento $\tau_{PR}$ de uma partícula em órbita planetocêntrica pode ser calculado analiticamente através da expressão (Burns et al. 1979)

$$\tau_{PR} = 9.3 \times 10^6 \frac{R^2 \rho r}{Q_{pr}} \quad \text{(anos)} \tag{2.12}$$

em que $R$ é a distância heliocêntrica do planeta. Os valores calculados através da equação 2.12 diferem menos de 5% dos valores obtidos através das simulações numéricas.

A etapa seguinte consistiu na análise de como a inclusão do achatamento de Urano altera os efeitos de cada componente da força de radiação solar.

No caso do arrasto de Poynting-Robertson, o achatamento adiciona uma oscilação de curto perído e pequena amplitude ($< 1$ km) no semi-eixo maior, mas a taxa de decaimento mantém-se inalterada e consistente com o valor estimado através da equação 2.12 (figura 2.16).

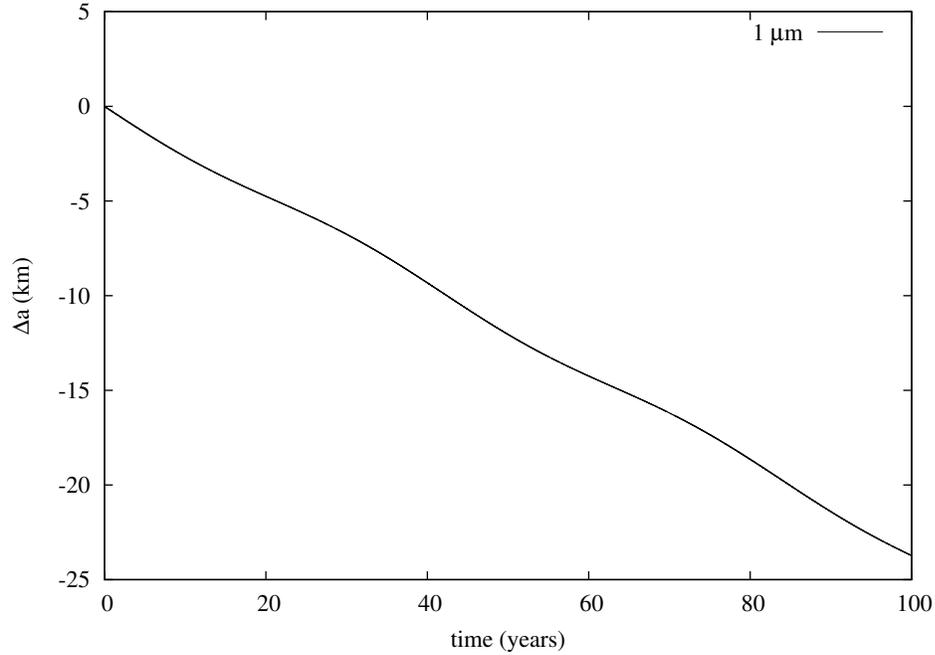

Figura 2.16: Variação do semi-eixo maior de partículas de 1 $\mu$m e 10 $\mu$m devido ao arrasto de Poynting-Robertson para o caso com achatamento. $\Delta a = 0$ corresponde ao semi-eixo maior inicial da partícula.

Os efeitos do achatamento são mais evidentes para a componente da pressão de radiação. A figura 2.17 mostra a variação da excentricidade para partículas de diferentes tamanhos e condições iniciais idênticas às da figura 2.13. A comparação entre as duas figuras mostra que o achatamento causa um amortecimento na variação da excentricidade e a diminuição na amplitude de oscilação é mais acentuada para partículas mais próximas ao planeta. Este comportamento é esperado uma vez que a intensidade da radiação solar diminui e os efeitos do achatamento de Urano aumentam quanto mais próximo do planeta (figura 2.12).

O perído de oscilação da excentricidade também é modificado pela inclusão do achatamento, passando a ser menor. O comportamento da excentricidade passa a ser modulado pela combinação de três frequências: $n_s$, $\varpi$ e uma frequência de curto perído relacionada à inclusão do termo $J_2$, onde $\dot{\varpi}$ é a taxa de precessão do pericentro da partícula e pode ser escrito como (Murray & Dermott 1999)

$$\dot{\varpi} \simeq \frac{3}{2} J_2 n \left( \frac{R}{a} \right)^2 \qquad (2.13)$$

$\dot{\varpi}$ varia entre 1.87 e 8.86 graus por ano para as partículas mais interna e externa, respectivamente, e estes valores correspondem aos obtidos através das simulações numéricas.

Dependendo do tamanho da partícula, do seu semi-eixo maior e do valor do coeficiente $J_2$, a inclusão do achatamento do planeta pode resultar em efeitos opostos. O achatamento de Marte aumenta a variação da excentricidade de partículas de poeira devido à pressão de radiação (Hamilton & Krivov 1996), enquanto para Netuno (Foryta & Sicardy 1996) e Saturno (Sfair et al. 2009) os efeitos são semelhantes aos encontrados para Urano.

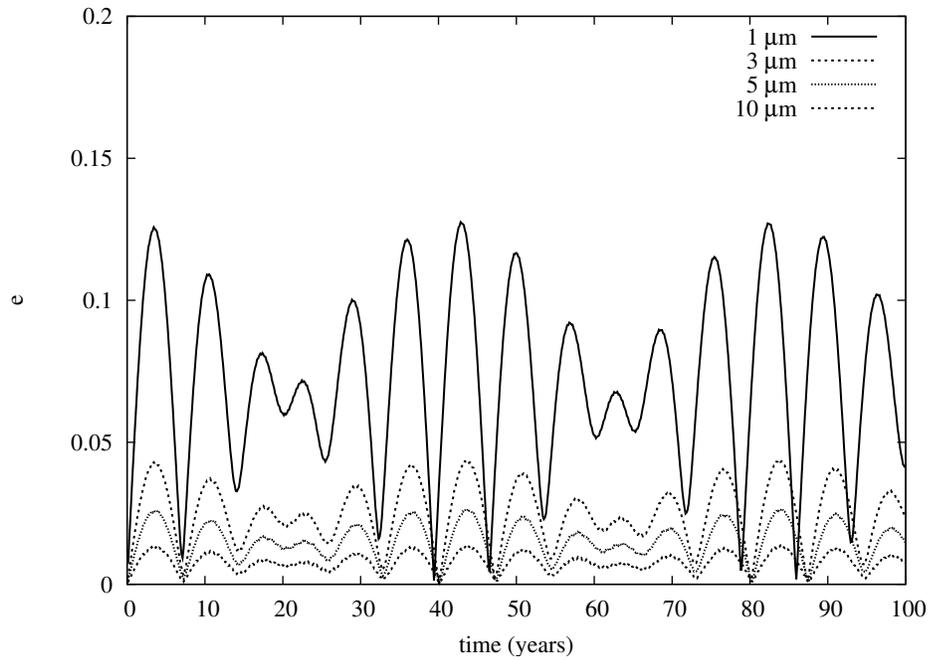

Figura 2.17: Variação da excentricidade devido à pressão de radiação para o caso com achatamento. As condições iniciais das partículas são idênticas às da figura 2.13.

### Evolução orbital

Além das perturbações devido à força de radiação solar e ao achatamento de Urano, foram incluídas nas simulações numéricas a interação gravitacional dos satélites próximos a cada anel: Puck e Mab no caso do anel $\mu$ e Portia e Rosalind no caso do anel $\nu$.

Devido ao grande número de partículas analisadas torna-se inviável apresentar graficamente a evolução orbital de todas elas. Por este motivo foram escolhidas algumas órbitas de partículas de 1 $\mu$m e 10 $\mu$m como exemplos representativos de todo o conjunto analisado.

As figuras 2.18 e 2.19 mostram a evolução do semi-eixo maior de partículas do anel $\mu$. Em cada gráfico, o eixo das ordenadas representa a variação do semi-eixo maior em relação a posição inicial da partícula, indicada por $\Delta a = 0$.

A evolução da órbita das partículas é dada pela combinação dos efeitos de todas as perturbações mencionadas anteriormente. A variação da excentricidade das partículas devido à interação gravitacional com os satélites é da ordem de $\mathcal{O}(10^{-4})$, o que é pelo menos duas vezes menor do que a causada pela pressão de radiação solar (figura 2.17). Desta forma que a evolução da excentricidade das partículas é dominada pela componente da radiação solar.

Mesmo com a redução causada pelo achatamento do planeta, a excentricidade das partículas atinge valores maiores que 0.01. Isto é suficiente para causar encontros próximos com os satélites e estes encontros podem causar variações ("saltos") no semi-eixo maior das partículas. As partículas de 1 $\mu$m apresentam maiores variações, algumas vezes chegando a centenas de quilômetros (figura 2.18), enquanto para as partículas de 10 $\mu$m os saltos possuem uma amplitude menor, da ordem de poucos quilômetros (figura 2.19).

Em alguns casos o semi-eixo maior da partícula aumenta, enquanto em outros ocorre uma redução. A direção de espalhamento depende da geometria do encontro partícula-satélite e mesmo partículas inicialmente distantes dos satélites podem sofrer encontros próximos, especialmente pequenas partículas que são mais sensíveis aos efeitos da pressão de radiação.

A tendência na redução do semi-eixo maior é causado pelo arrasto de Poynting-Robertson. Este efeito é mais nítido para partículas de 10 $\mu$m, as quais apresentam uma taxa de decaimento mais lenta e sofrem uma menor variação da excentricidade (figura 2.19b).

Nos casos apresentados nas figuras 2.18 e 2.19 o destino final das partículas é diferente. A variação do semi-eixo maior devido à encontros próximos com os satélites, combinada com a variação na excentricidade, faz com que a órbita das partículas cruze a órbita dos satélites e a sobreposição entre elas pode eventualmente resultar em uma colisão. No caso da partícula representada na figura 2.18a o semi-eixo maior varia mais de 300 km e após 365 anos a partícula colide com Puck.

As partículas do anel $\nu$ apresentam uma evolução orbital semelhante. As figuras 2.20 e 2.21 mostram a evolução do semi-eixo maior de partículas do anel $\nu$ sob influência do achatamento de Urano, da força de radiação solar e da perturbação gravitacional dos satélites Portia e Rosalind.

Os efeitos do achatamento de Urano são maiores na região do anel $\nu$ pois o anel está mais próximo ao planeta (figura 2.12), fazendo com que o amortecimento na variação da excentricidade seja maior do que no caso do anel $\mu$. Contudo, o anel $\nu$ não é tão extenso radialmente quanto o anel $\mu$, de forma que mesmo valores menores de excentricidade das partículas podem causar encontros próximos com os satélites. Novamente, estes encontros são responsáveis pelos saltos que podem ser observados nas figuras 2.20 e 2.21.

Ao contrário do que foi verificado para o anel $\mu$, o semi-eixo maior de algumas partículas de 10 $\mu$m do anel $\nu$ apresentam variações de centenas de quilômetros (figura 2.21b). Esta diferença entre o comportamento encontrado para cada anel está relacionada à proximidade entre os satélites, uma vez que a região compreendida entre as órbitas de Portia e Rosalind é menor que a região entre os satélites Puck e Mab.

A evolução das partículas de tamanhos intermediários (3 $\mu$m e 5 $\mu$m) dos dois anéis é semelhante aos casos que foram apresentados aqui. A excentricidade varia sob a combinação dos efeitos da pressão de radiação e do achatamento de Urano, enquanto o semi-eixo maior sofre perturbações devido à encontros próximos com os satélites.

Na seção seguinte será apresentada a análise do destino final das partículas analisadas.

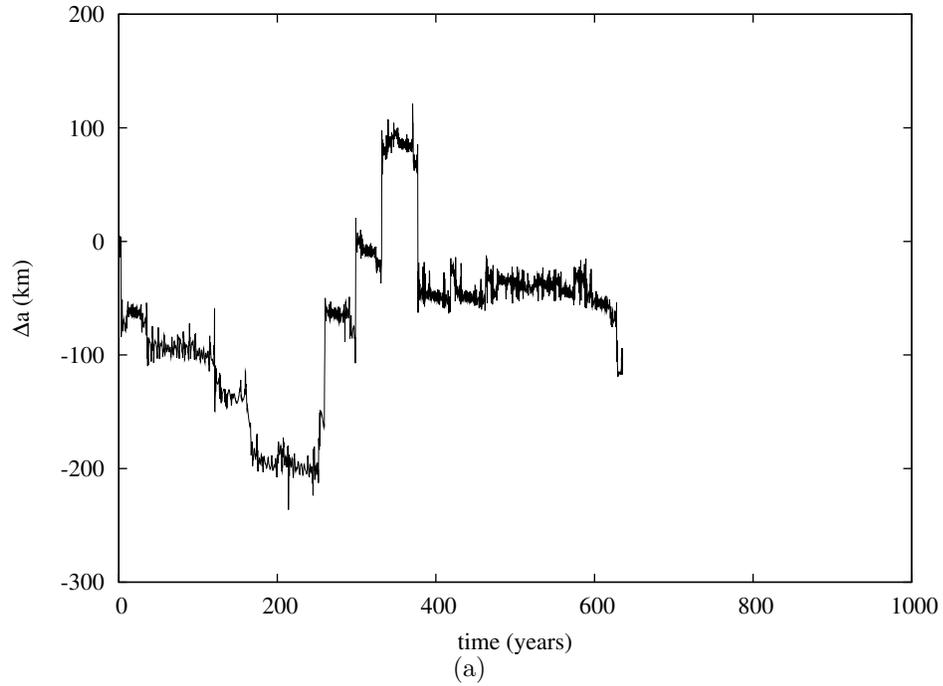

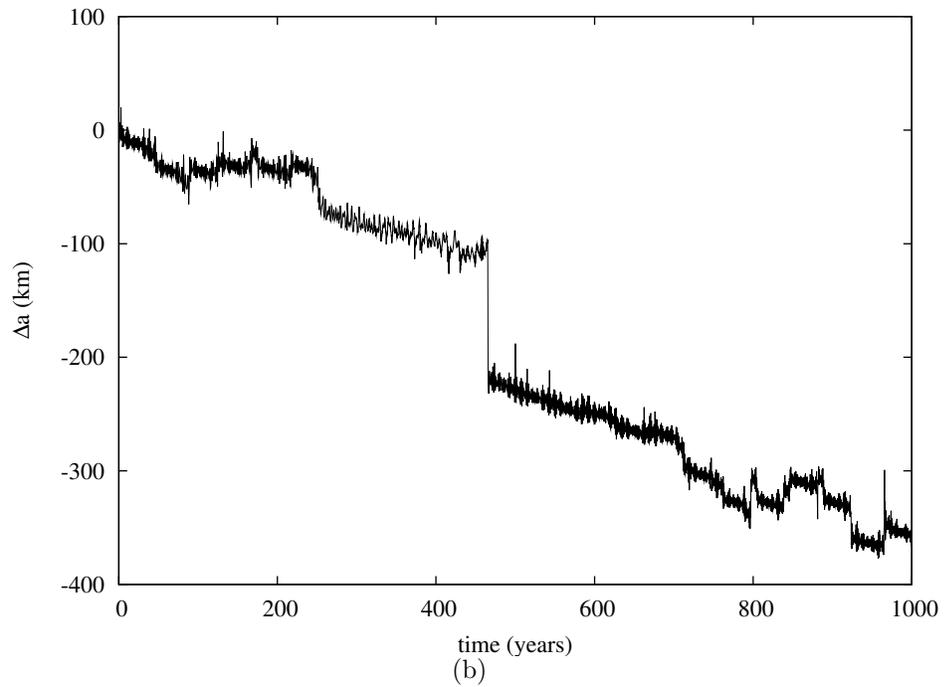

Figura 2.18: Evolução do semi-eixo maior de duas partículas de 1 $\mu$m do anel $\mu$ perturbadas pela força de radiação solar, pelo achatamento de Urano e pela interação gravitacional com os satélites Puck and Mab. Em (a) a partícula colide com Puck após 635 anos e a partícula representada em (b) permanece na região do anel durante todo o período analisado. Em cada gráfico $\Delta a = 0$ corresponde ao semi-eixo maior inicial da partícula. A largura do anel é 17000 km.

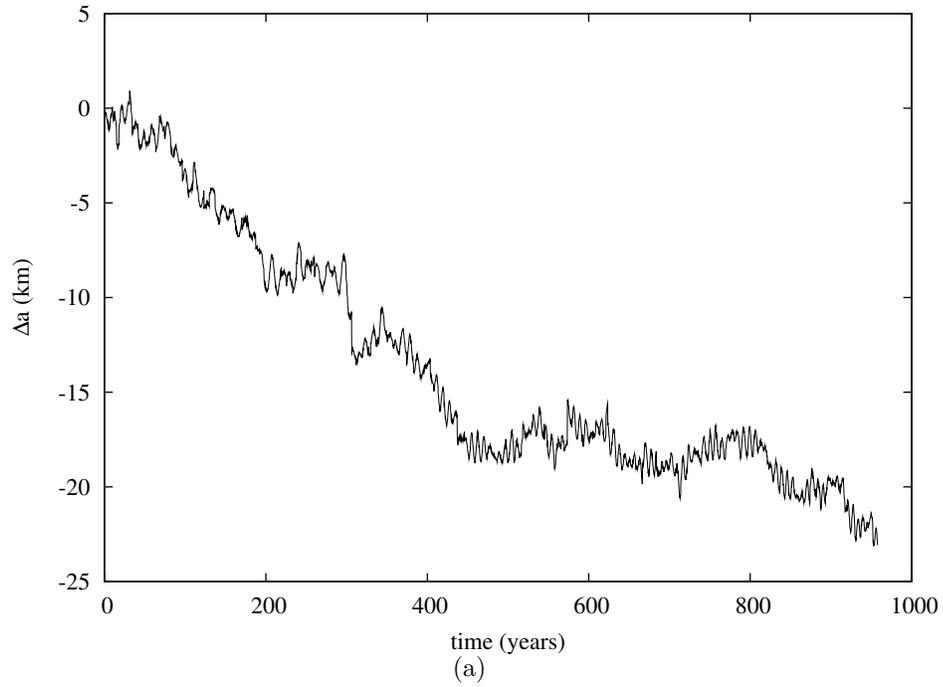

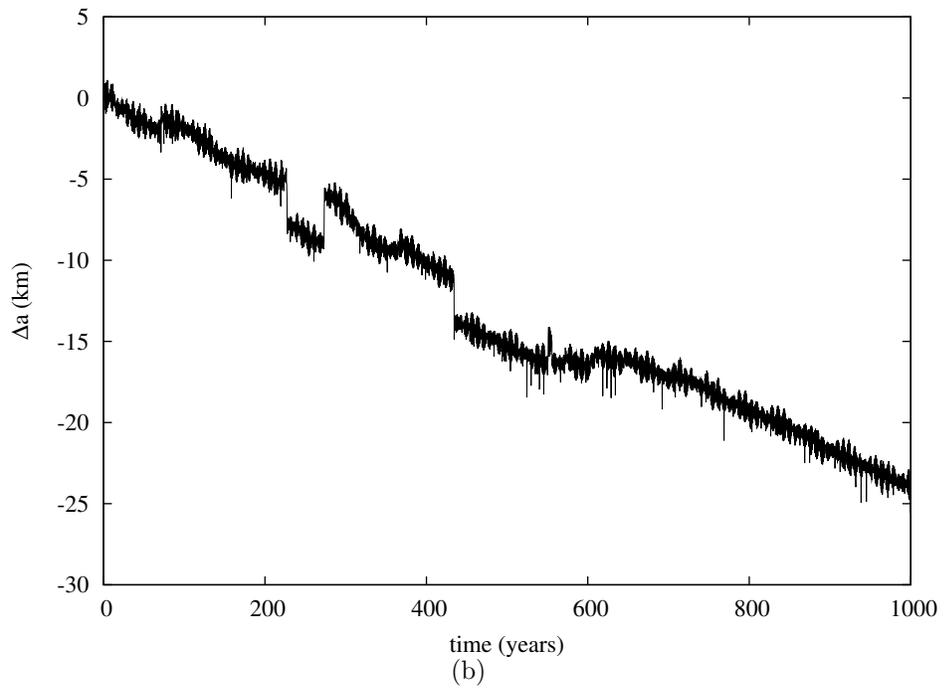

Figura 2.19: Evolução do semi-eixo maior de partículas de 10 $\mu$m do anel $\mu$ perturbadas pela força de radiação solar, pelo achatamento de Urano e pela interação gravitacional com os satélites Puck and Mab. A partícula representada em (a) colide com Mab após 950 anos e a partícula representada em (b) permanece na região do anel durante todo o período analisado.

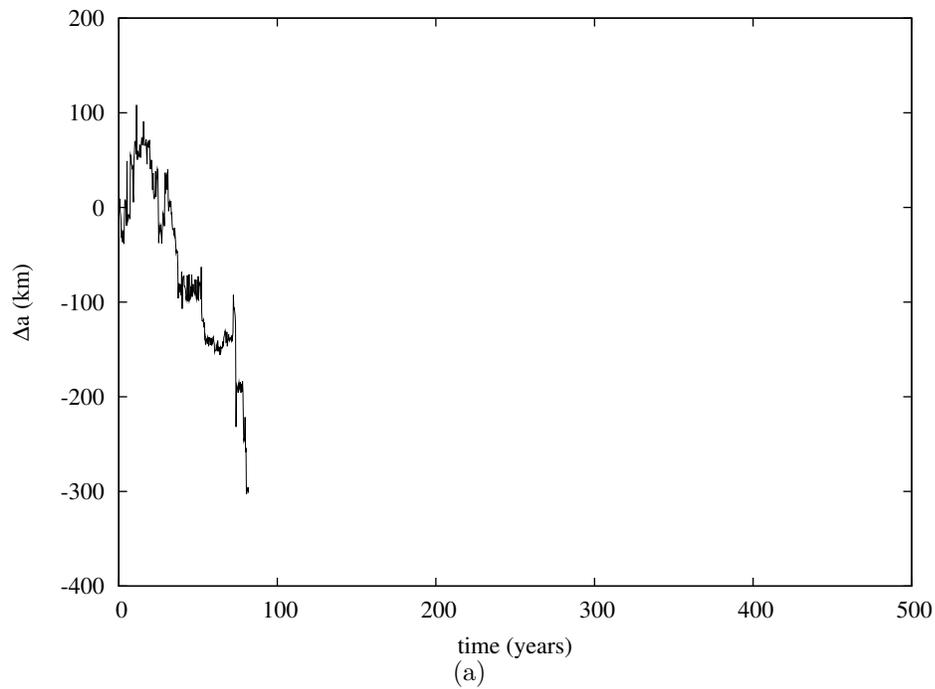

(a)

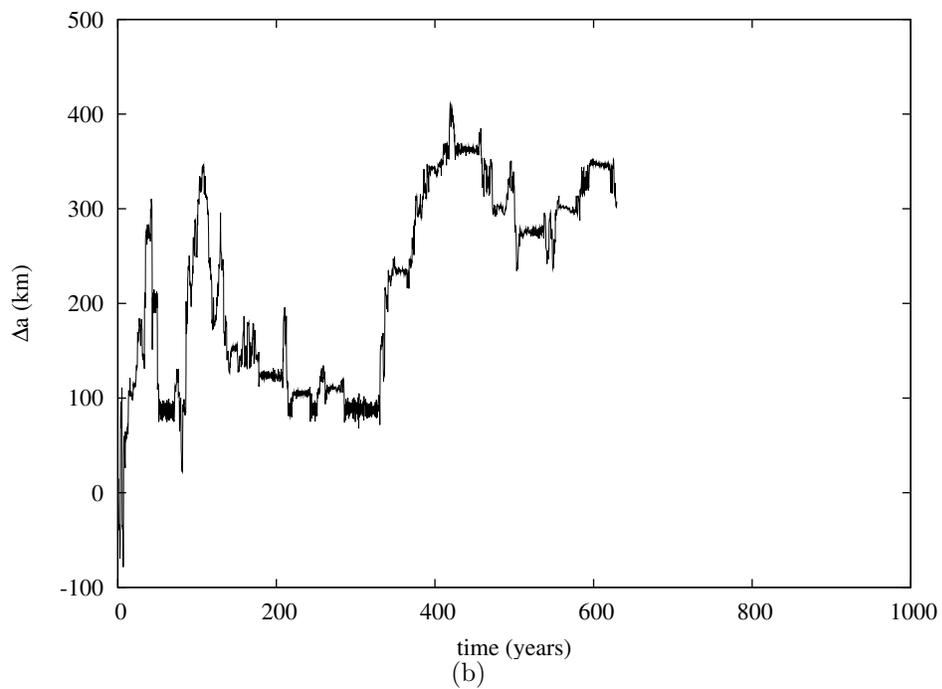

(b)

Figura 2.20: Evolução do semi-eixo maior de partículas de 1 $\mu$m do anel $\nu$ perturbadas pela força de radiação solar, pelo achatamento de Urano e pela interação gravitacional com os satélites Portia e Rosalind. A partícula representada em (a) colide com Portia após 80 anos e a partícula representada em (b) colide com Rosalind em aproximadamente 690 anos. Em cada gráfico $\Delta a = 0$ corresponde ao semi-eixo maior inicial da partícula. A largura do anel corresponde à 3800 km.

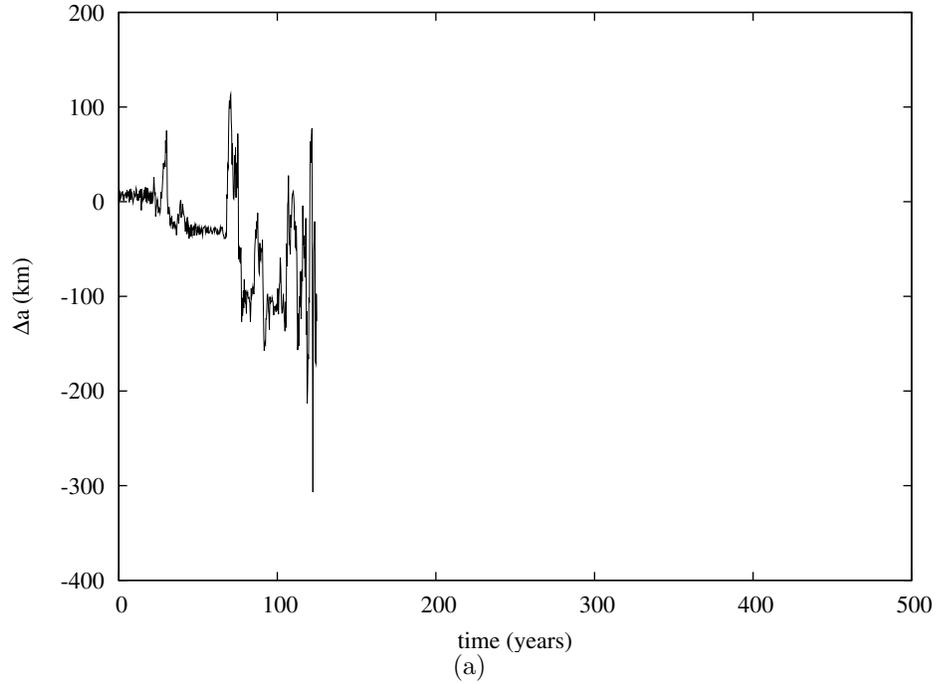

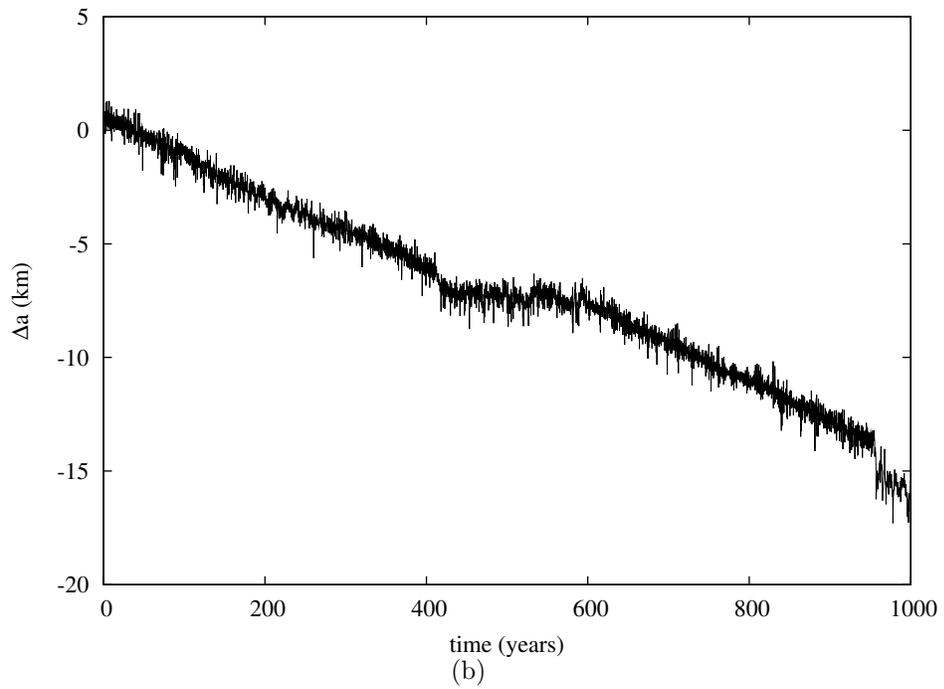

Figura 2.21: Evolução do semi-eixo maior de partículas de 10 $\mu$m do anel $\nu$ perturbadas pela força de radiação solar, pelo achatamento de Urano e pela interação gravitacional com os satélites Portia e Rosalind. A partícula representada em (a) colide com Portia após 125 anos, enquanto a partícula representada em (b) permanece na região do anel durante todo o período analisado.

**Colisões com os satélites**

A inclusão do achatamento de Urano evita o grande aumento na excentricidade das partículas. Desta forma todo o conjunto simulado permanece confinado na região dos anéis durante o intervalo de tempo analisado (1000 anos).

Não há evidência de uma relação direta entre o destino de uma partícula e sua posição inicial. Mesmo partículas inicialmente localizadas nas vizinhanças de um satélite podem ser espalhadas e permanecer na região do anel.

Entretanto, mesmo ficando na região dos anéis, a órbita de uma partícula pode eventualmente cruzar a órbita de um satélite e colidir com ele. As tabelas 2.6 e 2.7 resumem as informações sobre o destino das partículas dos anéis $\mu$ e $\nu$, respectivamente, ao final das integrações numéricas. Para cada satélite são apresentados a porcentagem do número total de partículas de cada tamanho que colide com o satélite ($N_\%$), a velocidade média de impacto ($\overline{v}$) e o tempo médio no qual ocorre a colisão ($\overline{T}$).

| Tamanho da | Puck | | | | Mab | | |
|:---:|:---:|:---:|:---:|:---:|:---:|:---:|:---:|
| partícula ($\mu m$) | $N_\%$ | $\overline{T}$ (anos) | $\overline{v}$ (m/s) | $N_\%$ | $\overline{T}$ (anos) | $\overline{v}$ (m/s) |
| 1 | 60 | 154 | 468 | 2 | 443 | 718 |
| 3 | 17 | 113 | 182 | 3 | 474 | 379 |
| 5 | 12 | 151 | 162 | 3 | 445 | 354 |
| 10 | 9 | 189 | 158 | 3 | 454 | 344 |

Tabela 2.6: Destino das partículas do anel $\mu$ ao final da simulação numérica. $N_\%$ é a porcentagem no número total de partículas que colidiu com o satélite, $\overline{T}$ é o tempo médio de colisão e $\overline{v}$ a velocidade média de impacto.

| Tamanho da | Rosalind | | | | Portia | | |
|:---:|:---:|:---:|:---:|:---:|:---:|:---:|:---:|
| partícula ($\mu m$) | $N_\%$ | $\overline{T}$ (anos) | $\overline{v}$ (m/s) | $N_\%$ | $\overline{T}$ (anos) | $\overline{v}$ (m/s) |
| 1 | 42 | 31 | 164 | 52 | 12 | 163 |
| 3 | 23 | 131 | 73 | 31 | 120 | 88 |
| 5 | 19 | 147 | 69 | 29 | 176 | 80 |
| 10 | 16 | 196 | 58 | 27 | 234 | 79 |

Tabela 2.7: Destino das partículas do anel $\nu$ ao final da simulação numérica. $N_\%$ é a porcentagem no número total de partículas que colidiu com o satélite, $\overline{T}$ é o tempo médio de colisão e $\overline{v}$ a velocidade média de impacto.

A análise dos valores apresentados nas tabelas 2.6 e 2.7 mostra que, tanto para o anel $\mu$ quanto para o anel $\nu$, existe uma relação entre o número de colisões $N_\%$ e o tamanho das partículas. O tempo médio de colisão $\overline{T}$ também mostra uma dependência com o raio das partículas. Isto é uma consequência das diferentes amplitudes de variação da excentricidade devido à pressão de radiação. Partículas pequenas são mais sensíveis aos efeitos da pressão de radiação e apresentam maiores variações de excentricidade. Com isso a possibilidade de que a órbita da partícula cruze a órbita de um satélite é maior.

A comparação para um dado tamanho de partícula mostra que as partículas do anel $\nu$ tem um tempo de vida menor do que as do anel $\mu$. Após 1000 anos, 94% das partículas com raio de 1 $\mu$m colidem com Portia ou Rosalind, e no mesmo intervalo de tempo pouco mais de 60% do número total das partículas deste tamanho do anel $\mu$ colidem com Puck ou Mab. Esta diferença resulta da diferença da extensão radial dos anéis, pois anel $\mu$ é $\sim 4.5$ vezes mais largo do que o anel $\nu$.

O maior número de colisões com os satélites internos aos anéis (Puck e Portia) deve-se principalmente a dois fatores: o tamanho e a localização destes satélites. A redução contínua do semi-eixo maior causada pelo arrasto de Poynting-Robertson faz com que haja uma tendência das órbitas das partículas à migrarem em direção ao planeta, aumentando o número de colisões com os satélites internos. Além disso, os satélites Puck e Portia são maiores que seus companheiros Mab e Rosalind, respectivamente. Isto é particularmente válido para o par Puck e Mab, pois o primeiro é 6.7 vezes maior que o segundo e com isso a área superficial de Puck chega a ser 45 vezes superior a de Mab, aumentando a possibilidade de impacto.

Quando uma partícula atinge a superfície de um satélite ela pode ser absorvida ou pode gerar novas partículas de poeira. Para cada colisão ocorrida durante as simulações numéricas foi calculada a velocidade relativa entre a partícula e o satélite, o que permite calcular a velocidade de impacto. A velocidade de impacto pode ser utilizada como medida da energia carregada pela partícula. Dependendo desta energia, o resultado da colisão pode ser distinto (Burns et al. 2001).

Uma forma de analisar o resultado da colisão é comparar a velocidade de impacto com a velocidade de escape $v_{esc}$ de cada satélite. A tabela 2.8 apresenta o valor de $v_{esc}$ para os quatro satélites próximos aos anéis $\mu$ e $\nu$.

| Satélite | $v_{esc}$ (m/s) |
|----------|-----------------|
| Puck | 69.0 |
| Mab | 10.2 |
| Portia | 59.7 |
| Rosalind | 30.7 |

Tabela 2.8: Velocidade de escape ($v_{esc}$) dos satélites próximos aos anéis de poeira.

A eficiência do processo de ejeção de material está, até certo ponto, relacionada à dife-rença entre a velocidade de colisão e a velocidade de escape. A maior parte das colisões com velocidade menor ou comparável à velocidade de escape pode resultar na deposição de material. Já colisões que ocorrem com maior velocidade relativa, sendo portanto mais energéticas, podem causar a ejeção de partícula para a região do anel (dependendo do ângulo de impacto).

A velocidade das colisões que ocorrem com os satélites Puck, Rosalind e Portia, para todos os tamanhos de partícula, são comparáveis às velocidades de escape. Desta forma, a maior parte destas colisões podem resultar na deposição de material sobre a superfície dos satélites.

Mab é uma exceção. Devido ao seu menor tamanho ($r = 12$ km), a velocidade de escape é apenas 10.2 m/s e a velocidade de colisão varia entre 344 m/s e 718 m/s para partículas de 10 $\mu$m e 1 $\mu$m, respectivamente, e em alguns casos o impacto ocorre com velocidades superiores à 1 km/s. Estas colisões podem produzir poeira para o anel, e a massa das partículas ejetadas pode ser maior do que a massa da partícula que impactou o satélite. Contudo, esse processo é pouco eficiente, uma vez que o número de colisões com o Mab é pequeno, em média 3% do número total de partículas analisadas.

### 2.4.4 Considerações finais

O estudo apresentado nesta seção mostrou que as partículas de poeira dos anéis $\mu$ e $\nu$ de Urano são fortemente perturbadas pela força de radiação solar e pelo achatamento do planeta. O achatamento de Urano reduz a variação da excentricidade das partículas devido à componente da pressão de radiação e desta forma as partículas permanecem na região dos anéis. Com isso evita-se que as partículas colidam com o anel $\epsilon$.

A combinação destas duas perturbações com a interação gravitacional com os satélites faz com que ocorram encontros próximos entre estes satélites e as partículas dos anéis. Os encontros resultam em variações do semi-eixo maior das partículas, que eventualmente colidem com algum dos satélite.

Dependendo da velocidade de impacto e da velocidade de escape do satélite, a colisão pode resultar na deposição de material sobre a superfície do satélite. Este mecanismo provavelmente ocorre com os satélites Puck, Portia e Rosalind, para os quais as duas velocidades são comparáveis. No caso de Mab a velocidade de impacto é superior à velocidade de escape e as colisões podem gerar mais material para o anel. Entretanto, este processo é pouco eficiente devido ao pequeno número de colisões com o satélite.

De fato, a análise das colisões e do destino das partículas depende de vários outros fatores. A geometria do impacto, propriedades físicas (densidade, composição, etc.) e mesmo a rotação dos satélites podem alterar o efeito resultante de uma colisão e a eficiência na produção de poeira. Há também a possibilidade que partículas colidam com corpos maiores imersos nos anéis, como os que supõem-se existir no anel $\nu$ (de Pater et al. 2006b). Outro processo possível é o impacto de micrometeoritos com os satélite, que pode ser uma fonte importante de material para o anel, o que será discutido na próxima seção. A combinação de todos estes mecanismos pode explicar a formação e manutenção dos anéis $\mu$ e $\nu$.

Os resultados aqui apresentados fazem parte do artigo "*Orbital evolution of the $\mu$ and $\nu$ dust ring particles of Uranus*", publicado no volume 505 da revista *Astronomy & Astrophysics*. Uma cópia do artigo pode ser encontrada no apêndice B.

## 2.5 Colisões de projéteis interplanetários com Mab

Como mostrado na seção anterior, a força de radiação solar é um mecanismo eficiente na remoção de partículas micrométricas da região dos anéis $\mu$ e $\nu$. Em uma escala de poucas centenas de anos mais de 60% das partículas de 1 $\mu$m analisadas colidem com algum dos satélites próximos ao anel $\mu$ (tabela 2.6).

Assim, é necessário buscar um mecanismo que alimente o anel de forma contínua e mantenha a população de partículas em um estado estacionário. No caso do anel $\mu$ a coincidência entre o pico do perfil radial do anel e a órbita de Mab faz com que o satélite seja um cantidado a fonte de partículas.

Foi mostrado que as colisões das partículas do próprio anel não ocorrem com energia suficiente para que o impacto gere mais poeira e as poucas colisões que acontecem com maior velocidade são raras, fazendo com que esse processo seja pouco eficiente.

Um método alternativo que possibilita a criação de novas partículas para o anel é a colisão de micrometeoros – ou IDPs (*Interplanetary Dust Particles*), com Mab. Neste seção será calculado o fluxo de IDPs na órbita de Urano e a quantidade de material que os impactos pode gerar. Também será apresentado um breve estudo da evolução orbital das partículas após o processo de ejeção.

### 2.5.1 Fluxo de IDPs

A taxa de produção de novas partículas para o anel é diretamente proporcional da quantidade de impactos de micrometeoros com a superfície do satélite, o que depende do fluxo de projéteis no espaço interplanetário.

A densidade de IDPs é difícil de ser medida observacionalmente, assim a melhor maneira de efetuar medições é localmente através de sondas espaciais. Alguns estudos foram realizados para determinar a quantidade de partículas sólidas a diferentes distâncias heliocêntricas. Grun et al. (1985), fazendo uso de dados provenientes em sua maior parte de sondas próximas a Terra, calcularam o fluxo de IDPs com até 1 g a uma distância de 1UA. Estes dados podem ser extrapolados para outros valores de massa e outras distâncias heliocêntricas, como foi feito por Colwell & Esposito (1990).

Um modelo mais elaborado foi desenvolvido por Divine (1993). Além dos dados de Grun et al. (1985), foram adicionadas informações enviadas por outras sondas (e.g. Pionner 10 e 11, Galileo e Ulysses) e medidas de radar. Para ajustar todos os dados são necessárias cinco populações distintas de partículas, com massas variando entre $10^{-18}$ e $1.0$ g na região compreendida entre 0.1 e 20 UA.

Porém, quanto mais afastado do Sol, menor é a quantidade de informações disponíveis sobre o meio interplanetário, tornando complexa a tarefa de calcular o fluxo de IDPs na órbita de Urano. Dados enviados pela sonda Pionner 10 foi detectado uma densidade espacial de partículas de $\geq 10^{-12}$ kg praticamente constante até 18 UA (Humes 1980). Por este motivo é esperado que o fluxo em Urano seja bastante similar ao valor de $1.8 \times 10^{-16}$ kgm$^{-2}$s$^{-1}$ encontrado para Saturno (Krivov et al. 2003). Desta forma, Porter et al. (2010) estabeleceram que o fluxo à distância heliocêntrica de Urano equivale a

$$F_{\mathrm{imp}}^{\infty} = 1.2 \times 10^{-16} \quad \text{kg m}^{-2}\text{s}^{-1} \tag{2.14}$$

sendo que o índice $\infty$ indica que este é o valor medido distante do planeta.

## 2.5.2 Focalização gravitacional

Quando um projétil interplanetário se aproxima de Urano e entra na sua esfera de Hill ele é acelerado em direção ao planeta. Essa focalização faz com que a velocidade e o número de IDPs aumente consideravelmente, sendo tanto maior quanto mais próximo do planeta.

Através da integral da energia é possível relacionar velocidade da partícula antes ($v_{\mathrm{imp}}$) e depois ($v_{\mathrm{imp}}^{\infty}$) da focalização gravitacional através da equação (Colombo et al. 1966)

$$\frac{v_{\mathrm{imp}}}{v_{\mathrm{imp}}^{\infty}} = \sqrt{1 + \frac{2GM_p}{a\left(v_{\mathrm{imp}}^{\infty}\right)^2}} \tag{2.15}$$

onde $G$ é a constante gravitacional e $M_p$ a massa do planeta.

Dadas as características da população de IDPs, a velocidade das partículas longe do planeta pode ser escrita como

$$v_{\mathrm{imp}}^{\infty} = V_p\sqrt{e^2 + i^2} \tag{2.16}$$

em que $V_p$ é a velocidade orbital média do planeta e $e \approx i \approx 0.3$. Substituindo estes valores obtém-se $v_{\mathrm{imp}}^{\infty} = 2.9$ km/s.

É importante salientar que não é levada em conta a velocidade orbital do satélite, o que alteraria a velocidade relativa do projétil e consequentemente a velocidade de impacto. Neste caso o processo resultante é dependente do tempo, gerando uma nuvem de poeira assimétrica e com densidade variável (Colwell 1993), o que está além do escopo deste trabalho.

A presença do planeta altera a densidade de projéteis $n_{\mathrm{imp}}^{\infty}$, que passa a ser $n_{\mathrm{imp}}$. Estas duas grandezas podem ser relacionadas através da equação 2.17, que é a versão corrigida por Spahn et al. (2006a) da equação 11 de Colombo et al. (1966).

$$\frac{n_{\mathrm{imp}}}{n_{\mathrm{imp}}^{\infty}} = \frac{1}{2}\frac{v_{\mathrm{imp}}}{v_{\mathrm{imp}}^{\infty}} + \frac{1}{2}\left[\left(\frac{v_{\mathrm{imp}}}{v_{\mathrm{imp}}^{\infty}}\right)^2 - \left(\frac{R_p}{a}\right)^2 + \left(1 + \frac{2GM_p}{R_p\left(v_{\mathrm{imp}}^{\infty}\right)^2}\right)\right]^{1/2} \tag{2.17}$$

O cálculo dos valores das equações 2.16 e 2.17 está apresentado na tabela 2.9.

### 2.5.3 Caracterização do satélite e produção de partículas

A eficiência do processo de produção de poeira também depende de características da superfície do satélite. Para quantificar esta eficiência pode ser utilizado o *yield* $Y$, parâmetro definido como a razão entre a massa total ejetada e a massa do impactor.

Vários modelos para $Y$ podem ser encontrados na literatura (Lange & Ahrens 1987, Koschny & Grün 2001, e. g.). Com base em ensaios experimentais, Koschny & Grün (2001) encontraram uma expressão para $Y$ em função da fração de silicato no material $G_{\mathrm{sil}}$ (0%= puro gelo, 100%= puro silicato), da velocidade de impacto $v_{\mathrm{imp}}$ e da massa do impactor (equação 2.18).

$$Y = 2.85 \times 10^{-8} \times 0.0149^{G_{\mathrm{sil}}/100} \times \left(\frac{1 - G_{\mathrm{sil}}/100}{927} + \frac{G_{\mathrm{sil}}/100}{2800}\right)^{-1} m_{\mathrm{imp}}^{0.23} v_{\mathrm{imp}}^{2.46} \tag{2.18}$$

Comparando a densidade de Mab com os satélites analisados por Krivov et al. (2003) pode-se assumir que $G_{\mathrm{sil}} = 0$ para Mab, simplificando a equação 2.18 para

$$Y = 2.64 \times 10^{-5} m_{\text{imp}}^{0.23} v_{\text{imp}}^{2.46}. \tag{2.19}$$

que está de acordo com o trabalho de Spahn et al. (2006a).

Conhecendo todos estes parâmetros pode-se calcular $M^+$, a taxa de produção de material por um satélite com raio $r$ e área da secção transversal $S = \pi r^2$:

$$M^+ = F_{\text{imp}} Y S. \tag{2.20}$$

em que $F_{\text{imp}}$ é o fluxo de IDPs após a focalização gravitacional dado por

$$F_{\text{imp}} = F_{\text{imp}}^{\infty} \left( \frac{v_{\text{imp}}}{v_{\text{imp}}^{\infty}} \right) \left( \frac{n_{\text{imp}}}{n_{\text{imp}}^{\infty}} \right) \tag{2.21}$$

### 2.5.4 Cálculo dos parâmetros

A determinação dos valores das equações 2.16-2.21 não apresenta maiores dificuldades. Os resultados obtidos estão apresentados na tabela 2.9. Para efetuar os cálculos assumiu-se que o impactor típico tem uma massa $m_{\text{imp}} = 10^{-8}$ kg, equivalente a uma partícula de $\sim 100~\mu$m em raio.

| | | |
|---|---|---|
| $v_{\text{imp}}/v_{\text{imp}}^{\infty}$ | 3.90 | |
| $n_{\text{imp}}/n_{\text{imp}}^{\infty}$ | 3.65 | |
| $F_{\text{imp}}$ | $1.70 \times 10^{-15}$ | kg m$^{-2}$ s$^{-1}$ |
| $M^+$ | $2.7 \times 10^{-3}$ | kg s$^{-1}$ |

Tabela 2.9: Parâmetros do modelo de impactos de IDPs.

### 2.5.5 Simulações numéricas

A evolução orbital das partículas após a ejeção pode ser tratada como um processo independente do mecanismo que as produziu. Nesta seção serão analisadas as órbitas das partículas criadas pelo impactos de IDPs na superfície de Mab sujeitas a combinação das perturbações devido à força de radiação solar, do achatamento do planeta e da perturbação gravitacional de Puck e de Mab.

## Modelo e condições iniciais

Assim como realizado na seção 2.4, foi analisado um conjunto de partículas esféricas com 1, 3, 5 e 10 $\mu$m de raio como amostra da população do anel. Partículas maiores podem ser geradas após o impacto, porém a probabilidade é bem menor (Krivov et al. 2003).

Um parâmetro importante a ser determinado é a velocidade $v_{\rm ej}$ com que as partículas são ejetadas após a colisão. Diferentemente das distribuições contínuas propostas na literatura (Kruger et al. 2000, Krivov et al. 2003) foi adotado os valores discretos de 10, 50, e 100 m/s para $v_{\rm ej}$, o que corresponde a aproximadamente 1, ,5 e 10 vezes a velocidade de escape $v_{\rm esc}$ de Mab. O limite inferior é a própria velocidade de escape, uma vez que partículas com $v < v_{\rm esc}$ não conseguem deixar as vizinhanças da superfície do satélite. O limite superior foi escolhido de forma a cobrir o intervalo de valores significativos das distribuições contínuas.

Para cada tamanho e velocidade de ejeção, um conjunto de 360 partículas foi uniformemente distribuido ao redor da superfície de Mab e em todos os casos não foi considerada nenhuma dependência entre a velocidade e o tamanho do grão (Nakamura & Fujiwara 1991). Apesar da colisão espalhar partículas em um cone com abertura de 53° (Nakamura & Fujiwara 1991, Spahn et al. 2006b) as trajetórias iniciais foram assumidas como linhas retas normais à superfície. Estas considerações simplificam o problema evitando a introdução de parâmetros adicionais no modelo, porém sem perder a visão global do processo de geração de poeira.

Para integrar as equações do movimento foi utilizado o pacote Mercury (Chambers 1999) com todas as adaptações que foram mencionadas na seção 2.4. O tempo total de integração foi de 1000 anos, exceto para as partículas que colidiram com algum dos satélites.

## Evolução orbital

Após a ejeção, a órbita inicial das partículas acompanha a órbita de Mab. Após poucos períodos orbitais as partículas se espalham formado um toróide ao redor do satélite, e a dispersão das partículas é tanto maior quanto maior for a velocidade inicial.

O estágio seguinte da evolução orbital é determinado pelos efeitos da radiação solar, que é responsável pelo aumento da excentricidade das partículas. Mesmo considerando os efeitos do achatamento a excentricidade pode atingir o valor de 0.13 para a partícula de 1 $\mu$m com velocidade inicial igual a velocidade de escape de Mab. Como comparação, com a mesma velocidade de ejeção, a excentricidade de uma partícula de 10 $\mu$m chega a 0.01.

Quando a excentricidade aumenta, as partículas são espalhadas radialmente e suas órbitas podem cruzar as órbitas de Puck ou Mab resultando em colisões. A tabela 2.10 resume o destino das partículas ao final das simulações para todos os tamanhos e valores de $v_{\rm ej}$ analisados, assim como o tempo médio ($\overline{T}$) para que ocorra a colisão com algum dos satélites.

| Tamanho da partícula | $v_{ej}$ | % de colisões | | | $\overline{T}$ |
|:---:|:---:|:---:|:---:|:---:|:---:|
| ($\mu$m) | ($\times\, v_{esc}$) | Puck | Mab | Total | (anos) |
| | 1 | 66 | 27 | 93 | 272 |
| 1 | 5 | 47 | 35 | 82 | 284 |
| | 10 | 44 | 36 | 80 | 280 |
| | 1 | 0 | 54 | 54 | 457 |
| 3 | 5 | 0 | 54 | 54 | 462 |
| | 10 | 0 | 44 | 44 | 464 |
| | 1 | 0 | 60 | 60 | 403 |
| 5 | 5 | 0 | 51 | 51 | 437 |
| | 10 | 0 | 44 | 44 | 437 |
| | 1 | 0 | 66 | 66 | 422 |
| 10 | 5 | 0 | 50 | 50 | 415 |
| | 10 | 0 | 52 | 52 | 420 |

Tabela 2.10: Destino das partículas ejetadas de Mab ao final da simulação numérica e o tempo médio das colisões.

O destino principal das partículas ejetadas é a colisão com algum dos satélites, sendo que, exceto para as partículas de 1 $\mu$m, as colisões sempre ocorrem com Mab. Isto é uma consequência da amplitude que a excentricidade pode atingir: apenas as menores partículas possuem valores de $e$ grandes o suficiente de forma que o pericentro da órbita cruze com Puck. Para as partículas que chegam próximas de Puck, o maior número de colisões com este satélite é esperado pois ele é aproximadamente 7 vezes maior que Mab.

Nos outros casos, mesmo considerando a oscilação da excentricidade e os saltos do semi-eixo maior causados pelos encontros próximos com Mab, as partículas permanecem próximas ao satélite da qual foram ejetadas. Este comportamento explica a semelhança do número de colisões para todos os valores de $v_{esc}$.

Os mesmos argumentos podem ser utilizados para entender o comportamento do tempo médio de colisão $\overline{T}$, que não apresenta uma dependência acentuada com a velocidade de ejeção. É importante ressaltar que o tempo médio em todos os casos é da ordem de algumas centenas de anos e que, principalmente para partículas maiores, a maioria das das colisões ocorre em $t \geq 100$ anos.

O espalhamento das partículas e a sobrevivência de acordo com o tamanho podem explicar o espectro azul do anel (de Pater et al. 2006b). As partículas de 1 $\mu$m são espalhadas por uma região radial maior e sobrevivem menos tempo no sistema, de forma que próximo da órbita de Mab permanecem as partículas maiores. Este fenômeno pode ser visto na figura 2.22, onde estão apresentadas as órbitas de todas as partículas durante o tempo de integração. As partículas de 1 $\mu$m (figura 2.22a) são espalhadas por toda a extensão radial do anel e chegam à órbita de Puck, enquanto as partículas de 10 $\mu$m (figura 2.22b) permanecem nas vizinhanças de Mab.

Assim, considerando que o processo de geração e remoção de partículas é contínuo, há uma separação acentuada entre os tamanhos de partículas, o que resulta em um espectro azul para o anel.

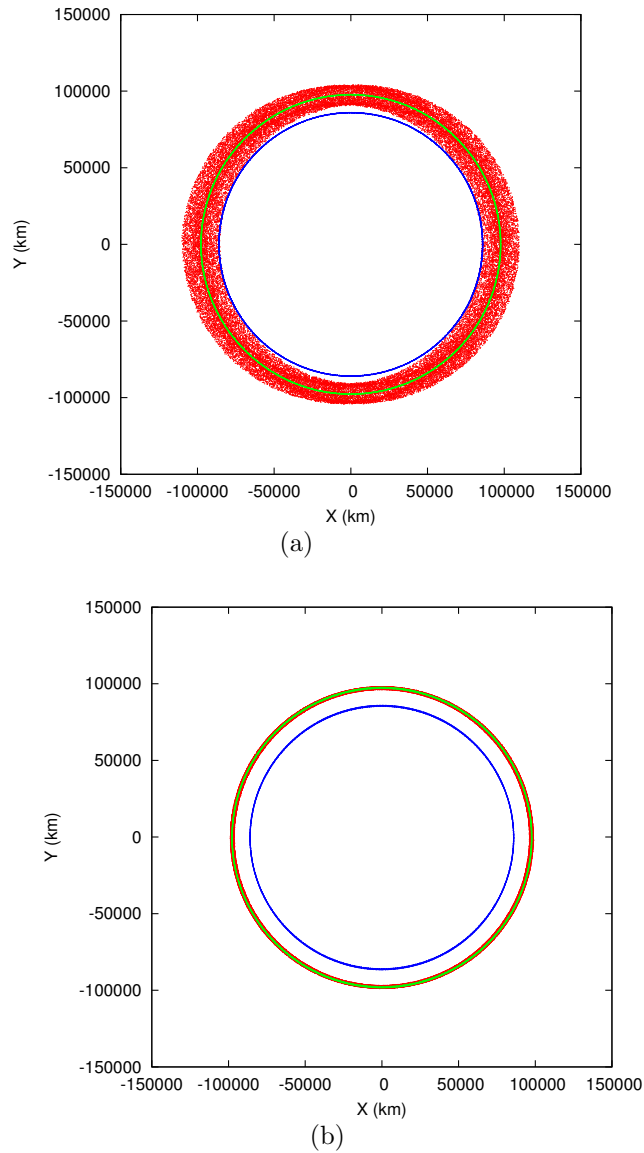

Figura 2.22: Sobreposição das órbitas das partículas (em vermelho) de (a) 1 $\mu$m e (b) 10 $\mu$m após a ejeção da superfície de Mab. Nos dois casos apresentados as simulações foram realizadas assumindo $v_{ej} = v_{esc}$. Também estão representadas as órbitas de Mab (em verde) e Puck (em azul).

### 2.5.6 A massa do anel

Para determinar se Mab é uma fonte eficiente para popular o anel é preciso calcular a massa total das partículas presentes no anel $\mu$ e comparar com a taxa de produção de poeira através dos impactos de IDPs. Devido às poucas informações disponíveis, esta tarefa é complicada e está sujeita à várias incertezas, de forma que todos os valores devem ser tratados como estimativas.

O anel $\mu$ apresenta um perfil triangular bem definido cujo pico coincide com a órbita de Mab. Como está é a região onde a maior parte das partículas permanece após ser ejetada (seção 2.5.5), considerou-se que a maior parte da massa do anel está distribuída em uma área $S$ dada por

$$S = 2\pi R \ \Delta R \qquad (2.22)$$

em que $R$ é o raio orbital e $\Delta R$ a largura da região.

Quando a distribuição do tamanho de partículas do anel, vários estudos relacionados a anéis de poeira (Colwell & Esposito 1990, Kruger et al. 2000, Showalter et al. 2009) considerram uma distribuição da forma

$$dN = Cs^{-q} \ ds \qquad (2.23)$$

em que $dN$ é o número de partículas, $s$ o tamanho do grão e $C$ uma constante. O valor do expoente $q$ varia de acordo com o sistema analisado e foi assumido como 3.5, igual ao encontrado para os anéis internos de poeira de Urano (Colwell 1993).

A profundidade óptica $\tau$ do anel é uma medida da sua transparência e como tal é proporcional à superfície ocupada pelas partículas (Burns et al. 2001). Assim, pode-se relacionar a densidade superficial de partículas $dN$ com $\tau$ através da relação

$$\tau = \int_{s1}^{s2} d\tau = \int_{s1}^{s2} \pi s^2 \ dN \qquad (2.24)$$

onde $s1$ e $s2$ são os tamanhos da menor e da maior partícula considerada, respectivamente. Como não existem informações precisas sobre o tamanho dos grãos presentes no anel $\mu$, assumimos que $s1 = 1 \ \mu$m e $s2 = 10 \ \mu$m.

Em Showalter & Lissauer (2006) pode ser encontrado o valor da profundidade óptica do pico do anel $\mu$, justamente a região onde permanece a maior parte das partículas ejetadas de Mab. Considerando partículas com densidade $\rho = 1$ g/cm$^{-3}$ em uma região com largura de 100 km localizada a 97700 km de Urano e profundidade óptica $\tau = 8.5 \times 10^{-6}$, a massa $m$ do anel é

$$m = \frac{4}{3}\pi \rho S \int_{s1}^{s2} s^3 \, dN = 2.2 \times 10^6 \text{ kg} \tag{2.25}$$

A título de comparação, esta massa corresponde a um satélite com apenas 7.4 m de raio e densidade igual a Mab.

É importante ressaltar que está massa é apenas uma estimativa, que dependendo dos parâmetros assumidos pode variar entre $10^5$ e $10^7$ kg. A incerteza provém da largura $\Delta R$ do anel, do expoente $q$ da distribuição de partículas e principalmente do tamanho dos grãos ($s1$ e $s2$).

### 2.5.7 Considerações finais

Considerando a taxa de produção de partículas devido à escavação de material da superfície de Mab por impactos de micrometeoritos (seção 2.5.4), são necessários 30 anos para produzir uma massa acumulada de $2.2 \times 10^6$ kg. Este tempo pode chegar a 250 anos quando considerado um cenário com o anel mais massivo, ou apenas 3 anos para para o limite inferior da estimativa da massa do anel $\mu$.

Dentro do intervalo assumido para a massa do anel, o tempo necessário para a geração das partículas é menor do que $\overline{T}$, exceto para os grãos de 1 $\mu$m que são removidos mais rapidamente do sistema (tabela 2.10). Assim, mesmo com a força de radiação solar atuando como um mecanismo destrutivo, Mab pode gerar partículas a uma taxa suficiente para manter uma população do anel em um estado estacionário e compatível com o que é observado. Por outro lado, a radiação solar é responsável pela segregação das partículas de acordo como tamanho do grão, o que resulta em uma distribuição característica de anéis com espectro azul.

Porém deve-se sempre ter em mente todas as incertezas do modelo apresentado. A estimativa da massa que pode ser gerada por Mab está intrinsecamente ligada ao fluxo e velocidade dos projéteis interplanetários na região da órbita de Urano, fatores para os quais não há uma grande confiabilidade. Já o fato do anel $\mu$ ser bastante tênue dificulta sua observação e consequentemente a determinação da profundidade óptica, refletindo na incerteza do cálculo da massa do anel.

Assim, dentro das limitações do modelo e dos parâmetros assumidos, Mab pode ser considerado a principal fonte de partículas para o anel $\mu$. Contudo, não deve-se excluir a possibi-lidade de que outros corpos ainda não detectados atuem como fontes adicionais de material.

Simulações numéricas mostraram que existem regiões onde satélites hipotéticos de até 10 km poderiam permanecer por até 1000 anos sem perturbar as órbitas de Puck e Mab (figura 2.23). Pode-se ver que praticamente toda a extensão radial do anel é estável durante o tempo analisado, com exceção das regiões próximas aos satélites. Estes resultados envolvendo as regiões de estabilidade, juntamente com uma análise semelhante feita para o anel $\nu$, foram apresentados no *42th DPS Meeting* (Sfair & Giuliatti Winter 2010).

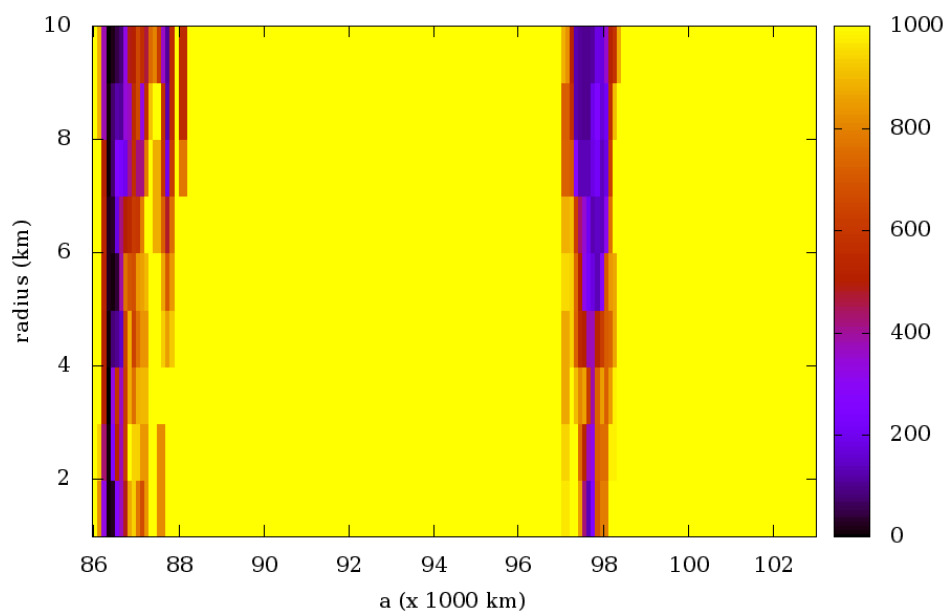

Figura 2.23: Tempo de sobrevivência em função do raio e do semi-eixo maior inicial dos satélites hipotéticos no anel $\mu$. Extraído de Sfair & Giuliatti Winter (2010).

O limite observacional elimina a possibilidade de existência de corpos > 5 km, mas um conjunto de corpos macroscópicos menores que este limite pode atuar como fonte complementar de partículas. Apesar de poder estar distribuidos ao longo do anel, para explicar o perfil triangular este conjunto deve estar localizado próximo ou na região coorbital de Mab, o que pode ter influência na órbita do satélite (Kumar et al. 2010).

# Referências Bibliográficas

# Capítulo 3

# O anel F de Saturno

## 3.1  Introdução

A primeira observação dos anéis de Saturno foi realizada em 1610 pelo italiano Galileo Galilei e foi uma das primeiras descobertas feitas com o recém inventado telescópio. Inicialmente ele interpretou a sua observação como duas luas orbitando o planeta e apenas em 1655 Christiaan Huygens apresentou a hipótese de que o sistema visto por Galileo era na verdade um anel sólido, eliptico.

No ano de 1859 James Clerk Maxwell provou matematicamente que um anel ao redor de um planeta seria estável somente se fosse composto por pequenas partículas independentes, contrariando as hipóteses de um anel (ou vários pequenos anéis) rígido. Em 1895, através de observações espectroscópias, Keller e Campell mostraram que os anéis de Saturno são formados por partículas, comprovando a idéia de Maxwell.

Até o advento das sondas espaciais o estudo de anéis planetários estava limitado a resolução dos telescópios localizados na superfície da Terra; além disso fatores como perturbações na atmosfera dificultavam as observações. Hoje sabe-se que o sistema de anéis de Saturno é vasto e complexo, formado por várias estruturas e com inúmeros satélites.

A sonda interplanetária Pioneer 11 (ou Pionner G), lançada em 1973, tinha como alvos Júpiter e Saturno e foi responsável pelo envio das primeiras imagens obtidas próximas a um anel planetário. Ao passar por Saturno ela fez uma das suas maiores descobertas: fotografou um anel estreito além da borda do anel A, posteriormente chamado anel F (Gehrels et al. 1980). As imagens enviadas pela sonda mostraram características peculiares do anel F, como uma região com maior concentração de partículas (*clumps*) e tranças no anel.

Estas particularidades do anel F fizeram com que ele se tornasse um dos objetivos principais da sonda Voyager II (que já estava a caminho de Saturno quando as imagens da Pionner 11 foram recebidas (Morrison 1982)). A figura 3.1 mostra imagens enviada pelas sondas

Voyager onde é possível identificar as estruturas peculiares do anel F.

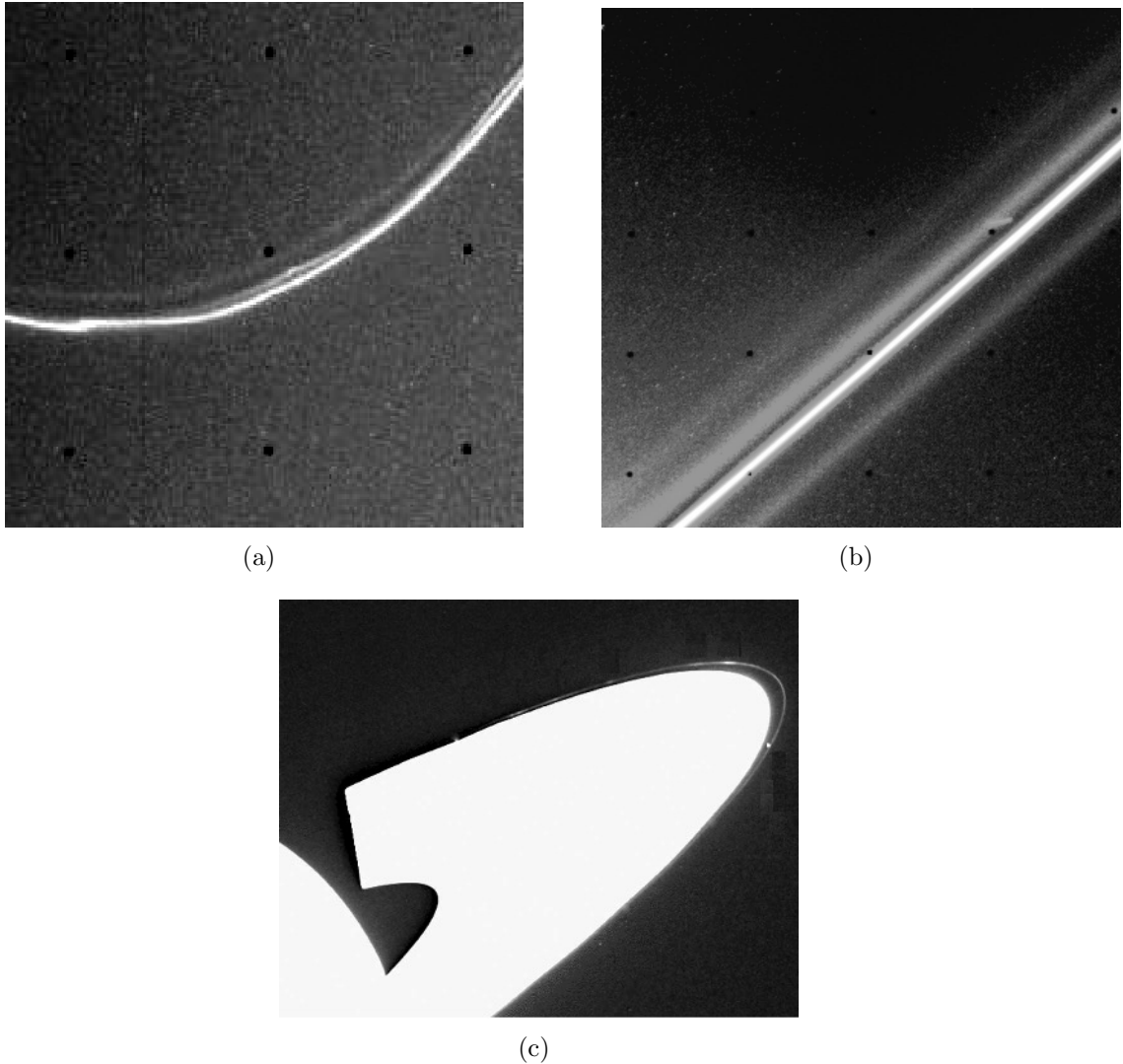

(a)

(b)

(c)

Figura 3.1: Algumas estruturas peculiares presentes no anel F: (a) imagem enviada pela Voyager I das tranças no anel (FDS34930.48); (b) imagem da estrutura múltipla enviada pela sonda Voyager II (FDS440005.1); (c) imagem enviada pela sonda Voyager II (FDS43408.10) onde é possível ver várias regiões com aglomeramento de partículas.

Essa nova visita ao sistema saturniano permitiu um estudo mais detalhado de todo o sistema de anéis devido à maior resolução das imagens. A sonda Voyager I também passou por Saturno, nove meses antes da Voyager II, mas suas câmeras eram de menor resolução e os objetivos principais desta missão estavam relacionados ao estudo da atmosfera do planeta (Morrison 1982). Mesmo assim, a partir das imagens da sonda Voyager I Collins et al. (1980)

encontraram dois pequenos satélites margeando o anel F, sendo um interno e outro externo ao anel. Prometeu, o satélite interno ao anel é mais massivo que Pandora, o satélite externo. Além disso a órbita de Prometeu é mais próxima ao anel F.

Atualmente a sonda Cassini está em órbita de Saturno e teve sua missão estendida até 2017. Suas câmeras, com resolução muito maior que as das sondas que passaram por Saturno anteriormente, revelaram com maiores detalhes os estruturas presentes no anel F, além de permitir uma melhor determinação dos elementos orbitais do anel e dos satélites próximos (Porco et al. 2005). Também foi descoberto um envelope de poeira com largura de 700 km ao redor do anel F, porém até o momento não foi publicado nenhum trabalho com mais informações sobre o tamanho das partículas que formam este envelope.

A evolução dinâmica do anel F é um assunto extensivo na literatura. Como exemplo, Murray et al. (2005) realizaram simulações numéricas da interação entre Prometeu e o anel F, considerando que o anel é formado por três faixas distintas e um envelope de poeira. Entre os resultados está a formação de falhas ou canais devido à aproximação do satélite, como havia sido previsto por Murray et al. (1997).

Porém, apenas a interação gravitacional não é suficiente para explicar todos os aspectos da formação e evolução do anel, principalmente tendo em vista a grande quantidade de partículas de poeira. Neste caso a evolução pode ser dominada por uma combinação de forças não gravitacionais que dependem fortemente do tamanho das partículas.

A tarefa de determinar precisamente a distribuição de partículas do anel F tem sido um desafio. Showalter et al. (1992) propuseram um modelo usando observações da Voyager e dados de ocultações estelares. Neste modelo o anel F é composto por um núcleo com $\sim 1$ km de largura formado por partículas centimétricas envolto por um envelope de poeira com $\sim 500$ km de largura formado por partículas micrométricas. Este núcleo do anel teria massa equivalente à um pequeno satélite com raio de $15 - 70$ km distribuída em um grande número de corpos menores.

Relacionados as forças não-gravitacionais e a fotometria do anel F, serão descritos neste capítulo os resultados obtidos em dois estudos. Na seção 3.2 é apresentado um estudo numérico sobre a influência da pressão de radiação solar nas partículas que formam o envelope de poeira do anel F, enquanto o desenvolvimento de um modelo fotométrico para o anel baseado nos dados enviados pela sonda Cassini é apresentado na seção 3.3.

## 3.2 Efeitos da pressão de radiação solar

As partículas de poeira que formam o anel F sofrem a ação de forças dissipativas. Uma destas forças, a pressão de radiação solar, pode ser dividida em duas componentes: a pressão

de radiação e o arrasto de Poynting-Robertson (P-R)(Burns et al. 1979). O objetivo deste trabalho é analisar os efeitos da radiação solar na evolução orbital das partículas de poeira do anel F, dando continuidade ao estudo iniciado em Sfair (2007).

### 3.2.1   Força de radiação solar

Como mostrado anteriormente, em um sistema de referência inercial cuja origem coincide com o centro de massa de um planeta (sistema planetocêntrico) a expressão vetorial para a força devido à radiação solar que atua sobre uma partícula que orbita este planeta é dada por (Mignard 1984)

$$\vec{F} = \beta \left[ \frac{\vec{r}_{sp}}{r_{sp}} \left[ 1 - \frac{\vec{r}_{sp}}{r_{sp}} \left( \frac{\vec{v}_p}{c} + \frac{\vec{v}}{c} \right) \right] - \left( \frac{\vec{v}_p}{c} + \frac{\vec{v}}{c} \right) \right] \tag{3.1}$$

em que $c$ é a velocidade da luz, $\vec{r}_{sp}$ é o raio vetor que liga o planeta ao Sol e $r_{sp} = |\vec{r}_{sp}|$, $\vec{v}$ é o vetor velocidade da partícula em relação ao planeta, $\vec{v}_p$ o vetor velocidade do planeta ao redor do Sol. Para partículas esféricas que obedecem uma óptica geométrica o valor da constante $\beta$ pode ser calculado através da equação (Burns et al. 1979)

$$\beta = 5.7 \times 10^{-5} \frac{Q_{\mathrm{pr}}}{\rho s} \tag{3.2}$$

sendo $s$ e $\rho$ o raio e a densidade da partícula no sistema CGS, respectivamente, e $Q_{pr}$ é uma constante que depende das propriedades ópticas do grão.

Na expressão (3.1) os termos proporcionais à velocidade correspondem ao arrasto de Poynting-Robertson (PR) enquanto o outro termo corresponde à pressao de radiação solar (RP). Por ser uma força sempre contrária à velocidade o PR causa uma diminuição contínua na energia da partícula, ocasionando uma redução no semi-eixo maior da órbita, fazendo com que a mesma decaia em direção ao planeta. O tempo de decaimento $\tau_{PR}$ de uma partícula pode ser estimado através da expressão (Burns et al. 1979)

$$\tau_{PR} = 9.3 \times 10^6 R^2 \rho s / Q_{pr} \quad \text{(anos)} \tag{3.3}$$

Já a RP é responsável pela oscilação da excentricidade da partícula, sendo que o período da variação em $e$ é igual ao movimento médio do planeta ao redor do Sol (Hamilton & Krivov 1996).

### 3.2.2  Modelo numérico e condições iniciais

Utilizando os parâmetros adimensionais definidos anteriormente (seção 2.4) para comparar as perturbações sofridas por uma partícula de 1 $\mu$m ao redor de Saturno (figura 3.2) fica evidente que, assim como ocorre em Urano, é necessário incluir no modelo tanto a perturbação devido ao achatamento do planeta (representado pelo parâmetro $W$ – equação 2.6) quanto pela pressão de radiação solar ($C$ – equação 2.4).

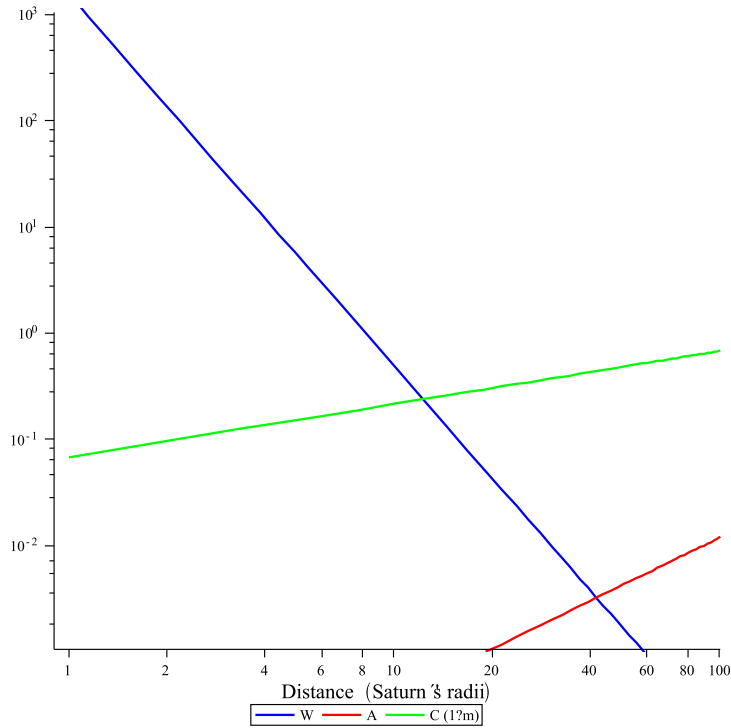

Figura 3.2: Parâmetros adimensionais das perturbações para uma partícula ao redor de Saturno em função da distância ao planeta.

Assim, as equações do movimento foram integradas com a versão adaptada do pacote Mercury (Chambers 1999), com a inclusão das componentes da força de radiação solar (equação 3.1) e do coeficiente gravitacional $J_2$ (Jacobson et al. 2006) para representar a assimetria na distribuição da massa do planeta.

Foram simuladas partículas de poeira localizadas em um envelope de 700 km de largura, com semi-eixo maior médio e excentricidade iguais ao núcleo do anel F (Murray et al. 2005). Os elementos orbitais dos satélites foram Extraídos de Jacobson et al. (2008) e para o cálculo da massa assumiu-se uma densidade média de 0.5 g/cm$^{-3}$ (Murray et al. 2005). A tabela 3.1 apresenta os valores utilizados nas simulações numéricas.

Foram simuladas partículas com raios de 1, 3, 5 e 10 $\mu$m. Para cada tamanho foi analisado um conjunto inicial de 100 partículas com semi-eixos maiores distribuídos uniformemente na região do envelope de poeira. O tempo de integração adotado foi de 100 anos, sendo que a partícula é removida do sistema caso seja detectada a colisão com o planeta ou algum dos satélites.

| Parâmetro | Prometeu | Pandora | Anel F (núcleo) |
|-----------|----------|---------|-----------------|
| m (kg) | $1.59 \times 10^{17}$ | $1.37 \times 10^{17}$ | – |
| $a$ (km) | 139380 | 141710 | 140224 |
| $e$ ($\times 10^{-3}$) | 2.2 | 4.2 | 2.6 |
| $\varpi$ (graus) | 161.0 | 83.4 | $0.0^*$ |
| $M$ (graus) | 242.3 | 202.5 | $0.0^*$ |

Tabela 3.1: Parâmetros utilizados nas simulações numéricas: massa (m), semi-eixo maior ($a$), excentricidade ($e$), longitude do pericentro ($\varpi$) e anomalia média ($M$). Os valores com $^*$ foram assumidos por simplicidade.

### 3.2.3 Resultados

**Efeitos sem o achatamento**

Como esperado, o estudo numérico mostrou que cada componente da força de radiação solar é responsável por um efeito distinto na órbita da partícula. A figura 3.3 mostra o decaimento do semi-eixo maior das partículas devido ao arrasto de Poynting-Robertson, quando não é considerado o efeito do achatamento do planeta. Todas as partículas localizadas no núcleo do anel ($\Delta a = 0$) apresentadas possuem elementos orbitais iniciais idênticos, diferindo apenas pelo tamanho.

A taxa de decaimento depende do tamanho da partícula, sendo que uma partícula de 1 $\mu$m decai 100 km em aproxiadamente 60 anos. Com uma simples extrapolação vê-se que o tempo necessário para que uma partícula deste tamanho colida com o planeta é de $5 \times 10^4$ anos, enquanto uma partícula de 10 $\mu$m demora aproximadamente $4 \times 10^5$ anos para colidir com Saturno. Estes dois valores são da mesma ordem de grandeza daqueles obtidos através da equação 3.3.

A variação na excentricidade das partículas causada pela radiação solar (figura 3.4) pode resultar na rápida colisão com o planeta, como ocorre para as partículas de 1 $\mu$m, cuja excentricidade atinge o valor de 0.6 em menos de 10 anos. Já para o caso das partículas maiores, o aumento de $e$ não é grande o suficiente para que o pericentro da partícula seja menor que o raio do planeta e a excentricidade oscila com período igual ao período orbital de Saturno.

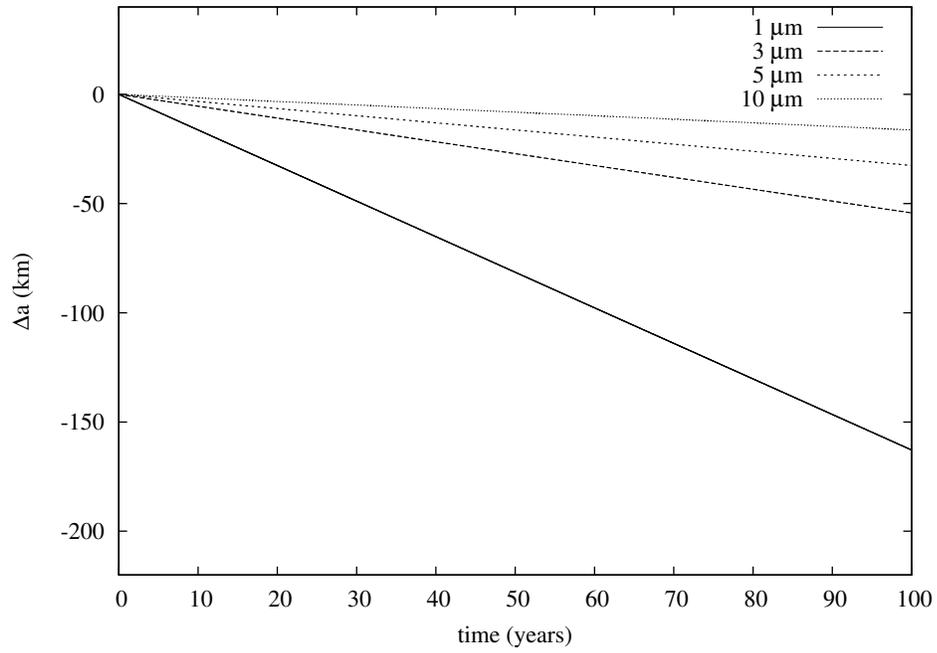

Figura 3.3: Variação do semi-eixo maior devido ao arrasto de Poynting-Robertson para o caso sem achatamento. $\Delta a = 0$ corresponde ao núcleo do anel.

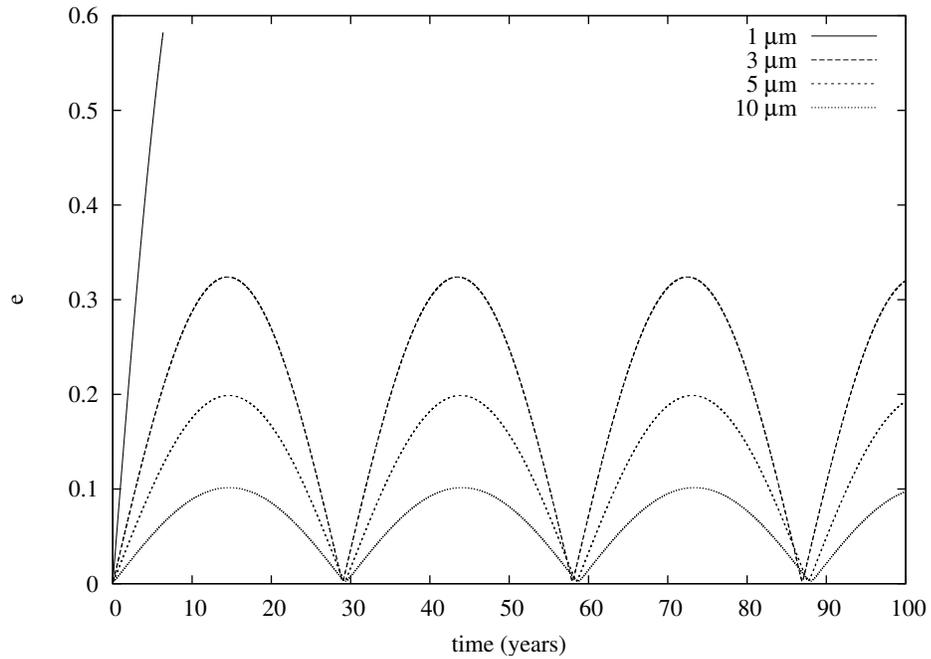

Figura 3.4: Variação da excentricidade devido à pressão de radiação para o caso sem achatamento.

Quando ambas as componentes da força de radiação solar são consideradas juntamente com a perturbação gravitacional de Prometeu e Pandora, a evolução do semi-eixo maior das partículas é alterada. A figura 3.5 mostra a evolução do semi-eixo maior e da excentricidade de partículas com 1 $\mu$m e 10 $\mu$m com condições iniciais às partículas das figuras 3.3 e 3.4.

A oscilação de curto-período do semi-eixo maior (figura 3.5a) é causada pela pressão de radiação. A amplitude da oscilação é de aproximadamente 10 km para as partículas de 1 $\mu$m e diminui conforme aumenta o tamanho do grão. Também é possível observar "saltos" do semi-eixo maiores, que estão associados à encontros próximos com os satélites (Winter et al. 2007).

A similaridade entre as figuras 3.4 e 3.5b mostra que, de acordo com as simulações numéricas, os efeitos gravitacionais dos satélites sobre a excentricidade das partículas é pequeno [$\mathcal{O}(10^{-5})$] quando comparado com os efeitos da pressão de radição.

Mesmo com o semi-eixo maior variando dentro da região do anel, o grande aumento da excentricidade faz com que as partículas analisadas cruzem a borda externa do anel A, consequentemente colidindo com as partículas do anel. Nos casos apresentados na figura 3.5 a colisão da partícula de 1 $\mu$m ocorre em menos de 3 meses e para partícula de 10 $\mu$m o tempo de colisão é aproximadamente 2,3 anos. A partícula mais externa de todo o conjunto analisado colide com o anel A quando sua excentricidade atinge o valor aproximado de 0.025.

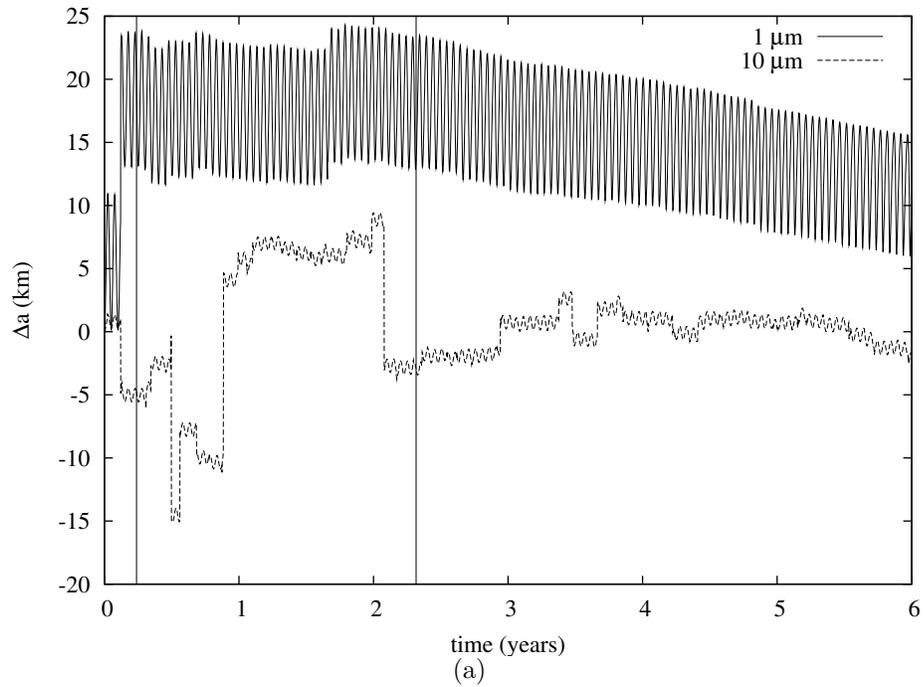

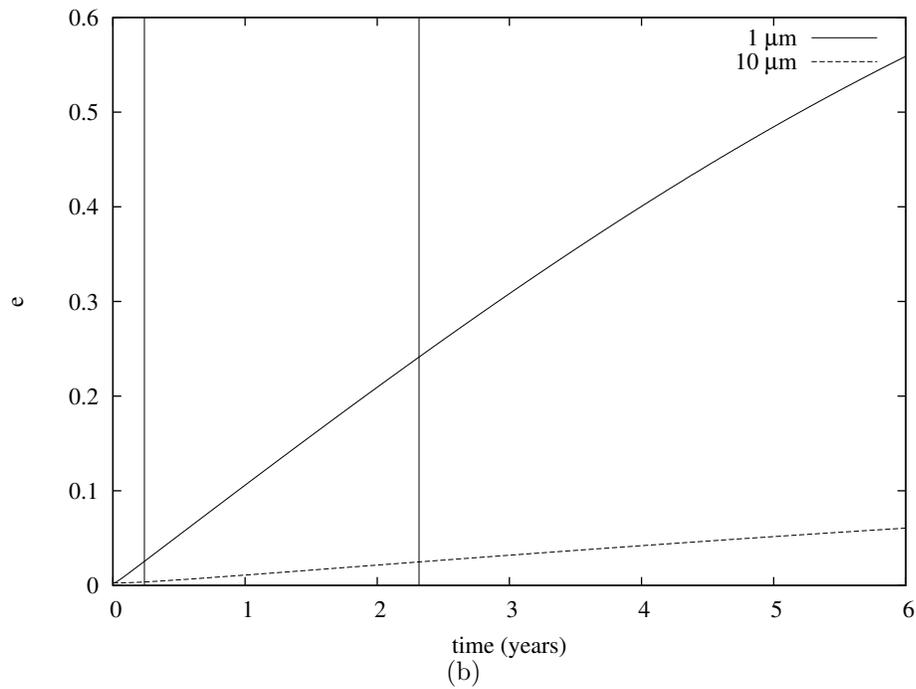

Figura 3.5: Evolução do (a) semi-eixo maior e (b) da excentricidade devido à força de radiação solar e a perturbação gravitacional de Prometeu e Pandora. Para cada caso foram analisadas duas partículas com raios de $1\mu m$ e $10\mu m$ com condições iniciais idênticas. O núcleo do anel está localizado em $\Delta a = 0$. As linhas verticais representam o instante de colisão das partículas com a borda externa do anel A.

**Efeitos com o achatamento**

A inclusão do achatamento do planeta faz com que haja um amortecimento na oscilação da excentricidade devido à pressão de radiação, reduzindo a amplitude máxima de 0.011 (figura 3.6). O período de oscilação da excentricidade, que antes estava relacionado ao movimento médio do planeta, passa a ser função também de $\varpi$, a taxa de variação do argumento do pericentro da partícula, que por sua vez depende do valor do coeficiente gravitacional $J_2$ e do semi-eixo maior da partícula.

A excentricidade induzida pela pressão de radiação pode variar entre 0.6 e 0.1 para partículas de 1 e 10 $\mu$m, respectivamente, mas devido ao aumento do momento angular – e consequentemente da velocidade orbital da partícula – a excentricidade máxima quando considerado o achatamento do planeta é de $\sim 0.01$ para ambas as partículas. O fato da excentricidade máxima não depender do tamanho da partícula pode parecer estranho, mas é um resultado esperado a partir da figura 3.2, onde fica evidente que o efeito do achatamento do planeta é pelo menos duas ordens de grandeza maior que a perturbação causada pela pressão de radiação.

A figura 3.7 apresenta a evolução do semi-eixo maior de duas partículas de diferentes tamanhos como exemplos representativos de todo o conjunto analisado. Nos dois casos as partículas de 3 $\mu$m e 5 $\mu$m estão sujeitas à combinação dos efeitos do achatamento de Saturno, da força de radiação solar e da perturbação dos satélites próximos.

O arrasto de Poynting-Robertson é o principal responsável pelo decaimento orbital da partícula, mas encontros próximos com os satélites fazem com que existam "saltos", algumas vezes aumentando e outras diminuindo o semi-eixo maior. Em todos os casos o semi-eixo maior das partículas analisadas permanecem na região do anel F, apesar da grande amplitude de variação que pode chegar a mais de 300 km em alguns casos.

Os dois casos apresentados na figura 3.7 resultaram na colisão das partículas com algum dos satélites: no caso (a) a partícula colide com Prometeu depois de $\sim 23$ anos e no caso (b) há a colisão com Pandora após 17 anos.

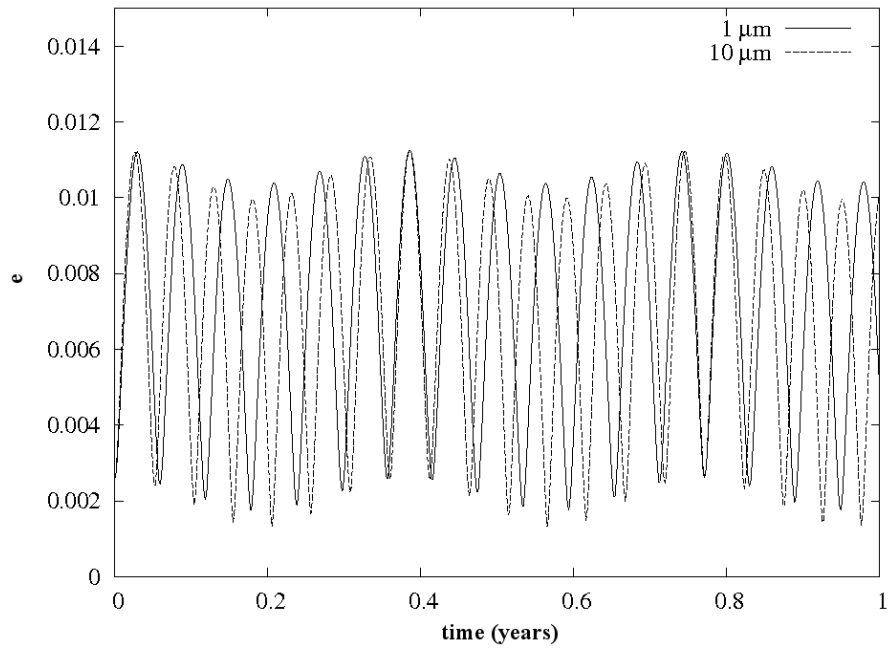

Figura 3.6: Evolução da excentricidade devido apenas à pressão de radiação solar e ao achatamento de Saturno de duas partículas com raios de $1\mu$m e $10\mu$m com condições iniciais idênticas às da figura (3.5).

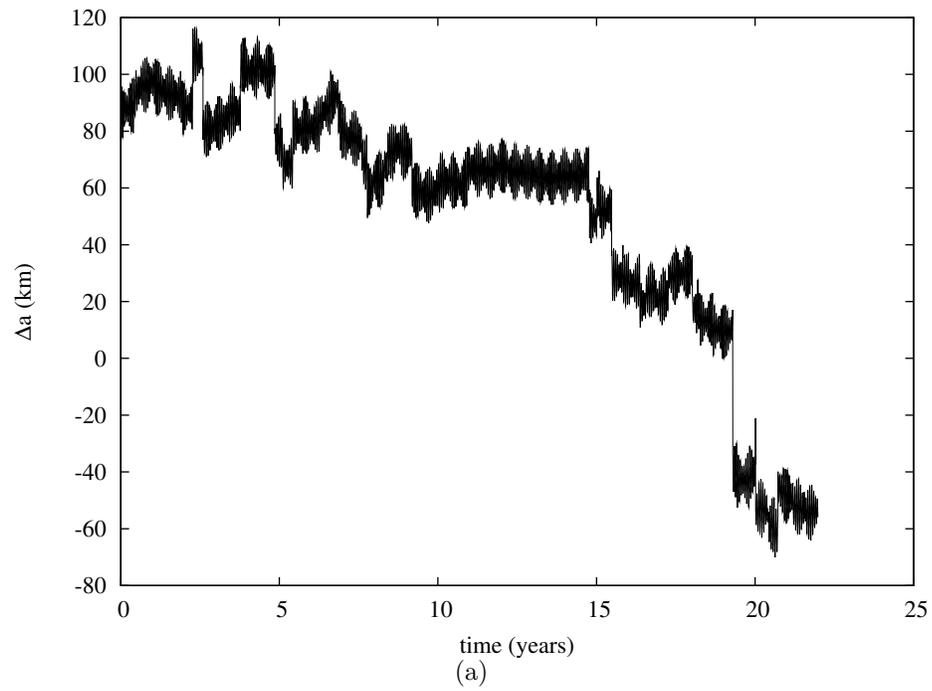

(a)

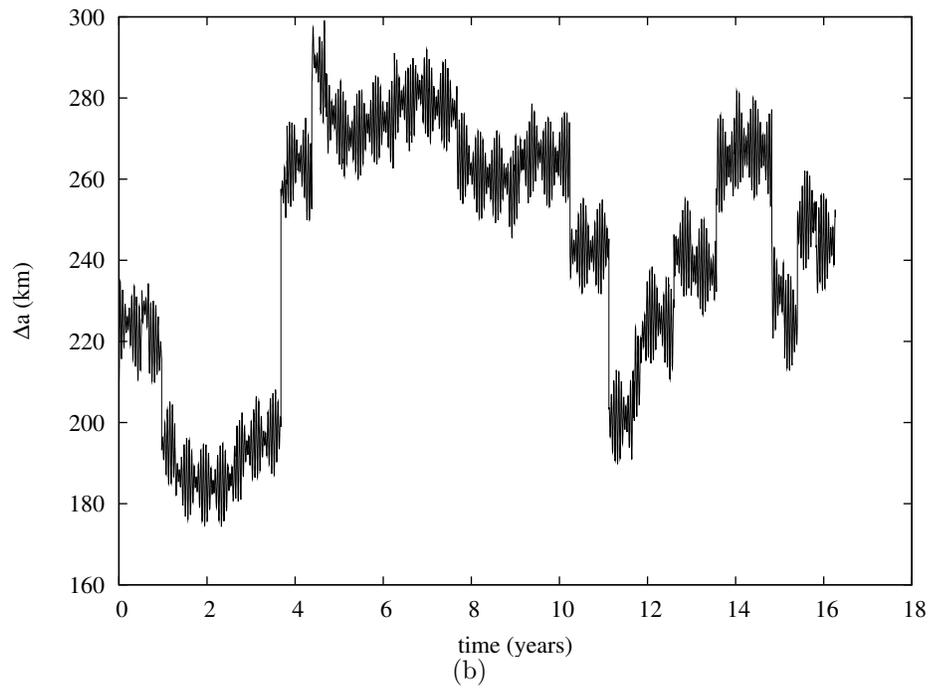

(b)

Figura 3.7: Evolução do semi-eixo maior de partículas de (a) 5 $\mu$m e (b) 3 $\mu$m sob influência da força de radiação solar, da perturbação gravitacional dos satélites próximos e do achatamento de Saturno.

**Colisões**

Para todas as partículas analisadas foi registrado seu destino ao final das simulações numéricas. Esses resultados estão apresentados nas tabelas 3.2 (caso sem achatamento) e 3.3 (caso com achatamento), assim como o tempo médio no qual ocorre a colisão ($< t_c >$).

| Partícula | Colisão com | | | $< t_c >$ |
|---|---|---|---|---|
| ($\mu$m) | Prometeu | Pandora | Anel A | (anos) |
| 1 | 3 | 1 | 96 | 0,25 ($\pm$ 0,03) |
| 3 | 2 | 6 | 92 | 0,70 ($\pm$ 0,10) |
| 5 | 9 | 10 | 81 | 1,10 ($\pm$ 0,26) |
| 10 | 20 | 15 | 65 | 1,90 ($\pm$ 0,60) |

Tabela 3.2: Destino das partículas sob influência da radiação solar e da perturbação gravitacional dos satélites próximos. Os números indicam a quantidade de colisões com cada satélite e com o anel A. $t_c$ indica o tempo médio de colisão para cada tamanho de partícula, com o respectivo desvio padrão.

| Partícula | Colisão com | | | $< t_c >$ |
|---|---|---|---|---|
| ($\mu$m) | Prometeu | Pandora | Anel A | (anos) |
| 1 | 57 | 43 | 0 | 6,5 ($\pm$ 5,8) |
| 3 | 74 | 26 | 0 | 5,2 ($\pm$ 4,8) |
| 5 | 76 | 24 | 0 | 7,3 ($\pm$ 6,2) |
| 10 | 68 | 32 | 0 | 7,7 ($\pm$ 6,5) |

Tabela 3.3: Destino das partículas sob influência da radiação solar, da perturbação gravitacional dos satélites próximos e do achatamento do planeta.

Compararando os resultados apresentados nas tabelas 3.2 e 3.3 fica evidente o efeito do achatamento na evolução orbital e no destino das partículas. No caso sem achatamento a taxa de colisão com o anel A varia entre 65% e 100% de acordo com o tamanho da partícula e em todos os casos a colisão ocorre em menos de 2 anos. Isso se deve ao aumento da excentricidade causado pela pressão de radiação (figura 3.5b).

O achatamento de Saturno evita o aumento da excentricidade, que passa a ser da ordem de $10^{-2}$. Este valor está abaixo da excentricidade mínima que uma partícula do anel F deve ter para que o pericentro de sua órbita cruze a borda do anel A. Desta forma a partícula permanece confinada na região do anel F, e aumentando a probabilidade de colisão com algum dos satélites próximos (Prometeu e Pandora), assim como o tempo médio para que ocorra uma colisão.

O maior número de colisões ocorre com Prometeu ($> 70\%$), uma vez que este satélite é maior que Pandora e está mais próximo ao anel. A diferença notada para partículas de 1 $\mu$m, para as quais o número de colisões com Prometeu e com Pandora é mais próximo, explica-se pelo maior decaimento do semi-eixo maior destas partículas que são mais sensíveis aos efeitos do arrasto de Poynting-Robertson do que as partículas maiores.

As figuras 3.8 e 3.9 mostram o tempo de colisão ($t_c$) em função da posição inicial das partículas contada a partir do centro do anel F (localizado em $\Delta a = 0$). São apresentados os resultados obtidos para partículas de 1 $\mu$m, 3 $\mu$m, 5 $\mu$m e 10 $\mu$m nos casos em que é levado em conta o efeito do achatamento do planeta e no caso onde o termo $J_2$ é desconsiderado.

Em todos os casos apresentados nas figuras 3.8 e 3.9 fica evidente o aumento no tempo médio de vida das partículas com a inclusão do achatamento de Saturno, especialmente para partículas de 1 $\mu$m. É possível ver que o valor médio de $t_c$ não difere muito para os vários tamanhos de partículas analisados. Esta similaridade entre os tempos médios está também relacionada à inclusão do achatamento que faz com que a amplitude de variação da excentricidade seja praticamente a mesma tanto para as menores quanto para as maiores partículas (figura 3.6). A similaridade na variação da excentricidade e o fato dos encontros próximos causarem variações no semi-eixo maior faz com que não exista um destino preferencial para as partículas de acordo com sua posição inicial. Por este mesmo motivo não é possível notar nenhuma relação entre o valor de $t_c$ e a posição inicial da partícula.

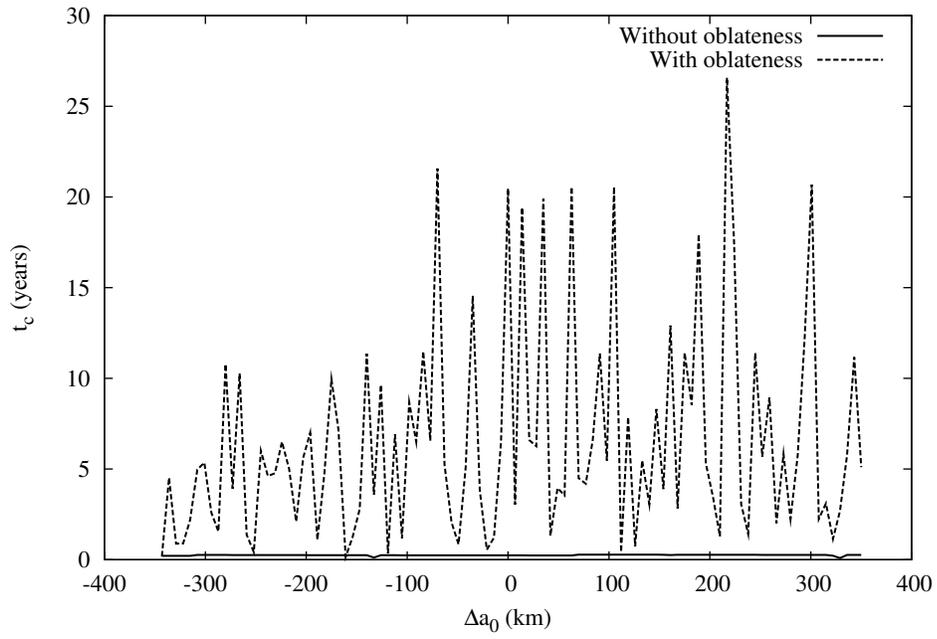

(a)

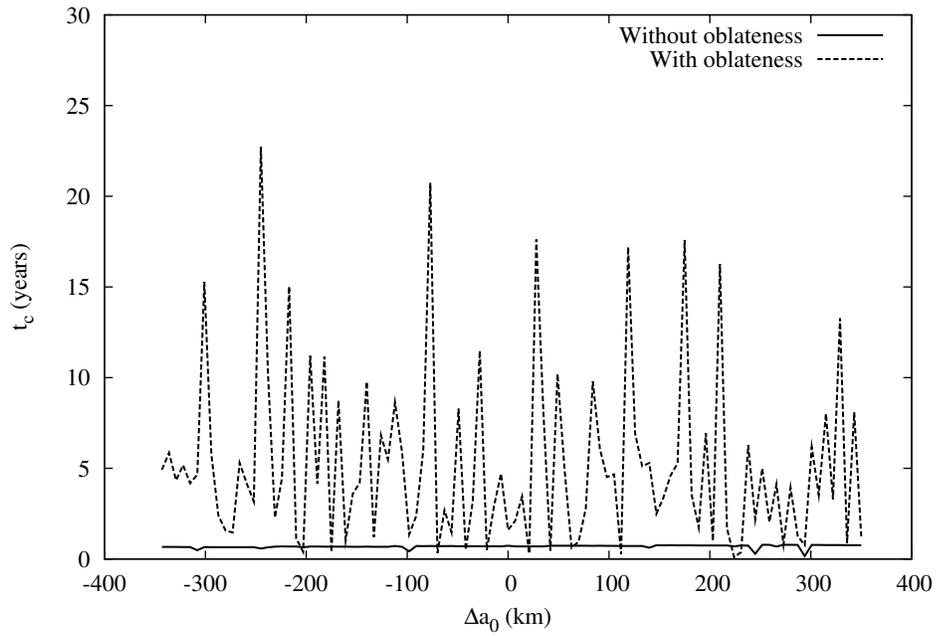

(b)

Figura 3.8: Tempo de colisão $t_c$ para partículas de (a) 1 $\mu$ e (b) 3 $\mu$m em função da posição inicial. A posição é contada em relação ao centro do anel localizado em 0. São apresentados os valores de $t_c$ para os casos com ( *With oblateness* ) e sem achatamento ( *Without oblateness* ).

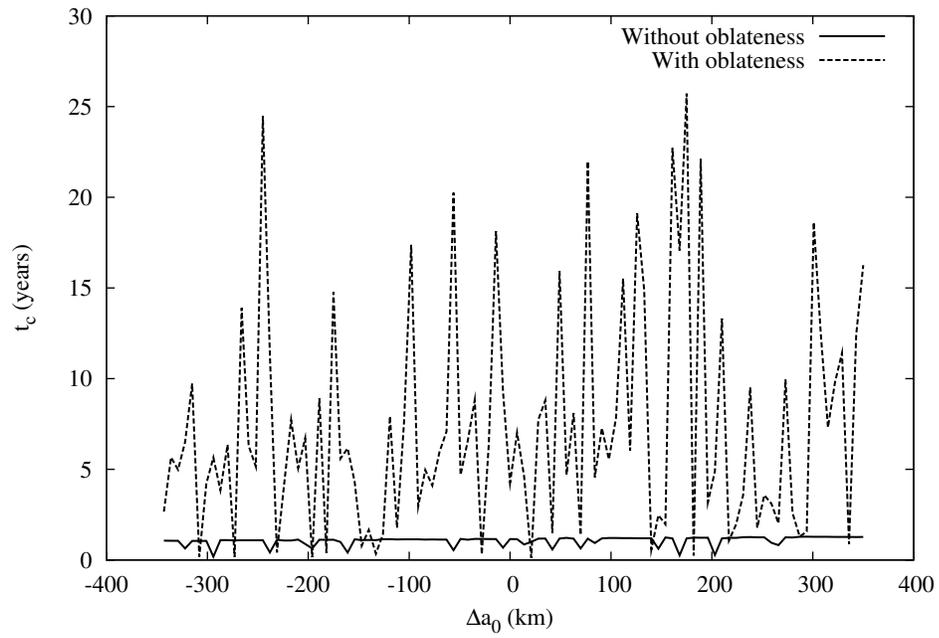

(a)

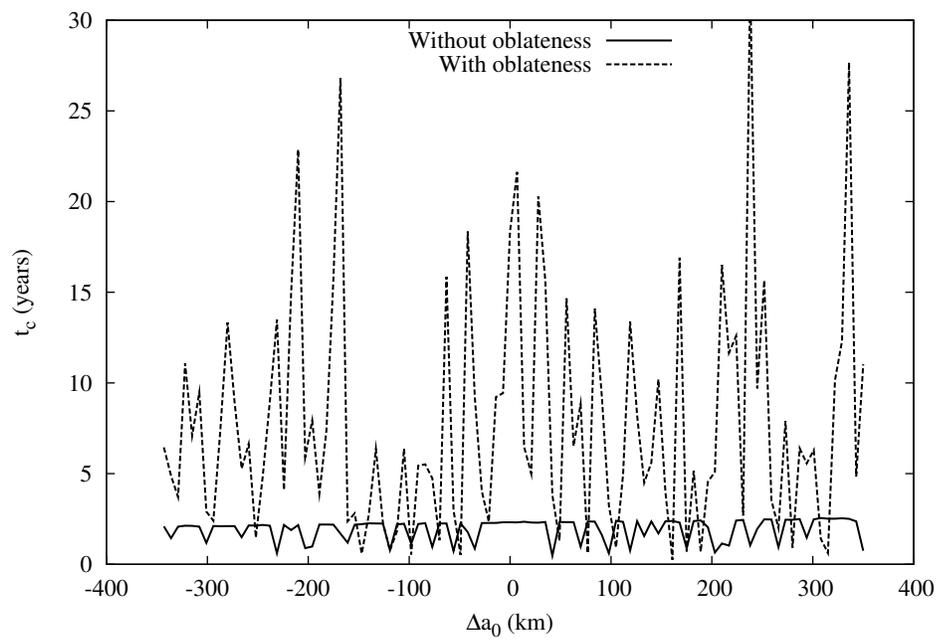

(b)

Figura 3.9: O mesmo caso apresentado na figura 3.8 para partículas de (a) 5 $\mu$ e (b) 10 $\mu$m.

### 3.2.4 Considerações finais

Apesar dos efeitos da força de radiação solar serem conhecidos na literatura, este estudo mostrou que o achatamento de Saturno é responsável pelo amortecimento na variação da excentricidade das partículas. Desta forma, as partículas de poeira do anel F permanecem na região compreendida entre as órbitas de Prometeu e Pandora, e acabam por colidir com algum dos satélites ao invés de colidir com o anel A.

O maior tempo de permanência na região do anel faz com que aumente a probabilidade da ocorrência de encontros próximos com os satélites. Estes encontros causam variações no semi-eixo maior das partículas, de forma que as partículas oscilam na região do anel e colidem com Prometeu ou Pandora em menos de uma década. Entretanto, neste período uma partícula cruza várias vezes o núcleo do anel e existe a possibilidade de que uma partícula de poeira, afetada pela força de radiação solar, colida com corpos maiores imersos no anel (Winter et al. 2007). Dependendo da configuração e da velocidade da colisão pode ocorrer a deposição de material na superfície deste corpo maior ou a ejeção de mais material para o anel (Burns et al. 2001). Mesmo a colisão das partículas com Prometeu ou Pandora pode ocorrer de forma a causar a ejeção de partículas em direção ao anel F.

Desta forma, mesmo sendo uma força dissipativa, a radiação solar não deve ser encarada apenas no sentido de reduzir o tempo de vida do anel. O processo de colisão poderá servir como fonte contínua e um mecanismo auxiliar na produção de material para o anel F.

Os resultados apresentados nesta seção fazem parte do artigo intitulado "*Dynamical Evolution of Saturn's F Ring Dust Particles*" publicado no volume 395 da revista "*Monthly Notices of the Royal Astronomical Society*". Uma cópia do artigo pode ser encontrada no apêndice C.

## 3.3  Análise fotométrica do anel F

Como mostrado na seção anterior a evolução do anel F está relacionada à uma combinação de forças que podem depender do tamanho das partículas. Desta forma, para melhor entender a evolução orbital é necessário conhecer a distribuição dos tamanhos das partículas.

Com este objetivo teve início durante o *COSPAR Capacity Building Workshop on Planetary Science* o projeto *"Photometry of the F-ring"*, cuja proposta é analisar o brilho no anel F através de imagens enviadas pela sonda Cassini e desenvolver um modelo fotométrico baseado na distribuição do tamanho das partículas que compõem o anel.

### 3.3.1  Obtenção dos dados, calibração e navegação

A primeira etapa do trabalho consistiu na busca e obtenção das imagens, o que foi realizado através da interface online OPUS (*Outer Planets Unified Search*) disponível no PDS Ring Node. Neste sistema é possível realizar a busca através de inúmeros parâmetros e devido à maior quantidade disponíveis de imagens optou-se por utilizar as duas câmeras do ISS (*Imaging Science Subsystem*): WAC (*Wide Angle Camera*) e NAC (*Narrow Angle Camera*), sendo que para as duas câmeras foram buscadas imagens obtidas com o filtro *clear* para simplificar a análise.

Uma vez escolhidos a câmera e o filtro, o próximo passo é selecionar opções para limitar apenas às imagens onde é possível ver o anel F. Para as imagens cujo cabeçalho já havia sido atualizado utilizou-se um parâmetro adicional para restringir o alvo de interesse (FRING) e no caso de observações dedicadas exclusivamente ao anel F a busca foi reduzida à época da obtenção das imagens.

Com estes critérios foram encontradas aproximadamente 2500 imagens (até o início de 2009) e a figura 3.10 mostra um exemplo das imagens obtidas. Porém nem todas as imagens são apropriadas para a análise, pois várias delas apresentam problemas como saturação ou apontamento incorreto. Por esta razão foi necessário inspecionar as imagens uma a uma, eliminando aquelas que apresentavam defeitos. Como esta etapa foi realizada manualmente aproveitou-se para classificar as imagens de acordo com a posição do anel F, facilitando a determinação da região sobre a qual será feita a integração (seção 3.3.2).

Após a seleção das imagens foi realizada a redução dos dados através da ferramenta CISSCAL (*Cassini Imaging Science Subsystem Calibration*), um aplicativo próprio para este fim escrito em IDL, e do volume COISS_0011 que contém todas as imagens de calibração. Ao final do processo obteve-se imagens com as contagens de cada pixel convertidas em valores de I/F.

Teve início então o processo de navegação com o intuito de determinar a resolução de cada imagem, o ângulo de fase (formado entre o Sol, o alvo e a câmera) e o ângulo de abertura do anel no instante de observação. Todos os cálculos geométricos foram realizados através das rotinas do sistema SPICE, que armazena os dados de telemetria da sonda e informações sobre os seus intrumentos, além das efemérides dos planetas e satélites, na forma de kernels. Com o uso das ferramentas do SPICE e do conhecimento da geometria da observação foi desenvolvido um código numérico que permitiu determinar a posição e orientação da Cassini durante a obtenção de cada imagem e consequentemente os parâmetros necessários.

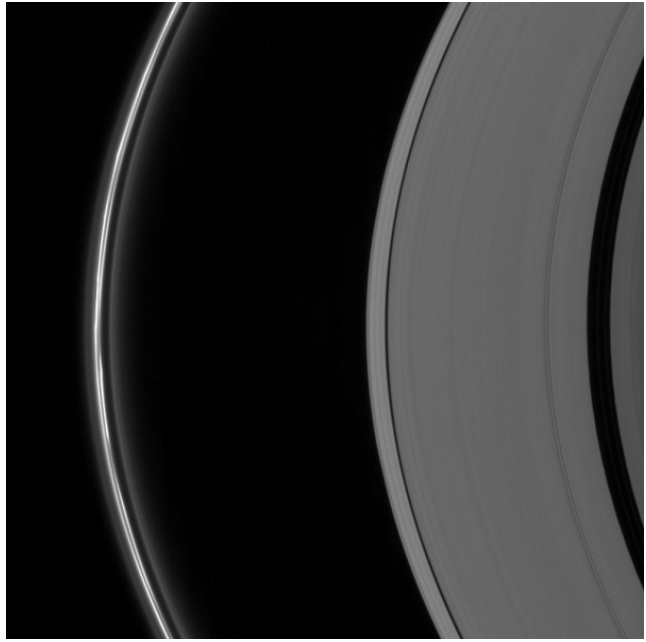

Figura 3.10: Imagem do anel F obtida pela câmera NAC da sonda Cassini utilizando o filtro *clear* sem nenhum tipo de tratamento.

### 3.3.2 Largura equivalente

Para uniformizar os resultados e facilitar a análise optou-se pelo cálculo da largura equivalente normal $W$[1], uma medida do total de luz refletida pelo anel para um observador situado perpendicularmente ao anel. $W$ pode ser definido como (Burns et al. 2001)

$$W = \sin \beta \int I(a)/F \mathrm{d}a \qquad (3.4)$$

em que I é a intensidade da luz refletida (potência por área por intervalo de comprimento de onda por esferorradiano), $\pi F$ é a densidade do fluxo solar incidente (potência por área por intervalo de comprimento de onda), $a$ é a distância ao centro do planeta medida no plano do anel e $\beta$ representa o ângulo de abertura do anel. $W$ apresenta como características a independência da resolução de cada imagem e de $\beta$ para pequenos valores de profundidade óptica, sendo assim um parâmetro confiável para realizar a comparação com dados de outras sondas.

---

[1]Apenas neste capítulo $W$ denota a largura equivalente normal. Nos demais é válida a definição dada pela equação 2.6.

O cálculo da largura equivalente é feito através da integração sob a curva obtida pela conversão de uma linha da imagem em um perfil radial de $I(a)/F$, tomando-se o cuidado de realizar a integração somente no intervalo de interesse, no caso o anel F. A figura 3.11 apresenta um perfil radial extraído de uma imagem obtida pela WAC, onde o anel F corresponde a estrutura proeminente localizada ao redor do pixel horizontal 450 e a região plana situada além do pixel 900 corresponde ao anel A que saturou a imagem.

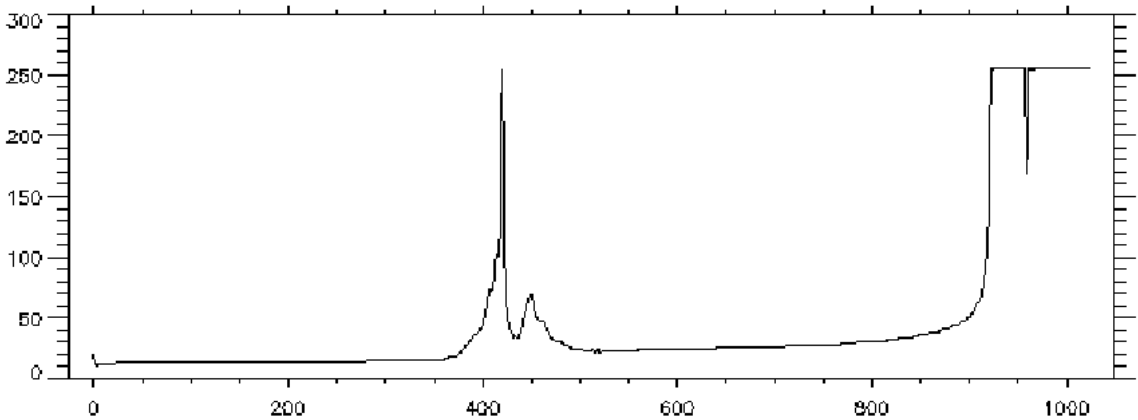

Figura 3.11: Exemplo de perfil radial obtido a partir da linha 512 de uma imagem do anel F. O gráfico representa a contagem em função da posição do pixel horizontal e a região integrada corresponde ao intervalo 300-550.

Como a posição do anel muda em cada imagem foi escrito um código em IDL para calcular automaticamente limites de integração. Este programa, aproveitando a classificação das imagens de acordo com a orientação, identifica a região do anel F através do cálculo das derivadas do perfil radial e assim determina os limites a serem utilizados.

### 3.3.3 Resultados

O conjunto total analisado compreende aproximadamente 1500 imagens obtidas tanto pela NAC quanto pela WAC, cobrindo vários ângulos de fase. A figura 3.12 mostra o valor da largura equivalente normal (denotada por NEW nas figuras) integrada sob a região do anel F em função do ângulo de fase.

A dispersão dos pontos obtidos para menores valores de $|\beta|$ indica que a profundidade óptica do anel não é pequena o suficiente para que seus efeitos sejam desprezados. Desta forma é necessário um ajuste mais elaborado que leve em conta o sombreamento mútuo das partículas. Com este novo modelo obtém-se a figura 3.13, na qual é possível ver que a dependência sistemática de $\beta$ foi eliminada. O melhor ajuste é obtido considerando uma profundidade óptica $\tau = 0.1$, que é assumido como o valor médio para o anel F.

O ajuste polinomial obtido para comportamento da curva de fase mostra que o anel é mais brilhante para grandes ângulos de fase, indicando a predominância de partículas micrométricas. Este resultado é compatível com uma distribuição da forma

$$n(a)\ a^{-p} \tag{3.5}$$

sendo $a$ o raio da partícula e $p = 3.5 - 4$.

Uma análise semelhante com imagens da sonda Voyager resulta em um valor de profundidade óptica $\tau = 0.02 - 0.04$ após a correção para $|\beta|$ (figura 3.14). A diferença de um fator 2-3 entre este valor e o determinado a partir das imagens da Cassini é sistemática e não é resultado de um efeito instrumental, como foi verificado através de comparações com imagens do anel A.

Em colaboração com M. M. Hedman e P. Nicholson, ambos pesquisadores da Cornell University, foram analisados perfis de ocultação obtidos através do fotopolarímetro da Voyager (PPS) e do instrumento VIMS da Cassini. A partir da medida da quantidade de luz bloqueada pelo anel durante uma ocultação estelar foi possível determinar a profundidade normal equivalente $N$. Apesar da possibilidade de um erro estatístico, o valor de $N$ derivado da única ocultação disponível para o PPS é menor que todos os 21 valores obtidos a partir do VIMS.

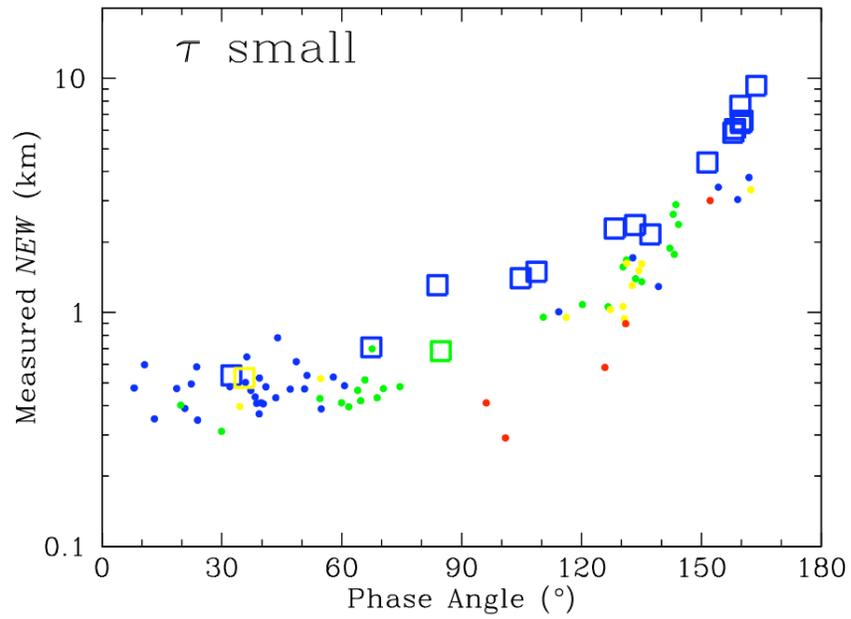

Figura 3.12: Largura equivalente (NEW) em função do ângulo de fase. Os quadrados são valores médios dos conjuntos da NAC e os pontos representam dados da WAC. As cores vermelho, amarelo, verde e azul indicam intervalos de 5° em $|\beta|$. Extraído de (Showalter et al. 2009).

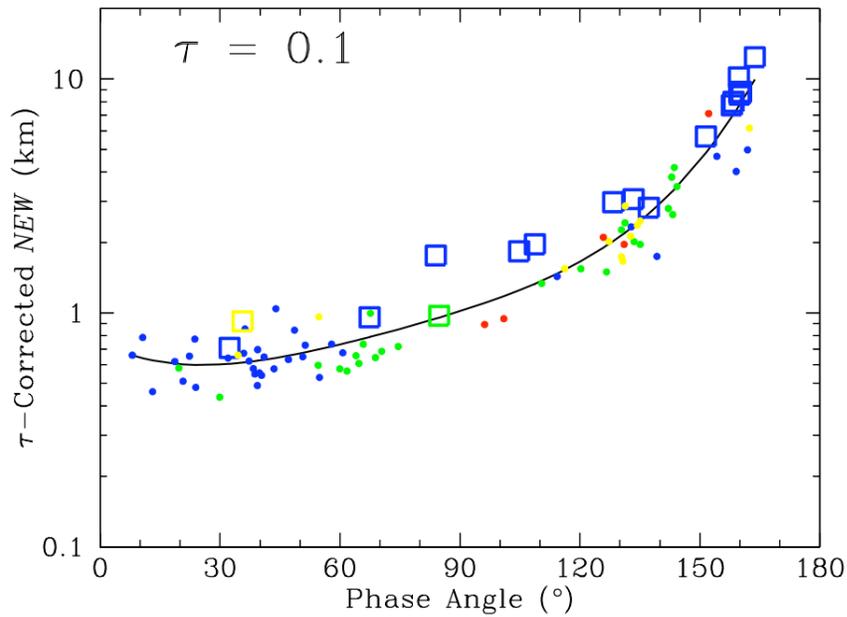

Figura 3.13: Largura equivalente (NEW) em função do ângulo de fase, levando em conta a correção da fotometria para $\tau = 0.1$. A curva mostra um ajuste polinomial e os quadrados são valores médios dos conjuntos da NAC, enquanto os pontos representam dados da WAC. As cores vermelho, amarelo, verde e azul indicam intervalos de 5° em $|\beta|$. Extraído de (Showalter et al. 2009).

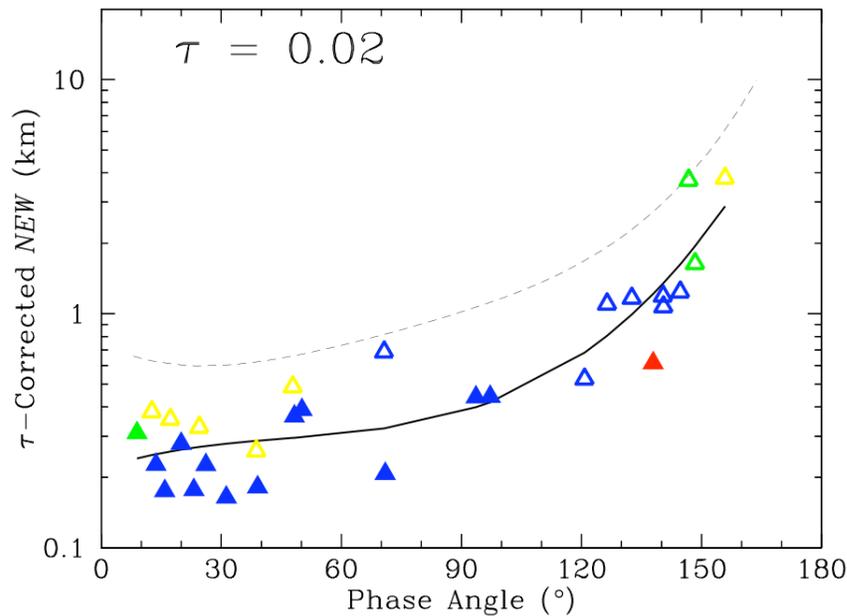

Figura 3.14: Comparação com os dados da Voyager 1 (triângulos abertos) e da Voyager 2 (triângulos preenchidos). As linhas indicam a função de fase ajustada para observações da Voyager (linha contínua) e da Cassini (linha tracejada). Extraído de (Showalter et al. 2009).

### 3.3.4 Considerações finais

A mudança da profundidade óptica do anel e o aumento do seu brilho indicam que a quantidade de poeira no anel sofreu um aumento desde a época da Voyager, contudo a distribuição do tamanho de partículas permaneceu praticamente o mesma.

A variação observada pode estar relacionada a proximidade anti-alinhamento das órbitas do anel e de Prometeu, o que altera a dinâmica e a interação entre as partículas do anel, o satélite e as moonlets (Winter et al. 2007).

Estes resultados foram apresentados em vários congressos, como por exemplo no *2009 European Planetary Science Congress*, cujo resumo pode ser encontrado no apêndice D. Eles também fazem parte de um artigo (autores: M. Showalter, R. S. French, R. Sfair, C. Aguelles, M. Pajuelo, P. Becerra, M. M. Hedman & P. D. Nicholson) que está em fase final de redação e será submetido para publicação em breve.

# Referências Bibliográficas

# Capítulo 4

# Análise de imagens dos anéis de Urano

## 4.1 Introdução

A maior parte das informações disponíveis sobre anéis planetários é proveniente de imagens obtidas através de telescópios (terrestres ou espaciais) e sondas interplanetárias. Ao contrário de outros conjuntos de dados que armazenam informações sobre um único ponto de um anel em um instante de tempo específico, uma imagem pode guardar um conjunto maior de informações como, por exemplo, valores de I/F (razão entre a quantidade de luz refletida pelo anel e o total de luz incidente) em função do raio orbital e longitude.

Tecnologias como a óptica adaptativa permitem que observações realizadas com telescópios terrestres resultem em imagens com qualidade comparável (por vezes superior) às obtidas com telescópios espaciais, porém com custo muito menor. Isto é particularmente útil para o caso de Urano, uma vez que o planeta foi visitado apenas pela sonda Voyager II em 1986 e desde então todas as observações tem sido realizadas com telescópios.

Além disso, com a utilização de diferentes filtros e com observações em diferentes geometrias e épocas, é possível cobrir um amplo espectro de frequências e resoluções, permitindo a caracterização física e orbital dos anéis, além de acompanhar a evolução temporal das estruturas presentes.

Todas estas características fazem com que a análise de imagens seja uma ferramenta fundamental para o estudo de anéis planetários e constituem um meio moderno de obtenção de dados.

Neste capítulo são apresentados os métodos utilizados e os resultados preliminares obtidos através da análise de imagens de Urano, seus satélites e anéis. Parte deste trabalho foi desenvolvida durante os períodos de estágio no Observatoire de Meudon.

## 4.2 Observação dos anéis de Urano

O sistema de anéis de Urano foi descoberto em 1977 através de uma ocultação estelar, que revelou a existência de quatro anéis estreitos e bem delimitados (Elliot et al. 1977). Posteriormente, entre os dados enviados após a passagem da sonda Voyager II por Urano em 1986 estavam a descoberta de mais seis anéis estreitos (Smith et al. 1986), que em conjunto com os quatro descobertos anteriormente formam o sistema principal de anéis de Urano.

Apesar de serem em sua maioria opticamente espessos, alguns destes anéis ($\eta$, $\delta$ e $\lambda$) possuem uma componente de baixa profundidade óptica indicando a presença de poeira. A exceção é o tênue anel $\zeta$, o mais extenso radialmente e próximo às camadas mais altas da atmosfera de Urano. Sua existência foi inferida a partir dos dados da Voyager (de fato, o anel é visível em apenas uma imagem) e observações recentes feitas com o telescópio Keck confirmaram a existência do anel que se estende gradualmente de 37850 km até 41350 km (de Pater et al. 2006).

A observação destas regiões de poeira é difícil pois elas são ofuscadas pelo brilho do planeta e dos anéis principais. Porém, devido à inclinação do planeta um observador na Terra observa o plano dos anéis de Urano em diferentes configurações durante o período orbital do planeta. A figura 4.2 mostra a variação das latitudes da Terra ($B$) e do Sol ($B'$) em relação a Urano no período de junho de 2007 a junho de 2008.

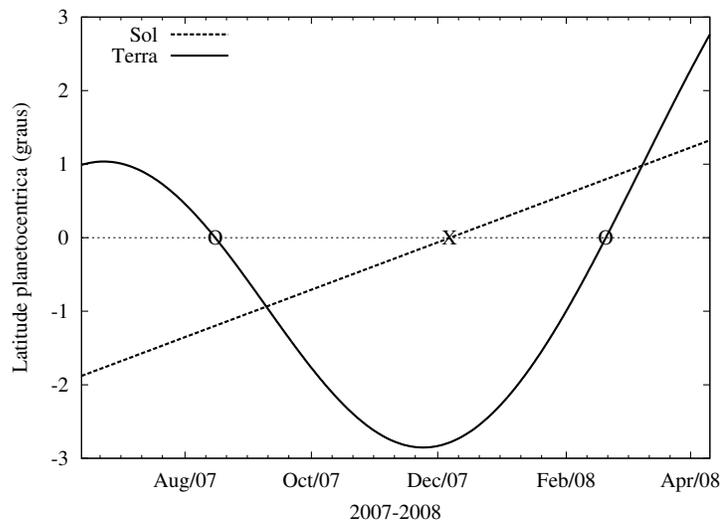

Figura 4.1: Variação das latitudes da Terra ($B$) e do Sol ($B'$) em relação a Urano

A partir dos ângulos das latitudes terrestre e solar é possível definir dois parâmetros $\mu$ e $\mu'$ como

$$\mu \equiv \sin B$$
$$\mu' \equiv \sin B'$$

e com isso a geometria da observação pode ser separada em duas configurações de acordo com a iluminação solar:

- quando o Sol e a Terra estão do mesmo lado do plano dos anéis ($\mu\mu' > 0$) o lado iluminado dos anéis é visível

- quando o Sol e a Terra estão em lados opostos do plano dos anéis ($\mu\mu' < 0$) o lado escuro dos anéis pode ser observado.

De forma geral, as passagens pelo plano dos anéis são denominadas de RPX ou *Ring plane crossing*. Como é possível ver na figura 4.2, nos anos de 2007 e 2008 ocorreram dois instantes (agosto de 2007 e fevereiro de 2008) nos quais tem-se que $\mu = 0$, correspondendo a passagem da Terra pelo plano dos anéis. Já em dezembro de 2007 tem-se que $\mu' = 0$, definido como o equinócio de Urano, quando o Sol cruza o equador do planeta.

O equinócio de Urano ocorre aproximadamente a cada 42 anos, metade do período orbital do planeta. Nesta configuração a incidência dos raios solares é paralela ao plano dos anéis e com isso o brilho dos anéis principais é bastante atenuado devido à auto ocultação das partículas, tornando propícia a observação das regiões de poeira. A figura 4.2 mostra o perfil radial dos anéis de Urano em julho de 2004, agosto de 2006 e logo após o RPX pela Terra em maio de 2007. Em 2004 a estrutura mais brilhante é o anel $\epsilon$, o anel mais denso do sistema e que consequentemente reflete a maior parte quantidade da luz incidente. Conforme se aproxima o RPX da Terra, a intensidade do anel diminui chegando a praticamente desaparecer no perfil obtido em maio de 2007. Neste mesmo perfil a região mais evidente é o anel $\zeta$, que não era visível anteriormente. Isto é um indicativo que este anel é composto majoritariamente por partículas de poeira.

de Pater *et al.* (2007) realizaram a comparação entre os perfis radiais obtidos pela Voyager em 1986 (French et al. 1991) com os dados referentes ao RPX da Terra em 2007. Algumas estruturas podem ser identificadas nos dois conjuntos de dados, como por exemplo o anel $\eta$ e uma região de poeira localizada em 43.000 km. Porém, em alguns casos há diferenças entre os dois perfis, como pode ser visto na região em 45.000 km que é bastante evidente em 2007 mas parece ser completamente desprovida de poeira nos dados da Voyager, como pode ser visto na figura 4.3. Isso indica que houve uma alteração na distribuição das partículas de poeira ao redor de Urano.

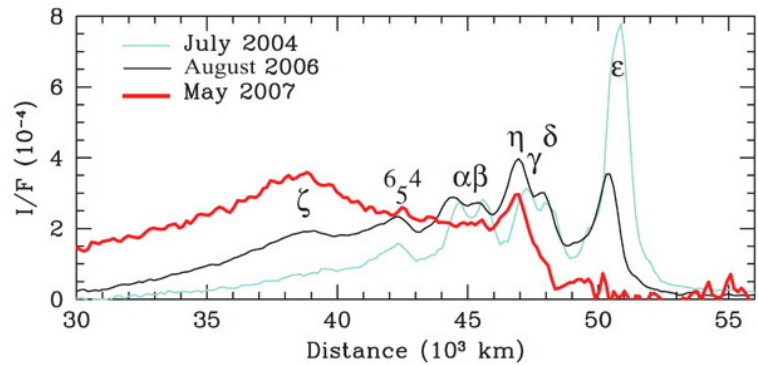

Figura 4.2: Perfil radial dos anéis de Urano obtidos com o telescópio Keck em diferentes épocas próximas ao RPX da Terra em maio de 2007. Adaptado de de Pater et al. (2007).

Outra diferença bastante evidente é a defasagem do anel ζ entre a posição determinada pela Voyager (linha azul) e os dados recentes (linha vermelha) (figura 4.3). Como as observações foram feitas para diferentes ângulos de fase não é possível fazer nenhuma inferência sobre a distribuição de tamanho das partículas e com isso foram propostas duas possíveis explicações para a defasagem encontrada: ou o anel é composto por duas populações de partículas com tamanhos diferentes e espacialmente separadas, algo difícil de ser explicado, ou ocorreu uma mudança na distribuição radial do anel entre 1986 e 2007, o que é mais provável.

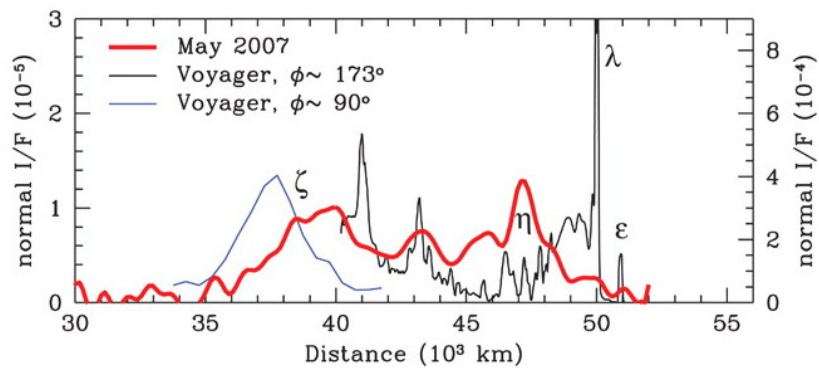

Figura 4.3: Comparação entre os perfis radiais após a deconvolução dos dados da Voyager (em preto, com escala no lado direito) e do telescópio Keck (em vermelho, com escala no lado esquerdo). Também é apresentado o perfil do anel ζ obtido na imagem 26846.50 pela Voyager (em azul). Adaptado de de Pater et al. (2007).

Aproveitando o RPX ocorrido no final do ano de 2007, o Prof. Bruno Sicardy do Obser-

vatoire de Paris organizou uma missão na qual foram realizadas observações de Urano e seus
anéis utilizando o VLT (*Very Large Telescope*) nos dias 07, 08, 09 e 10 de dezembro, obtendo
imagens tanto do lado iluminado quando do lado escuro dos anéis. Neste período a latitude
do Sol variou entre $0.015°$ e $-0.027°$, enquanto a latitude da Terra manteve-se praticamente
constante em $2.77°$. A figura 4.4 mostra um diagrama do planeta para um observador situado
no VLT no dia 07/12/07 às 23h50h0s UTC, sendo praticamente idêntico para os demais dias
(exceto pela posição dos satélites).

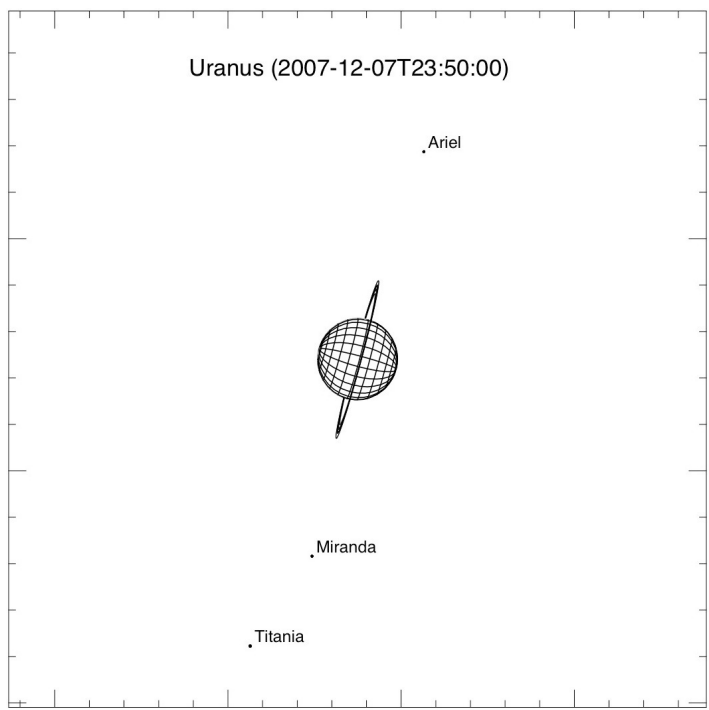

Figura 4.4: Diagrama do sistema de Urano para o dia 07/12/2009, onde estão identificados
os satélites visíveis em um campo com 30 arcsec.

Nas seções seguintes serão descritas as principais características do telescópio (seção 4.3),
do processo de obtenção e características das imagens (seção 4.4), além da calibração (seção 4.5),
centragem (seção4.7) e os resultados obtidos.

## 4.3   VTL

O VLT situa-se em Cerro Paranal, Chile, e é operado pelo *European Southern Obser-
vatory* (ESO). O Observatório conta com quatro telescópios de 8.2 m (UTs) e mais quatro
telescópios móveis auxiliares de 1.8 m, que podem fazer observações na região do espectro
visível e também no infravermelho próximo e médio ($0.3 - 20$ $\mu$m). Os telescópios podem ser operados individualmente, o que ocorre na maior parte do tempo, ou em conjunto para formar um interferômetro. As observações de Urano durante o RPX foram realizadas com o telescópio UT4 (Yepun) com o instrumento NACO (NAOS-CONICA), um sistema para óptica adaptativa no infravermelho.

Quando a luz de um objeto atravessa a atmosfera da Terra ela cruza regiões com turbulência, causando uma perturbação nas frentes de onda que resulta na distorção da imagem. Para corrigir isso, o sistema NAOS (*Nasmyth Adaptive Optics System*) de óptica adaptativa (OA) possui um sensor que analisa em uma escala de microssegundos as frentes de onda distorcidas e com o auxílio de 185 atenuadores introduz pequenas deformações em um espelho para compensar a perturbação da atmosfera (E. Pompei 2009).

Para que possa ser aplicada a técnica de OA é necessário um objeto de referência para comparação, geralmente uma estrela brilhante em um campo próximo ao observado. Caso não seja possível encontrar uma estrela para referência existe a possibilidade de utilizar um laser para criar uma estrela fictícia. A eficiência da óptica adaptativa depende do brilho e morfologia deste objeto de referência, de fatores ambientais como a velocidade da turbulência atmosférica, da massa de ar que a luz atravessa e fatores instrumentais.

Acoplado ao sistema NAOS está a câmera CONICA, que conta com um detector CCD (*charged-coupled device*) de $InSb$ com $1026 \times 1024$ pixels, dos quais apenas $1024 \times 1024$ são utilizáveis. Cada pixel do CCD mede 27 $\mu$m e é sensível a comprimentos de onda no intervalo $0.8 - 5.5$ $\mu$m, que corresponde ao infravermelho próximo.

A CONICA pode operar em 5 diferentes modos, com diferentes valores de escala de placa, campo de visão e comprimentos de onda. As observações de Urano foram realizadas no modo S27, com escala de 27.15 mas/pixel, campo de visão de $28 \times 28$ segundos de arco e sensível ao espectro na região de 1.0 $\mu$m-2.5 $\mu$m. A escala de placa será discutido em mais detalhes na seção 4.6.

## 4.4 Características das imagens

### Quantidade, seleção, tamanho e rotação

As observações foram realizadas nos dias 07, 08, 09 e 10 de dezembro de 2007, por simplicidade denotados como N07, N08, N09 e N10, respectivamente. A figura 4.5 mostra a secção de uma imagem obtida em cada noite, com a identificação dos satélites e da orientação das imagens.

Nem todas as imagens obtidas são úteis para a análise pois algumas apresentam problemas principalmente devido à imprecisão durante as correções introduzidas pela óptica adaptativa. Isso ocorre principalmente quando um satélite está próximo ou sobre o disco do planeta, resultando em uma imagem aparentemente "fora de foco".

Desta forma, antes de prosseguir a análise é necessário fazer uma seleção das imagens apropriadas para cada noite, descartando aquelas que apresentam problemas. Um resumo da quantidade de dados obtidos é apresentado na tabela 4.1, que contém o número total de imagens obtidas em cada noite, a quantidade de imagens adequadas e os satélites regulares de Urano visíveis em pelo menos uma parte do conjunto de cada noite.

| Noite | Total de imagens | Imagens adequadas | Satélites visíveis |
|-------|------------------|-------------------|--------------------|
| N07   | 99               | 83                | A, M, T, U         |
| N08   | 72               | 59                | A, M, U            |
| N09   | 136              | 97                | A, M               |
| N10   | 144              | 102               | A, M, T, U         |

Tabela 4.1: Quantidade de imagens obtidas e de imagens aproveitáveis em cada noite, além dos satélites regulares visíveis no campo: (A)riel, (M)iranda, (T)itânia e (U)mbriel.

O conjunto de imagens foi obtido utilizando toda a área disponível do CCD, resultando em imagens com dimensão de $1024 \times 1024$ pixels.

Para facilitar a determinação dos perfis radiais dos anéis é conveniente que eles estejam posicionados vertical ou horizontalmente na imagem. Por este motivo as imagens foram rotacionadas de aproximadamente 15° (note os eixos N-E na figura 4.5) de modo que os anéis estejam orientados verticalmente com o norte apontando para cima. Na seção 4.6 a rotação da imagem será discutida em mais detalhes.

**Filtro**

As observações foram realizadas no infravermelho próximo (NIR[1]) com a utilização do filtro Ks, cujo comprimento de onda central é igual a 2.18 $\mu$m e a largura é de 0.35 $\mu$m. A transmitância deste filtro está apresentada na figura 4.6 e corresponde a uma janela de infravermelho da atmosfera, ou seja, uma região do espectro que consegue atravessar a atmosfera sem que haja absorção e re-emissão (Lord 1992).

A escolha deste filtro deu-se pelo fato de que na banda Ks a luz do Sol é consideravelmente absorvida pelo metano e hidrogênio presentes na atmosfera de Urano (Fink & Larson 1979, Karkoschka 1994). Com isso a quantidade de luz espalhada pelo planeta é bastante reduzida, tornando possível analisar as regiões mais internas dos anéis que geralmente são ofuscadas pelo brilho do planeta.

---

[1] *Near InfraRed*

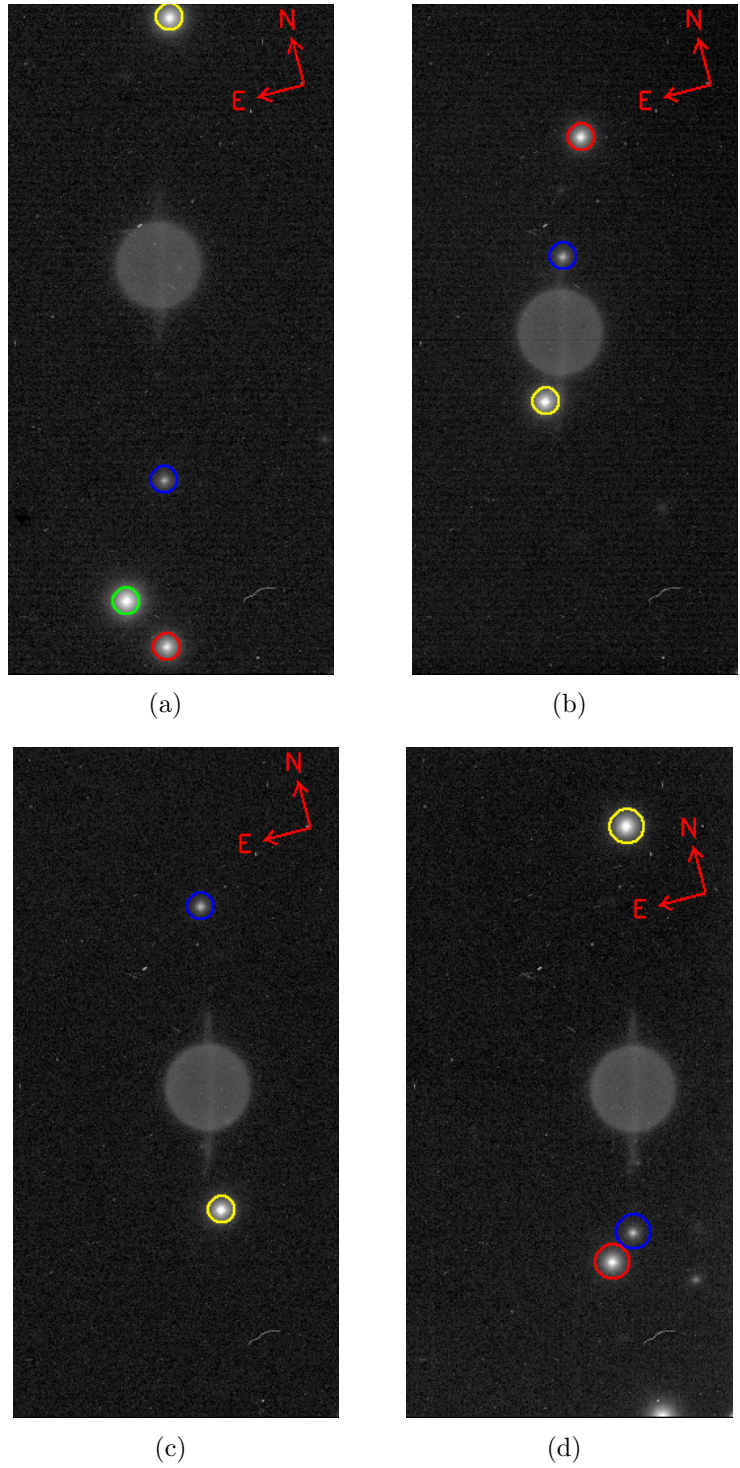

Figura 4.5: Secção de imagens obtidas em (a) N07, (b) N08, (c) N09 e (d) N10. Em cada caso estão identificados os satélites Ariel (amarelo), Miranda (azul), Titânia (verde) e Umbriel (vermelho) e a orientação Norte-Leste.

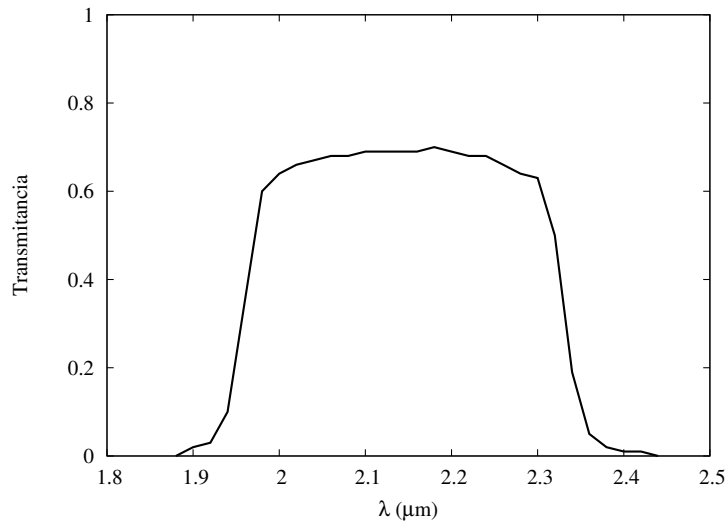

Figura 4.6: Curva de transmitância do filtro Ks utilizado nas observações.

**Tempo de exposição**

O tempo de exposição anunciado para todas as imagens de todas as noites é igual a 60 segundos. Este valor é suficientemente grande para obter uma boa razão sinal/ruído sem causar a saturação dos satélites.

Utilizando o valor presente no cabeçalho das imagens, a fotometria dos satélites e do planeta resulta em valores muito diferentes do esperado. Uma possível explicação é que o tempo de exposição seja diferente dos 60 segundos anunciados. Isto será discutido em mais detalhes na seção 4.10.

***Jitter***

Por razões técnicas a observação é feita em bloco de doze exposições divididas em ciclos de três imagens, sendo que após cada ciclo ocorre a alternância da posição do planeta no CCD. Essa alternância é conhecida *jitter* e é utilizada para efetuar a descarga do CCD. Uma consequência desta limpeza da carga é a formação de uma imagem negativa "fantasma", como pode ser visto na figura 4.7.

Devido ao *jittering* e ao fato de que todos os objetos de interesse concentram-se em apenas metade da área do CCD, as imagens foram cortadas pela metade no sentido da largura preservando somente a região onde se encontra o planeta. Com isso a dimensão das imagens foi reduzida de $1024 \times 1024$ pixels para $512 \times 1024$ pixels, diminuindo o espaço necessário para o armazenamento e aumentando a velocidade da análise computacional.

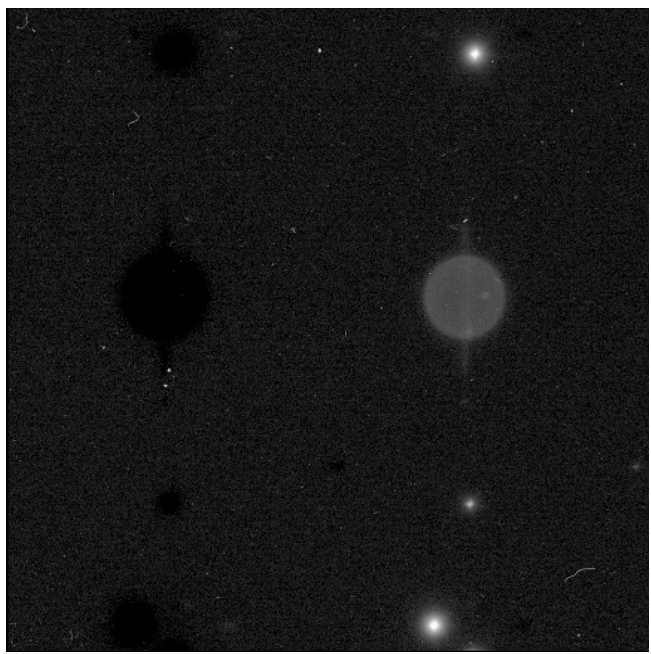

Figura 4.7: Imagem obtida em N07 na qual é possível ver a imagem do planeta no lado direito e o fantasma negativo criado devido ao *jitter* no lado esquerdo.

## 4.5 Calibração das imagens

As imagens obtidas pelo telescópio estão inicialmente em um estado bruto, ou seja, estão sujeitas a uma série de efeitos que alteram a forma como cada pixel do CCD registra os dados. Desta forma, para obter os dados corretos dos objetos astronômicos as imagens precisam ser tratadas e esses efeitos indesejados corrigidos.

Os fótons provenientes do objeto observado atingem os pixels do sensor e serão absorvidos, criando uma corrente elétrica devido ao efeito fotoelétrico e os diferentes valores da corrente para cada pixel serão traduzidos em diferentes intensidades na imagem resultante. Porém existem vários fenômenos eletrônicos e instrumentais que produzem pequenas correntes, causando um ruído na imagem.

Alguns destes efeitos que geram ruído na imagem e devem ser corrigidos são:

- *Dark current*: a temperatura do dispositivo pode criar um ruído aleatório devido à excitação dos elétrons do CCD, criando uma pequena corrente. Este efeito pode ser reduzido mantendo-se o CCD resfriado.

- Fundo de céu (*sky background* ou simplesmente *sky*): a contribuição do céu aumenta as contagens do CCD e varia de acordo com vários fatores, tais como as condições de observação e a presença de objetos brilhantes próximos. No caso das observações de Urano com o VLT o campo das imagens é consideravelmente pequeno, o que reduz a contaminação por outras fontes.

- Sensibilidade: a eficiência em transformar a energia do fóton incidente em corrente elétrica varia para cada pixel do CCD.

- *Bad pixels*: alguns pixels do CCD podem estar permanentemente danificados, resultando em contagens não confiáveis.

- Raios cósmicos: a incidência de raios cósmicos durante a exposição causa uma elevação na contagem, podendo resultar na saturação do pixel e regiões adjacentes.

A diferença da sensibilidade de cada pixel pode ser corrigida através das imagens de *flat-field* (ou simplesmente *flat*). As imagens de *flat* são obtidas apontando para uma região uniformemente iluminada, como por exemplo o céu durante o dia ou uma tela dentro da cúpula iluminada por uma fonte artificial. Dividindo-se a imagem de ciência pelo *flat* é compensado o efeito devido à eficiência de cada pixel.

Já o efeito devido a *dark current* pode ser compensado subtraindo da imagem de ciência uma outra imagem com mesmo tempo de exposição obtida com o obturador fechado (imagem de *dark*).

Para a determinação da variação do céu existem diferentes opções. Uma delas é a técnica de *microscanning*, que permite separar a contribuição do céu no sinal das imagens supondo que o ruído segue uma distribuição de Poisson (Devillard 1999). Outra opção é utilizar a metade da imagem que não contém o planeta devido ao *jittering* para criar uma imagem do céu com a mediana de cada série. Uma vez conhecida a contribuição do céu, ela pode ser subtraida das imagens de ciência.

As correções de *flat-field*, *dark current* e *sky* para as imagens de Urano foram realizadas por Benoit Carry seguindo o *pipeline* desenvolvido pela equipe do ESO para o VLT-NaCo (Devillard 1999, N. Ageorges 2007). As imagens de *flat* e *dark* são fornecidas pela equipe do ESO para cada noite de observação e a estimativa do céu foi obtida utilizando o *jittering* e a mediana das séries de imagens.

O processo de remoção de pixels defeituosos e raios cósmicos foi realizado posteriormente e compreende duas fases. Primeiro é necessário identificar os pixels que apresentam o problema, ou seja, aqueles pixels cuja contagem está acima de um dado limite calculado em função do valor do céu e do desvio padrão. Estes pixels com defeito irão formar uma máscara com a mesma dimensão da imagem.

Conhecendo os pixels com defeito, eles precisam ser substituídos por valores apropriados. Para isso, a contagem de cada pixel da mascára é substituído pela mediana dos pixels vizinhos, suavizando a imagem e eliminando as regiões defeituosas ou atingidas por raios cósmicos.

A tentativa inicial para determinar os pixels foi criar uma máscara aproveitando as metades das imagens que não contém o planeta devido ao *jittering*. Porém esse processo pode ser otimizado com a utilização de um filtro mediano implementado pela função `sigma_filter`, que faz parte da *IDL Astronomy User's Library* (Landsman 1993). Com isso a detecção e remoção de pixels problemáticos são feitas de forma automática para cada imagem, dispensando a necessidade de criar uma máscara. A sintaxe utilizada para a rotina `sigma_filter` aplicada sobre um vetor *array* que contém a imagem é

IDL > sigma_filter ( array , box , N_sigma=n , / iterate )

Esta função calcula a mediana dos pixels de uma região móvel com tamanho *box* centrada em cada pixel da imagem, excluindo o valor deste pixel central. Se a contagem do pixel central foi maior que $n$ vezes o desvio padrão da mediana calculada para a região, o valor é substituído pela mediana. O parâmetro */iterate* faz com que a rotina seja aplicada de forma recursiva até que não aja mais alterações no valor dos pixels.

Os parâmetros utilizados correspondem a uma região com 10 pixels de largura e limite de corte como sendo 4 vezes o desvio padrão da mediana. Estes valores precisam ser determinados com cuidado, pois se for escolhida uma região muito grande alguns satélites pequenos com poucos pixels de largura (e.g. Miranda) podem ser removidos da imagem, enquanto uma região pequena é ineficiente para remover os *bad pixels*. Algo semelhante ocorre com $N\_sigma$, pois um valor alto impede a remoção dos raios cósmicos, enquanto valores menores causam a suavização da imagem em regiões indesejadas, como no centro dos satélites por exemplo.

A figura 4.8 ilustra os resultados obtidos com o uso da função `sigma_filter`. A imagem inicial (a) mostra uma região com $150 \times 350$ onde é possível observar os satélites Ariel e Umbriel. Esta imagem foi corrigida para compensar o *flat-field*, *dark-current* e o fundo de céu, porém é possível identificar vários pontos brancos (raios cósmicos) e uma região contínua de pixels defeituosos na região superior esquerda da imagem. Após a aplicação do filtro mediano praticamente todos os raios cósmicos e *bad pixels* foram removidos, como pode ser visto em (b). A diferença entre a imagem original e a imagem após o filtro é mostrada em (c), indicando que a contagem dos satélites não foi alterada com a aplicação do filtro.

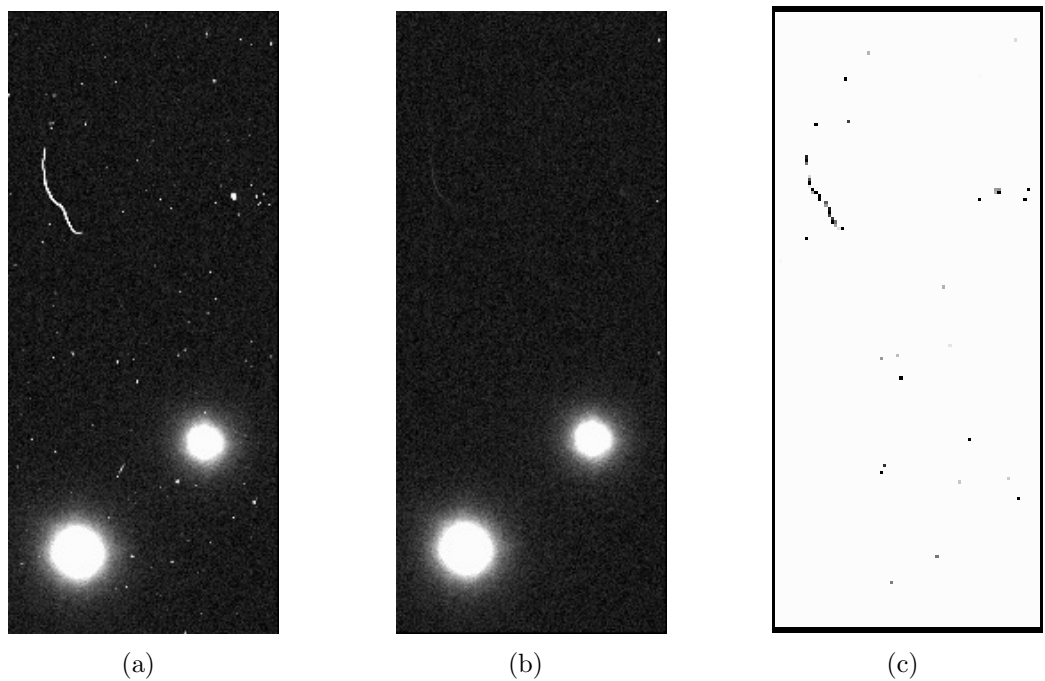

Figura 4.8: Resultado da aplicação do filtro mediano: (a) imagem original contendo *bad pixels* e raios cósmicos, (b) resultado após a aplicação do filtro mediano e (c) diferença entre as imagens anteriores mostrando os pixels que foram corrigidos.

## 4.6 Escala e rotação

Para determinar o perfil radial dos anéis é necessário conhecer a escala da imagem em mas/pixel e a rotação da imagem em relação ao norte, uma vez que as imagens foram rotacionadas para que o plano dos anéis ficasse disposto verticalmente. Estas informações também são necessárias para centralizar as imagens utilizando as efemérides dos satélites, como será mostrado na seção 4.7.3.

Apesar dos valores da escala e da rotação estarem presentes no cabeçalho das imagens, a precisão não é suficiente e não há uma estimativa do erro. Por esta razão o cálculo destes valores foi realizado através de um programa em FORTRAN escrito pelo Prof. Bruno Sicardy. Este programa efetua uma série de operações para ajustar um conjunto de $N$ pontos $(p_n, q_n)$ sobre um outro conjunto de $N$ pontos $(u_n, v_n)$. O ajuste é obtido através da minimização de uma função $S$ dada por

$$S = \sum_{n=1}^{N} |\alpha Z_n + \beta - W_n|^2 \qquad (4.1)$$

em que

$$Z = p + iq$$
$$W = u + iv \tag{4.2}$$
$$i \equiv \sqrt{-1}$$

As variáveis $\alpha$ e $\beta$ podem ser escritas como

$$\alpha = \rho e^{\varphi} \tag{4.3}$$
$$\beta = \rho_0 e^{\varphi_0} \tag{4.4}$$

sendo

$$\rho = \frac{\left| \sum_n ZW^* \right|}{\sum_n ZZ^*} \tag{4.5}$$

$$\varphi = -\operatorname{Im}\left( \ln \sum_n ZW^* \right) \tag{4.6}$$

em que $^*$ denota o complexo conjugado da variável. De maneira semelhante é possível escrever as variáveis $\rho_0$ e $\varphi_0$.

As operações realizadas para minimizar a função $S$, em ordem de execução, são:

1. Cálculo e alinhamento dos vetores dos baricentros das nuvens de pontos $Z = Z(p, q)$ e $W = W(u, v)$

2. Expansão e rotação de $Z$ para ajustar sobre $W$

3. Translação de $Z$ sobre $W$

4. Determinação da barra de erro

Após estas etapas, a escala $E$ e a rotação $R$ da imagem são dadas por

$$E = |\alpha| \tag{4.7}$$
$$R = |\ln(\alpha)| \tag{4.8}$$

e também são informadas pelo programa as barras de erro de $E$ e $R$.

Para obter a escala da imagem em unidades de mas/pixel e a rotação em graus é preciso informar as efemérides dos satélites em relação ao centro do planeta e suas posições na imagem. As efemérides são dadas nas componentes $x, y$ medidas em mas/pixel e irão formar os vetores $W$ no programa. Já as posições dos satélites são dadas em pares $(x, y)$ medidos em pixel e irão formar os vetores $Z$. Desta forma, somente será possível determinar a escala se existirem satélites visíveis e por esta razão o presente método não pode ser aplicado à todas as imagens dos conjuntos.

Os fatores que influenciam na precisão da escala e da rotação das imagens obtidas são:

- Quantidade de imagens e de satélites visíveis

- Determinação do centróide do satélite

- Precisão das efemérides

Quanto maior for o número de imagens e satélites visíveis, maior será a quantidade de pontos a serem ajustados e consequentemente maior será a precisão obtida. Desta forma deu-se preferência às imagens com pelo menos dois satélites visíveis, especialmente Ariel e Umbriel por serem mais brilhantes.

A posição dos satélites na imagem é dada através das coordenadas do centróide obtidas pela rotinas `gctrd` ou `ctrd` (Landsman 1993). A rotina `gctrd` faz o ajuste 2D de uma gaussiana, porém é bastante sensível à presença de bad pixels ou qualquer contaminação de fontes próximas. Por este motivo foi dada preferência para a função `ctrd`, que possui um método mais robusto baseado no algoritmo `FIND-DAOPHOT` para identificação de perturbações no brilho de imagens (i.e. estrelas). Contudo, deve-se evitar imagens nas quais os satélites estejam muito próximos, quando o brilho de um pode se sobrepor a luz refletida pelo outro, assim como satélites próximos ao planeta ou aos anéis.

A sintaxe básica da rotina `ctrd` aplicada em uma imagem armazenada no vetor $img$ é

IDL > ctrd , img , x , y , xcen , ycen , [ fwhm ]

em que $x, y$ são os valores em pixel da posição inicial aproximada do centro do satélite obtidos visualmente e $xcen, ycen$ são as saídas contendo as coordenadas do centróide, também em pixels. Optou-se por realizar a busca do centróide em uma região com aproximadamente 4 pixels ao redor de $x, y$, o que corresponde a $fwhm = 6$ (meia altura igual a 1.5 $\sigma$).

As efemérides dos satélite podem ser encontradas no sistema *Horizons*, mantido pelo JPL (*Jet Propulsion Laboratory*), e no sistema de efemérides de satélites naturais desenvolvido pelo IMCCE (*Institut de Mécanique Céleste et de Calcul des Ephémérides*). A preferência de um sobre o outro depende do modelo adotado para o cálculo e o período de atualização das efemérides. No caso dos satélites de Urano, os valores fornecidos pelo JPL e pelo IMCCE não diferem muito, mas devido ao fato de ter sido atualizado mais recentemente foi dada preferência para o *Horizons*.

Os valores das efemérides foram exportados em intervalos regulares de 60 segundos para todas as noites de observação, porém o tempo entre valores sucessivos pode não corresponder ao tempo registrado para cada imagem. Desta forma é necessário utilizar um programa em FORTRAN que faz a interpolação das efemérides para obter o valor em um instante $t = (t_0 + t_{exp})/2$, em que $t_0$ e $t_{exp}$ são o instante inicial de obtenção da imagem e o tempo de exposição nominais informados no cabeçalho do arquivo.

Deve-se apenas tomar cuidado com a orientação do sistema de eixos escolhido. Ao exportar as efemérides através do *Horizons* o eixo das abcissas (eixo vertical da imagem), o valor é positivo na direção Oeste, enquanto no programa (e também no IDL) é assumida como positiva a direção Leste.

Os valores obtidos para a escala e para a rotação, assim como a precisão, após o processo de seleção das imagens apropriadas, da obtenção da posição e das efemérides dos satélites, estão apresentados na tabela 4.2.

|      | Satélites   | N° imagens | Rotação (graus)    | Escala (mas/pixel) |
|------|-------------|------------|--------------------|--------------------|
| N07  | A, T, U, M  | 5          | $-14.72 \pm 0.02$  | $27.08 \pm 0.02$   |
| N08  | A, U        | 5          | $-14.70 \pm 0.04$  | $27.11 \pm 0.02$   |
| N09  | A, M        | 10         | $-14.74 \pm 0.05$  | $27.09 \pm 0.02$   |
| N10  | A, U, M     | 7          | $-14.70 \pm 0.01$  | $27.01 \pm 0.05$   |

Tabela 4.2: Valores obtidos para a escala e rotação em cada uma das noites. Estão apresentados o número de imagens e os satélites utilizados para cada noite: (A)riel, (T)itânia, (M)iranda e (U)mbriel.

Os valores anunciados para a escala e para a rotação são 27.15 mas/pixel e $-14.97°$, respectivamente. Já os os valores obtidos através de uma média ponderada dos dados apresentados na tabela 4.2 correspondem à **27.06 mas/pixel** e **$-14.72°$**. Estes valores estão de acordo com medidas anteriores (Sicardy 2009, Souami 2009) derivadas de outras observações com o mesmo telescópio e instrumentos (câmera, filtro,...) e serão adotados como padrão em todos os cálculos neste trabalho.

## 4.7 Centragem

Para que estruturas tênues sejam visíveis é necessário aumentar a quantidade de luz recebida pelo telescópio para que a intensidade da imagem seja maior.

Uma forma de se obter este resultado é através do aumento do tempo de exposição da imagem. Porém o aumento no tempo de exposição eleva a quantidade de ruído na imagem devido à vários fatores como, por exemplo, variações na atmosfera e raios cósmicos.

Um alternativa para aumentar a razão sinal/ruído é obter várias imagens com menor tempo de exposição e então adicioná-las para produzir o resultado equivalente a uma única imagem com o tempo de exposição igual a soma dos tempos individuais.

Como o interesse é ressaltar os anéis, ao somar as imagens é necessário que o planeta permaneça sempre na mesma posição. Porém a posição de um objeto no céu varia com o passar do tempo e essa alteração pode não ser devidamente compensada pelo acompanhamento do telescópio.

Desta forma, antes de somar as imagens é preciso garantir que a posição do centro do planeta coincida em todas elas. Caso exista uma diferença é preciso determinar exatamente a defasagem entre as imagens e então move-las para uma origem em comum.

A técnica padrão para a determinação da diferença entre duas imagens é feita considerando um ponto fixo comum nas duas imagens, usualmente uma estrela. Porém o campo das imagens obtidas no VLT é pequeno e não há nenhuma estrela presente, então este procedimento não pode ser aplicado. Nas seções seguintes serão descritos alguns dos métodos alternativos que podem ser utilizados para calcular o deslocamento entre duas imagens obtidas.

### 4.7.1 Função de *cross-correlation*

Uma das técnicas que pode ser utilizada para determinar o deslocamento entre duas imagens é a função de *cross-correlation*, que oferece uma estimativa do grau de correlação entre dois vetores.

Se considerarmos dois vetores unidimensionais $x_i$ e $y_i$, sendo $i = 0, 1, 2, \ldots, N$, o coeficiente de correlação $r$ é definido como (Hoel 1966)

$$r = \frac{\sum\limits_i \left(x_i - \bar{x}\right)\left(y_i - \bar{y}\right)}{\sqrt{\sum\limits_i \left(x_i - \bar{x}\right)^2}\sqrt{\sum\limits_i \left(y_i - \bar{y}\right)^2}} \tag{4.9}$$

em que $\bar{x}$ e $\bar{y}$ são as médias dos vetores $x$ e $y$, respectivamente. Se os vetores $x$ e $y$ diferem apenas por uma translação, o valor de $r$ indica quanto o vetor $y$ deve ser deslocado para que os dois vetores fiquem idênticos.

Este conceito pode ser generalizado para vetores de duas dimensões, como é o caso das imagens analisadas. A função de *cross-correlation* para duas imagens $I_1$ e $I_2$ é definida como

o produto (Devillard 1999)

$$\sum_{p_1 \in w_1} \sum_{p_2 \in w_2} p_1 \otimes p_2 \qquad (4.10)$$

em que $p_1$ é o índice do pixel que percorre a região de interesse $w_1$ da imagem $I_1$ e $p_2$ é o índice do pixel que percorre a região de interesse $w_2$ da imagem $I_2$. O produto de *cross-correlation* indicado por $\otimes$ pode ser definido por várias funções como, por exemplo,

$$p_1 \otimes p_2 = \sum_{w_1, w_2} (p_1 - p_2)^2 \qquad (4.11)$$

ou

$$p_1 \otimes p_2 = \sum_{w_1, w_2} (p_1 p_2) \qquad (4.12)$$

ou ainda através de uma relação semelhante à equação 4.10.

Se a função a ser utilizada for a soma do quadrado da diferença entre as regiões de interesse das imagens (equação 4.11), deve-se buscar o mínimo da correlação entre $I_1$ e $I_2$ (figura 4.9). Por outro lado, se a função escolhida for a soma do produto dos valores dos pixels ou for da forma da equação (4.9), o valor que corresponde ao melhor ajuste entre as imagens é aquele que maximiza o produto $p_1 \otimes p_2$.

Para aplicar o método de *cross-correlation* foi escrito um programa que contém as rotinas `correl_optmize`, `correl_analyze` e `correl_images` (Landsman 1993). O método utilizado assume que não há alteração na escala das imagens, assim como não há rotação entre elas, ou seja, a diferença entre duas imagens deve-se somente a uma translação nas direções $x$ e $y$. Com isso, o programa para determinar o deslocamento entre duas imagens segue as seguintes etapas:

- A rotina `correl_optmize` recebe como parâmetros de entrada os vetores de duas imagens, sendo que a primeira ($I_1$) servirá como imagem de referência em relação a qual será dado o *offset* das demais imagens ($I_2$). Usualmente a imagem inicial de cada conjunto foi utilizada como referência.

- Um laço percorre todo o conjunto de imagens e armazena em $I_2$.

- A rotina `correl_optmize` faz o deslocamento de $I_2$ de forma iterativa nas direções $x$ e $y$. Para cada deslocamento $(\Delta x, \Delta y)$ a rotina `correl_images` calcula o produto de correlação entre $I_1$ e $I_2$ utilizando como função a generalização bidimensional da equação (4.9).

- A rotina `correl_analyze` determina o valor ótimo de $(\Delta x, \Delta y)$ que maximiza a função de *cross-correlation*.

A precisão padrão deste método é de 1 pixel em cada direção, porém é possível definir um parâmetro de magnificação para determinar a diferença entre as imagens com uma precisão de uma fração de pixel. A sintaxe completa utilizada para o método de *cross-correlation* é

IDL > c o r r e l _ o p t i m i z e , I1 , I2 , dx , dy , /mag=5

em que $dx$ e $dy$ são as variáveis que armazenam o deslocamento entre $I_1$ e $I_2$, com precisão de até 1/5 de pixel.

Para aumentar a confiabilidade do resultado limitou-se a região de interesse (ROI[2]) de cada imagem somente a um quadrado com 80 pixels de lado no qual todo o planeta está inscrito, como mostrado na figura 4.10. É importante notar que as coordenadas desta região devem permanecer fixas em relação à imagem e o planeta deve estar aproximadamente no centro para evitar problema com condições de contorno na borda da ROI.

Com isso o processo computacional torna-se mais eficiente e elimina-se a possibilidade de que a função de correlação tente ajustar duas imagens através da posição dos satélites ao invés do centro do planeta. Contudo, imagens nas quais há presença de *bright spots*, como nuvens ou satélites sobre o disco do planeta, devem ser rejeitadas.

Foram realizados dois testes para analisar a precisão deste método. Inicialmente foram criadas imagens fictícias contendo uma estrela (seguindo uma distribuição gaussiana de intensidade) e um ruído de fundo aleatório. Para este caso a precisão alcançada varia entre 1/25 e 1/4 pixel, dependendo da razão sinal/ruído. Para o outro teste foi escolhida uma imagem qualquer do conjunto e então foi criada uma cópia desta imagem deslocada nas direções $x$ e $y$ com auxílio da função `fshift`. Aplicando o método de *cross-correlation* descrito anteriormente é possível recuperar o deslocamento da imagem com uma precisão entre 1/10 e 1/2 pixel. Desta forma, assumiu-se como precisão máxima para o método o valor de 0.5 pixel.

A figura 4.11 mostra dois instantes do processo de *cross-correlation*: em (a) é mostrada uma etapa intermediária com a sobreposição de duas imagens da noite N07 deslocadas pela função `correl_optimize`, e em (b) está a diferença entre estas duas imagens para o melhor ajuste entre elas, que corresponde ao máximo da função de correlação com $\Delta x = 1.0$ e $\Delta y = 0.5$.

---

[2]*Region Of Interest*

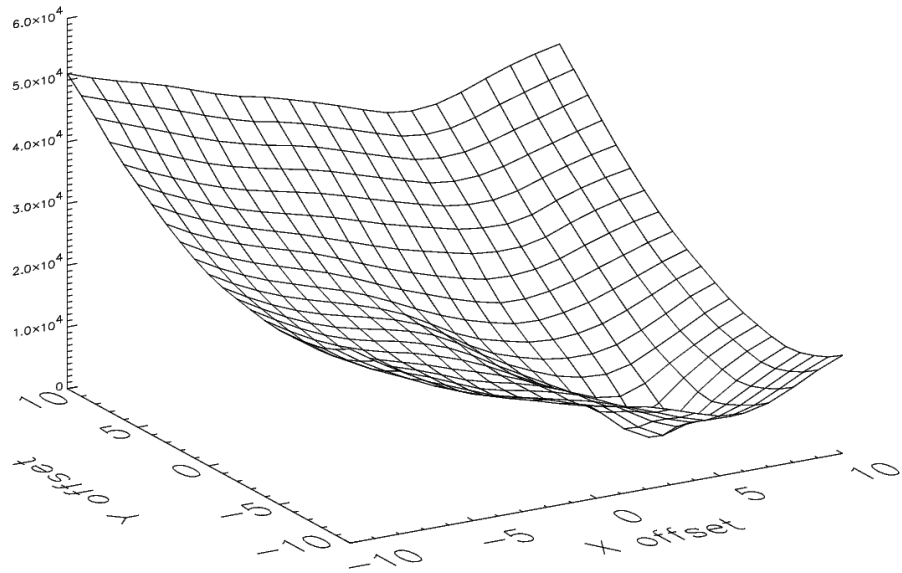

Figura 4.9: Esboço da função de correlação em função do deslocamento nas direções $X$ e $Y$. Neste exemplo busca-se o mínimo da função de *cross-correlation*, que ocorre em $(5, -5)$. Extraído de (Devillard 1999).

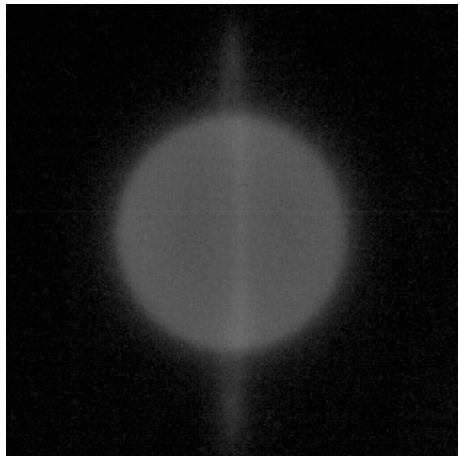

Figura 4.10: Exemplo de ROI limitando uma região ao redor do planeta onde será aplicado o método de *cross-correlation*.

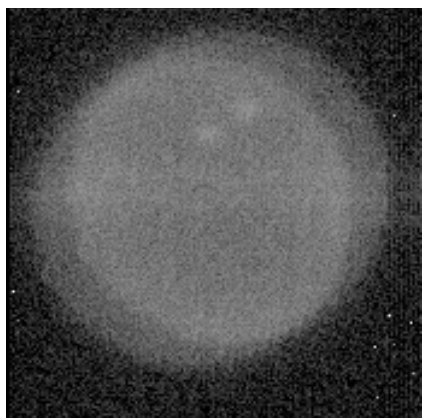 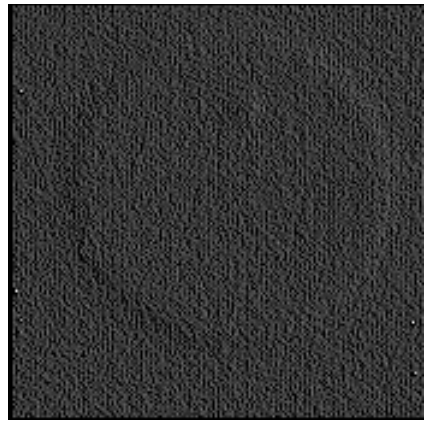

(a)                 (b)

Figura 4.11: Etapas do processo de *cross-correlation*: (a) duas imagens sobrepostas durante deslocamento na busca do máximo da funçao de correlação e (b) a diferença entre as duas imagens para o deslocamento que resulta no melhor ajuste.

### 4.7.2 Interspectrum

Um método bastante flexível e robusto para determinar o deslocamento entre duas imagens é o interspectrum, baseado em operações em um espaço de Fourier.

A transformada de Fourier $\mathcal{F}$ de uma função $f(x)$ é definida como (Arfken & Weber 2001)

$$\mathcal{F}[f(x)] \equiv \mathcal{F}(k) = \int_{-\infty}^{\infty} f(x)e^{-ikx}dx \tag{4.13}$$

enquanto a reconstrução da função $f(x)$ é obtida através da transformada inversa de Fourier $\mathcal{F}^{-1}$, definida como

$$f(x) = \mathcal{F}^{-1}[\mathcal{F}(k)] \equiv \frac{1}{2\pi}\int_{-\infty}^{\infty} \mathcal{F}(k)e^{ikx}dk \tag{4.14}$$

Partindo da definição da transformada de Fourier (equação 4.13) e fazendo um deslocamento $a$ no espaço real tem-se que

$$\begin{aligned}\mathcal{F}(k) = \quad & \mathcal{F}[f(x-a)] = \int_{-\infty}^{\infty} f(x-a)e^{-ik(x-a)}dx \\ = \quad & \mathcal{F}(x)e^{-ika} \end{aligned} \tag{4.15}$$

em que se fez a tranformação $(x-a) \to x$ para obter a equação (4.15). Desta forma, um deslocamento no espaço real representa uma alteração de fase no espaço de Fourier.

No caso de uma função discreta $x_n$ ($n = 0, 1, \dots, N-1$), a versão contínua da transformada de Fourier (equação 4.13) e a transformada inversa (equação 4.14) são substituídas pelas transformadas discreta de Fourier (DTF[3]) e sua correspondente inversa, cujas definições são (Arfken & Weber 2001)

$$X_k = \mathcal{F}(x_n) \equiv \sum_{n=0}^{N-1} x_n e^{-\frac{2\pi i}{N}kn}, \qquad k = 0, 1, \dots, N-1 \tag{4.16}$$

$$x_n = \mathcal{F}^{-1} \equiv \frac{1}{N}\sum_{k=0}^{N-1} X_k e^{-\frac{2\pi i}{N}kn}, \qquad n = 0, 1, \dots, N-1 \tag{4.17}$$

O cálculo da DFT é realizado através da função `fft` que faz parte da biblioteca padrão do IDL. Esta rotina nada mais é do que uma implementação de algoritmo que aplica transformada rápida de Fourier (FFT[4]).

Sabendo que um deslocamento de uma imagem é traduzido como uma mudança de fase no espaço de Fourier, o método do interspectrum propõe determinar a distância relativa entre duas imagens $I_1$ e $I_2$ através de uma medida da fase $\Phi$ do interspectrum entre elas. Esta fase é definida como (Gaulme et al. 2008)

$$\Phi = \arg \left[ \mathcal{F}(I_1) \times \mathcal{F}^{-1}(I_2) \right]$$
$$\propto (x_2 - x_1) + (y_2 - y_1) \tag{4.18}$$

onde $(x_1, y_1)$ e $(x_2, y_2)$ são as coordenadas do centro do planeta nas imagens $I_1$ e $I_2$, respectivamente.

---

[3]*Discrete Fourier Transform*
[4]*Fast Fourier Transform*

O programa utilizado para aplicar este método (`interspectrum6`) é uma versão adaptada e melhorada do código desenvolvido por alguns membros do LESIA. A rotina recebe como parâmetros de entrada dois vetores contendo as imagens (ou as ROIs) – *array*1 e *array*2 – que devem ter obrigatoriamente as mesmas dimensões, e um parâmetro $D$ que será utilizado durante o ajuste na determinação de $\Phi$. A sintaxe da função `interspectrum6` é

IDL > interspec6 , array1 , array2 , D, dx, dy onde $dx$ e $dy$ são as variáveis de armazenam o deslocamento entre as imagens.

Assim como no caso da função de *cross-correlation*, o método do interspectrum traz melhores resultados se for aplicado à uma região específica da imagem. Novamente optou-se por limitar a ROI em uma pequena área ao redor do planeta, porém a escolha do tamanho desta região deve ser feita com cuidado. O cálculo do interspectrum requer uma quantidade razoável de pontos para obter bons resultados, então deve-se evitar regiões muito pequenas. Por outro lado, regiões muito grandes resultam na elevação do ruído na imagem e consequentemente diminuem a precisão obtida, apesar da habilidade do método em tratar casos com baixa razão S/N. Após vários testes optou-se por manter a ROI como uma região quadrada com 80 pixels de lado e em uma posição fixa em relação a imagem.

Uma vez determinada a ROI que será utilizada, a rotina `interspectrum6` aplica um filtro *lowpass* para remoção de uma parte do ruído das imagens. Sabendo que o ruído é representado principalmente por altas frequências no espaço de Fourier (Gonzalez & Woods 1992), é calculada a FFT dos vetores *array*1 e *array*2 e sobre a transformada de cada ROI é ajustada uma curva gaussiana. Esse ajuste tem como consequência a atenuação do espectro eliminando as componentes de alta frequência e a reconstrução através da transformada inversa resulta em imagens suavizadas.

A etapa seguinte é o cálculo da transformada de Fourier das imagens reconstruídas e o interspectrum entre elas, para então determinar o plano de fase $\Phi$ da transformação. Para melhorar a precisão do método e reduzir o custo computacional, a determinação de $\Phi$ não é feita utilizando todos os pontos resultantes da transformação, mas apenas aqueles cujo módulo do intespectrum está acima de um parâmetro de corte ($D$) ao longo de uma função peso.

O resultado do método é bastante sensível a escolha da função peso e do parâmetro $D$. Foram realizados inúmeros testes fixando uma função peso e variando $D$ em um intervalo $[0.1, 5]$. Para as imagens de Urano os melhores resultados foram obtidos utilizando $\log_{10}$ como função peso e $D = 2$, parâmetros que foram mantidos para todas as imagens de todas as noites.

Uma vez conhecida a função peso e o plano do interspectrum é utilizada a rotina `mpfit2dfun` para obter as equações paramétricas deste plano, uma vez que o deslocamento entre as imagens é função da fase deste plano (equação 4.18). Se a equação do plano é da forma

$$\Phi = Ax + By + C \qquad (4.19)$$

os deslocamentos $\Delta x$ e $\Delta y$ entre as imagens são dados por

$$\Delta x = -\frac{SA}{2\pi} \qquad (4.20)$$

$$\Delta y = -\frac{SB}{2\pi} \qquad (4.21)$$

em que o fator $S$ está relacionado com o tamanho da ROI escolhida.

A precisão dos resultados obtidos através do interspectrum foi testada seguindo o mesmo procedimento aplicado ao método da função de *cross-correlation*. No primeiro teste foram criadas imagens fictícias com uma estrela (seguindo uma distribuição gaussiana de intensidade) e um ruído de fundo aleatório, variando a razão S/N no intervalo $[5, 30]$. Nestes casos foi possível determinar a diferença entre a posição das imagens com a excelente precisão de $10^{-2}$ pixel, mesmo para os menores valores de S/N. Este teste também mostrou que o método do interspectrum obtém melhores resultados se a separação inicial das imagens não for muito grande, tipicamente da ordem de 10 pixels ou menor.

Testes da aplicação do método para a ROI extraída de uma imagem real e uma cópia desta região deslocada com a função `fshift` resultam em uma precisão da ordem de 1/10 pixel ou menor, dependendo da imagem escolhida. Desta forma, apesar de conservador o valor de 0.1 pixel foi considerado o limite da precisão do interspectrum.

### 4.7.3 Efemérides

Outra técnica possível para determinar o centro do planeta (e consequentemente o deslocamento entre as imagens) é utilizar as efemérides dos satélites, uma vez que elas são dadas em relação ao centro de Urano.

O programa desenvolvido para determinar a escala e rotação das imagens também pode ser utilizado para determinar a posição do centro do planeta. Partindo das definições dos vetores complexos $Z$ e $W$ (equação 4.2), e da função de minimização $S$ (equação 4.1) pode-se escrever

$$\alpha Z + \beta - W = 0$$
$$\alpha\,(Z_p - Z_s) + \beta - (W_p - W_s) = 0 \tag{4.22}$$

em que índices $p$ e $s$ referem-se aos dados do planeta e do satélite, respectivamente. Como a efeméride de Urano em relação ao próprio o planeta é zero o vetor $W_p$ é nulo, de forma que o vetor com as coordenadas do centro do planeta $Z_p$ é obtido através da relação

$$Z_p = Z_s + \frac{W_s - \beta}{\alpha} \tag{4.23}$$

Com isso, sabendo-se as efemérides de um satélite, a escala e a rotação da imagem é possível determinar o centro do planeta. A precisão deste método está sujeita aos mesmos fatores que afetam a determinação da escala: a presença de satélites, a determinação dos seus centróides e da confiabilidade das efemérides. Levando estes fatores em conta, a precisão deste método é tipicamente 0.5 pixel.

## 4.8   Deslocamento e adição das imagens

A aplicação de qualquer um dos métodos anteriores (ou de uma combinação deles) para uma sequência de imagens de uma determinada noite de observação resulta em uma lista que contém o nome de cada arquivo e a diferença em pixels $\Delta x$ e $\Delta y$ na posição do planeta em relação à primeira imagem da sequência.

A escolha da técnica de centragem depende de vários fatores. A presença de nuvens ou satélites sobre o disco do planeta prejudica a utilização da função de correlação, pois neste caso o ajuste seria feito sobre estes pontos brilhantes ao invés do centro do planeta.

Algo semelhante ocorre com o método do interspectrum. Apesar de mais robusto e preciso, se existirem muitas regiões brilhantes que dificultem a determinação do contorno do planeta não é possível realizar o ajuste da gaussiana necessário. A figura 4.12 mostra a comparação entre as imagens de Netuno e Urano no filtro Ks obtidas no VLT em 2007 e as respectivas superfícies. Apesar do ruído, os limites da superfície de Urano podem ser identificados, o que não ocorre no caso de Netuno devido à presença de inúmeras nuvens que elevam a contagem dos pixels em determinadas regiões.

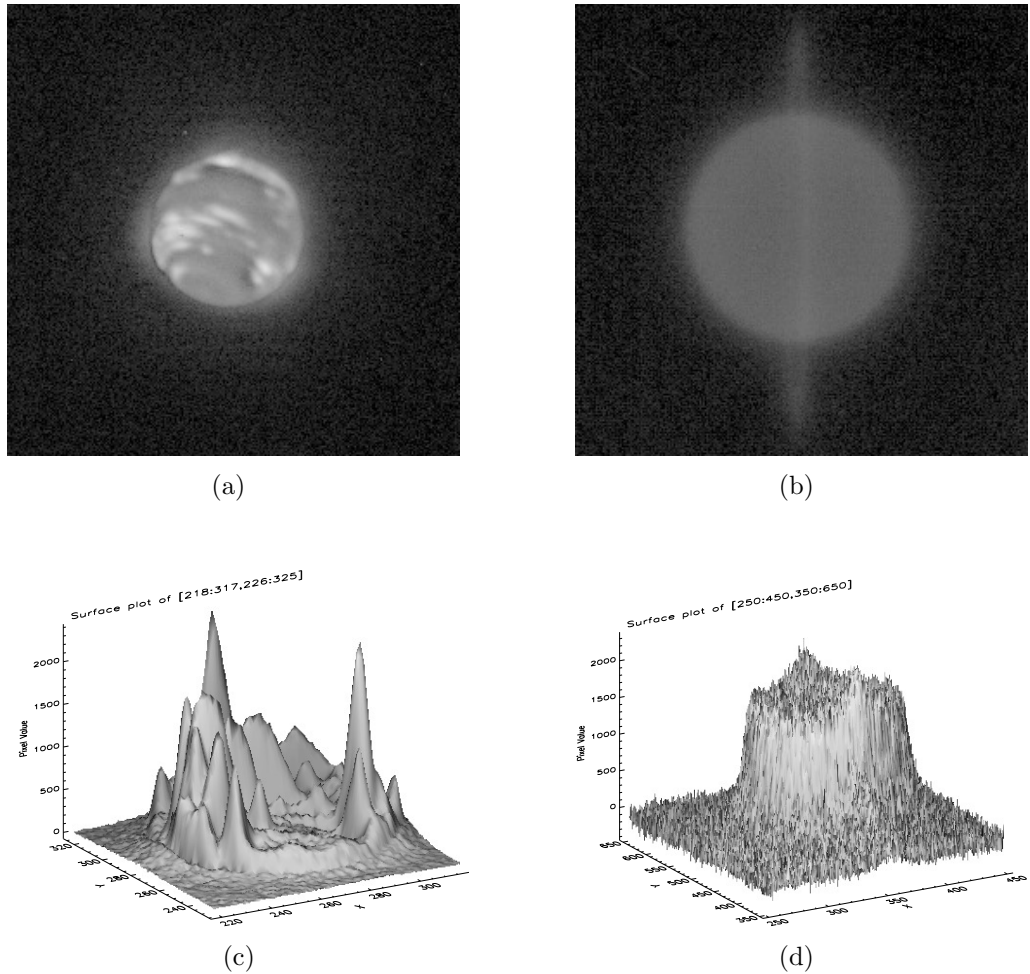

(a)                       (b)

(c)                       (d)

Figura 4.12: Comparação entre as imagens de (a) Netuno e (b) Urano obtidas em 2007 com o filtro Ks no VLT. Em (c) e (d) estão apresentados as superfícies de Netuno e Urano, respectivamente, onde fica evidente a presença das nuvens em Netuno.

O método das efemérides é particularmente útil nos casos em que há dificuldades na utilização da imagem do próprio planeta para realizar a centragem, como ocorre com a presença de muitas nuvens. A figura 4.13 mostra o resultado obtido por este método após a adição de 72 imagens de Netuno. Outro fator que contribuiu para a escolha desta técnica para as imagens de Netuno foi a atualização recente das efemérides (Jacobson 2009), o que torna o método mais preciso.

Para determinar a diferença da posição de Urano nas imagens optou-se pelo método do interspectrum que, apesar do maior custo computacional, apresenta a melhor precisão. Os resultados obtidos mostram que o acompanhamento do telescópio foi efetivo e a diferença da posição do centro do planeta não variou mais que poucos pixels para o conjunto de imagens de cada noite.

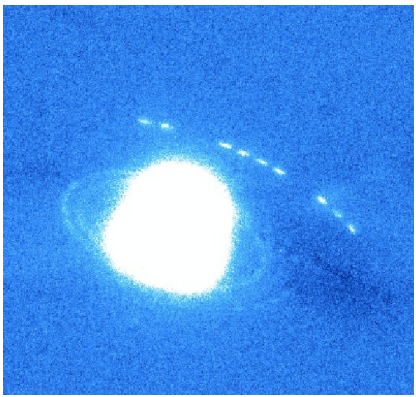

Figura 4.13: Resultado da adição de 72 imagens de Netuno no filtro Ks obtidas no VTL com tempo de exposição de 90 segundos. É possível identificar os anéis Le Verrier (interno) e Adams (externo), além do satélite Proteus. A imagem do planeta parece totalmente brilhante devido às nuvens. Adaptado de (Souami 2009).

Uma vez conhecida a diferença de posição de cada imagem é utilizada a função `fshift` para fazer o deslocamento segundo os valores de $\Delta x$ e $\Delta y$. Após este processo o centro de Urano possui as mesmas coordenadas em todas as imagens, que podem então ser combinadas resultando em uma única imagem para cada noite de observação.

A figura 4.14 mostra a imagem resultante para cada noite após a soma de todas as imagens adequadas (tabela 4.1), correspondendo a imagens com tempo de exposição igual a 83, 59, 97 e 102 minutos para N07, N08, N09 e N10, respectivamente.

Além de simplesmente somar as imagens outra forma possível de combiná-las é tomando a mediana do conjunto. Para tanto, as imagens de cada noite são enfileiradas criando um cubo e então a resultante será o valor da mediana calculada ao longo de cada pixel do cubo. Com isso os pequenos satélites são removidos da imagem resultante pois eles movem-se rapidamente, assim como suas posições nas imagens.

A figura 4.15 mostra a superfície da imagem na região de Ariel para uma combinação utilizando a soma (a) e a mediana (b) do conjunto. O processo de combinação mediana é bastante eficiente para pequenos satélites, mas é possível ver em (b) que mesmo para um satélite brilhante com Ariel a intensidade é consideravelmente reduzida.

A desvantagem deste tipo de combinação é que o brilho total da imagem é menor do que no caso da soma, pois a intensidade resultante de cada pixel é somente o valor da mediana do pixel correspondente no cubo. Assim, para o conjunto de imagens de cada noite a resultante terá a intensidade equivalente ao tempo de exposição de uma única imagem. Uma alternativa para aumentar a intensidade é dividir o cubo e aplicar este método à subconjuntos contendo de 10 a 15 imagens e então adicionar estas imagens intermediárias, resultando em uma imagem final com maior intensidade e sem a presença dos pequenos satélites.

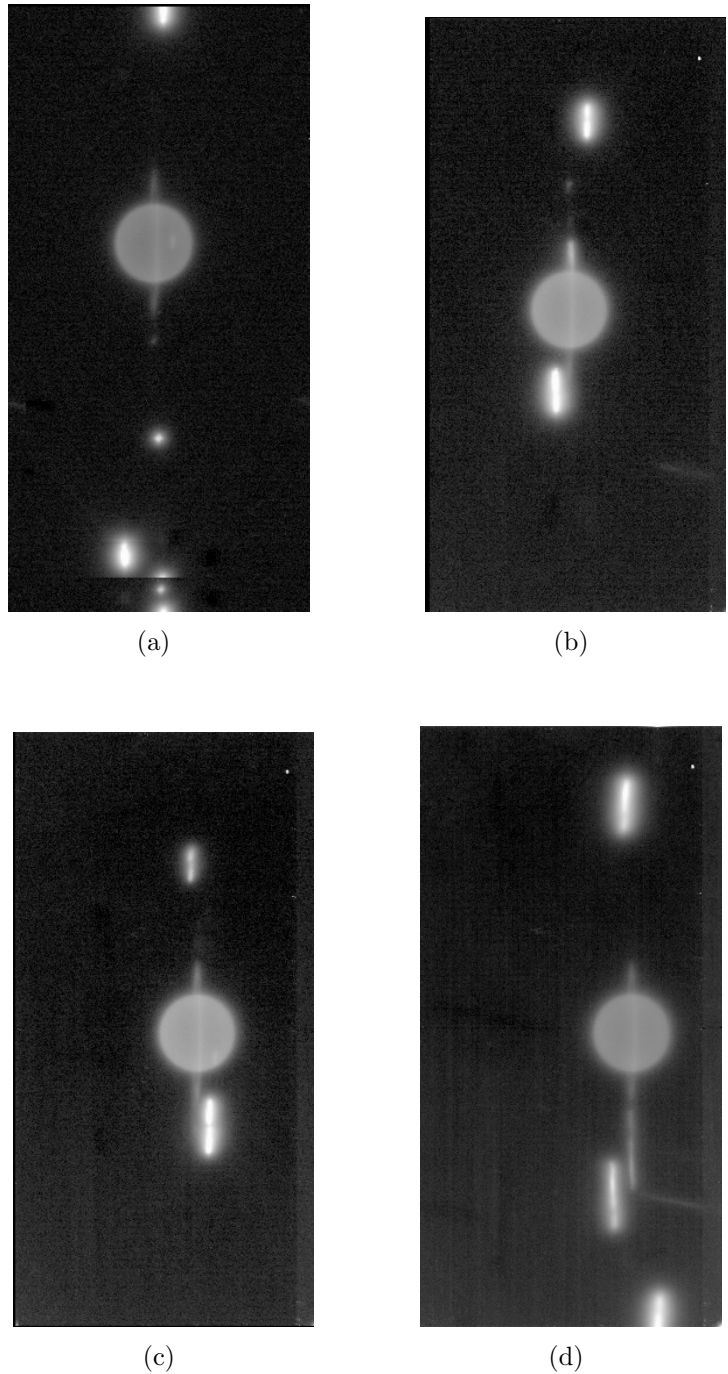

(a)

(b)

(c)

(d)

Figura 4.14: Imagens resultantes para cada noite de observação: (a) N07, (b) N08, (c) N09 e (d) N10.

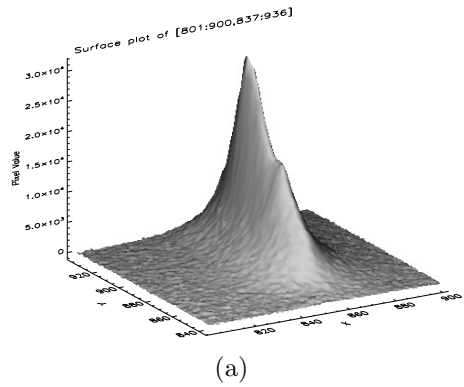

(a)

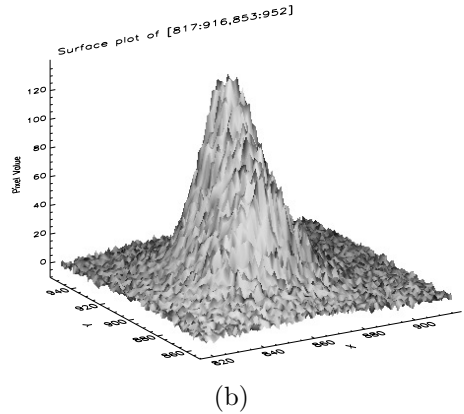

(b)

Figura 4.15: Superfície da imagem na região de Ariel após uma combinação com a (a) soma e a (b) mediana do conjunto de imagens.

## 4.9   Perfis radiais

Com as imagens resultantes para cada noite de observação é possível obter a distribuição radial dos anéis de poeira de Urano, que corresponde ao perfil vertical das imagens. Cada perfil é obtido através da contagem da intensidade sobre os pixels de uma coluna que obviamente deve interceptar a região dos anéis (figura 4.16).

Nesta figura, o pico localizado aproximadamente no pixel 300 corresponde ao satélite Titânia, enquanto a região do planeta está compreendida entre 550 e 693, aproximadamente. Imediatamente a esquerda e a direita do planeta é possível ver pequenas elevações que correspondem aos anéis de poeira.

Para determinar a distância radial dos anéis é preciso conhecer a posição do centro do planeta e converter a escala de pixels para quilômetros. Como mostrado anteriormente (seção 4.7.3), o centro do planeta pode ser determinado a partir das efemérides dos satélites, assim como a escala da imagem em mas/pixel (seção 4.6). Para converter a resolução angular de cada pixel em km/pixel é necessário saber a distância entre o observador e Urano.

Assumindo que esta distância é igual a 20.108 UA (Giorgini et al. 1996) e constante para todas as noites de observação, após um cálculo geométrico simples tem-se que a escala das imagens corresponde a 394 km/pixel.

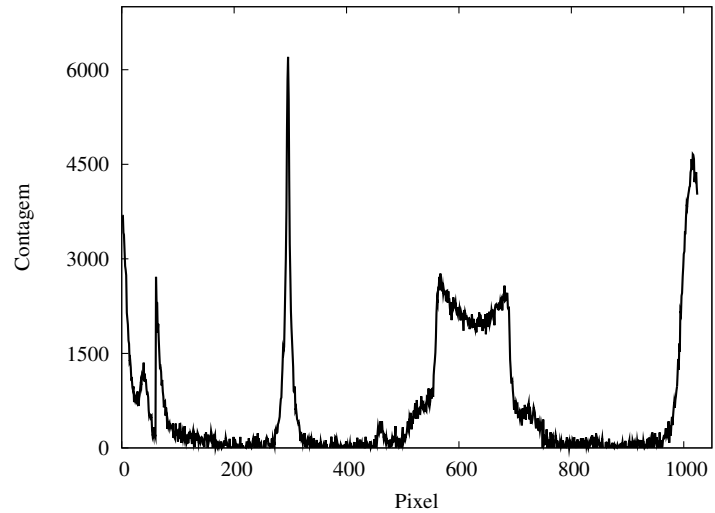

Figura 4.16: Exemplo de perfil radial obtido para a imagem resultante de N07. Neste caso a contagem foi feita sobre a coluna 623, que corresponde a coordenada $x$ do centro de Urano.

Com estas informações é possível obter o perfil radial dos anéis tanto ao norte quanto ao sul, como mostrado nas figuras 4.17 e 4.18. Nestas figuras a distância é contada em relação ao centro do planeta e é positiva no sentido norte. Em cada gráfico são apresentados os perfis obtidos sobre a coluna central, que passa pelo centro do Urano, e também a soma do perfil de dez colunas ao redor da coluna central.

O brilho devido à proximidade de satélites contamina o perfil radial dos anéis, como pode ser visto para a região sul das imagens de N08 e N09 e de maneira menos acentuada para N10 (compare com as imagens resultantes de cada noite na figura 4.14).

É impossível identificar os anéis e outras estruturas nestes perfis sem antes fazer algumas correções. Por estarem muito próximos ao planeta é necessário remover a contribuição do brilho do planeta, o que pode ser feito aplicando alguma técnica de coronografia digital; de maneira mais simples pode ser estimada a contribuição do brilho de Urano após uma rotação (supondo que o brilho do planeta é uniforme, uma hipótese bastante razoável).

Também é necessário fazer a deconvolução das imagens e dos perfis radiais. Este processo envolve transformações e operações no espaço de Fourier e resulta na redução do ruído da imagem e diminuindo o aspecto "borrado"(Starck et al. 2002), consequentemente melhorando a resolução e tornando possível a identificação dos anéis principais e das regiões de poeira.

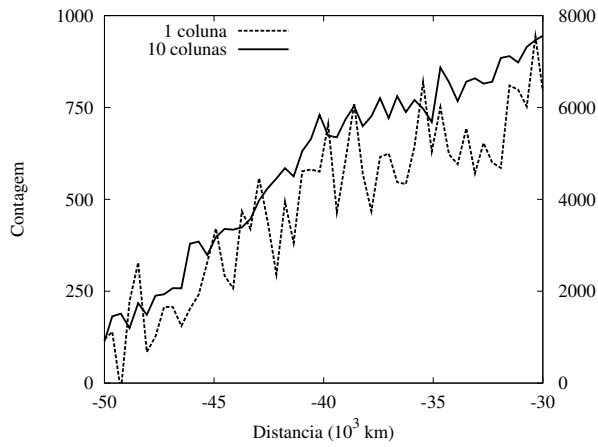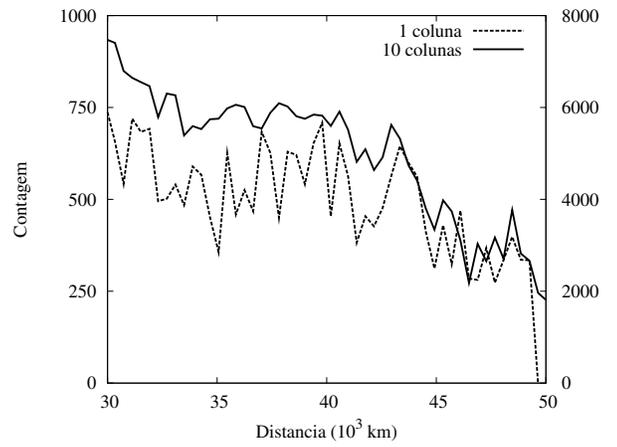

(a)

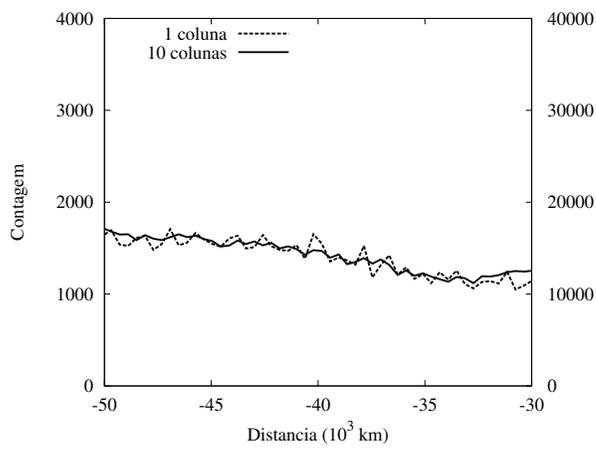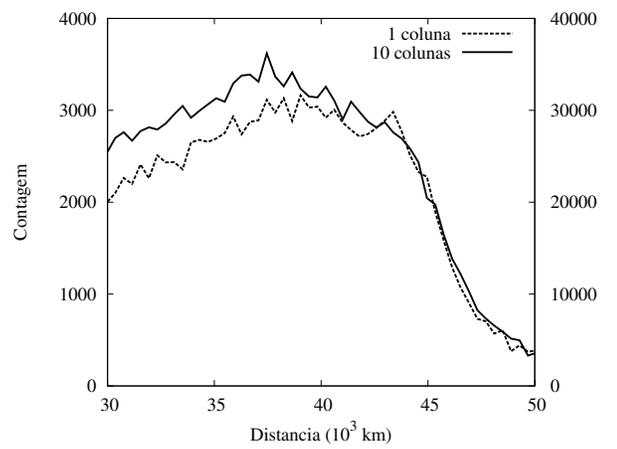

(b)

Figura 4.17: Perfis radiais dos anéis em (a) N07 e (b) N08 integrados sobre uma e dez colunas, cujas escalas estão representadas nos lados esquerdo e direito, respectivamente. A distância é contada a partir do centro do planeta e positiva no sentido norte.

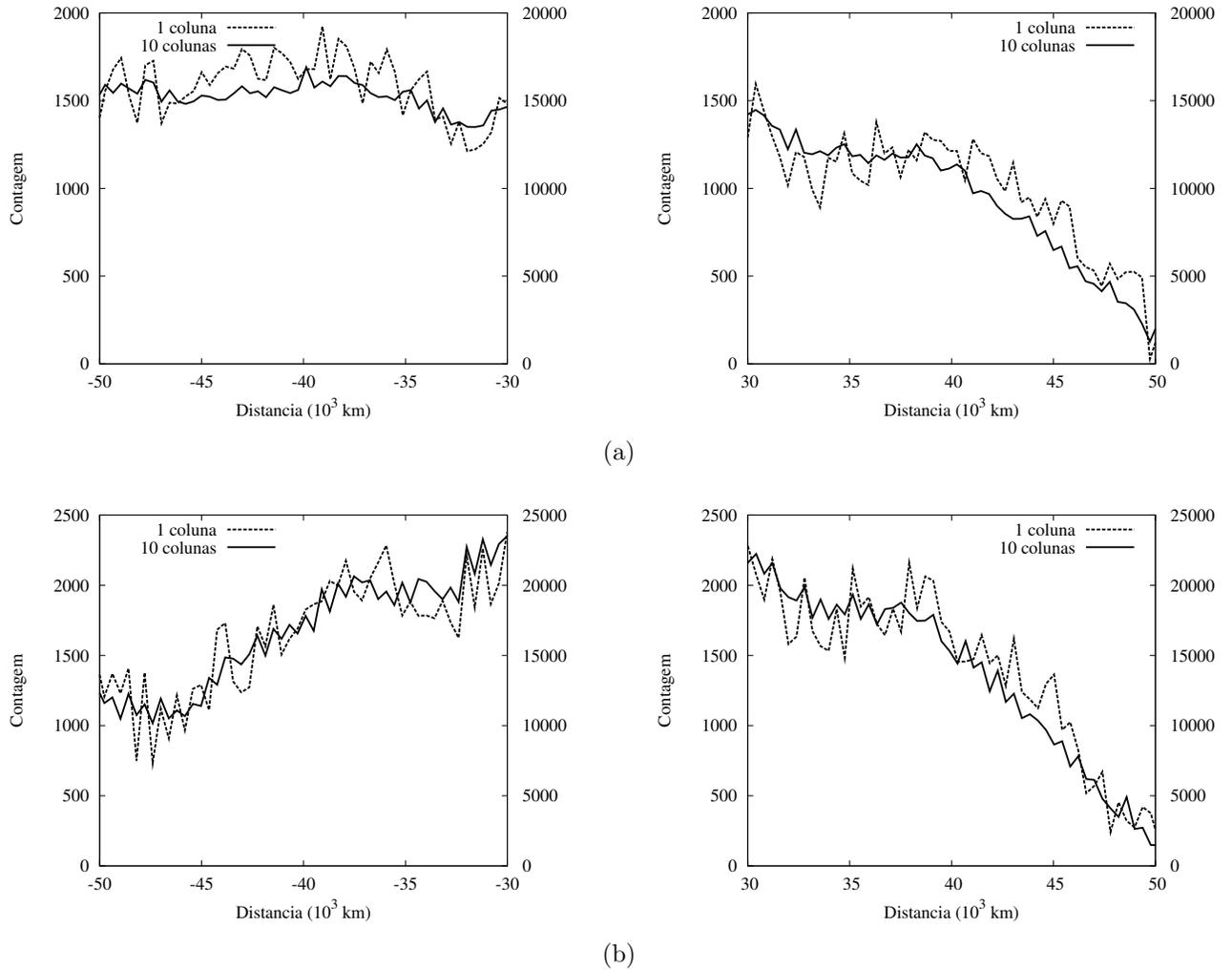

Figura 4.18: Perfis radiais dos anéis em (a) N09 e (b) N10 integrados sobre uma e dez colunas, cujas escalas estão representadas nos lados esquerdo e direito, respectivamente. A distância é contada a partir do centro do planeta e positiva no sentido norte.

## 4.10 Fotometria

Para inferir as propriedades dos anéis é necessário converter a contagem dos pixels em magnitudes ou em unidades de $I/F$, em que I é a intensidade da luz refletida pelo anel e $\pi F$ é a densidade do fluxo solar incidente.

Como a intensidade dos anéis não é conhecida, optou-se por iniciar o cálculo da fotometria com a determinação da magnitude dos satélites, uma vez que estes valores são, a princípio, bem determinados. Primeiramente é necessário escolher o método a ser usado para a fotometria, que pode ser de abertura ou diferencial (Currie et al. 2000).

A fotometria de abertura consiste basicamente na definição de uma região que contém o objeto e então realizar a contagem da intensidade (fluxo) nesta região. Já a fotometria diferencial realiza um ajuste gaussiano sobre o perfil do objeto para definir a região na qual será feita a contagem. Devido ao ajuste que é feito, a fotometria diferencial é útil em campos contendo vários objetos próximos, para os quais é difícil a identificação individual (Diolaiti et al. 2000). Porém nas imagens de Urano os satélites estão usualmente distantes e optou-se por realizar a fotometria de abertura devido à sua simplicidade.

O primeiro passo para realizar a fotometria de abertura é determinar o tamanho das regiões nas quais serão feitas a contagem e a estimativa da contribuição do céu. Para isso são definidos três raios: raio de abertura ($r$), raio interno ($r_{in}$) e o raio externo ($r_{out}$). O raio de abertura determina a região onde será feita a contagem e deve conter toda a luz refletida pelo satélite. Por este motivo $r$ não pode ser muito pequeno, mas também não pode ser arbitrariamente grande para que não sejam contadas contribuições de outras fontes. Os raios interno e externo delimitam uma coroa na qual será feita a contagem do céu, que deve ser subtraida da contagem do satélite. É importante que a área compreendida entre $r_{in}$ e $r_{out}$ não esteja contaminada com a luz proveniente de nenhum outro objeto.

Inicialmente foi escolhida uma imagem de N10 na qual os satélites Miranda, Ariel e Umbriel estão visíveis e distantes um do outro e do planeta. A contagem foi realizada com uma versão modificada da função `aper` (Landsman 1993), cuja sintaxe é

$$\mathrm{IDL} > \mathrm{aper\_mod} \,,\; \mathrm{image} \,,\; \mathrm{xc} \,,\; \mathrm{yc} \,,\; \mathrm{m0} \,,\; /\mathrm{exact}$$

em que *image* é o vetor contendo a imagem, *xc* e *yc* são as coordenadas em pixel do centróide do satélite obtidas com a rotina `ctrd` e *m0* é a magnitude de ponto zero.

Ao ser executada, a função `aper_mod` requer os valores dos raios das regiões de interesse. Para determinar o raio de abertura ótimo foram realizados vários testes alterando $r$ entre 5 e 45 pixels em intervalos de 5 pixels e em todos os casos assumiu-se $r_{in} = r + 5$ e $r_{out} = r_{in} + 10$. A figura 4.19 apresenta o fluxo obtido para cada satélite com os diferentes raios de abertura.

Para os três satélites o fluxo difere bastante para pequenos raios de abertura, pois a região escolhida não é grande o suficiente para conter toda a luz do satélite. Já para $r \geq 30$ a contagem é praticamente estável para os três satélites, ou seja, o fato de aumentar a abertura não irá resultar no aumento do fluxo.

Conhecendo o fluxo $F$ do satélite é possível calcular sua magnitude $Z$ através da relação

$$Z = -2.5 \log_{10} \left( \frac{F}{F_0} \right) \tag{4.24}$$

em que $F_0$ é o fluxo de referência que deve ser obtido com o mesmo filtro. Esta relação pode ser reescrita como

$$Z = Z_0 - 2.5 \log_{10}(F) \qquad (4.25)$$

onde $Z_0$ é chamado magnitude de ponto zero e, a princípio, é fornecida para cada noite de observação.

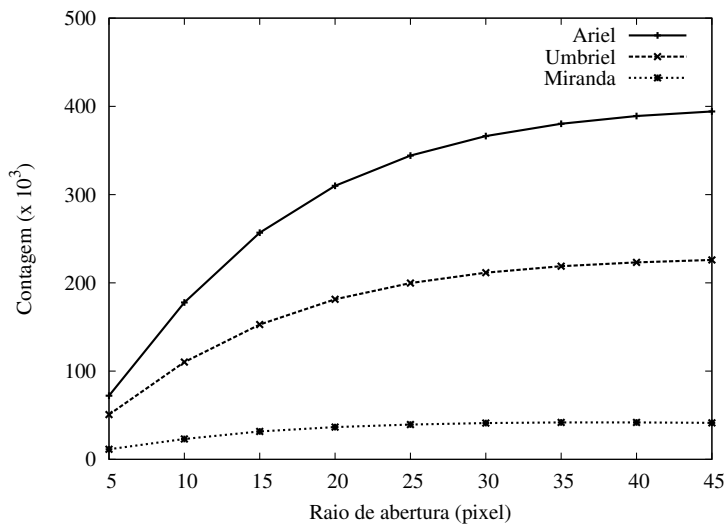

Figura 4.19: Contagem em função do raio de abertura para os satélites Ariel, Umbriel e Miranda para uma imagem de N10.

A figura 4.20 apresenta o cálculo da magnitude em função do raio de abertura, considerando o fluxo apresentado na figura 4.19 e $Z_0 = 23.01$ para N10. Como esperado, a magnitude mantém-se praticamente constante para $r \geq 30$ pixels.

Este procedimento foi repetido para várias imagens com estes satélites em diferentes noites e os valores obtidos para os fluxo e para as magnitudes foram os mesmos. Estes valores estão apresentados na tabela 4.3, juntamente com os valores esperados.

| Satélite | Magnitude derivada do fluxo | Magnitude esperada (Kesten et al. 1998) |
|---|---|---|
| Ariel | 9.02 | 13.05 |
| Umbriel | 9.62 | 13.45 |
| Miranda | 11.47 | 15.29 |

Tabela 4.3: Valores obtidos para a magnitude dos satélites e os valores esperados para a banda K derivados de Kesten et al. (1998)

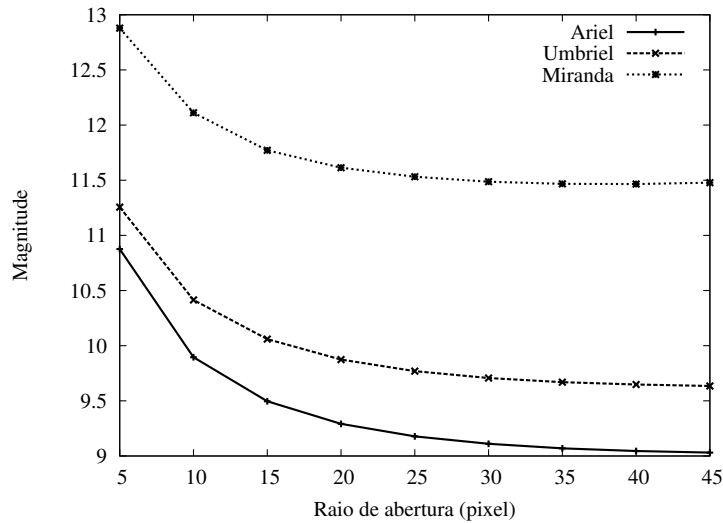

Figura 4.20: Magnitude de Ariel, Umbriel e Miranda em função do raio de abertura.

Apesar da razão entre as magnitudes calculadas para os satélites ser próxima da razão entre os valores esperados, em valores absolutos há uma grande diferença entre as magnitudes, com os satélites aparecendo muito mais brilhantes do que o esperado. Como a magnitude relativa entre os satélites é compatível com os valores anunciados, a causa da discrepância provavelmente deve-se a algum efeito sistemático. Foram levantadas algumas hipóteses para tentar entender esta diferença:

- *Determinação do fluxo*: inicialmente suspeitou-se da existência de um problema na determinação do fluxo do satélite. Porém foram realizados testes com diferentes métodos e softwares (IRAF, MIDAS e o próprio IDL com rotinas alternativas como a `atv`) e em todos a medida do fluxo é praticamente a mesma;

- *Alteração de albedo*: o albedo dos satélites de Urano não é homogêneo (Karkoschka 2001), mas esta diferença de albedo não é grande o suficiente para explicar a discrepância da magnitude;

- *Distância*: a magnitude dos satélites varia conforme a distância entre eles e o observador. Contudo, a compensação devido à distância não altera significativamente a magnitude;

- *Magnitude de ponto zero*: além de não ter sido fornecida para todas as noites, existe um erro na estimativa da magnitude de ponto zero. Este erro pode compensar uma pequena parte da diferença encontrada;

- *Confiabilidade do valor anunciado*: outra possibilidade é que o valor anunciado para as magnitudes apresente algum erro. Porém foram comparados dados de duas fontes distintas (catálogo 2MASS (Sicardy 2009) e Kesten et al. (1998)) e ambos mostraram-se consistentes;

- *Tempo de exposição*: a diferença no fluxo, que resulta na diferença da magnitude encontrada pode ser consequência do maior tempo de exposição da imagem. Segundo a equipe de engenharia do VLT-NaCo, apesar do tempo de exposição de cada imagem ser anunciado como 60 segundos, cada imagem é formada pela média de seis exposições com dez segundos cada. Com isso o fluxo é reduzido de um fator 6 e, por consequência, a magnitude do satélite aumenta.

## 4.11 Etapas futuras

Os resultados apresentados aqui fazem parte de um projeto que está em desenvolvimento.
A continuidade deste trabalho iniciará com a determinação da razão instrumental que explique a discrepância encontrada para a magnitude dos satélites e com o cálculo do fator necessário para realizar a correção. Tendo esta informação será possível utilizar os satélites como referência para obter a intensidade da luz refletida pelos aneis em função do fluxo incidente e assim determinar a quantidade de luz espalhada para as diferentes condições de iluminação.

Para remover a contribuição do brilho do planeta estuda-se aplicar uma técnica de coronografia, com o objetivo de "limpar" o perfil radial dos aneis. Entre os métodos que poderão ser aplicados estão um algoritmo de coronografia e a utilização da simetria das imagens. De maneira semelhante está sendo procurado um método para isolar o brilho dos satélites na porção sul das imagens de N08 e N09, de forma a poder utilizar os perfis desta região.

Em seguida será realizada a deconvolução que permita melhorar a resolução da imagem e extrair exatamente a posição dos aneis estreitos de Urano, além da posição radial das regiões de poeira. Também será desenvolvido um modelo para o espalhamento de luz através do qual poderá ser determinada a profundidade óptica dos aneis e, em última instância, a distribuição do tamanho das partículas (Chandrasekhar 1960).

Conhecendo a posição dos anéis e regiões de poeira será possível comparar os dados com valores obtidos anteriormente (Gibbard et al. 2005, de Pater et al. 2006). Isso permitirá uma análise sobre a evolução dinâmica do sistema de aneis, uma vez que estudos anteriores (de Pater et al. 2007) mostraram uma alteração considerável na posição dos aneis em uma escala de décadas.

Este estudo será feito através de simulações numéricas considerando efeitos como o achatamento do planeta, interações com satélites próximos e *moonlets*, além de forçar dissipativas como a pressão de radiação solar, que pode ser determinante na evolução orbital das partículas (Sfair & Giuliatti Winter 2009).

Por fim, como subproduto desta análise espera-se construir um modelo médio que, ao ser subtraído das imagens, permita identificar a posição dos satélites que formam a família de Portia. Estes dados poderão ser úteis para realizar a astrometria e melhorar as efemérides disponíveis.

# Referências Bibliográficas

# Apêndices

# Apêndice A

# Plutão

## A.1  Introdução

Plutão e Caronte formam um sistema binário cujo centro de massa está localizado fora do corpo principal. A descoberta de dois sátelites Weaver et al. (2005, 2006), Nix e Hydra, tornou o sistema ainda mais interessante.

Relacionados a este sistema foram desenvolvidos três estudos. O primeiro, apresentado na seção A.2, trata sobre a interação mútua entre os objetos e as alterações nas órbitas. Em seguida (seção A.3) foi realizado um estudo sobre a possibilidade dea existência de partículas e corpos maiores na região dos novos sátelites e a influência que eles teriam nas órbitas de Nix e Hydra.

Ainda em andamento, está sendo realizada a análise da influência da radiação solar sobre partículas microscópicas que podem existir ao redor de Plutão, Caronte e dos sátelites. Uma breve descrição deste estudo pode ser encontrada na seção A.4.

## A.2  Determinação das órbitas

A primeira determinação das órbitas dos sátelites foi realizada por Buie et al. (2006). Eles utilizaram um sistema baricêntrico levando em conta as perturbações devido à Plutão e Caronte, mas sem considerar perturbações mútuas entre Nix e Hydra. Por não considerar todas as interações, os valores obtidos para as massas de Plutão e Caronte não estavam de acordo com estudos anteriores (Lee & Peale 2006).

Tholen et al. (2008) analisaram a dinâmica do sistema Plutão-Caronte-Nix-Hydra através de simulações numéricas considerando a interação entre todos os corpos. As integrações foram realizadas utilizando dados astrométricos de diferentes fontes (Weaver et al. 2006, Buie et al.

2006, Stern et al. 2007, Tholen & Buie 1997, Beletic et al. 1989) obtidas em diferentes épocas, permitindo verificar a auto-consistência do modelo.

Em conjunto com o Prof. Bruno Sicardy teve início um projeto cujo objetivo era determinar as alterações nas posições esperadas de Plutão e Caronte devido à perturbações de Nix e Hydra e também determinar as órbitas destes satélites.

A colaboração neste projeto consistiu na comparação de simulações numéricas com os resultados de Tholen et al. (2008). Para realizar as simulações, após alguns testes optou-se pelo Mercury (Chambers 1999), um pacote que oferece vários métodos de integração (RADAU, Bulirsch-Stoer, mapa simplético e um integrador híbrido) para o problema de n-corpos que interagem gravitacionalmente. O método escolhido foi o Bulirsch-Stoer que, apesar de executar um número maior de passos, possibilita uma precisão melhor.

Como resultado, os elementos orbitais de Caronte, Nix e Hydra obtidos através de simulações mostraram-se de acordo com os resultados apresentados Tholen et al. (2008).

Um subproduto desta análise mostrou-se particularmente útil em diversos estudos posteriores. Os elementos orbitais dos satélites são comumente dados em relação ao baricentro do sistema, porém o pacote Mercury realiza todas as integrações em um referencial planetocêntrico. Desta forma é necessário efetuar a conversão das posições e velocidades entre os dois sistemas de referência antes de iniciar a integração numérica.

Várias rotinas, em diferentes linguagens de programação, foram desenvolvidas para realizar as conversões. Isso permitiu a rápida transformação das condições iniciais para o programa, assim como o cálculo automático dos valores resultantes das simulações numéricas. Exemplos de trabalhos que fizeram uso destas rotinas podem ser encontrados em dos Santos (2010a) e Pires Dos Santos et al. (2011).

## A.3   Estabilidade e satélites hipotéticos

Através de simulações numéricas foi analisado o comportamento de partículas sob o efeito gravitacional de Plutão, Caronte, Nix e Hydra. Apesar da presença dos satélites reduzir a região estável onde as partículas podem sobreviver, existem configurações onde alguns aglomerados permanecem durante todo o intervalo de integração ($10^5$ períodos orbitais do binário).

Verificou-se também que satélites ainda não detectados podem existir no sistema sem perturbar a excentricidade de Nix e Hydra de um fator maior que $10^{-3}$. Os locais propícios para procurar estes satélites são as regiões coorbitais dos satélites, a região entre eles ou ainda além da órbita de Hydra.

Estes resultados estão apresentados no artigo *Gravitational effects of Nix and Hydra in*

*the external region of the Pluto–Charon system* publicado no volume 410 da revista *Monthly Notices of the Royal Astronomical Society.* Uma cópia pode ser encontrada no apêndice E.

## A.4  Efeitos da força de radiação solar

O sistema de Plutão é constantemente bombardeado por objetos provenientes do cinturão de Kuiper. As colisões destes objetos, tanto com os satélites quanto com os corpos principais, podem causar a ejeção de poeira (Stern 2009).

Estas partículas de poeira presentes no sistema são bastante afetadas pela força de radiação solar, apesar da grande distância ao Sol. O parâmetro adimensional $C$ (definido na equação 2.4) calculado para Plutão considerando partículas de diferentes tamanhos (figura A.1), e a comparação com os valores obtidos para Urano (figura 2.12) e Saturno (figura 3.2) mostra que esta perturbação não pode ser desconsiderada. Também é apresentado o parâmetro $A$ referente a maré solar (equação 2.3), que pode ser desprezado na região de Nix e Hydra.

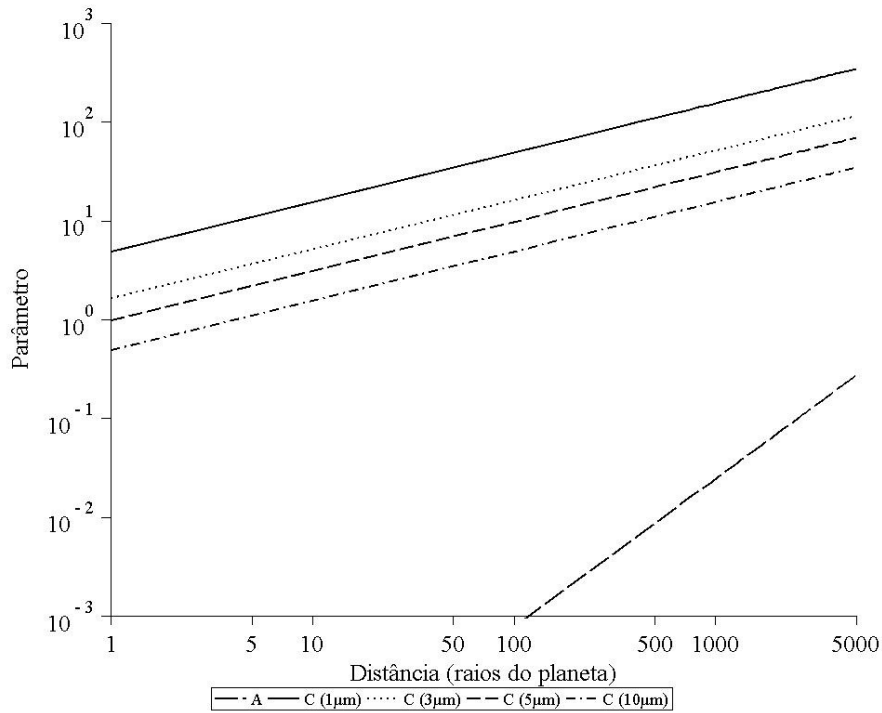

Figura A.1: Parâmetros adimensionais das forças para uma partícula ao redor de Plutão em função da distância ao planeta. O coeficiente $C$ foi calculado para diferentes tamanhos de partículas.

A princípio este efeito poderia ser inesperado, uma vez que Plutão está duas vezes mais distante do Sol que Urano, e portanto o fluxo solar é aproximadamente 4 vezes menor. Porém a pressão de radiação não depende somente do fluxo solar, mas também do movimento médio da partícula (c.f. equação 2.4). Assim, partículas ao redor de Plutão movem-se mais lentamente do que em Urano, de forma que o trabalho realizado devido à troca de *momentum* acaba sendo mais eficiente. A título de comparação foi realizada uma simulação numérica com um planeta hipotético semelhante a Plutão, porém a uma distância heliocêntrica igual a de Urano. Os resultados mostraram que, devido à menor massa do planeta, o efeito sentido devido à força de radiação solar por uma partícula ao redor de Plutão é maior do que o verificado para Urano (dos Santos 2010b).

Para realizar as simulações das partículas ao redor de Plutão foi necessário adaptar o código utilizado anteriormente, alterando as equações para o movimento médio do Sol $n_s$ que era considerado constante. Esta não é uma aproximação válida para Plutão, uma vez que a órbita do planeta possui uma excentricidade de 0.25 (Murray & Dermott 1999). Este estudo está em andamento e uma versão preliminar dos resultados pode ser encontrada em Pires Dos Santos et al. (2010).

# Referências Bibliográficas

# Apêndice B

# Artigo A&A - Urano

Cópia da versão publicada do artigo *Orbital evolution of the μ and ν dust ring particles of Uranus* publicado na revista *Astronomy and Astrophysics*, 505.

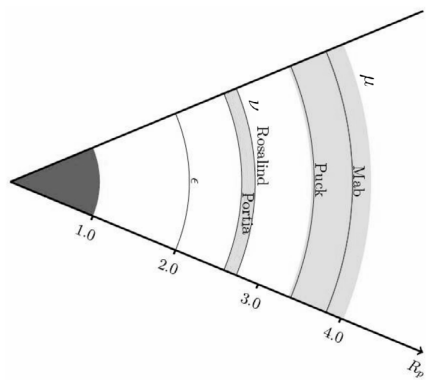

**Astronomy & Astrophysics**

# Orbital evolution of the μ and ν dust ring particles of Uranus

R. Sfair and S. M. Giuliatti Winter

UNESP-São Paulo State University, C.P. 205, Guaratinguetá, CEP 12516-410, SP, Brazil
e-mail: rsfair@feg.unesp.br

## ABSTRACT

The μ and ν rings of Uranus form a secondary ring-moon system with the satellites Puck, Mab, Portia, and Rosalind. These rings are tenuous and dominated by micrometric particles, which can be strongly disturbed by dissipative forces, such as the solar radiation pressure. In the region of these rings, the solar radiation force and the planetary oblateness change the orbital evolution of these dust particles significantly. In this work, we performed a numerical analysis of the orbital evolution of a sample of particles with radii of 1, 3, 5, and 10 μm under the influence of these perturbations, combined with the gravitational interaction with the close satellites. As expected, the Poynting-Robertson component of the solar radiation force causes the collapse of the orbits on a timescale between $3.1 \times 10^5$ and $3.6 \times 10^6$ years, while the radiation pressure causes an increase in the eccentricity of the particles. The inclusion of Uranus's oblateness prevents a large variation in the eccentricity, confining the particles in the region of the rings. The encounters with the close satellites produce variations in the semimajor axis of the particles, leading them to move inward and outward within the ring region. These particles can either remain within the region of the rings or collide with a neighbouring satellite. The number of collisions depends on the size of both the particles and the satellites, and the radial width of the ring. For the time span analysed, the percentage of particles that collide with a satellite varies from 43% to 94% for the ν ring, and from 12% to 62% for the μ ring. Our study shows that all collisions with Portia and Rosalind have the value of impact velocity comparable to the escape velocity, which could result in the deposition of material onto the surface of the satellite. Collisions between Puck and particles larger than 1 μm also occur at an impact velocity comparable to the value of the escape velocity. The exception is Mab, which is hit by particles with velocities several times larger than the escape velocity. These collisions are energetic enough to eject material and supply material to the μ ring. However, only a few particles (3%) collide with the surface of the satellite Mab at such a velocity.

**Key words.** planets: rings – planets and satellites: individual: Uranus – methods: N-body simulations

## 1. Introduction

Until 1977, the only known planet surrounded by rings was Saturn. In this year, during the occultation of the star SAO 158687, four narrow rings around Uranus were discovered (Elliot et al. 1977), and further observations from Voyager II found six more rings and faint dust bands between the rings (Smith et al. 1986). The ε ring orbits at 51 149 km (2 $R_p$), the outer edge of the main ring system.

Showalter & Lissauer (2006) obtained several images with long exposure times of the Uranian system using the Hubble Space Telescope (HST), which discovered two rings, μ and ν, which represented a secondary ring-moon system orbiting outside the main rings. Both rings are very faint and their radial profiles, defined in terms of the brightness of the ring as a function of the distance from the planet's centre, show a distinguished triangular profile. Their orbits are closely related with satellites: the peak, which corresponds to the distance from the planet's centre of maximum normal value of $I/F$ (measurement of the brightness of the ring that would be observed from a perpendicular point of view), of μ ring is almost aligned with the orbit of the satellite Mab, while the inner edge of the ring coincides with the orbit of Puck. The ν ring peak does not match the orbit of any satellite, but the ring orbits between the satellites Rosalind and Portia. Figure 1 shows a schematic view of the μ and ν rings and the closest satellites. The radial extension of each ring is represented by shadow areas, and the solid lines correspond to the orbits of Puck, Mab, Rosalind, and Portia. All distances are represented in units of Uranus' radius ($R_p$), and the solid line at

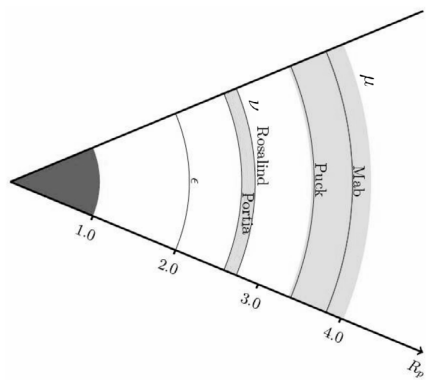

**Fig. 1.** Schematic view of the secondary ring-moon system of Uranus. The solid lines correspond to the satellites orbits and the rings are represented by shadow areas. The planet (shadow area until 1 $R_p$) and the ε ring are also indicated. All distances are scaled in units of Uranus' radius ($R_p$).

2 $R_p$ indicates the position of the ε ring, the outermost ring of the main system.

The μ and ν rings are radially wide and predominantly forward scattering, indicating that they are dominated by micrometric dust (de Pater et al. 2006). Dust particles can be disturbed by several non-gravitational forces, such as the solar radiation force, the solar tidal, and the planetary oblateness. These forces are much weaker than the gravity force of the planet. However, they change the orbital energy of the particle, which can be important to its long-term evolution.

In this paper, we analyse the orbital evolution of a sample of dust particles, located at the secondary ring system of Uranus, affected by the solar radiation pressure, the oblateness of the planet, and the gravitational perturbations of the satellites. In Sect. 2, we analyse the strength of the solar radiation pressure and the oblateness of the planet, and in Sect. 3, we describe the numerical method. The effects of these perturbations on the ring particles and the orbital evolution of them are analysed in Sects. 4 and 5, respectively. Section 6 deals with the collisions between the particles and the satellites. Our final conclusions are presented in the last section.

## 2. Force parameter

Hamilton & Krivov (1996) presented a way to compare the strength of the perturbation forces using dimensionless parameters that depend on the dust particles' semimajor axis, and the physical properties of the dust and the planet. Following their notation, we calculated the values of the solar tidal ($A$), oblateness ($W$), and radiation pressure ($C$) parameters. The solar tidal parameter $A$ is defined as

$$A = \frac{3}{4}\frac{n_s}{n},\tag{1}$$

where $n_s$ is the mean motion of the planet around the Sun and $n$ is the mean motion of the particle.

The parameter $W$, which describes the strength of the planetary oblateness, is written as

$$W = \frac{3}{2}J_2\left(\frac{R}{a}\right)^2\frac{n_s}{n},\tag{2}$$

where $J_2$ denotes the second zonal harmonic coefficient, $R$ is the radius of the planet, and $a$ is the semimajor axis of the particle.

The ratio $\sigma$ of the radiative force to the planets' gravity of a particle with radius $r$ and density $\rho$ is

$$\sigma = \frac{3}{4}Q_{pr}\frac{F_s a}{GM_s c\rho r}.\tag{3}$$

In this equation, $Q_{pr}$ is the radiation pressure coefficient, $F_s$ is the incoming solar flux at the planet, $GM_s$ is the planet's gravitational constant, and $c$ is the speed of light.

Using the previous definition of $\sigma$, the perturbation caused by radiation pressure is written as

$$C = \frac{3}{2}\frac{n}{n_s}\sigma.\tag{4}$$

All of these parameters were calculated for a dust particle around Uranus, and the results are presented in Fig. 2.

As can be seen in Fig. 2, the solar tide is relevant to particles located faraway from the planet ($>50\ R_p$), thus it can be safely ignored in the region of the $\mu$ and $\nu$ rings. On the other hand, for these rings the effects of radiation pressure and oblateness are appreciable and must be taken into account. In the following sections, we present our results for a numerical study of the evolution of a sample of particles disturbed by the gravitational effects of the closest satellites to each ring, the solar radiation pressure, and the oblateness of Uranus.

### 2.1. Solar radiation force

Most of the planetary ring systems are coplanar to the orbital plane of the planet, that is, the inclination $i$ of the ring particle is $i \leq \gamma$, where $\gamma$ is the obliquity of the planet (Hamilton 1993). In these cases, the problem can be treated as bidimensional, when the Sun and the ring particle are located in the same plane.

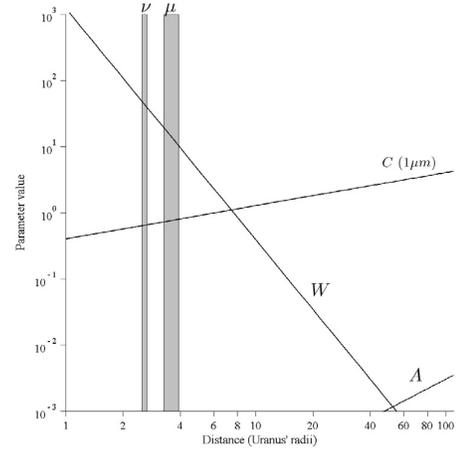

**Fig. 2.** Parameters values as a function of the distance of the planet: $A$ (solar tide), $W$ (planetary oblateness), and $C$ (radiation pressure). The vertical bars denote the radial extent of the $\mu$ and $\nu$ rings. We assumed a spherical grain with radius of 1 $\mu$m and uniform density of 1.0 g cm$^{-3}$.

Although Uranus's dusty rings describe equatorial orbits, the obliquity of Uranus is 97.8° (Murray & Dermott 1999). Therefore, this problem requires a three-dimensional approach, since the Sun and the ring particle do not share the same orbital plane.

For a particle that describes an orbit around a planet with obliquity $\gamma$, the components of the solar radiation force, in a Cartesian inertial frame centred on the planet, can be written as (Mignard 1984)

$$F_x = \frac{\beta GM_s}{r_{sp}^2}\left[\cos\left(n_s t\right) - \left(\frac{x_s}{r_{sp}}\right)^2\left(\frac{v_{xs}}{c} + \frac{v_x}{c}\right) - \left(\frac{v_{xs}}{c} + \frac{v_x}{c}\right)\right],\tag{5}$$

$$F_y = \frac{\beta GM_s}{r_{sp}^2}\left[\cos(\gamma)\sin\left(n_s t\right) - \left(\frac{y_s}{r_{sp}}\right)^2\left(\frac{v_{ys}}{c} + \frac{v_y}{c}\right)\right.$$
$$\left. - \left(\frac{v_{ys}}{c} + \frac{v_y}{c}\right)\right],\tag{6}$$

$$F_z = \frac{\beta GM_s}{r_{sp}^2}\left[\sin(\gamma)\sin\left(n_s t\right) - \left(\frac{z_s}{r_{sp}}\right)^2\left(\frac{v_{zs}}{c} + \frac{v_z}{c}\right)\right.$$
$$\left. - \left(\frac{v_{zs}}{c} + \frac{v_z}{c}\right)\right],\tag{7}$$

where $r_{sp}$ is the module of the Sun-planet vector, and $(x_s, y_s, z_s)$ are the components of $r_{sp}$. The components of the velocity of the particle are $(v_x, v_y, v_z)$, and $(v_{xs}, v_{ys}, v_{zs})$ are the components of the velocity of the planet around the Sun. In these equations, the term independent of the velocities is called the radiation pressure (RP hereafter) and the remaining terms correspond to the Poynting-Robertson drag (PR drag).

If we assume that a spherical particle obeys geometrical optics, the parameter $\beta$ defined earlier as the ratio of the radiation force to the solar gravitational force can be written as (Burns et al. 1979)

$$\beta = 5.7 \times 10^{-5}\frac{Q_{pr}}{\rho r},\tag{8}$$

where $Q_{pr}$, the radiation pressure coefficient, is assumed to be equal to the unit to represent the value of an ideal material.

## 3. Numerical simulations

We performed a numerical study of the evolution of a sample of particles located at the $\mu$ and $\nu$ rings influenced by the gravity field of Uranus, the solar radiation force, and the gravitational perturbation of the closest satellites.

The numerical integrations were carried out using the variable timestep Bulirsch-Stoer algorithm from the Mercury package (Chambers 1999). The Mercury package allows the inclusion of a user-defined force such as Eqs. (5)–(7). It can also handle the case of a non-spherical central body including terms for the multipole expansion of the gravity field. We restricted our analysis to the first term of the expansion ($J_2$) for Uranus, a reasonable approximation when higher terms of the expansion are at least two orders of magnitude smaller.

In our model, the planet is in a circular orbit around the Sun, thus $n_s$ is constant, as well as the solar flux. We did not take into account the reflection from the planet, when the contribution of this effect was at least an order of magnitude weaker than those produced by the direct radiation (Hamilton & Krivov 1996). For the same reason, we ignored the effects of the grain rotation (Yarkovsky effect).

The semimajor axis ($a$), eccentricity ($e$), inclination ($i$), pericentre ($\omega$), ascending node ($\Omega$), and the radius of the satellites used in the numerical simulations are listed in the Table 1. All orbital elements were derived from Showalter & Lissauer (2006), and the density of all satellites was assumed to be 1.3 g cm$^{-3}$, equivalent to those for the satellite Miranda. Table 2 (adapted from Showalter & Lissauer 2006) summarizes the information for the $\mu$ and $\nu$ rings orbital radii. The rings were assumed to be in circular and equatorial orbits. Uranus's parameters used in the numerical simulations (radius, mass, semimajor axis, and $J_2$ coefficient) were derived from Murray & Dermott (1999).

The gravitational effects of the other satellites of the Portia family were not taken into account since these satellites are small and far from the $\mu$ and $\nu$ rings.

No information has so far regarding the particle size distribution of the $\mu$ and $\nu$ rings. Since de Pater et al. (2006) showed that these rings are dominated by micrometric dust, we assumed spherical particles of size 1, 3, 5, and 10 $\mu$m with density of 1 g cm$^{-3}$ (pure solid ice). For each ring, an ensemble of 10 000 particles of each size were radially and azimuthally distributed within the ring region. The initial conditions of the particles were chosen following a random normal distribution.

Each particle was numerically integrated for a time span of 1000 years. When the distance between the particle and a satellite was less than the satellite's radius, a collision was detected. In this case, the particle is removed from the system and the impact parameters (position and velocity) were recorded.

## 4. Solar radiation force and planetary oblateness effects

Firstly, we analysed the effects of each component of the solar radiation force on the orbital elements of the particle, and how accounting for the planetary oblateness modified these effects.

As expected, if the oblateness of the planet is not taken into account, the radiation pressure component causes a variation of a few kilometres in the semimajor axis and induces large variations in the eccentricity of a dust particle. Figure 3 shows how

**Table 1.** Satellite orbital elements and physical parameters used in the numerical simulations.

|  | Mab | Puck | Rosalind | Portia |
|---|---|---|---|---|
| $a$ (km) | 97 735 | 86 004 | 69 926 | 66 097 |
| $e$ ($\times 10^{-3}$) | 2.54 | 0.39 | 0.58 | 0.51 |
| $i$ (°) | 0.134 | 0.321 | 0.093 | 0.18 |
| $\omega$ (°) | 240.30 | 331.88 | 257.28 | 84.41 |
| $\Omega$ (°) | 350.74 | 199.48 | 157.64 | 77.71 |
| $r$ (km) | 12 | 81 | 36 | 70 |

**Table 2.** The $\mu$ and $\nu$ rings radii.

| Ring | Inner edge (km) | Outer edge (km) | Peak radius (km) | Width (km) |
|---|---|---|---|---|
| $\mu$ | 86 000 | 103 000 | 97 700 | 17 000 |
| $\nu$ | 66 100 | 69 900 | 67 300 | 3800 |

RP changes the eccentricity of four particles with identical initial conditions but different sizes. Figure 3a shows the evolution in the eccentricity of particles located at the $\mu$ ring, and Fig. 3b for those particles at the $\nu$ ring.

The overall behaviour of the eccentricity variation caused by the radiation pressure component for both $\mu$ and $\nu$ rings is the same. For larger particles, the period of oscillation is approximately equal to the orbital period of the planet ($\sim$84 years). The concavities seen in Fig. 3 are related to a highly inclined Uranus' equatorial plane. A local minimum appears approximately at the half of the orbital period of the planet, as the Sun crosses the ring plane.

It can be seen that the eccentricity of the 1 $\mu$m particle reaches a value higher than 0.5, which is enough to cause a collision with the $\epsilon$ ring. Before colliding with the main ring system, the particles of the $\mu$ and $\nu$ rings should cross the orbits of several satellites of the Portia family and may collide with one of these satellites.

In contrast to the radiation pressure, the Poynting-Robertson component causes no secular change in the eccentricity. The main effect of the PR drag is a decrease in the semimajor axis caused by the energy loss experienced by the particle. A 1 $\mu$m particle initially located at $a = 97\,500$ km (approximately at the peak of the $\mu$ ring) decays 23 km in about 100 years, while the variation in the semimajor axis of a 10 $\mu$m particle is roughly 2 km (Fig. 4). If we assume a constant decay rate, the time until the orbits collapse varies from $3.1 \times 10^5$ to $3.6 \times 10^6$ years, depending on the size of the particle. Burns et al. (1979) provide an analytical expression to calculate the lifetime of a circumplanetary particle in an equatorial orbit and the values derived from our numerical simulations are accurate to about 5% for particles of all sizes.

The next step was to analyse the effects of the solar radiation force and a non-spherical central body. The oblateness of Uranus adds a short-period variation in the semimajor axis disturbed by the Poynting-Robertson component, although the decay rate remains the same (Fig. 5).

The effect of the oblateness is more evident in the eccentricity of the particle under the effects of the radiation pressure component. Figure 6 shows the variation in the eccentricity for particles of different sizes and identical initial conditions as presented in Fig. 3a. A comparison between Figs. 3a and 6 shows that the oblateness causes a "damping" in the eccentricity variation.

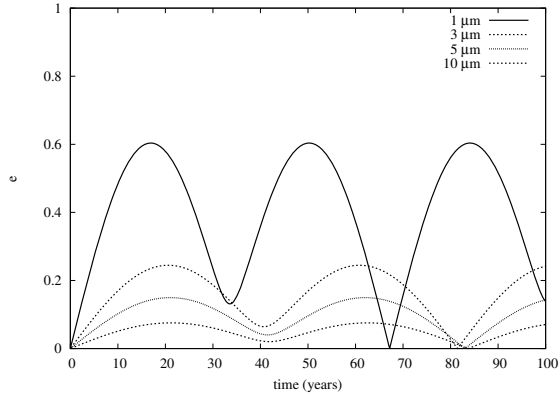

(a)

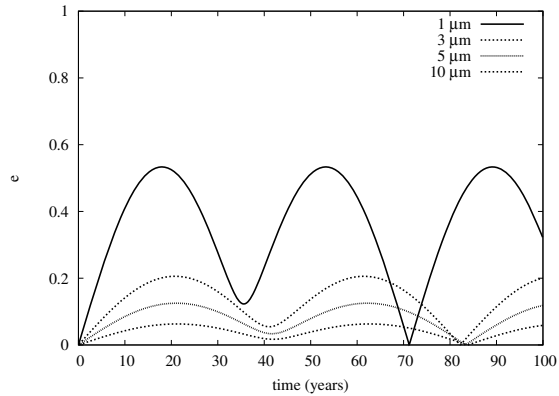

(b)

**Fig. 3.** Evolution of the eccentricity due to the radiation pressure component for ring particles with identical initial conditions: **a)** for the $\mu$ ring and **b)** for the $\nu$ ring.

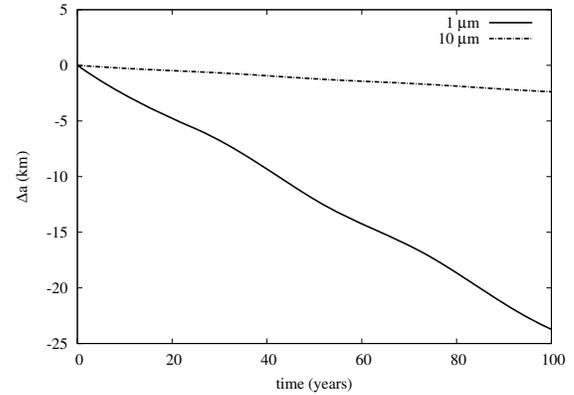

**Fig. 4.** Evolution of the semimajor axis due to the PR drag for 1 $\mu$m and 10 $\mu$m sized particles of $\mu$ ring with identical initial conditions. $\Delta a = 0$ corresponds to the initial semimajor axis of the particle.

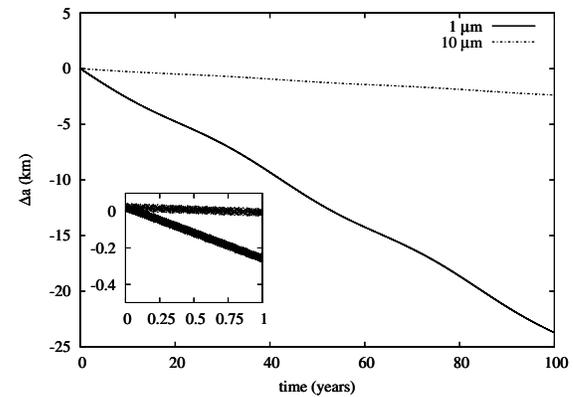

**Fig. 5.** Evolution of the semimajor axis under the effects of the PR drag and the planetary oblateness. The initial conditions of the particles are the same as in Fig. 4.

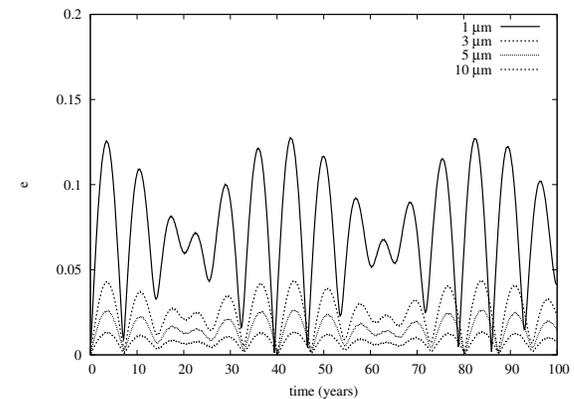

**Fig. 6.** Evolution of the eccentricity due to the radiation pressure and the planetary oblateness ($J_2$ term) for those particles located at the $\mu$ ring. The initial conditions are identical to those presented in Fig. 3a.

The decrease in the amplitude is enhanced as the distance to the planet decreases, when the rate of precession can be described by an inverse square law (cf. Eq. (9)). The reduction is more evident for the $\nu$ ring particles. Figure 7a and b illustrates the variation in the eccentricity for 1 $\mu$m and 10 $\mu$m sized particles of the $\nu$ ring with the same initial conditions as the particles shown in Fig. 3b.

This decrease in the eccentricity keeps the particle within the region of the rings. A similar process occurs in the Saturn's F-ring, where the damping in the eccentricity avoids the particles leaving the ring region (Sfair et al. 2009).

The effects of the planetary oblateness on a particle, already disturbed by the solar radiation pressure, can lead to different outcomes. For dust particles located close to Phobos and Deimos (Mars' satellites), the oblateness of the planet enhances the variation in the eccentricity produced by RP component (Hamilton & Krivov 1996), while for those particles at the ring system of Neptune (Foryta & Sicardy 1996) and Saturn (Sfair et al. 2009) the effect is the opposite. The values of the eccentricity are "damped" when the planetary oblateness is considered.

The damping in the eccentricity occurs when the precession due to the oblateness is faster than the planet's mean motion. In this case, the forced eccentricity induced by the radiation pressure has insufficient time to develop fully and tends to disappear (Mignard 1984).

The period of oscillation also changes and becomes shorter. The period of the eccentricity is modulated by the coupling of three frequencies: $n_s$, $\varpi$, and a short-period frequency related to the inclusion of the $J_2$ term, where $\varpi$ is the pericentre precession rate, which can be written as (Murray & Dermott 1999)

$$\varpi \simeq \frac{3}{2} J_2 n \left(\frac{R}{a}\right)^2.$$ (9)

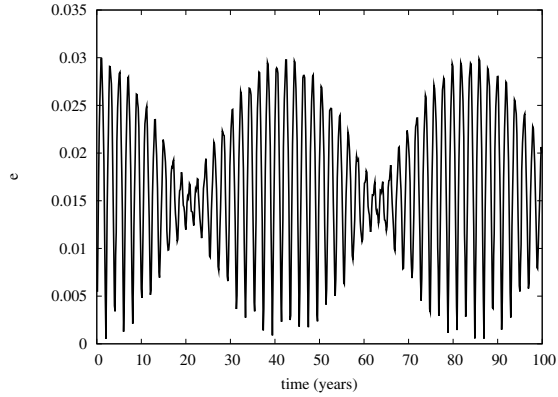

(a)

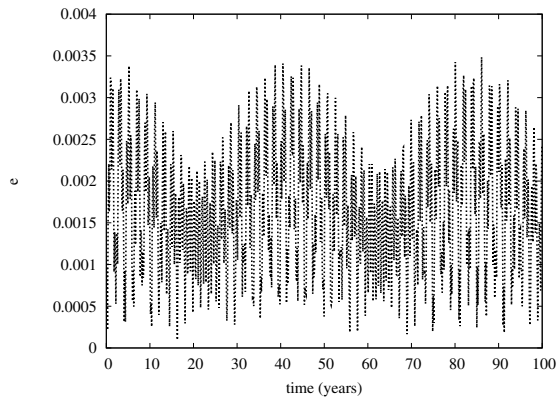

(b)

**Fig. 7.** Evolution of the eccentricity due to the radiation pressure component and the planetary oblateness for $\nu$ ring particles. The size of the particles are **a)** 1 $\mu$m and **b)** 10 $\mu$m, and the initial conditions are the same as in Fig. 3b.

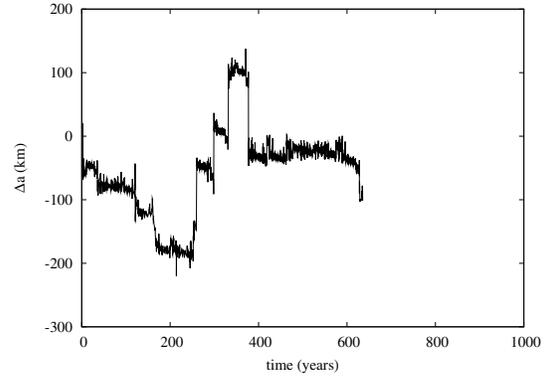

(a)

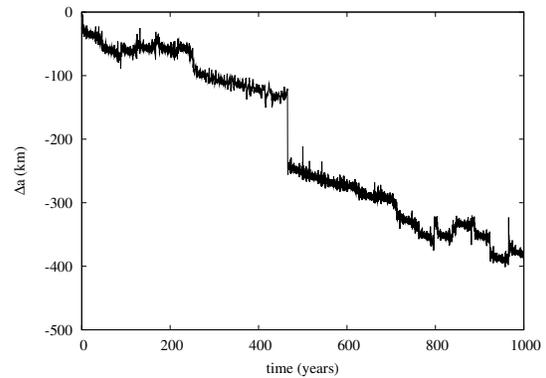

(b)

**Fig. 8.** Temporal evolution of the semimajor axis of two 1 $\mu$m sized particles located in $\mu$ ring under the effects of the solar radiation force, the planetary oblateness, and the satellites Puck and Mab. In **a)** the particle collides with Puck after 635 years. The outer and inner edges of the ring are located at 5485 km and $-11\,515$ km, respectively; the particle represented in **b)** remains within the ring region for the time span of 1000 years. The outer and inner edges of the ring are located at 5027 km and $-11\,970$ km, respectively. In each plot, $\Delta a = 0$ corresponds to the initial semimajor axis of the particle.

The time rate of $\varpi$ varies from 1.87 to 8.86 degrees per year for the innermost and outermost particles, respectively. These values agree fairly well with those obtained by means of numerical simulations.

## 5. Orbital evolution

In addition to the effects of the solar radiation pressure and the oblateness of Uranus, the dust ring particles are also perturbed by the closest satellites. The $\mu$ ring particles are perturbed by the satellites Puck and Mab, and the $\nu$ ring particles suffer the gravitational effects of Portia and Rosalind. Because of the large number of particles analysed, we present some representative plots of the 1 $\mu$m and 10 $\mu$m sized particles.

Figures 8 and 9 show the evolution of the semimajor axis of two $\mu$ ring particles. In each plot, the $y$-axis represents the semimajor axis displacement from its initial value as a function of time ($x$-axis). All particles were integrated for a time span of 1000 years, unless a collision occurs.

The amplitude of the variation in the eccentricity, caused by the gravitational perturbation of the satellites, is $O(10^{-4})$. It is, at least, two orders of magnitude smaller than the variation produced by the RP component. Therefore, the evolution of the eccentricity is dominated by this component of the solar radiation force. The eccentricity can reach values higher than 0.01. Such values are high enough to cause close encounters with the satellites, which cause sudden variations (*jumps*) of the semimajor axis. The 1 $\mu$m particles exhibit amounts of these axis large variations, sometimes of hundreds of kilometres (Fig. 8), while the majority of the 10 $\mu$m particles exhibit "jumps" of a few kilometres (Fig. 9). Even those particles initially faraway from the neighbourhood of the satellites can experience close encounters with them, especially small particles, which are more sensitive to the effects of the radiation pressure.

In the cases shown in Figs. 8 and 9, the particles have different orbital evolution. The variation in the semimajor axis of the particles, due to close encounters with the satellites, in combination with the eccentricity oscillation, allow the orbit of the particle to cross the orbit of the satellite, leading to an eventual collision. For example, the semimajor axis of the particle represented in Fig. 8 has a variation larger than 300 km, implying that a collision with Puck will occur in ∼360 years.

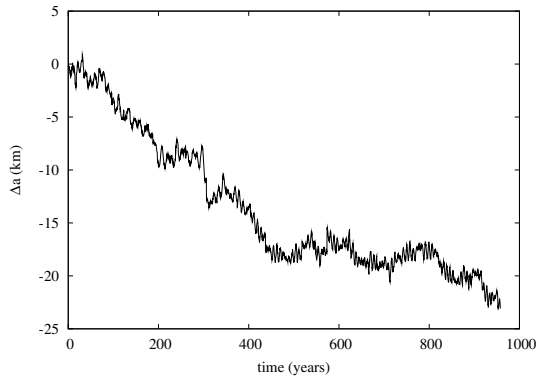

(a)

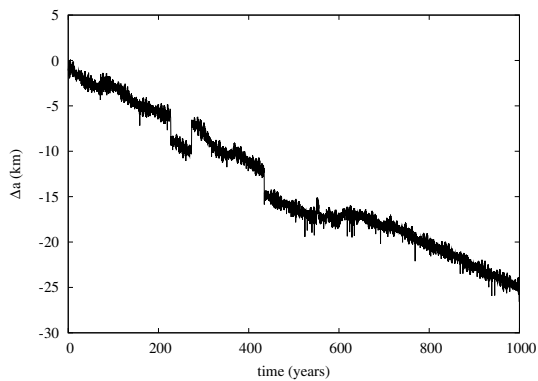

(b)

**Fig. 9.** Temporal evolution of the semimajor axis of two 10 $\mu$m particles located in $\mu$ ring under the effects of the solar radiation force, the planetary oblateness, and the satellites Mab and Puck. The particle represented in **a)** collides with Mab after 950 years. The outer and inner edges of the ring are located at 8700 km and $-8300$ km, respectively. The particle represented in **b)** survived within the ring region for the time span of 1000 years. The outer and inner edges of the ring are located at 1392 km and $-10\,708$ km, respectively.

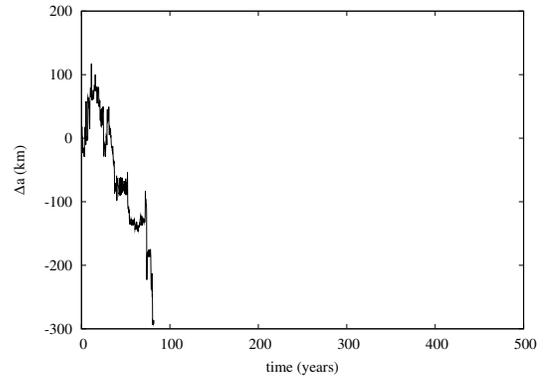

(a)

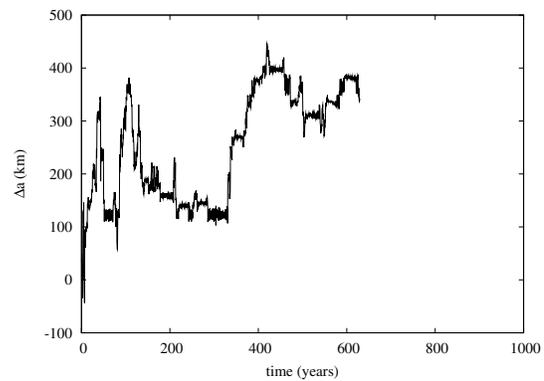

(b)

**Fig. 10.** Evolution of the semimajor axis of 1 $\mu$m particles of the $\nu$ ring disturbed by the solar radiation force, planetary oblateness, and gravitational influence of Portia and Rosalind. In **a)** the particle collides with Portia after 80 years. The outer and inner edges of the ring are located at 682 km and $-3118$ km, respectively. The particle represented in **b)** collides with Rosalind in approximately 690 years. The outer and inner edges of the ring are located at 3053 km and $-747$ km, respectively. In each plot, $\Delta a = 0$ corresponds to the initial semimajor axis of the particle.

A similar behaviour was found for the $\nu$ ring particles. Figures 10 and 11 show the evolution of the semimajor axis of four $\nu$ ring particles, orbiting an oblate Uranus, affected by the solar radiation force and the gravitational perturbations of Portia and Rosalind.

As expected, the strength of the planetary oblateness is stronger in the $\nu$ ring region (Fig. 2), which makes the eccentricity damping more effective. However, even a small $a$ variation in the eccentricity leads to close encounters with one of the satellites (Fig. 10). Some 10 $\mu$m sized particles of the $\nu$ ring undergo deviations of some hundreds of kilometres in semimajor axis (Fig. 11a), as a consequence of the proximity with Portia and Rosalind.

The global behaviour of the representative sample of the particles with 3 $\mu$m and 5 $\mu$m sizes do not differ from the cases presented here. The eccentricity varies according to the combination of the $J_2$ term and the RP component. The close encounters between the particle and the satellite produce variations (jumps) in the semimajor axis within the ring region.

## 6. Collisions between satellites and particles

Since the oblateness prevents the overstated increase in the eccentricity, the complete ensemble of particles remains within the region of the rings for the time span of 1000 years. However, some particles cross the orbit and can collide with the close satellite. Tables 3 and 4 summarize the number of collisions between the particles and the satellites in the numerical simulations. $N_\%$ is the percentage of the particles that collide with the satellite, $\overline{T}$ is the mean time of the collision, and $\overline{v}$ is the mean impact velocity.

For both rings, the percentage of the number of collisions ($N_\%$) and the mean time ($\overline{T}$) exhibit a correlation with the size

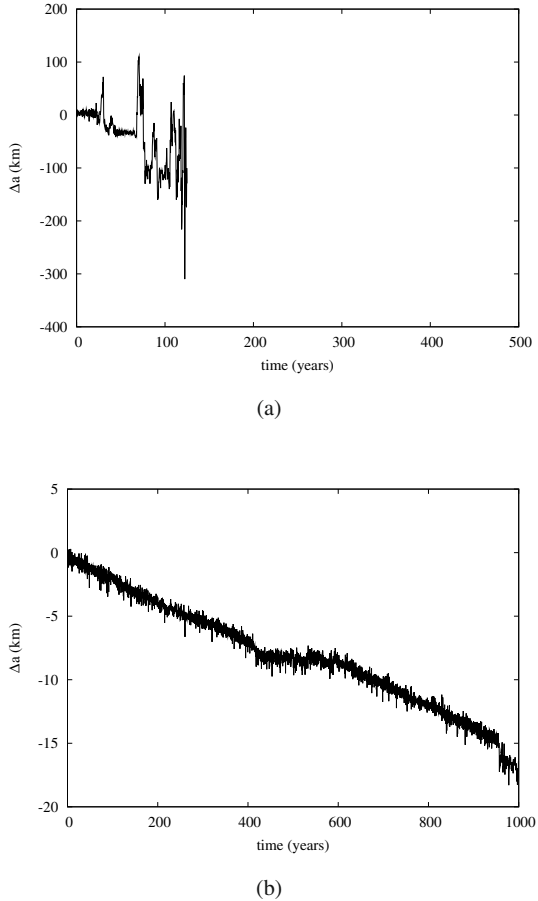

(a)

(b)

of the particle. This is because of the different values of the eccentricity, produced by the effect of the RP component. Small particles, which are more sensitive to this effect, exhibit large variations in their eccentricity. The orbits of 1 $\mu$m sized particles cross a broad region of the ring, increasing the probability of collision with a close satellite.

By comparing both ring systems, we can verify that the particles located at the $\nu$ ring have a shorter lifetime. After 1000 years, 94% of the 1 $\mu$m sized particles collided with Portia or Rosalind. Meanwhile, the number of collisions with Puck and Mab is roughly 60% of the total number of particles. This difference is caused mainly by the radial extension of the $\nu$ ring, which is ∼4.5 times narrower than the $\mu$ ring. All particles that did not suffer any collision with the satellites remained within the region of the rings.

The number of collisions with the satellites Portia and Puck depends, primarily, on the size and location of the satellites. The decrease in the semimajor axis, due to the PR drag, increases the collision between the satellites, located close to the inner edge of the ring, and the particles. Portia and Puck are both larger than the satellites Mab and Rosalind. It can be seen from Table 1 that Puck is ∼7 larger than Mab, which increases the superficial area of Puck and consequently the probability of collision with the particles. More than 50% of the $\mu$ ring particles collide with Puck.

When a particle hits the surface of the satellite, it may be absorbed or it can produce dust ejecta. At the time of the collision, we computed the relative velocities of the satellite and the particle, which allows us to calculate the impact velocity. The impact velocity is a reliable measurement of the energy carried by the particle. Depending on the impact velocity, the particle can be absorbed or dust ejecta can be produced.

The escape velocities ($v_{esc}$) of Puck, Mab, Portia, and Rosalind are approximately 70 m s$^{-1}$, 10 m s$^{-1}$, 60 m s$^{-1}$, and 30 m s$^{-1}$, respectively. Most of the collisions where $\bar{v} \gg v_{esc}$ can generate ejecta debris that can escape from the parent satellite. The fate of these debris depends strongly on the size of the satellite and the geometry during impact. Tables 3 and 4 show that most of the values of the impact velocity of the ring particles are comparable to the escape velocity of the satellite. Collisions that could produce dust ejecta are those between 1 $\mu$m sized particles and the satellites. All dust particles that collide with Mab have $\bar{v} \gg v_{esc}$. However, this means of suppling dust particles to the ring is insufficient, since the number of collisions is less than 3% of the total amount of analysed particles.

## 7. Final discussion

We have performed a numerical analysis to determine the orbital evolution of a sample of dust particles, located at the $\mu$ and $\nu$ rings, under the combined effects of the solar radiation force, the oblateness of Uranus, and the gravitational perturbation of the closest satellites.

Each component of the solar radiation force is responsible for a distinctive effect in the orbit of the particle. The semimajor axis of the particles decays due to PR drag, and the RP component causes large variations in the eccentricity. The effects on the semimajor axis of the particles does not change considerably when the oblateness of Uranus is taken into account. However, the damping of the eccentricity, due to the $J_2$ term, keeps the particles within the region of the rings and prevent the collision with the $\epsilon$ ring.

The close encounters with the satellites induce sudden variations in the semimajor axis of the particles, although the particles still do not leave the ring region. Consecutive encounters increase the probability of collisions between the particles and the satellites, especially for 1 $\mu$m particles, which are more sensitive to the effects of RP component. The number of collisions and the mean time of the collision depend on both the size of the particle and the width of the rings.

The total number of collisions is higher for the inner satellites Puck and Portia. All collisions with Portia and Rosalind have an impact velocity comparable to the value of the escape velocity, which could deposit material onto the surface of the satellite. Collisions between Puck and particles larger than 1 $\mu$m also have an impact velocity comparable to the value of the escape velocity. However, the impact velocity of those 1 $\mu$m particles that collide with Puck is almost seven times higher than the escape velocity. Further analysis is required in order to verify whether these collisions can produce dust ejecta. Only a few particles (3%) hit the surface of the satellite Mab at a velocity that can cause the ejection of material in the $\mu$ ring.

Other mechanisms of dust production, such as bombardment by micrometeoroids, will have to be introduced to explain the origin of the micrometric particles in the $\mu$ and $\nu$ rings. This study is under investigation.

**Table 3.** Results of the collision between $\mu$ ring particles and the close satellites (Mab and Puck) after 1000 years.

| Particle size | Puck | | | Mab | | |
|---|---|---|---|---|---|---|
| ($\mu$m) | $N_\%$ | $\overline{T}$ (years) | $\overline{v}$ (m s$^{-1}$) | $N_\%$ | $\overline{T}$ (years) | $\overline{v}$ (m s$^{-1}$) |
| 1 | 60 | 154 | 468 | 2 | 443 | 718 |
| 3 | 17 | 113 | 182 | 3 | 474 | 379 |
| 5 | 12 | 151 | 162 | 3 | 445 | 354 |
| 10 | 9 | 189 | 158 | 3 | 454 | 344 |

**Table 4.** Results of the collision between $\nu$ ring particles and the close satellites (Rosalind and Portia) after 1000 years.

| Particle size | Rosalind | | | Portia | | |
|---|---|---|---|---|---|---|
| ($\mu$m) | $N_\%$ | $\overline{T}$ (years) | $\overline{v}$ (m s$^{-1}$) | $N_\%$ | $\overline{T}$ (years) | $\overline{v}$ (m s$^{-1}$) |
| 1 | 42 | 31 | 164 | 52 | 12 | 163 |
| 3 | 23 | 131 | 73 | 31 | 120 | 88 |
| 5 | 19 | 147 | 69 | 29 | 176 | 80 |
| 10 | 16 | 196 | 58 | 27 | 234 | 79 |

*Acknowledgements.* The authors thank FAPESP, CAPES, and CNPq for the financial support. We also thank the anonymous referee for its helpful comments.

# Apêndice C

# Artigo MNRAS - Saturno

Cópia da versão publicada do artigo *Dynamical evolution of Saturn's F ring dust particles* publicado na revista *Monthly Notices of the Royal Astronomical Society*, 395.

# Dynamical evolution of Saturn's F ring dust particles

R. Sfair,★ S. M. Giuliatti Winter, D. C. Mourão and O. C. Winter

*UNESP – São Paulo State University, Guaratinguetá, CEP 12516-410, SP, Brazil*

## ABSTRACT

Saturn's F ring has been the subject of study due to its peculiar structure and the proximity to two satellites, named Prometheus (interior) and Pandora (exterior to the ring), which cause perturbations to the ring particles. Early results from Voyager data have proposed that the ring is populated with centimetre- and micrometre-sized particles. The Cassini spacecraft also detected a less dense part in the ring with width of 700 km. Small particles suffer the effects of solar radiation. Burns et al. showed that due to effects of one component of the solar radiation, the Poynting–Robertson drag, a ring particle will decay in the direction of the planet in a time much shorter than the age of the Solar system. In this work, we have analysed a sample of dust particles (1, 3, 5 and 10 μm) under the effects of solar radiation, the Poynting–Robertson drag and the radiation pressure components and the gravitational effects of the satellites Prometheus and Pandora. In this case, the high increase of the eccentricity of the particles leads almost all of them to collide with the outer edge of the A ring. The inclusion of the oblateness of Saturn in this system significantly changes the outcome, since the large variation of the eccentricity is reduced by the oblateness effect. As a result, there is an increase in the lifetime of the particle in the envelope region. Our results show that even the small dust particles, which are very sensitive to the effects of solar radiation, have an orbital evolution similar to larger particles located in the F ring. The fate of all particles is a collision with Prometheus or Pandora in less than 30 years. On the other hand, collisions of these particles with moonlets/clumps present in the F ring could change this scenario.

**Key words:** Solar system: general.

# 1 INTRODUCTION

The region which encompasses the orbits of the F ring and the close satellites Prometheus and Pandora has been studied since the Voyager spacecraft encountered the Saturnian System in 1980 and 1981. The unique structure of the F ring can be seen in details through the data sent by the Cassini spacecraft (Porco et al. 2005).

Modelling the F ring population has also been a challenge. Showalter et al. (1992) have proposed a model for the F ring population by using photometric observations and occultation data. Their model consists of a ring composed by centimetre-sized particles in a small core of about 1-km-wide and micrometre-sized dust in a much wider 'envelope' (about 500 km wide). They proposed that this core has an equivalent mass to a moonlet of 15–70 km radius distributed along the ring in a large number of small bodies.

This population of dust particles located in the envelope region of the F ring suffers the effects of perturbative forces. As has been analysed by Burns, Lamy & Soter (1979), the effects of the solar radiation can be divided into two components: the radiation pressure

(RP) and the Poynting Robertson (PR) drag. If the two components of the solar radiation could be separate, we could expect two different responses in the orbital motion of a dust particle. Under the effects of the RP, the eccentricity of the particle will oscillate and under the effects of the PR drag, which removes the orbital angular momentum from the particle through the absorption and the re-emission of the solar radiation, the semimajor axis of the particle will decrease until collision with the planet in a time less than the age of the Solar system (Burns et al. 1984).

The Cassini spacecraft sent several F ring images showing a 700-km-wide envelope which could be seen through the enhanced images (Porco et al. 2005). However, no work has been published so far to provide any further information regarding the size of these particles located in the envelope region.

Murray et al. (2005) have included in their numerical simulations of the F ring an envelope composed of dust particles. They showed that at some configurations Prometheus can enter the envelope region.

The goal of this paper is to analyse the orbital evolution of the dust particles located at the Saturnian F ring taking into account the solar radiation, the gravitational perturbations of Prometheus and Pandora and the oblateness of Saturn. In the next section, we derive the components of the solar radiation force; and in Section 3 we present our results. The conclusions are summarized in the last section.

★E-mail: rsfair@feg.unesp.br

## 2 SOLAR RADIATION FORCE

The dynamical system considered here is within the framework of the planar four-body problem, planet–satellites–ring particle and a time-dependent force. We have also included the planet's non-spherical mass distribution taking into account the quadruple term $J_2$, all other terms have been neglected. The vectorial expression for the solar radiation force on a particle around a planet is given by (Mignard 1984)

$$\boldsymbol{F} = \beta \left\{ \frac{\boldsymbol{r}_{sp}}{r_{sp}} \left[ 1 - \frac{\boldsymbol{r}_{sp}}{r_{sp}} \left( \frac{\boldsymbol{v}_p}{c} + \frac{\boldsymbol{v}}{c} \right) \right] - \left( \frac{\boldsymbol{v}_p}{c} + \frac{\boldsymbol{v}}{c} \right) \right\}, \quad (1)$$

where $c$ is the speed of light, $\boldsymbol{v}$ is the velocity vector of the particle relative to the planet, $\boldsymbol{r}_{sp}$ is the Sun–planet position vector, $\boldsymbol{v}_p$ is the velocity vector of the planet relative to the Sun and $r_{sp} = |\boldsymbol{r}_{sp}|$. The first term corresponds to the RP component and the other terms to the PR drag component. We assumed spherical particles which obey geometrical optics. The value of $\beta$ is given by (Burns et al. 1979)

$$\beta = 5.7 \times 10^{-5} \frac{Q_{PR}}{\rho s}, \quad (2)$$

where $s$ and $\rho$ are the radius and the density of a spherical particle in cgs units and $Q_{PR}$ is a constant value which depends on the optical properties of the grain.

The components of the solar radiation force can be given by

$$F_x = \frac{\beta G M_s}{r_{sp}^2} \left[ \cos(n_s t) - \left( \frac{x_s}{r_{sp}} \right)^2 \left( \frac{v_{xs}}{c} + \frac{v_x}{c} \right) \right. $$
$$\left. - \left( \frac{v_{xs}}{c} + \frac{v_x}{c} \right) \right] \quad (3)$$

$$F_y = \frac{\beta G M_s}{r_{sp}^2} \left[ \sin(n_s t) - \left( \frac{y_s}{r_{sp}} \right)^2 \left( \frac{v_{ys}}{c} + \frac{v_y}{c} \right) \right. $$
$$\left. - \left( \frac{v_{ys}}{c} + \frac{v_y}{c} \right) \right], \quad (4)$$

where $G$ is the gravitational constant, $M_s$ and $n_s$ are the mass and the mean motion of the Sun, respectively, $(x_s, y_s)$ are the coordinates of the Sun–planet position, $(v_x, v_y)$ are the components of the velocity of the particle. The subscript '$s$' in the velocity components refers to the Sun. The first term corresponds to the RP component and the other terms to the PR drag component. We have neglected the planetary shadow and the light reflected from the planet, since both effects are at least one order of magnitude weaker than the effect caused by the solar radiation (Hamilton & Krivov 1996).

## 3 NUMERICAL SIMULATIONS

We have numerically simulated a sample of dust particles initially located in the envelope region of the F ring. This region is 700 km wide and has a mean semimajor axis and eccentricity equal to the core of the F ring (Murray et al. 2005). These dust particles are disturbed by the gravitational perturbations of Prometheus and Pandora, the oblateness of Saturn ($J_2$ term (Jacobson et al. 2006)) and the RP and PR drag components.

The initial positions of the satellites adopted in the numerical simulations were derived from Jacobson et al. (2008). The values

| Parameters | Prometheus | Pandora | F ring core |
|---|---|---|---|
| $m$ (kg) | $1.59 \times 10^{17}$ | $1.37 \times 10^{17}$ | – |
| $a$ (km) | 139 380 | 141 710 | 140 224 |
| $e(\times 10^{-3})$ | 2.2 | 4.2 | 2.6 |
| $\varpi(°)$ | 161.0 | 83.4 | – |
| $M(°)$ | 242.3 | 202.5 | – |

of the mass ($m$), semimajor axis ($a$), eccentricity ($e$), longitude of pericentre ($\varpi$) and mean anomaly ($M$) of Prometheus and Pandora are listed in Table 1.

The sizes of the particles were chosen to be 1, 3, 5 and 10 μm. For each particle size, we have simulated 100 particles with initial semimajor axis uniformly distributed in the envelope region of the F ring (700 km). The other initial orbital parameters of the particles were $e = 2.6 \times 10^{-3}$, $M = 0$ and $\varpi = 0$.

The numerical simulations were performed using the Bulirsh–Stöer numerical integrator from the Mercury package (Chambers 1999). Some modifications were necessary in order to include the solar radiation force.

### 3.1 Without the effects of Saturn's oblateness

The effect of each component of the solar radiation was analysed separately. Fig. 1 shows the time evolution of the semimajor axis and eccentricity of four particles with size of 1, 3, 5 and 10 μm with the same initial conditions. The semimajor axis of the F ring's core is located at 0. As expected, the PR drag component is responsible for the decrease in semimajor axis of each particle leading to a collision with the planet. The rate of decrease of the semimajor axis depends on the size of the particle, the 1-μm-sized particle decays 100 km in about 60 years. An extrapolation of the decay time obtained in Fig. 1a shows that the particles collide with Saturn in $5 \times 10^4$ and $4 \times 10^5$ yr, considering 1- and 10-μm-sized particles, respectively. These values have the same magnitude of the estimation derived by Burns et al. (1979) (cf. their equation 55).

The RP component provokes an oscillation in the semimajor axis and a large variation in the eccentricities (Fig. 1b) of the particles. For 1-μm-sized particle, the eccentricity reaches large values (about 0.6), leading to a collision with the planet in less than 10 years. For larger particles, the eccentricity oscillates with a period equal to the orbital period of Saturn.

When both components of the solar radiation and the gravitational influence of the satellites Prometheus and Pandora are taken into account, the behaviour of the semimajor axis of the particles changes. Fig. 2 shows the orbital evolution of two particles, of 1 and 10 μm in size, initially located at the semimajor axis of the core of the F ring. By comparing Figs 1(a) and 2(a), we verified that the semimajor axis of both particles (fig. 2a) oscillates due to RP component. The smaller particle has an amplitude of oscillation of about 10 km. The temporal evolution of the semimajor axis of both particles also shows 'jumps' (Fig. 2a) which are associated with close approaches with one of the satellites (Winter et al. 2007). However, the temporal variations of the eccentricities presented in Figs 1(b) and 2(b) are similar, since, according to other numerical simulation that we have performed, the gravitational effects of the satellites on the eccentricities of the particles are small [$\mathcal{O}(10^{-5})$] when compared with the effects of the RP component. The vertical lines in Fig. 2 show the collision time between the particles and the A ring. The smaller particle collides with the A ring in less than ~3 months and the larger particle in less than 2.3 years. We will discuss the collision time in the next section.

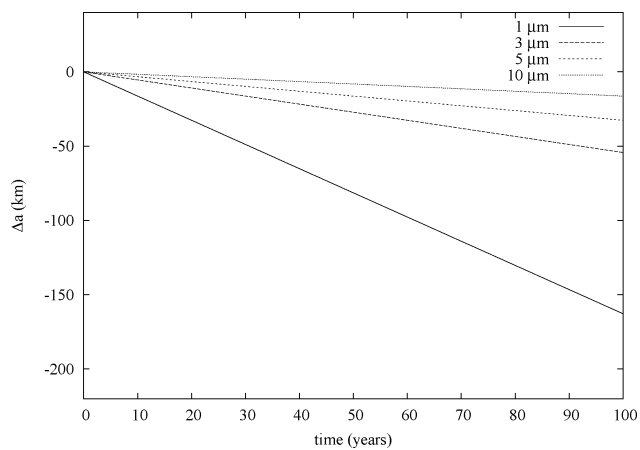

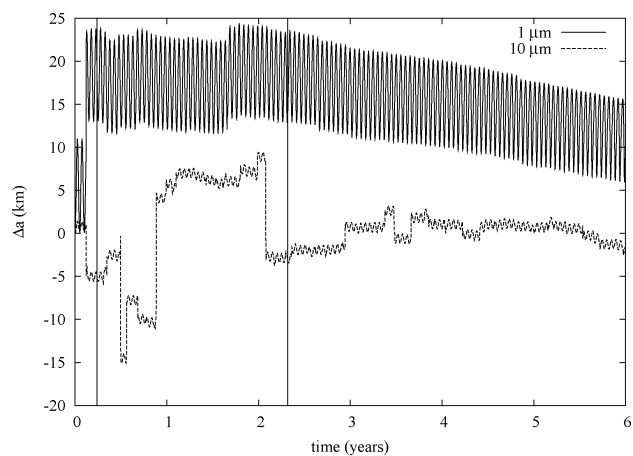

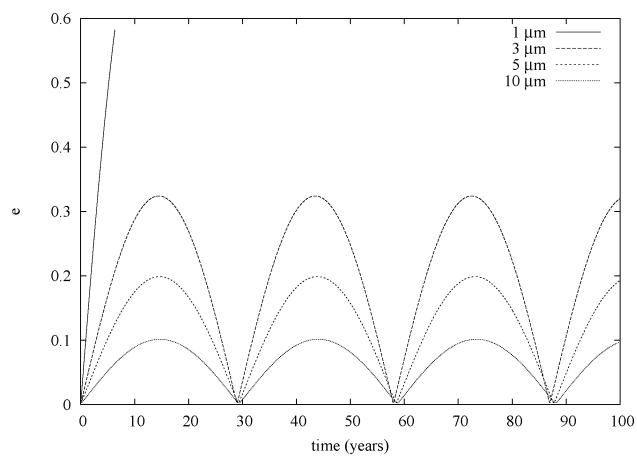

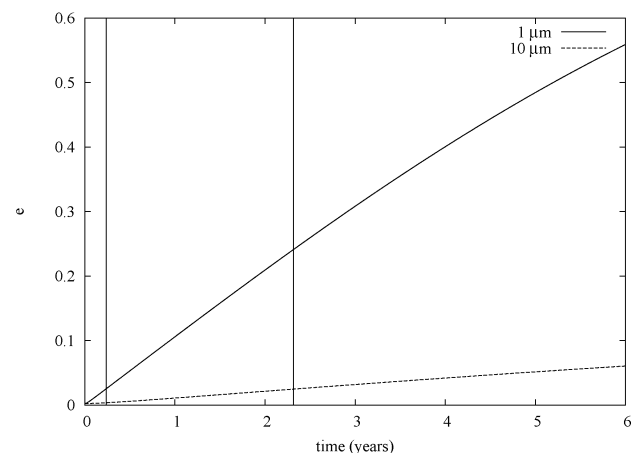

**Figure 1.** Time evolution of the (a) semimajor axis due to the effects of the PR drag only and (b) the eccentricity due to the RP component of four particles of different sizes. The semimajor axis of the F ring's core is located at 0.

**Figure 2.** Evolution of the (a) semimajor axis and (b) eccentricity of 1- and 10-μm-sized particles under the effect of the solar radiation and the gravitational perturbation of Prometheus and Pandora. The semimajor axis of the F ring's core is located at 0. Vertical lines represent the collision time between the particles and the A ring.

### 3.2 With the effects of Saturn's oblateness

In this section, we compare the effects caused by the inclusion of the gravity coefficient $J_2$ in the orbital evolution of a dust F ring particle. Fig. 3 shows the variation of the eccentricities of two particles, sizes of 1 and 10 μm, with identical initial conditions. As can be seen in Fig. 1(b), the RP component induces larger variation in the eccentricity of the smallest particle. For example, the eccentricity of the 1-μm-sized particle reaches the value 0.6, and for a 10-μm-sized particle the eccentricity increases up to 0.1. When the effect of Saturn's oblateness is considered, the amplitude of the eccentricity decreases to a value of ~0.01 for both particles, as shown in Fig. 3. This result was explained by Hamilton & Krivov (1996). They analysed the effects on a dust particle when the planetary oblateness and the RP component act simultaneously. For the 1-μm-sized particle located up to 9 $R_p$ (see their fig. 1), where $R_p$ is the Saturnian radius, the effect of the planetary oblateness is stronger than the perturbation caused by the PR drag component.

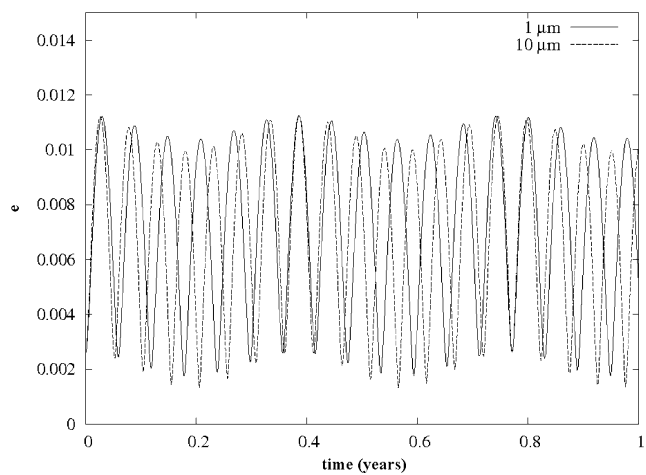

**Figure 3.** Variation of the eccentricity of two particles of size 1 and 10 μm disturbed only by the solar radiation and the oblateness of Saturn. The initial conditions of the two particles are identical.

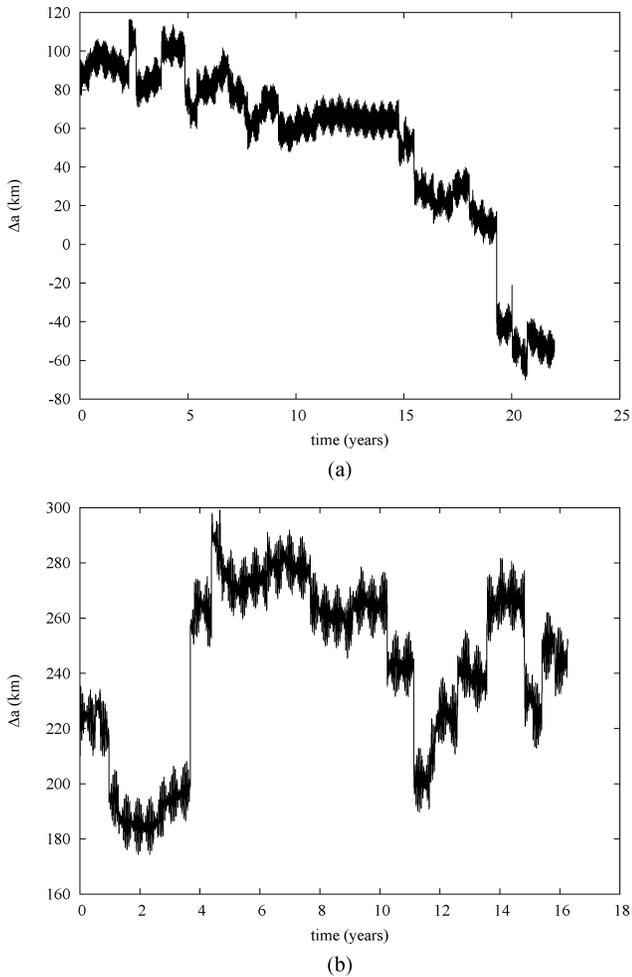

**Figure 4.** Variation of the semimajor axis of (a) a 5-μm-sized particle and (b) a 3-μm-sized particle under the effects of the solar radiation, the gravitational perturbation of the satellites and the oblateness of Saturn. The particle represented in (a) collides with Prometheus, while the particle in (b) has a collision with Pandora. The semimajor axis of the F ring's core is located at 0.

The strength of the planetary oblateness decreases with the increase of the distance from the planet. On the other hand, the strength of the PR drag component is weaker for those particles located near the planet. The period of the variation of the eccentricities (less than 0.1 year) is much smaller than the period showed in our Fig. 1(b) (about 30 years).

In Fig. 4, the evolution of the semimajor axis of two representative examples of our numerical simulations for 5- and 3-μm-sized particles under the effects of the solar radiation force, the gravitational perturbation of Prometheus and Pandora and the oblateness of Saturn are presented. As can be seen in Fig. 1(a), a particle of 5 μm size decays ∼30 km in 100 years due to the PR drag component. The sudden changes in the semimajor axis of the particles, seen in Fig. 4, are due to close encounters between the particle and one of the satellites, which induces an increase or decrease of its value (Winter et al. 2007). Due to the oblateness of Saturn, the collision with one of the satellites will also depend on the variation of the semimajor axis, since larger variations of the eccentricity, due to the RP component, are damped by the effects caused by the oblateness.

We have computed the number of the particles that collide with one of the satellites or with the outer edge of the A ring, located at ∼136 780 km from Saturn. We have also calculated the mean collision time, $\langle T_c \rangle$, and its standard deviation. Table 2 presents these results for those particles under the gravitational effects of the satellites and the solar radiation. As can be seen, most of the particles collide with the outer edge of the A ring in less than 3 years. Table 3 presents the results for the numerical simulations of the system that also took into account the oblateness of Saturn. A comparison of Tables 2 and 3 shows a remarkable difference on the fate of the particles. The effects of the oblateness of the planet are such that the particles collide with one of the satellites before reaching the outer edge of the A ring. This is due to the fact that the oblateness damps the variation of the eccentricity. A ring particle located at the outer edge of the envelope needs an eccentricity of ∼0.025 to cross the outer edge of the A ring. However, the variation of the eccentricity, when the oblateness is taking into account, reaches a maximum value ∼0.012 (Fig. 3), avoiding the collision with the outer edge of the A ring. By comparing Tables 2 and 3, it can be seen that there is an increase of the collision time, when the oblateness of the planet is included, independently of the size of the particle.

**Table 2.** The fate of the dust particles under the gravitational effects of the satellites and the solar radiation.

| Particle (μm) | Prometheus collision | Pandora collision | A ring collision | $\langle T_c \rangle$ (yr) |
|---|---|---|---|---|
| 1 | 3 | 1 | 96 | 0.25(±0.03) |
| 3 | 2 | 6 | 92 | 0.7(±0.1) |
| 5 | 9 | 10 | 81 | 1.1(±0.25) |
| 10 | 20 | 15 | 65 | 1.9(±0.6) |

**Table 3.** The fate of the dust particles when the oblateness of Saturn is also taken into account.

| Particle (μm) | Prometheus collision | Pandora collision | A ring collision | $\langle T_c \rangle$ (yr) |
|---|---|---|---|---|
| 1 | 57 | 43 | 0 | 6.5(±5.8) |
| 3 | 74 | 26 | 0 | 5.2(±4.8) |
| 5 | 76 | 24 | 0 | 7.3(±6.2) |
| 10 | 68 | 32 | 0 | 7.7(±6.5) |

From Table 3 we have also noted that about 70 per cent of the particles collide with Prometheus, since this satellite is closer to the ring. The only exception occurs for 1-μm-sized particle, because its semimajor axis presents a larger oscillation than the semimajor axis of the other particles, which is caused by the PR drag component.

Fig. 5 shows the collisional time $(T_c)$ as a function of $\Delta a_0$ for a sample of particles of 1, 3, 5 and 10 μm in size. The F ring's core is located at 0. The full lines, in Fig. 5, represent those particles under the gravitational effects of Prometheus and Pandora and the solar radiation pressure. Almost all particles collide with the outer edge of the A ring in a time less than 3 years. The increase of the eccentricities, due to the RP component, of these tiny particles leads to a large variation in the orbital radius which allows the orbits of the particles to cross the outer edge of the A ring. In Fig. 5, those particles represented by the dashed line are also perturbed by the oblateness. When this effect is included, the collision time increases for all particles.

The difference in the collision time is due to the fact that the planetary oblateness damps the eccentricity caused by the RP component. Our results show that when the oblateness of Saturn is present almost 70 per cent of the particles collide with Prometheus. For this case, the variation of the orbital radius is smaller

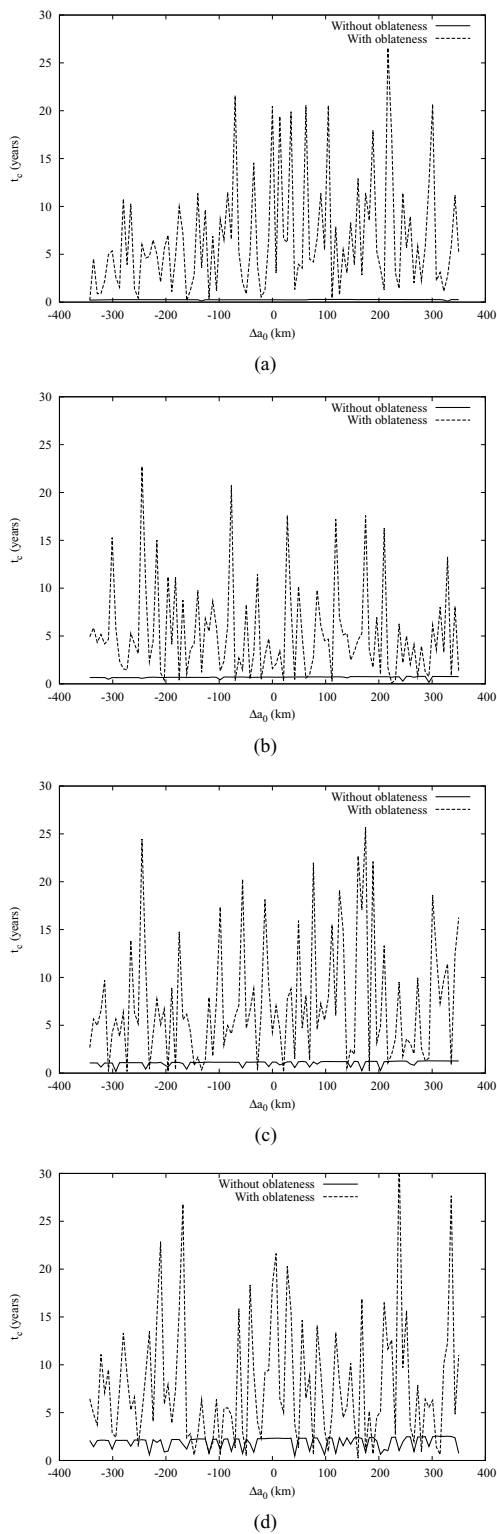

**Figure 5.** Collision time ($t_c$) for particles of different sizes: (a) 1-μm-, (b) 3-μm-, (c) 5-μm- and (d) 10-μm-sized particles as a function of the initial semimajor axis $a_0$. Dashed lines represent those particles perturbed by the solar radiation pressure and Prometheus and Pandora, and the full lines represent those particles perturbed also by the $J_2$ term. The semimajor axis of the F ring's core is located at 0.

allowing more collisions with Prometheus since it is closer to the envelope region.

## 4 FINAL COMMENTS

In this work, we have analysed the behaviour of a sample of F ring dust particles under the effects of the solar radiation force and the gravitational perturbations of Prometheus and Pandora, in orbit around an oblate Saturn.

When the oblateness of Saturn is not considered, the large variation of the eccentricity, induced by the RP component, leads most of the particles to a collision with the outer edge of the A ring. However, the variation in the eccentricity of the F ring particles is decreased by the effect of the planetary oblateness. As a result, there is an increase in the lifetime of the particle in the envelope region.

Our results show that the dust particles, which are very sensitive to the effects of the solar radiation pressure, have an orbital evolution similar to the population of large particles.

Most of these particles (about 70 per cent) collide with Prometheus in less than 30 years. It is important to point out that these dust particles spend most of the time, before colliding with the satellites, wandering in the F ring region. Cassini data showed that moonlets and clumps populate the F ring (Porco et al. 2005). The paper by Winter et al. (2007) has shown that these bodies are in confined chaotic orbits. Therefore, the dust particles can collide with these moonlets/clumps before leaving the F ring region.

## ACKNOWLEDGMENTS

The authors thank FAPESP, CAPES and CNPq for the financial support. We thank the Ubuntu-BR Community. We thank C. Murray for their critical reading of the manuscript.

This paper has been typeset from a TEX/LATEX file prepared by the author.

# Apêndice D

# Resumo Europlanet

Cópia do resumo enviado para o *2009 European Planetary Science Congress*. O congresso foi realizado em Potsdam, Alemanha, e o trabalho foi apresentado pelo Dr. Mark Showalter.

Além deste congresso, o trabalho *The Brightening of Saturn's F Ring* foi apresentado em outros eventos, tais como o 41th DPS Meeting e 40th DDA Meeting, ambos organizados pela American Astronomical Society.

# The Brightening of Saturn's F Ring

**M. R. Showalter** (1), R. S. French (1), R. Sfair (2), C. Arguelles (3), M. Pajuelo (3), P. Becerra (3), M. M. Hedman (4) and P. Nicholson (4)

(1) SETI Institute, Mountain View, California, USA, (2) Sao Paolo State University, Sao Paolo, Brazil, (3) Pontifica Universidad Catolica del Peru, Lima, Peru, (4) Cornell University, Ithaca, New York, USA (mshowalter@seti.org / Fax: +1-650-962-9419)

## Abstract

Saturn's F ring has brightened markedly in the last 25 years. It is twice as bright in the Cassini data as it was in the Voyager data from 1980 and 1981. This conclusion is based on photometric comparisons of Cassini and Voyager images, and is supported by occultation data. We attribute this change to increasing perturbations by nearby Prometheus, which passes closer to the ring now that it did in the Voyager era, yielding more dust.

## Image Analysis

Our analysis procedures follow those originated by Showalter et al. 1992 [1] for the Voyager images. Our primary Cassini data set consists of the narrow-angle, clear-filter images from 19 "FMOVIEs," obtained by staring at a ring tip while the ring rotated underneath. These sample many ring longitudes, enabling us to average out the large intrinsic variations within the ring. This data set is supplemented by ~ 300 wide-angle images, which capture more phase angles but have incomplete longitudinal coverage. We calibrated the images and obtained profiles of ring reflectivity versus radius from each. We measured the area under the curve to yield "normal equivalent width" *NEW*, defined as

$$NEW = \sin(B) \int I/F \, dr \quad,$$

where *I/F* is the reflectivity (intensity *I* normalized by solar flux density), *r* is orbital radius and *B* is the ring opening angle. *NEW* has the properties that it is independent of image resolution and is also independent of *B* in the limit where optical depth is small. Figure 1 shows our raw measurements. The color coding (red, yellow, green and blue for increasing |*B*|) shows that *NEW* is systematically smaller for observations taken closer to the ring plane. This suggests that optical depth τ is not

small enough for its effects to be ignored. Using standard models for mutual shadowing within the ring (but neglecting multiple scattering), we find that assuming τ = 0.1 eliminates any systematic *B*-dependence, and thereby reduces the scatter among the points (Fig. 2). We interpret this value as representing the mean optical depth of the ring. The phase curve increases toward high phase angles due to diffraction by the prevalent, micrometer-sized dust grains in the ring. Photometric modeling shows that this curve is consistent with a differential size distribution of the form $n(a) \sim a^p$, where *a* is the particle radius. The data imply *p* = 3.5–4; such a power-law model is consistent with that found in other dusty rings.

For the Voyager data, we have repeated the analysis of Showalter et al. [1] using the latest calibration and analysis techniques. Voyager measurements show the same *B*-dependence, which can be eliminated by assuming τ ~ 0.02–0.04 (Fig. 3). These measurements fall consistently below the Cassini measurements by a factor of 2–3, depending on the exact value of τ assumed. No plausible assumptions about τ can reconcile the two sets of measurements.

If the ring had a strong spectral slope, then the discrepancy might be explained by the different bandpasses of the Voyager and Cassini cameras. However, our measurements indicate that the ring color is neutral. We have also performed a few comparisons of the A ring edge, imaged by both spacecraft under similar lighting and viewing geometries, to confirm that our calibrations are consistent. We therefore conclude that we have detected an actual change in ring properties. However, note that the phase curves from Voyager and Cassini (Figs. 2 and 3) have a nearly constant ra-tio, suggesting that the distribution of particle sizes is essentially unchanged.

**Occultation Data**

We have investigated the optical depth of the ring directly using occultation profiles from Voyager's photopolarimeter (PPS) instrument and from Cassini's Visual and Infrared Mapping Spectrometer (VIMS). We measure the amount of light blocked as a star passes behind the ring, and infer a "normal equivalent depth" *NED*, representing the radial integral of $\tau$ after correction for ring opening angle *B*. *NED* from the single PPS occultation is smaller than that from *any* of the 21 VIMS occultations examined, and it is less than half of the VIMS mean. Because this one measurement could be a statistical outlier, we cannot draw firm conclusions; nevertheless, it provides additional support for the hypothesis that the ring has changed. Measured $\tau$ values are generally consistent with our inferences above based on the image analysis.

**Discussion**

The F ring is almost surely held together by an unseen population of meter- to km-sized bodies. These bodies contain most of the ring's mass, but are masked by the far more numerous fine dust particles. The recent change probably represents an increase in the dust production rate, rather than a change in the number of underlying source bodies. Stirring by Prometheus is probably the dominant driver for dust production. Both the moon and ring are on eccentric orbits, and their minimum separation distance varies as the pericenters go into and out of alignment. The pericenters are approaching anti-alignment, so we hypothesize that closer passages by Prometheus are driving the recent increase in dust.

**Acknowledgments**

This work was supported by NASA's Cassini Data Analysis Program through grant NNX07AJ76G. We acknowledge the contributions of the Cassini ISS and VIMS teams as well as those of their predecessors on the Voyager Project.

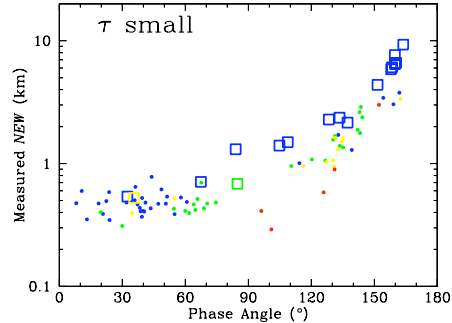

Figure 1: Raw photometry from the Cassini images. Large squares are averages from FMOVIEs; dots are measurements from wide-angle images. Colors red, yellow, green and blue indicate increasing ranges of |*B*| in 5° steps.

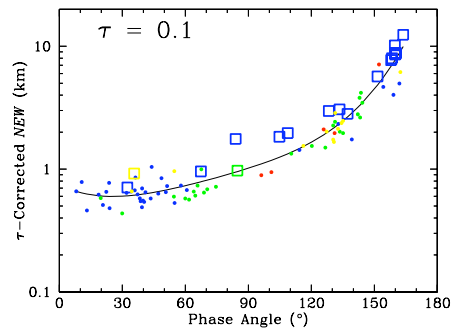

Figure 2. The same photometry as above, but corrected for $\tau = 0.1$. The scatter among the points is smaller and no dependence on |*B*| is evident. A quartic polynomial fit to the data is shown as an approximation to the underlying phase function.

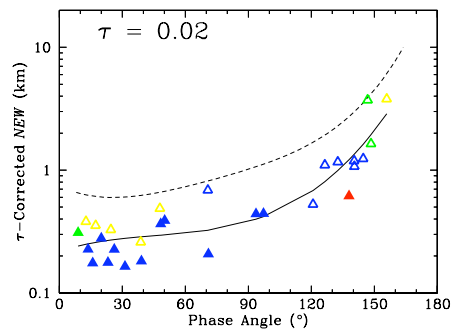

Figure 3. Measurements similar to those above but from the Voyager camera. Open triangles are from Voyager 1; closed triangles are from Voyager 2. A best-fit quartic phase curve is also shown (solid) and the Cassini fit from above is shown for comparison (dashed).

# Apêndice E

# Artigo MNRAS - Plutão

Cópia da versão publicada do artigo *Gravitational effects of Nix and Hydra in the external region of the Pluto–Charon system* publicado na revista *Monthly Notices of the Royal Astronomical Society*, 410.

# Gravitational effects of Nix and Hydra in the external region of the Pluto–Charon system

## P. M. Pires dos Santos,[★] S. M. Giuliatti Winter and R. Sfair

*UNESP-São Paulo State University, Guaratinguetá, CEP 12.516-410 SP, Brazil*

## ABSTRACT

Two new companions to the Pluto–Charon binary system have been detected in 2005 by Weaver et al. These small satellites, named Nix and Hydra, are located beyond Charon's orbit. Although they are small when compared to Charon, their gravitational perturbations can decrease the stability of the external region (beyond Charon's orbit). The dynamical structure of this external region is analysed by numerically simulating a sample of particles under the gravitational effects of Pluto, Charon, Nix and Hydra. As expected the effects of Nix and Hydra decrease the external stable region. Agglomerates of particles can survive even after $10^5$ orbital periods of the binary in some regions, such as coorbital to Nix and Hydra and between their orbits. We also analysed the effects of hypothetical satellites on the orbital evolution of Nix and Hydra in order to constrain an upper limit size. Some hypothetical satellites can be coorbital to Nix or Hydra without provoking any significant gravitational effects on them.

**Key words:** Kuiper belt: general – planetary systems.

## 1 INTRODUCTION

The first work to study the stability in the region of Pluto–Charon system was performed by Stern et al. (1994). They analysed the orbits of a sample of test particles under the gravitational effects of Pluto and Charon. Their results showed an unstable region interleaved between ∼1.8 A and 2.4A, where $A = 19\,600$ km, for test particles initially with inclination ($I$) and eccentricity ($e$) equal to zero. The integration time was $10^5$ orbital periods of the binary ($T_{P-C}$). An additional study was performed in order to verify how a massive hypothetical satellite can exist in the stable region of the system without causing any significant gravitational effects on Charon's eccentricity. Their results show that no satellites larger than $3 \times 10^{-4} m_{P-C}$, where $m_{P-C}$ is the mass of the binary, can exist in the internal (0.1–0.4A) and external (at 2A) regions without inducing an eccentricity on Charon's orbit larger than $10^{-3}$.

Nagy, Süli & Érdi (2006) analysed the dynamical structure of the phase space of the internal and external regions of the Pluto–Charon system through a set of numerical simulations of the spatial circular restricted three-body problem (Pluto–Charon particle). The gravitational effects of Nix and Hydra were not taken into account in the dynamical system. They numerically simulated a sample of particles initially located in the external region at semimajor axis $a$ (relative to the barycentric reference frame) between 0.55B and 5B, where $B = 19\,571.4$ km. The eccentricities of the particles were taken from 0 to 0.3. The results (see their fig. 3), concerning particles for the planar case ($I = 0$), showed that for $a < 2.15B$ the system is unstable for all values of the eccentricity $e$ for a time-span of $10^3$ orbital periods of Charon. However, for $a \geq 2.15B$ there is a stable region depending on the value of $e$. Between Nix's and Hydra's orbits the eccentricity can reach values of up to 0.31.

An extension of this work has been presented by Süli & Zsigmond (2009) through an analysis of the spatial elliptic restricted three-body problem for a time-span of $10^4$ orbital periods of the binary. They generated stability maps for the $(a - e)$ and $(a - I)$ orbital element spaces near the satellites Nix and Hydra (taken as massless bodies) using three different complementary methods. As a result they obtained an unstable zone larger than the unstable zone derived from the circular restricted three-body problem. Both satellites, Nix and Hydra, are in stable regions. Nix might be in the 4:1 mean motion resonance if the values of the argument of pericentre or longitude of node falls in a certain range, while Hydra is not in the 6:1 mean motion resonance for arbitrary values of the argument of pericentre or longitude of nodes.

The encounter between the New Horizons mission and the Pluto system in 2015 will help to reconnaissance this peculiar system. The entire Hill sphere will be searched for additional rings and satellites during the approach phase (Young et al. 2008). Stern et al. (2006) proposed the possibility of rings in the Pluto system with a characteristic ring optical depth of $\tau = 5 \times 10^{-6}$ between Nix and Hydra. Steffl et al. (2006) analysed a set of *Hubble Space Telescope* data in order to search for additional satellites in the external and internal regions. By assuming spherical satellites and no limb darkening they ruled out (at the 90 per cent level of confidence) additional satellites with diameters larger than ∼50 km located between the orbits of Nix and Hydra, and ∼36 km beyond Hydra's orbit. These values were calculated by assuming a very dark albedo of $\rho_v = 0.04$. If they assumed that these hypothetical satellites are as reflective as Charon, the values of the diameters decrease to 16 and 12 km, respectively.

The goal of this paper is to explore the external region taking into account the gravitational effects of Nix and Hydra on test particles. We also analysed the effects that a sample of small hypothetical satellites can induce on Nix's and Hydra's orbital eccentricities. These results can place an additional upper limit on the estimated size of these hypothetical satellites.

In the next section we present a sample of diagrams of semimajor axis versus eccentricity for particles in P-type orbits around the barycentre of Pluto–Charon system after approximately $10^5$ orbital periods of the binary. In Section 3 we discuss the gravitational effects of a sample of small hypothetical satellites on Nix's and Hydra's orbital eccentricity. Our results are discussed in the last section.

## 2 DIAGRAMS (A-E)

Holman & Wiegert (1999) divided the orbits of a binary into three classes, following the designation given by Dvorak (1986): in the P-type orbits the particle is around the barycentre of the system, in the S-type orbits the particle is around one of the massive bodies and the third class is related to those particles around the Lagrangian triangular points, $L_4$ and $L_5$ (Murray & Dermott 1999). In this section we analyse the gravitational perturbations due to Nix and Hydra on particles in P-type orbits around the barycentre of the two massive bodies Pluto and Charon.

Table 1 gives the orbital parameters, diameters and masses of Charon, Nix and Hydra. The orbital parameters $a, e, I, \omega, \Omega$ denote the semimajor axis, eccentricity, inclination, argument of pericentre and longitude of the ascending node, respectively. The orbital parameters of Charon are relative to Pluto and those of Nix and Hydra are relative to the barycentre of the system. The mass of the satellites was obtained assuming the density equals to 1.63 g cm$^{-3}$. Pluto's mass is taken to be equal to $1.304 \times 10^{22}$ kg (Tholen et al. 2008). The orbital period of the binary is 6.3872 d (Tholen et al. 2008).

The orbital plane of Pluto–Charon was taken to be the reference plane. The numerical integrations have been carried out using the variable time-step Bulirsch–Stoer algorithm from the Mercury package (Chambers 1999). The perturbation of the Sun was neglected since we have simulated particles located up to $10^5$ km.

**Table 1.** Satellite orbital elements and physical parameters used in the numerical simulations (derived from Tholen et al. 2008), at Epoch JD 245 2600.5. The orbital parameters $a, e, I, \omega, \Omega$ denote the semimajor axis, eccentricity, inclination, argument of pericentre and longitude of the ascending node, respectively. The orbital parameters of Charon are relative to Pluto and the elements of Nix and Hydra are relative to the barycentre of the system. The masses of Charon, Nix and Hydra were obtained by assuming the density equals to 1.63 g cm$^{-3}$.

| Parameters | Charon | Nix | Hydra |
|---|---|---|---|
| $a$ (km) | 19 570.3 | 49 240 | 65 210 |
| $e$ | 0.0035 | 0.0119 | 0.0078 |
| $I$ (°) | 96.168 | 96.190 | 96.362 |
| $\omega$ (°) | 157.9 | 244.3 | 45.4 |
| $\Omega$ (°) | 223.054 | 223.202 | 223.077 |
| Diameter (km) | 1212 | 88 | 72 |
| Mass (kg) | $1.520 \times 10^{21}$ | $5.8 \times 10^{17}$ | $3.2 \times 10^{17}$ |

### 2.1 Prograde orbits

To obtain the diagrams we use a set of 404 505 initial conditions varied in the following ways.

(i) the semimajor axis (relative to the barycentre of the system) was distributed from $1d$ to $5d$ ($d = 19\,570$. km), with $\Delta a = 0.005d$.

(ii) The eccentricity assumed values from 0 to 0.2, with $\Delta e = 0.05$.

(iii) A set of 101 values of the argument of pericentre was randomly chosen between $0°$ and $360°$ for particles in prograde orbits. All particles started at the true anomaly equals to zero. The inclination was assumed to be 0 (prograde orbits) relative to the Pluto–Charon orbital plane.

The orbital evolution of the test particles are represented in Figs 1(a)–(c). We considered that the particle leaves the system when the distance between the particle and the barycentre is $10d$. When the distance between the particle and a massive body was less than the body's radius a collision was detected.

Fig. 1 shows the final values of the semimajor axis and eccentricity for a sample of particles, in prograde orbits, which survived after (a) 10 000 d (about $10^3 T_{P-C}$), (b) 70 000 d (about $10^4 T_{P-C}$) and (c) 650 000 d (about $10^5 T_{P-C}$). The satellites Nix and Hydra are located at 2.516$d$ and 3.332$d$, respectively. Particles initially located at about $1d < a < 3d$ are represented in red, those particles initially at about $3d < a < 4d$ are in green and those at about $4d < a < 5d$ are represented in blue colour. Different colours can help to visualize the orbital evolution of the particles.

The instability region near Charon's orbit extends up to $2d$, as firstly verified by Stern et al. (1994). These particles leave the system in less than $10^3 T_{P-C}$. Some clumps of particles start forming interior to Nix's orbit, coorbital to Nix and Hydra and in a small region between the orbits of these two small satellites. In Fig. 1(c) these clumps are clearly visible at about 1.8$d$, 2$d$ and 2.3$d$ interior to Nix's orbit. As discussed by Stern et al. (1994) these clumps might be associated to mean motion resonances with Charon. An agglomerate of particles can also be seen at $2.49d \leq a \leq 2.54d$ and $3.27d \leq a \leq 3.38d$, for values of $e \leq 0.1$. Particles with $a$ ranging between 2.78$d$ and 3.01$d$ (between the orbits of Nix and Hydra) also survived for this time-span of the integration. Particles with eccentricities larger than 0.1 were scattered due to close encounters with Nix and Hydra.

As can be seen in Fig. 1 the gravitational perturbations of both satellites Nix and Hydra on those test particles located beyond Hydra's orbit are small. Some structures appear to be associated with resonances between the particles and the satellites. Two bodies are in mean motion resonance when $n/n' = p/(p + q)$, where $n$ and $n'$ are their mean motions, $p$ and $q$ are small integers and $q$ is of the order of the resonance. The mean motions are measured in the sidereal system (Sülli & Zsigmond 2009).

The locations of the mean motion resonance between the particle and the satellite Nix (top) and Hydra (bottom) are indicated in the figure. Particles near to a resonance had their eccentricities increased; this effect can be visualized, for example, for those particles located near the 3 : 2 mean motion resonance with Hydra.

Distinct regimes can be caused by the gravitational perturbations of an embedded satellite, in circular orbit around the primary, in a sample of massless particles. A chaotic zone, where the particles would be scattered by the gravitational perturbations of the satellite,

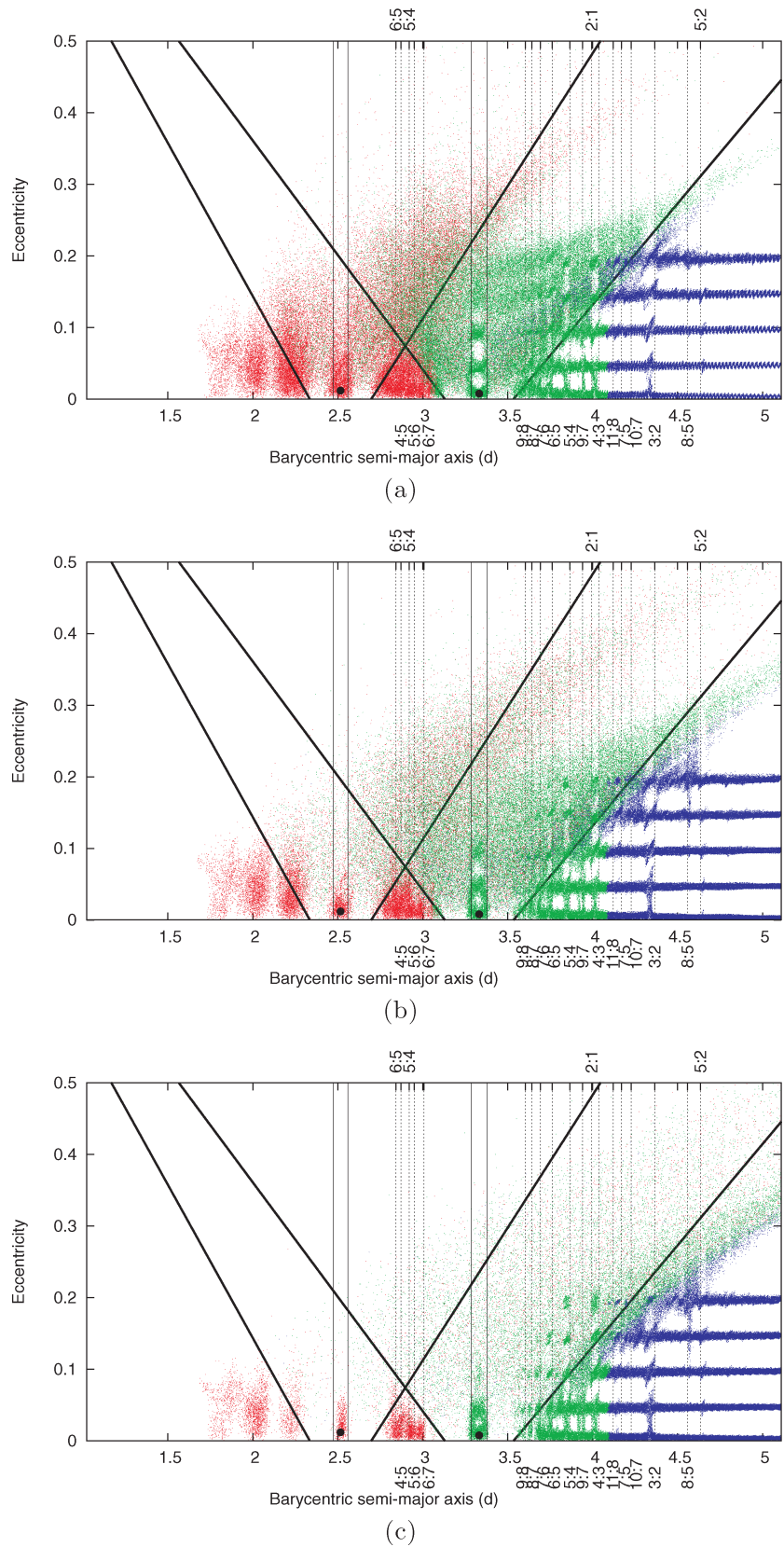

**Figure 1.** Diagram of the final semimajor axis and eccentricity for a set of test particles, in prograde orbits, after (a) 10 000 d, (b) 70 000 d and (c) 650 000 d. The satellites Nix and Hydra (represented by black circles) are located at 2.516*d* and 3.332*d*, respectively. The locations of the mean motion resonance between the particle and the satellite Nix (top) and Hydra (bottom) are indicated in this figure. Particles initially located at about 1*d* < *a* < 3*d* are represented in red, those particles initially at about 3*d* < *a* < 4*d* are in green and those at about 4*d* < *a* < 5*d* are represented in blue colour.

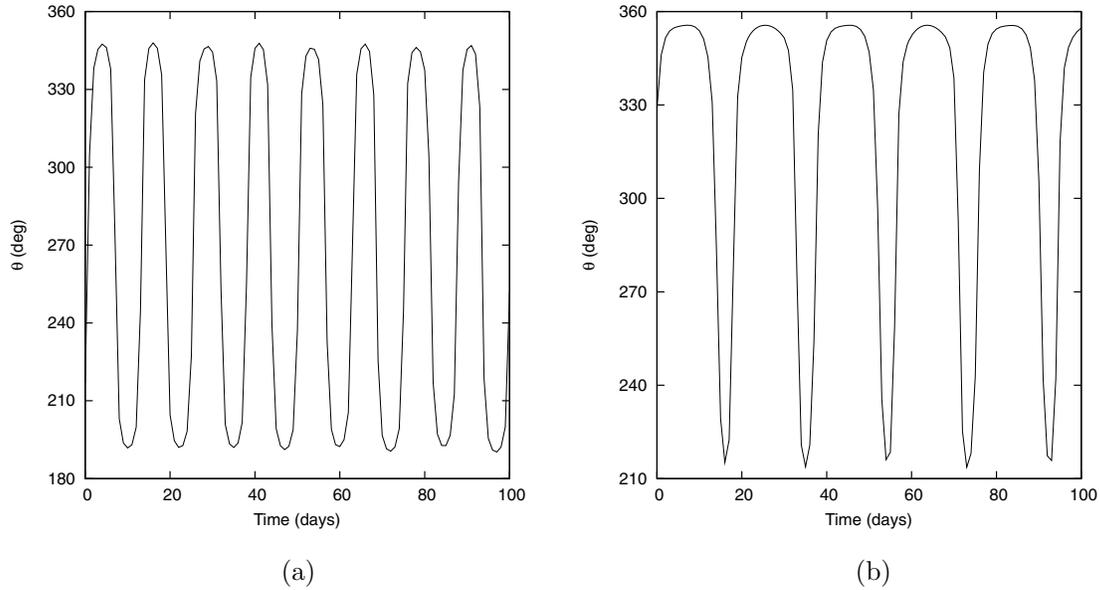

(a)                                            (b)

**Figure 2.** The angle $\theta$ versus time for one representative coorbital particle to (a) Nix located at $2.516d$ and (b) Hydra at $3.332d$.

has a width ($w_{ch}$) given by (Wisdom 1980)

$$w_{ch} = 2 \times 1.3\mu^{2/7}a_{sat}, \tag{1}$$

where $\mu$ is the mass ratio between the mass of the satellite ($m_{Nix}$ or $m_{Hydra}$) and the mass of the binary ($m_{P-C}$) and $a_{sat}$ is the semimajor axis of the satellite. The width of the gap, derived from equation (1), generated by Nix is $\sim7000$ km and by Hydra is $\sim7900$ km. These values derived from the theory fit well the values shown in Fig. 1(c) for the eccentricity equals to 0.

Coorbital particles to Nix and Hydra have also survived after $10^5$ $T_{P-C}$. From the circular restricted three-body problem the width of the coorbital region can be given by (Murray & Dermott 1999)

$$w \sim 0.5\mu^{1/3}a_{sat}. \tag{2}$$

From equation (2), the coorbital region generated by Nix is $\sim1700$ km and by Hydra is $\sim1800$ km. These values are represented by vertical lines, despite the fact that the width of the coorbital region, given by equation (2), is valid for the eccentricity of the particle equals to 0.

We have numerically simulated the agglomerates of particles located at $2.5d$ (near to Nix's orbit) and $3.3d$ (near to Hydra's orbit) in order to verify if the angle $\theta$, the angular separation between the particle and the satellite, is smaller than $1°$. Values of $\theta < 1°$ imply that the particle is no longer coorbital to the satellite. During the numerical integration, the majority of the particles remained coorbital to the satellites Nix and Hydra. As can be seen in Fig. 1, only a small amount of particles do not belong originally in the coorbital region. Fig. 2 shows the angle $\theta$ as a function of time of a representative coorbital particle to Nix (Fig. 2a) and to Hydra (Fig. 2b).

The lines shown in Fig. 1 represent the crossing lines. Those particles which enter the region limited by these lines are crossing the orbits of Nix or Hydra. Therefore, these particles will be ejected from the system or they will collide with the satellite.

Some particles cross the orbits of the satellites and can collide with them during the numerical simulation. Table 2 summarizes the numbers of collisions between the particles and the four massive bodies (Pluto, Charon, Nix and Hydra) for particles in prograde and retrograde (Section 2.2) orbits. Our results, concerning particles in prograde orbits, show that about 50 per cent of the initial set of the particles survived after $10^5$ $T_{P-C}$, most of them were initially located beyond Hydra's orbit. About 32 per cent of the particles were ejected from the system.

**Table 2.** Result of the collisions between the test particle, in prograde and retrograde orbits, with Pluto, Charon, Nix and Hydra after $10^5$ $T_{P-C}$, and the percentage of the particles ejected from the system.

|             |        | Prograde (per cent) | Retrograde (per cent) |
|-------------|--------|---------------------|-----------------------|
|             | Pluto  | 8                   | 0.6                   |
|             | Charon | 10                  | 1.5                   |
| Collisions  | Nix    | 5.5                 | 11                    |
|             | Hydra  | 5                   | 11                    |
|             | Total  | 29                  | 24                    |
| Escape      |        | 32                  | 5                     |

### 2.2 Retrograde orbits

To obtain the diagram presented in Fig. 3 we use a set of 40 050 initial conditions varied in the following ways.

(i) The values of the semimajor axis and the eccentricity were the same as assumed for those particles in prograde orbits (Section 2.1).

(ii) The argument of pericentre have been randomly distributed into 10 different values. All particles started at the true anomaly equals to zero. The inclination was assumed to be $180°$ (retrograde orbits) relative to the Pluto–Charon orbital plane.

Fig. 3 shows the final values of the semimajor axis and eccentricity of a sample particles, in retrograde orbits, after $10^5$ $T_{P-C}$. The lines across this figure represent the collision lines, as discussed in Section 2.1. Particles inside these lines may collide with one of the smaller satellites or may be ejected from the system, resulting in a less dense region.

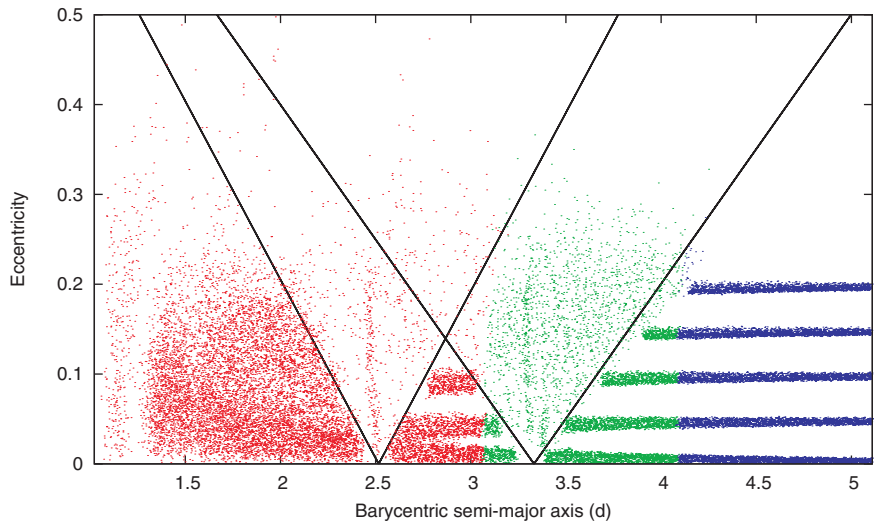

**Figure 3.** Diagram of the final semimajor axis and eccentricity for a set of test particles, in retrograde orbits, after $10^5 T_{P-C}$. The satellites Nix and Hydra are located at 2.516$d$ and 3.332$d$, respectively. Particles initially located at approximately $1d < a < 3d$ are represented in red, those particles initially at $3d < a < 4d$ are represented in green and those at $4d < a < 5d$ are represented in blue colour.

Comparison between Figs 1 and 3 shows that the stability region for those particles in retrograde orbits is larger than for particles in prograde orbits, as verified in Stern et al. (1994) for the region interior to Nix's orbit. Particles initially located interior to Nix, from $a \sim 1.31d$ to the inner boundary of Nix's collisional region, remained in this region. Their eccentricities reach values of up to 0.3.

An agglomerate of particles, initially between the orbits of Nix and Hydra, remained in this region. Beyond Hydra's orbit the stability region is larger than that obtained for particles in prograde orbits. Particles with $a \geq 3.4d$, this limit changes for different values of the eccentricity, suffer only a small gravitational perturbation of the satellites.

Table 2 presents the number of collisions between the particles and the massive bodies. Our results, concerning particles in retrograde orbits, show that about 24 per cent of the particles collided with one of the massive bodies and only 5 per cent are ejected from the system. Most of the particles survived after $10^5 T_{P-C}$.

## 3 HYPOTHETICAL SATELLITES

In this section we analyse the gravitational effects of a sample of hypothetical satellites on the orbits of Nix and Hydra. These satellites were initially in circular, coplanar and prograde orbits around the barycentre of the system. A similar analysis was carried out by Stern et al. (1994) in order to constrain the mass of a hypothetical satellite near Pluto and Charon, without inducing any eccentricity in Charon's orbit larger than $10^{-3}$.

The initial semimajor axis (relative to the barycentre of the system) of these hypothetical satellites (moonlets) were distributed from $2d$ to $5d$, with $\Delta a = 0.005d$. All moonlets started at the same true anomaly equals to zero. The orbital parameters of Pluto, Charon, Nix and Hydra are listed in Table 1. All bodies were treated as spherical.

The diameters of the hypothetical satellites were assumed to be between 2 km ($m = 6.8 \times 10^{12}$ kg) and 50 km ($m = 1.06 \times 10^{17}$ kg) for those satellites interior to Hydra's orbit, and between 2 and 36 km ($m = 3.9 \times 10^{16}$ kg) for those satellites exterior to Hydra's orbit.

The masses of the hypothetical satellites were obtained by assuming the density equals to 1.63 g cm$^{-3}$. The stepsizes of the diameters of the hypothetical satellites are 2 km. This range of values was chosen based on the results presented by Steffl et al. (2006). They have ruled out additional satellites with diameters larger than 50 km located between the orbits of Nix and Hydra, and 36 km beyond Hydra's orbit. These values were obtained by assuming a very dark albedo of $p_v = 0.04$.

Fig. 4(a) depicts the results of the numerical simulations, where the values of the time (in days) are plotted as a function of the barycentric semimajor axis and radius (in km) of the hypothetical satellites. Inspection of this figure shows that interior to Hydra's orbit most of the satellites collided with Charon, Nix or were ejected from the system. 75 per cent of the satellites coorbital to Nix and Hydra, and 33 per cent of those satellites located between their orbits survived during the time of the integration (650 000 d). Beyond 3.6$d$ all the hypothetical satellites remained in the system, since the gravitational effects of Charon, Nix and Hydra are less effective.

The available data are not sufficient to allow the precise definition of the orbital elements of Nix and Hydra. Precise orbital elements will be accomplished through the data sent by the New Horizons spacecraft during the period before and after its close approach with the Pluto system.

From Tholen et al. (2008), the osculating eccentricity variations as a function of time for Nix and Hydra are [0, 0.0272] and [0, 0.0179], respectively. Considering these results we constrained the size and position of the survived satellites (presented in Fig. 4a) to select only those satellites that did not induce any eccentricity larger than $10^{-3}$ in the orbits of Nix and Hydra.

During the numerical integrations the eccentricities of Nix and Hydra were monitored. Fig. 4(b) shows the location and size (in radius) of those hypothetical satellites that cause only a small increase (less than $10^{-3}$) in the eccentricities of Nix and Hydra. Therefore hypothetical satellites can reside coorbital to Nix and Hydra, in a small region between their orbits and beyond Hydra's orbit. Their presence will affect the eccentricities of these two small satellites, but this effect cannot be detected by current technology.

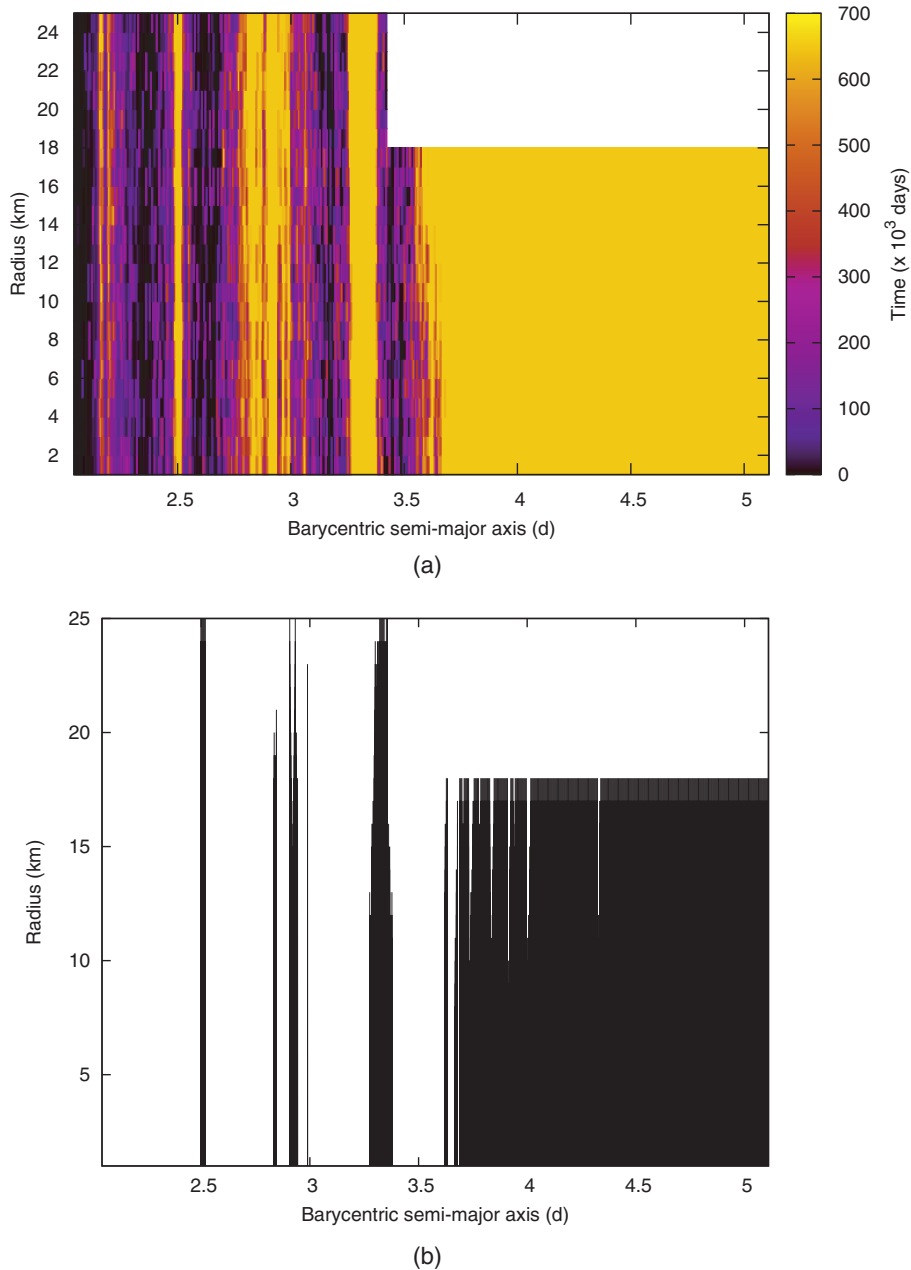

(a)

(b)

**Figure 4.** This figure presents the (a) values of the time (in days) as a function of the barycentric semimajor axis and radius (in km) of the hypothetical satellites and (b) those survived hypothetical satellites which cause only a negligible variation in the eccentricities of Nix and Hydra (less than $10^{-3}$).

## 4 FINAL COMMENTS

Two new companions to the Pluto–Charon binary system have been detected in 2005 by Weaver et al. (2006). These small satellites, named Nix and Hydra, are located beyond Charon's orbit. Although they are small when compared to Charon, their gravitational perturbations can decrease the stability of the external region. The dynamical structure of this external region is analysed by numerically simulating a sample of particles in P-type orbits under the gravitational effects of Pluto, Charon, Nix and Hydra. The particles are separate into two groups: prograde and retrograde orbits. The presence of the small satellites Nix and Hydra, as expected, decreases the stable region beyond Charon's orbit. A simple analysis, taking into account the circular restricted three-body problem, can explain the formation of the agglomerates of particles, for particles in prograde orbits, near and coorbital to Nix and Hydra. Comparison between particles in prograde and retrograde orbits shows that the stability region for those particles in retrograde orbits is larger than for particles in prograde orbits.

Small unseen satellites can also exist in this system without inducing an eccentricity in the orbits of Nix and Hydra larger than $10^{-3}$. These hypothetical satellites can be localized in the coorbital regions of Nix and Hydra, between their orbits and beyond Hydra's orbit.

The encounter between the New Horizons spacecraft and the Pluto system in 2015 will help to elucidate this peculiar quadruple system formed, up to now, by two massive bodies, Pluto and Charon, and two small satellites, Nix and Hydra.

**ACKNOWLEDGMENTS**

The authors thank Á. Sülli for his critical reading of the manuscript. This work was supported by FAPESP (Proc. 2007/06275-0 and Proc. 2006/04997-6).

This paper has been typeset from a T$_E$X/L$^A$T$_E$X file prepared by the author.